%% file: main.tex
\documentclass[12pt,twoside]{book}
\usepackage[T1]{fontenc}
\usepackage{xspace}
\usepackage{setspace}
\usepackage{url}

\usepackage{fancyhdr}
\usepackage{latexsym,amssymb,amsmath,amsfonts,amsthm}

\usepackage[ruled, vlined, linesnumbered]{algorithm2e}
\usepackage{algorithmic}

\usepackage{url}
\usepackage{lscape}
\usepackage{verbatim}
\usepackage{rotate}
\usepackage{placeins}
\usepackage{fancyvrb}

\usepackage[nocompress]{cite}
\usepackage[numbers,sort&compress]{natbib}

\usepackage{enumerate}
\usepackage{times}
\usepackage{moreverb} 
\usepackage{listings}

\usepackage[table]{xcolor}
\usepackage{mathrsfs}
\usepackage{paralist}
\usepackage{booktabs}
\usepackage{titlesec}
\usepackage{stmaryrd}

\usepackage{multirow,amsmath,booktabs,dcolumn}
\usepackage{graphicx}
\usepackage{caption}
\usepackage{subcaption}

\usepackage{pdfpages}
\usepackage[pdfpagemode={UseOutlines},bookmarks=true,bookmarksopen=true,
   pdfauthor={Vinayak Gupta},pdfcreator={Vinayak Gupta},
   bookmarksopenlevel=0,bookmarksnumbered=true,hypertexnames=false,
   colorlinks,linkcolor={black},citecolor={black},urlcolor={red},
   pdfstartview={FitV},unicode,breaklinks=true]{hyperref}
   
\theoremstyle{definition}
\newtheorem{definition}{Definition}

\newtheorem*{problem*}{Problem Statement}

\graphicspath{ {images/} }
\newcommand{\bm}[1]{\boldsymbol{#1}}
\newcommand{\cm}[1]{\mathcal{#1}}
\newcommand{\ds}[1]{\boldsymbol{\mathcal{#1}}}
\newcommand{\bs}[1]{\boldsymbol{#1}}
\newcommand{\eg}{\emph{e.g.}}
\newcommand{\ie}{\emph{i.e.}}
\newcommand{\etc}{\emph{etc.}}
\newcommand{\xhdr}[1]{\vspace{1mm} \noindent{{\bf #1.}}}

\newcommand{\cin}{check-in\xspace}
\newcommand{\cins}{check-ins\xspace}

\newcommand{\imtpp}{\textsc{IMTPP}\xspace}
\newcommand{\imtppp}{\textsc{IMTPP++}\xspace}
\newcommand{\axolotl}{\textsc{Axolotl}\xspace}
\newcommand{\reformd}{\textsc{Reformd}\xspace}
\newcommand{\revamp}{\textsc{Revamp}\xspace}
\newcommand{\nsr}{\textsc{NeuroSeqRet}\xspace}
\newcommand{\proactive}{\textsc{ProActive}\xspace}

\titlespacing{\section}{0pt}{\parskip}{-\parskip}
\titlespacing{\subsection}{0pt}{\parskip}{-\parskip}
\titlespacing{\subsubsection}{0pt}{\parskip}{-\parskip}

\date{}
\begin{document}
\pagestyle{empty}
\cleardoublepage
\onehalfspacing 
\pagenumbering{gobble}
\input{rest/cover_page}

\cleardoublepage
\input{rest/copyright}

\cleardoublepage
\input{rest/inner_cover_page}

\cleardoublepage
\pagenumbering{roman}
\input{rest/certificate}
\cleardoublepage
\input{rest/acknowledge}

\normalfont
\input{chapters/001_abstract}
\cleardoublepage

\includepdf[pages=-]{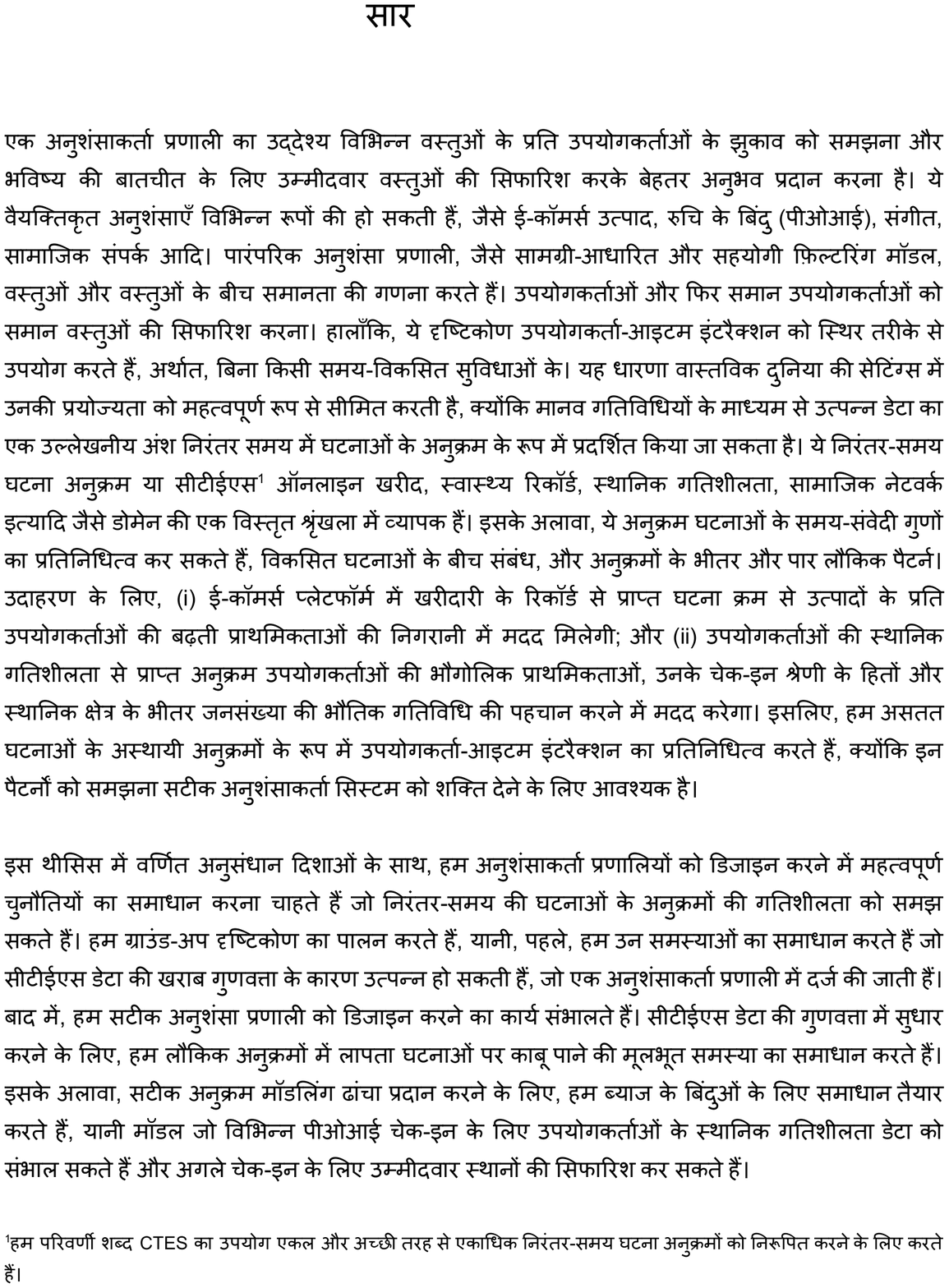}
\cleardoublepage

\pagestyle{fancy}
\tableofcontents

\listoffigures
\listoftables
\newpage
\newpage

\pagenumbering{arabic}
\pagestyle{fancy}
\onehalfspacing

\chapter{Introduction}
\input{chapters/002_introduction}

\chapter{Background}\label{chapter:background}
\input{chapters/003_background}

\part{Temporal Sequences with Missing Events}
\chapter{Overcoming Missing Events in Continuous-Time Sequences}\label{chapter:imtpp}
\input{chapters/004_imtpp}

\part{Recommendation in Spatio-Temporal Settings}
\chapter{Cross-Region Transfer for Spatial Features}\label{chapter:axolotl}
\input{chapters/005_axolotl}

\chapter{Sequential Recommendation using Spatio-Temporal Sequences}\label{chapter:reformd}
\input{chapters/006_reformd}

\chapter{Learning Spatial Behaviour using User Traces on Smartphone Apps}\label{chapter:revamp}
\input{chapters/007_revamp}

\part{Applications of Modeling Temporal Sequences}
\chapter{Large-Scale Retrieval of Temporal Sequences}\label{chapter:nsr}
\input{chapters/008_neuroseqret}

\chapter{Modeling Human Action Sequences}\label{chapter:proact}
\input{chapters/009_proactive}

\chapter{Conclusion and Future Work}\label{chapter:conclusion}
\input{chapters/010_conclusion}

\pagestyle{fancy}
\cleardoublepage
\addtocontents{toc}{\vspace{1em}}

\bibliographystyle{ACM-Reference-Format}
{\footnotesize
\bibliography{refs.bib}}
\pagestyle{plain}
\onehalfspacing
\cleardoublepage
\input{rest/pub}
\cleardoublepage
\input{rest/bio}
\cleardoublepage
\pagestyle{plain}


\end{document}

%% file: rest/cover_page.tex
\begin{titlepage}

\begin{center}


\LARGE 
\MakeUppercase{\textbf{Modeling Time-series and Spatial Data for Recommendations and other Applications}}\\

\vspace{3cm}

\LARGE

\MakeUppercase{\textbf{Vinayak Gupta}} 

\vspace{6cm}
\hspace{0cm}
\hbox{\includegraphics[width=8pc]{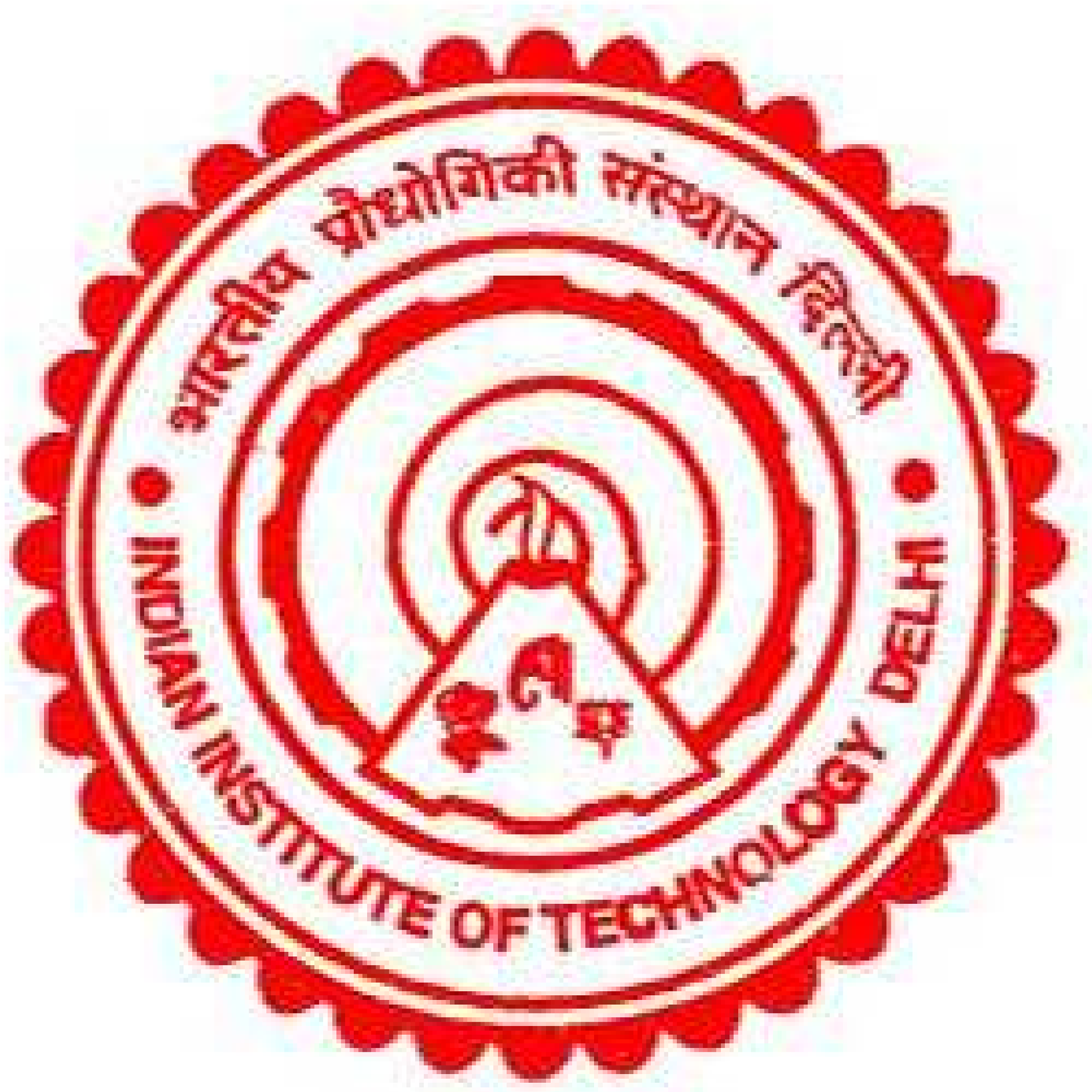}}
\vspace{1cm}

\large{DEPARTMENT OF COMPUTER SCIENCE \& ENGINEERING}\\
\large{INDIAN INSTITUTE OF TECHNOLOGY DELHI}\\
\large{2022}\\

\end{center}

\end{titlepage}

%% file: rest/copyright.tex
\begin{center}

\large

\ \\ \
\vspace{16cm}
\ \\ \
\copyright{Vinayak Gupta - 2022}\\
All rights reserved.

\end{center}

%% file: rest/inner_cover_page.tex

\begin{center}

\LARGE
\MakeUppercase{\textbf{Modeling Time-series and Spatial Data for Recommendations and other Applications}}\\
\vspace{1cm}

\large

{by}\\
\vspace{.3cm}
{VINAYAK GUPTA}\\
\vspace{.3cm}
{Department of Computer Science and Engineering}\\
\vspace{2cm}
{Submitted}\\
\vspace{0.3cm}
{in fulfillment of the requirements of the degree of {\bf  Doctor of Philosophy}}\\
\vspace{5cm}
{to the }\\
\vspace{1cm}

\hspace{0cm}
\hbox{\includegraphics[width=6pc]{rest/iitd-logo.pdf}}

\vspace{0.3cm}
{\bf
\large{Indian Institute of Technology Delhi}\\
\large{2022}\\
}

\end{center}


%% file: rest/certificate.tex
\chapter*{Certificate}
\addcontentsline{toc}{chapter}{Certificate}

This is to certify that the thesis titled {\small \textbf{MODELING TIME-SERIES AND SPATIAL DATA FOR RECOMMENDATIONS AND OTHER APPLICATIONS}} being submitted by {\small\textbf{Mr. VINAYAK GUPTA}} for the award of \textbf{Doctor of Philosophy} in \textbf{Computer Science and Engineering} is a record of bona fide work carried out by him under my guidance and supervision at the \textbf{Department of Computer Science and Engineering, Indian Institute of Technology Delhi}. The work presented in this thesis has not been submitted elsewhere, either in part or full, for the award of any other degree or diploma.

\vspace {10 pc}

\begin{flushright}
\noindent{Srikanta Bedathur} \\
\noindent{Associate Professor} \\
\noindent{Department of Computer Science and Engineering}\\
\noindent{Indian Institute of Technology Delhi} \\
\noindent{New Delhi- 110016}
\end{flushright}

%% file: rest/acknowledge.tex
\chapter*{Acknowledgements}
\addcontentsline{toc}{chapter}{Acknowledgements}
\setlength{\parindent}{0pt} 
\setlength{\parskip}{2ex}

I am highly indebted to my guide, Prof. Srikanta Bedathur, a brilliant advisor and an invaluable friend over the years. His `\textit{what are we trying to achieve here}' approach to all aspects of research and life will continue to inspire me. I will greatly cherish our discussions on topics ranging from sophisticated neural models trained on tens of GPUs to casual gigs on academic life. I am beyond belief grateful that he took me under his guidance and withstanding my never-ending tantrums around `\textit{what am I here?}'. This thesis is a tiny aspect of the immense knowledge I have gained by working under him.

A special thanks must go to Prof. Abir De for his invaluable encouragement and constant update meetings that always motivated me to work on my projects. Without his support, a significant part of the research presented in this thesis would have never begun. Never had I thought that a small interaction at CODS-COMAD 2019 would result in a collaboration of many years and multiple top-tier publications. I am also grateful to my research committee members Prof. Parag Singla and Prof. Rahul Garg, and Dr. L. V. Subramaniam, for their critical evaluation and suggestions that helped me shape most of the work in this thesis.

Special shout-outs to my SIT 309 gang -- Dishant Goyal, Sandeep Kumar, Omais Shafi, Dilpreet Kaur, Ovia Seshadri, and Arindam Bhattacharya for their constant support via poker nights, cricket games, food hunting trips, and innumerable coffee breaks. To Dishant, I would like to know if I'll ever be able to pay him back for being my go-to guy for many years. I am also thankful to have immense support from Apala Shankar Garg and Garima Gaur during my degree's initial and final phases, respectively. I'm also in massive debt to my friends from my IIIT days, specially Vijendra Singh, Ayush Srivastava, Mayur Mishra, Avashesh Singh, Ayushi Jain, and Ovais Malik, for enduring my never-ending cribs about grad school. Lastly, I thank my parents, my brother Kartik, and my cousins Surabhi and Ketan, for standing by me.

I am the author of this thesis, but all of them are surely the authors of me.

\vspace{0.1cm}
{ \begin{flushright}{\bf Vinayak Gupta}\end{flushright} }

%% file: chapters/001_abstract.tex
\chapter*{Abstract}
\addcontentsline{toc}{chapter}{Abstract}

A recommender system aims to understand the users' inclination towards the different items and provide better experiences by recommending candidate items for future interactions. These personalized recommendations can be of various forms, such as e-commerce products, points-of-interest (POIs), music, social connections, \etc\ Traditional recommendation systems, such as content-based and collaborative filtering models, calculate the similarity between items and users and then recommending similar items to similar users. However, these approaches utilize the user-item interactions in a \textit{static} way, \ie, without any time-evolving features. This assumption significantly limits their applicability in real-world settings, as a notable fraction of data generated via human activities can be represented as a sequence of events over a continuous time. These continuous-time event sequences or CTES~\footnote{We use the acronym CTES to denote a single and well as multiple continuous-time event sequences.} are pervasive across a wide range of domains such as online purchases, health records, spatial mobility, social networks, \etc\ Moreover, these sequences can implicitly represent the time-sensitive properties of events, the evolving relationships between events, and the temporal patterns within and across sequences. For example, (i) event sequences derived from the purchases records in e-commerce platforms can help in monitoring the users' evolving preferences towards products; and (ii) sequences derived from spatial mobility of users can help in identifying the geographical preferences of users, their \cin category interests, and the physical activity of the population within the spatial region. Therefore,  we represent the user-item interactions as temporal sequences of discrete events, as understanding these patterns is essential to power accurate recommender systems.

With the research directions described in this thesis, we seek to address the critical challenges in designing recommender systems that can understand the dynamics of continuous-time event sequences. We follow a \textit{ground-up} approach, \ie, first, we address the problems that may arise due to the poor quality of CTES data being fed into a recommender system. Later, we handle the task of designing accurate recommender systems. To improve the quality of the CTES data, we address a fundamental problem of overcoming \textit{missing} events in temporal sequences. Moreover, to provide accurate sequence modeling frameworks, we design solutions for points-of-interest recommendation, \ie, models that can handle spatial mobility data of users to various POI \cins and recommend candidate locations for the next \cin. Lastly, we highlight that the capabilities of the proposed models can have applications beyond recommender systems, and we extend their abilities to design solutions for large-scale CTES retrieval and human activity prediction. 

To summarize, this thesis includes three directions: (i) Temporal Sequences with Missing Events; (ii) Recommendation in Spatio-Temporal Settings; and (iii) Applications of Modeling Temporal Sequences. 
In the first part, we present an unsupervised model and inference method for learning neural sequence models in the presence of CTES with missing events. This framework has many downstream applications, such as imputing missing events and forecasting future events. 
In the second part, we design point-of-interest recommender systems that utilize the geographical features associated with the spatial \cins to recommend future locations to a user. Here, we propose solutions for \textit{static} POI recommendation, sequential recommendations, and recommendations models that utilize the similarity between physical mobility and the smartphone activities of users. 
Lastly, in the third part, we highlight the strengths of the proposed frameworks to design solutions for two tasks: (i) retrieval systems, \ie, retrieving relevance sequences for a given query CTES from a large corpus of CTES data; and (ii) understanding that different users take different times to perform similar actions in activity videos. Moreover, in each chapter, we highlight the drawbacks of current deep learning-based models, design better sequence modeling frameworks, and experimentally underline the efficacy of our proposed solutions over the state-of-the-art baselines. Lastly, we report the drawbacks and the possible extensions for every solution proposed in this thesis. 

A significant part of this thesis uses the idea of modeling the underlying distribution of CTES via \textit{neural} marked temporal point processes (MTPP). Traditional MTPP models are stochastic processes that utilize a fixed formulation to capture the generative mechanism of a sequence of discrete events localized in continuous time. In contrast, neural MTPP combine the underlying ideas from the point process literature with modern deep learning architectures. The ability of deep-learning models as accurate function approximators has led to a significant gain in the predictive prowess of neural MTPP models. In this thesis, we utilize and present several neural network-based enhancements for the current MTPP frameworks for the aforementioned real-world applications. 

%% file: chapters/002_introduction.tex
Recommender systems have become pervasive across various applications, including finance, social networks, healthcare, and spatial mobility. These recommendations can be from a wide range of sources, such as e-commerce products~\cite{ngcf,danser,sasrec,tisasrec}, music~\cite{van2013deep,mrecsys,nazari2020recommending}, news articles~\cite{lu2015content,newsrec}, points-of-interest~\cite{locate,deepmove,axolotl}, \etc\
With the widespread use of applications on the web, it is easier for people to provide feedback about their likes or dislikes to their application providers. These providers can use the collected feedback to understand the customers' preferences and provide better recommendations. These recommendations can enhance the customer experience, as the recommended content is generally preferable to the suggestions given at random~\cite{charurec}. In the terminology of recommender systems, the entities to which the recommendations are provided are called \textit{users}, and the products that can be recommended are referred to as \textit{items}. The working of a recommender system can be divided into two phases -- understanding the preferences of the users and recommending candidate items for future interactions. 
Traditional recommendation systems, such as content-based and collaborative filtering models, generate relationships between users and items based on their past interactions~\cite{charurec}. Specifically, content-based filtering uses the similarity between item features to recommend similar items to the users based on the items the users have interacted with in the past. In the same context, collaborative filtering uses similarities between users and items simultaneously to provide recommendations. 
Thus, in a collaborative filtering-based recommender system, it is possible to recommend an item to user A based on the interests of a similar user B. The similarity between users and items is calculated based on the features available for each entity. However, a significant drawback of these approaches is that they utilize the user-item interactions in a \textit{static} way, \ie, they cannot calculate similarities between users and items when the features evolve with time. This limitation drastically reduces their applicability in real-world applications, as most user-generated data is in the form of temporal sequences. In detail, the data extracted from a majority of online activities, physical actions, and natural phenomena can be represented as sequences of discrete events localized in continuous time, \ie, continuous-time event sequences (CTES). 
Thus, to design accurate recommender systems, it is necessary to understand the rich information encoded in these temporal sequences. To highlight a few examples, to recommend the future purchases of a user, we must capture the time-sensitive relationship between products of different types; to advertise meet-up places in the neighborhood, we must understand how user preferences change with time -- `coffee houses' during the day and `social joints' in the night; and in the particular case of recommending actions that people should take to prepare breakfast, we must capture the time they might take to complete independent actions, such whisking an egg or slicing vegetables. 

In recent years, deep learning frameworks have shown unmatched prowess in understanding visual data~\cite{krizhevsky2017, resnet, he2015delving}, natural language text~\cite{transformer, bert}, and audio~\cite{hershey2017cnn, nagrani17}. Thus, it is hardly surprising that designing recommender systems using deep learning tools has attracted significant attention and research efforts~\cite{sasrec, ngcf, pinsage, tisasrec, danser, bert4rec, jing2017neural, fpmc}. However, due to the rich information encoded in these CTES, it is challenging to learn the dynamics of these sequences with standard deep-learning models. The rich information encoded in every CTES data may include the ever-changing time intervals between interactions, high variance in the length of sequences for different users, additional attributes such as spatial features with every event, and the influence structure between events -- within and across different sequences. Encapsulating these features in a recommender system is necessary to perform a wide range of downstream tasks, such as forecasting future interactions for users and identifying the most-likely time when the user will interact with an item. 

In this thesis, we seek to address critical challenges in designing neural recommender systems that can understand the dynamics of continuous-time event sequences. To better address these challenges, we follow a \textit{ground-up} approach, \ie, first, we address the problems that may arise due to the poor quality of sequential data being fed into the systems. Then we address the task of designing systems that accurately understand these sequences. In the context of poor data quality, we address a real-world problem with CTES data, of overcoming missing events, \ie, the loss in data quality as the data-collection procedure could not record all events in the sequences due to constraints such as crawling or privacy restrictions. This data loss is a crucial problem as the performance of any recommender system, or deep-leaning in general is conditioned on the availability of high-quality data. Moreover, since most of the sequential models have considered only the settings where the training data is completely observed, \ie, there are no missing observations, their predictive performance deteriorates significantly as the sequence quality further degrades~\cite{rnn_miss,medical,traffic}. Later, we address the problems associated with designing accurate sequence modeling frameworks that can better understand the users' preferences and recommend candidate items. For this, we consider the task of points-of-interests (or POI) recommendation, \ie, given the past mobility records of a user via her \cins, the goal is to recommend the most likely POIs that the user will visit in her future \cins. Addressing this problem is necessary as recent research has shown that accurate advertisements on POI networks, such as Foursquare and Instagram, can achieve up to 25 times the return-on-investment~\cite{fsqstats}. However, since deep learning models require large quantities of training data, designing a POI recommender system can be challenging if the spatial data for the underlying region is insufficient to train large neural networks. The skew in geographical distribution is a result of the variability in the quantity of mobility data across spatial regions, \ie, a majority of the human population is located in urban and suburban regions~\cite{population}. This variability in data makes it challenging to design POI recommendation systems for regions with limited data. Thus, we overcome problems associated with designing POI systems with \textit{limited} training data.

Across different chapters of this thesis, we propose robust yet scalable sequence modeling frameworks. However, the ability to understand the dynamics of CTES better can have applications beyond recommender systems. Therefore, in the last part of this thesis, we utilize the predictive prowess of the proposed frameworks and design solutions for two real-world applications of large-scale sequence retrieval and human activity prediction. In detail, we show that the current approaches for both applications have a limited ability to model the temporal relationships between events in a sequence. To this extent, we propose deep-learning frameworks that understand the dynamics of CTES data and outperform the current approaches for the two applications and a wide range of downstream tasks. 

For a significant part of this thesis, we model the dynamics of CTES using \textit{neural} marked temporal point processes (MTPP\footnote{In this and the following chapters, we use MTPP to denote a single and well as multiple point processes.})~\cite{rmtpp, nhp, thp, sahp, fullyneural, intfree, xiaointaaai} -- probabilistic generative models that learn the latent interaction between the current and the past events in a CTES. Neural MTPP models bridge the gap between the universal approximation ability of deep learning models and the probabilistic formulation of point processes. In detail, the traditional MTPP models, like self-exciting processes, use a fixed mathematical function to model the interaction between events in a CTES~\cite{hawkes, daley2007introduction}. Thus, capturing complex relationships within the CTES requires the knowledge to represent these patterns using mathematical functions while being scalable for large datasets. In contrast, the neural MTPP combine the underlying ideas from the point process literature with deep-learning models' ability to be effective function approximators. Thus, these models have the flexibility of a neural network and can benefit from the sequence modeling ability of neural recurrent layers~\cite{dlbook} and transformers~\cite{transformer}.

\begin{table}[t]
  \caption{List of first-author research contributions made in this thesis. 
  }
  \vspace{-3mm}
  \centering
  \resizebox{\textwidth}{!}{
  \begin{tabular}{l|c|c|c}
  \toprule
  \textbf{Ch.} & \textbf{Ref.} & \textbf{Title} & \textbf{Venue} \\ \hline
  \multirow{3}{*}{\ref{chapter:imtpp}} & \cite{imtpp} & Learning Temporal Point Processes with Intermittent Observations & AISTATS 2021\\
  & \multirow{2}{*}{\cite{imtppp}}  & Modeling Continuous Time Sequences with Intermittent Observations & \multirow{2}{*}{ACM TIST 2022} \\  
  & & using Marked Temporal Point Processes & \\ \hline

  \multirow{2}{*}{\ref{chapter:axolotl}} & \multirow{2}{*}{\cite{axolotl}} & Doing More with Less: Overcoming Data Scarcity for POI & \multirow{2}{*}{ACM TIST 2022} \\ & & Recommendation via Cross-Region Transfer & \\ \hline

  \ref{chapter:reformd} & \cite{reformd} & Region Invariant Normalizing Flows for Mobility Transfer  & CIKM 2021 \\ \hline

  \ref{chapter:revamp} & \cite{revamp} & Modeling Spatial Trajectories using Coarse-Grained Smartphone Logs  & IEEE TBD 2022 \\ \hline

  \multirow{2}{*}{\ref{chapter:nsr}} & \multirow{2}{*}{\cite{neuroseqret}} & Learning Temporal Point Processes for Efficient Retrieval of  & \multirow{2}{*}{AAAI 2022} \\ & & Continuous Time Event Sequences & \\ \hline  
  
  \multirow{2}{*}{\ref{chapter:proact}} & \multirow{2}{*}{\cite{proactive}} & ProActive: Self-Attentive Temporal Point Process Flows for  & \multirow{2}{*}{KDD 2022} \\ & & Activity Sequences & \\
  \bottomrule
  \end{tabular}
  }
  \label{int_tab:papers}
  \vspace{-3mm}
\end{table}

\section{Our Research Contributions}
In this thesis, we address the problems associated with designing recommender systems that can understand user-item interactions in the form of temporal sequences. Firstly, we address problems associated with missing events that degrade the quality of the CTES data. Later, we design accurate POI recommendation frameworks that utilize temporal data and understand the physical mobility of users. We demonstrate that our proposed architectures can effectively overcome the drawbacks of the data-collection process and simultaneously outperform the state-of-the-art models for predicting and forecasting future events. Moreover, we highlight the strengths of the proposed models and propose solutions that are tailored to address a few real-world CTES applications that have been overlooked in the past -- large-scale CTES retrieval and human activity prediction. The research publications that constitute this thesis are listed in Table~\ref{int_tab:papers}, and can be organized into the following three parts:

\textbf{1. Temporal Sequences with Missing Events.} In the first part of the thesis, we address a data-related problem of overcoming missing events in temporal sequences and the limitations of traditional approaches in modeling these sequences. Addressing this problem is crucial as most of the existing models and inference methods in the neural MTPP literature consider only the \textit{complete} observation scenario. In a complete observation setting, the underlying event sequence is assumed to be observed entirely with no missing events -- an ideal setting that is rarely applicable in real-world applications. A recent line of work that considers missing events while training MTPP utilizes supervised learning techniques that require additional knowledge of \emph{missing} or \emph{observed} labels for each event in a sequence. The need for additional knowledge further restricts their practicability in real-world settings, as in several scenarios, the details of the missing events are not known apriori. To address these problems, in Chapter~\ref{chapter:imtpp}, we propose \textbf{\imtpp} (\textbf{I}ntermittently-observed \textbf{M}arked \textbf{T}emporal \textbf{P}oint \textbf{P}rocesses), a novel unsupervised model and inference method for learning MTPP in the presence of event sequences with missing events. Specifically, we first model the generative processes of observed and missing events using two MTPP, representing the missing events as latent random variables. Then, we devised an unsupervised training method that jointly learns the MTPP through variational inference. Such a formulation can impute the missing among the observed events, which enhances its predictive prowess and identify the optimal position of missing events.

\textbf{2. Recommendations in Spatio-Temporal Settings.} In the second part of the thesis, we design recommender systems for datasets with geographical features. In detail, we analyze the physical mobility data of users worldwide and highlight the problems associated with recommending spatial points of interest (POI) for users to visit in the future. These problems primarily arise due to the high variance in the volume of data collected across different spatial regions. In contrast to the concept of incomplete sequences in Chapter~\ref{chapter:imtpp}, we regard \textit{limited} data as the problem of data scarcity, \ie, the available data is assumed to be complete but in short supply. However, the volume is insufficient to train a deep neural network-based recommender system effectively. Thus, in this part, we focus on the task of points-of-interest (POI) recommendation, \ie, recommending candidate locations to a user based on her past visits. Unlike standard item recommendation, the task of POI recommendation is more challenging as a user's preference in the network is influenced by its geo-location and the distribution of nearby POIs. We address three real-world recommendation scenarios that are affected by limited data:
\begin{compactitem}
\item[\textbf{2 (a)}.]\textbf{Top-$k$ POI Recommendation with Limited Data.} Firstly, we address the problems arising from limited data in designing systems for top-$k$ recommendations, \ie, estimating the probability of a specific user checking into a candidate POI using their past \cins. Therefore, in Chapter~\ref{chapter:axolotl}, we present \textbf{\axolotl} (\textbf{A}utomated \textit{cross} \textbf{Lo}cation-network \textbf{T}ransfer \textbf{L}earning), a novel method aimed at transferring location preference models learned in a data-rich region to boost the quality of recommendations in a data-scarce region significantly. Precisely, we deploy two channels for information transfer, (i) a \emph{meta-learning} based procedure learned using location recommendation as well as social predictions, and (ii) a \textit{lightweight} unsupervised cluster-based transfer across users and locations with similar preferences. Both of these work together synergistically to achieve improved accuracy of recommendations in data-scarce regions without any prerequisite of overlapping users and with minimal fine-tuning. Our model is built on top of an \emph{twin graph-attention} neural network, which captures user- and location-conditioned influences in a user-mobility graph for each region.

\item[\textbf{2 (b)}.]\textbf{Sequential POI Recommendation with Limited Data.} In addition to top-$k$ recommendations, we design neural models to overcome regional data scarcity for sequential POI recommendations, \ie, continuous-time recommendations for the following spatial location. In contrast to top-$k$ recommendations, here, our goal is to recommend specifically the \textit{next} candidate POI and the probable \cin time using the past \cin sequence of the user. To overcome data scarcity for sequential POI recommendation, in Chapter~\ref{chapter:reformd}, we propose \textbf{\reformd} (\textbf{Re}usable \textbf{F}lows f\textbf{or} \textbf{M}obility \textbf{D}ata), a novel transfer learning framework for continuous-time location prediction for regions with sparse \cin data. Specifically, we model user-specific \cin sequences in a region using a marked temporal point process with \textit{normalizing} flows to learn the inter-\cin time and geo-distributions~\cite{shakir, flowbook}. Later, we transfer the model parameters of spatial and temporal flows trained on a data-rich \textit{source} region for the next \cin and time prediction in a \textit{target} region with scarce \cin data. We capture the evolving region-specific \cin dynamics for MTPP and spatial-temporal flows by maximizing the joint likelihood of \textit{next} \cin with three channels -- (i) \cin category prediction, (ii) \cin time prediction, and (iii) travel distance prediction. 

\item[\textbf{2 (c)}.]\textbf{User App-Usage to Physical Mobility.} Here, we overcome the problems associated with limited data by using the different features associated with users' physicla mobility. Specifically, we aim to capture the relationship between users' mobility and smartphone usage. This task is based on the intuition that as every user carries their smartphone wherever they go -- a crucial aspect ignored by the current models for spatial recommendations. Thus, in Chapter~\ref{chapter:revamp}, we present \textbf{\revamp} (\textbf{Re}lative position \textbf{V}ector for \textbf{A}pp-based \textbf{M}obility \textbf{P}rediction), a sequential POI recommendation approach that uses smartphone app-usage logs to identify the mobility preferences of a user. This work aligns with the recent psychological studies of online urban users that show that the activity of their smartphone apps largely influences their spatial mobility behavior. Specifically, our proposal for using coarse-grained data refers to data logs collected in a privacy-conscious manner consisting only of the following: (i) category of the smartphone app used, such as `retail',`social', \etc; and (ii) category of the \cin location. Thus, \revamp is not privy to precise geo-coordinates, social networks, or the specific app used. Buoyed by the efficacy of self-attention models, we learn the POI preferences of a user using two forms of positional encodings -- absolute and relative -- with each extracted from the inter-\cin dynamics in the mobility sequence of a user.
\end{compactitem}

\textbf{3. Applications}
In the third part of the thesis, we address the limitations of the current sequence models that restrict their modeling ability in many real-world applications. In detail, we highlight that the sequence modeling propose ability of MTPP models proposed in the previous chapters can have applications beyond recommender systems, and can be used to better learn the embeddings of CTES in the specific application settings. These embeddings are then tailored to outperform the existing approaches in two downstream tasks -- large-scale retrieval of CTES and modeling actions performed by humans in activity sequences.

\begin{compactitem}
\item[\textbf{3 (a)}.] \textbf{Large Scale CTES Retrieval.} The recent developments in MTPP frameworks have enabled an accurate characterization of temporal sequences for a wide range of applications. However, the problem of retrieving such sequences still needs to be addressed in the literature. Specifically, given a large corpus of temporal sequences and a query sequence, our goal is to retrieve all the sequences from the corpus that are \textit{relevant} to the query. Therefore, in Chapter~\ref{chapter:nsr}, we propose \nsr, a family of MTPP models to retrieve and rank a relevant set of continuous-time event sequences for a given query sequence from a large corpus of sequences. More specifically, we first apply a trainable unwarping function on the query sequence, which makes it comparable with corpus sequences, especially when a relevant query-corpus pair has individually different attributes. Next, we feed the unwarped query and corpus sequences into MTPP-guided neural relevance models. We develop two variants of the relevance model, which offer a tradeoff between accuracy and efficiency. We also propose an optimization framework to learn binary sequence embeddings from the relevance scores, suitable for locality-sensitive hashing, leading to a significant speedup in returning top-$k$ results for a given query sequence.

\item[\textbf{3 (b)}.] \textbf{Actions by Humans in Activity Sequences.} Unlike machine-made time series, the sequences of actions done by different humans are highly disparate, as the time required to finish a similar activity might vary between people. Therefore, understanding the dynamics of these sequences is essential for many downstream tasks such as activity length prediction, goal prediction, \etc\ Existing neural approaches that model an activity sequence are either limited to visual data or are task-specific, \ie, limited to the next action or goal prediction. In Chapter~\ref{chapter:proact}, we present \proactive (\textbf{P}oint P\textbf{ro}cess flows for \textbf{Activ}ity S\textbf{e}quences), a neural MTPP framework for modeling the continuous-time distribution of actions in an activity sequence while simultaneously addressing three real-world applications -- next action prediction, sequence-goal prediction, and \textit{end-to-end} sequence generation. Specifically, we utilize a self-attention module with temporal normalizing flows to model the influence and the inter-arrival times between actions in a sequence. For time-sensitive prediction, we perform a constrained margin-based optimization to predict the goal of the sequence with a limited number of actions.
\end{compactitem}

\section{Organization of Thesis}
We organize the rest of the thesis as follows: In Chapter~\ref{chapter:background}, we review some background on marked temporal point processes, graph neural networks, and self-attention models. Part 1 of the thesis presents a method for addressing missing data problems in temporal sequences (Chapter~\ref{chapter:imtpp}). In Part 2, we design POI recommendation systems that overcome limited data problems -- in a top-$k$ setting (Chapter~\ref{chapter:axolotl}) and for sequential recommendation (Chapter~\ref{chapter:reformd}) -- and identify the influence of app usage on the physical mobility (Chapter~\ref{chapter:revamp}). In Part 3, we present two novel applications of neural MTPP models in large-scale sequence retrieval (Chapter~\ref{chapter:nsr}) and modeling human actions in activity sequences (Chapter~\ref{chapter:proact}). Finally, in Chapter~\ref{chapter:conclusion} we summarize our contributions and discuss future avenues of research that this thesis offers.

%% file: chapters/003_background.tex
In this chapter, we provide a detailed overview of marked temporal point processes (MTPP), a crucial element of most approaches proposed in this work. In addition, we offer a brief introduction to graph neural networks (GNN). These neural network-based approaches constitute the necessary background for understanding this work's subsequent chapters. Specifically, MTPP are essential to understand the contributions made in Chapters~\ref{chapter:imtpp},~\ref{chapter:reformd},~\ref{chapter:nsr}, and~\ref{chapter:proact} and GNN are necessary for Chapter~\ref{chapter:axolotl}.

\section{Marked Temporal Point Process}
Marked temporal point processes~\cite{hawkes,daley2007introduction,rasmussen2018lecture} are probabilistic generative models for continuous-time event sequences. In recent years, MTPP have emerged as a powerful tool to model asynchronous events localized in continuous time~\cite{daley2007introduction,hawkes}, which have a wide variety of applications, \eg, information diffusion, disease modeling, finance, etc. Driven by these motivations, in recent years, MTPP have appeared in a wide range of applications in healthcare~\cite{lorch2018stochastic,rizoiu_sir}, traffic \cite{rmtpp,initiator}, web and social networks~\cite{gupta2021learning,du2015dirichlet,srijan,rmtpp,farajtabar2017fake,imtpp,colab}, and finance~\cite{bacry,bauwens}. Moreover, MTPP models are even applied to many applications, including seismology and neuroscience~\cite{survey_tpp}.

An MTPP represents an event using two quantities: (i) the time of its occurrence and (ii) the associated mark, where the latter indicates the category of the event and therefore bears different meanings for different applications. For example, in a social network setting, the marks may indicate users' likes, topics, and opinions on the posts; in finance, they may correspond to the stock prices and the category of sales; in healthcare, they may indicate the state of the disease of an individual. 
Mathematically, MTPP can be represented as a probability distribution over sequences of variable lengths belonging to a closed time interval $[0, T]$, and can be realized as an event sequence $\cm{S}_k=\{e_i=(c_i,t_i) | i \in[k]\}$, where $k$ is the number of events. Here, the times are ever-increasing, \ie, $0 < t_1 < t_2 < \cdots < t_N \le T$ and $c_i \in \mathcal{C}$ is the corresponding mark with $\mathcal{C}$ as the set of all categorical marks. Note that across the different chapters in this thesis, we denote the mark in an event $e_i$ using $c_i$ as well as $x_i$. However, both these notations mean the same and are used interchangeably. We also denote $\Delta_{t,i} = t_{i} - t_{i-1}$ as the inter-event time difference between events $e_i$ and $e_{i-1}$. 

\subsection{Conditional Intensity Function of an MTPP} 
For an MTPP, the time of each event is a random variable. Therefore, given the times of past events, $\{t_1, \cdots t_i-1\}$, we can determine $t_i$ using the following functions~\cite{de2019temporal}:
\begin{compactitem}
\item Conditional probability density function $f^*(t) = f(t|\cm{S}(t))$, that determines that the next event will occur in interval $[t, t+dt)$.
\item Cumulative distribution function $F^*(t) = F(t|\cm{S}(t)) = \int_{t_{i-1}}^{t} f^*(\tau) d\tau$, that determines the probability that the next event will occur before time $t$.
\item A secondary cumulative distribution function, called \textit{survival} function represented as, $S^*(t) = S(t|\cm{S}(t)) = 1 - F^*(t)$. This function represents the probability that the next event will not occur before time $t$.
\end{compactitem}

Using these functions, we can determine the characteristics of future events in a sequence. However, a major drawback of representing an MTPP using these functions is that we cannot combine multiple MTPP models together. Therefore, we resort to characterizing the event times of an MTPP using a conditional intensity function (CIF) or hazard function, denoted by $\lambda^*(t) = \lambda(t|\cm{S}(t))$, that represents the conditional probability that next event in a sequence has not happened before time $t$ and will happen during the interval $[t, t+dt)$. Mathematically, we can define the relationship between $\lambda^*(t)dt$, $f^*(t)$, and $S^*(t)$ as:
\begin{equation}
\lambda^*(t)dt = \frac{f^*(t) dt}{S^*(t)},
\end{equation}
Intuitively, $\lambda^*(t)$ is an instantaneous rate of events per unit of time. Using the conditional intensity function, we can easily combine multiple MTPP models. Specifically, for two MTPP models $\cm{S}_A(t)$ and $\cm{S}_B(t)$ with intensities $\lambda^*_1(t)$ and $\lambda^*_2(t)$ respectively, we can characterize the joint history $\cm{S}(t) = \cm{S}_A(t) \cup \cm{S}_B(t)$ as:
\begin{equation}
\lambda^*(t) = \lambda^*_1(t) + \lambda^*_2(t),
\end{equation}
\xhdr{Relationship between $\lambda^*(t)$ and $f^*(t)$}
By definition, we have the following:
\begin{equation}
\lambda^*(t) = \frac{f^*(t)}{S^*(t)} = -\frac{1}{S^*(t)} \frac{dS^*(t)}{dt} = - \frac{d \log S^*(t)}{dt},
\end{equation}
Thus, if we integrate the left and right-hand sides in the above equation and consider that $dS^*(t) = -f^*(t)dt$, we obtain the relationship between $\lambda^*(t)$ and $f^*(t)$ as:
\begin{equation}
f^*(t) = \lambda^*(t) \exp \bigg(-\int_{t_{i-1}}^{t} \lambda^*(\tau) d \tau \bigg),
\end{equation}
\xhdr{Relationship between $\lambda^*(t)$ and Log-likelihood}
We can compute the log-likelihood that $\lambda^*(t)$ will generate the sequence $\cm{S}(t)$ with parameters $\theta$ as:
\begin{equation}
\mathscr{L} = \bigg(\sum_{i=1}^{n} \log \lambda^*_{\theta}(t_i) - \int_{t_{i-1}}^{t_i} \lambda^*_{\theta} (\tau) d\tau \bigg) - \int_{t_n} ^{T} \lambda^*_{\theta}(\tau),
\end{equation}
This can be resolved in the following:
\begin{equation}
\mathscr{L} = \sum_{i=1}^{n} \log \lambda^*_{\theta}(t_i) - \int_{0}^{T} \lambda^*_{\theta} (\tau) d\tau, \, T \ge t_n.
\end{equation}


\subsection{Normalizing Flows}
Normalizing flows~\cite{shakir,ppflows} (NF) are generative models used for density estimation and event sampling. They work by mapping simple distributions to complex ones using multiple bijective, \ie, reversible functions. For \eg, the function $r(x)=x+1$ is a reversible function because, for each input, a unique output exists and vice-versa, whereas the function $r(x) = x^2$ is not a reversible function. In detail, let $\bs{Z} \in \mathbb{R}^D$ be a random variable with a known probability density function $p_{\bs{Z}} : \mathbb{R}^D \rightarrow \mathbb{R}$. Let $g$ be an invertible function and $\bs{Y} = g(\bs{Z})$. Then, via the change of variables formula~\cite{kobyzev2020normalizing}, the probability density function of $\bs{Y}$ is:
\begin{equation}
p_{\bs{Y}} (\bs{y}) = p_{\bs{Z}} \big( f(\bs{y}) \big) \, \Big | \mathrm{det} \, D f(\bs{y}) \Big |,
\end{equation}
where $f(\cdot)$ is the inverse of $g$ and $D f(\bs{y}) = \frac{\delta f}{\delta \bs{y}}$ is the Jacobian of $f$. Here, the above function $g(\cdot)$ (a generator) projects the base density $p(\bs{Z})$ to a more complex density, and this projection is considered to be in the \textit{generative} direction. Whereas the inverse function $f(\cdot)$ moves from a complicated distribution towards the simpler one of $p(\bs{Z})$, referred to as the \textit{normalizing} direction. Since in generative models, the base density $p(\bs{Z})$ is considered as Normal distribution, this formulation gives rise to the name \textit{normalizing} flows. To sample a point $\bs{y}$, one can sample a point $\bs{z}$ and then apply the generator $\bs{y} = g(\bs{z})$. Such a procedure supports closed-form sampling. Moreover, modern approaches for normalizing flows approximate the above functions using a neural network~\cite{autoregressive,dhaliwal,shakir}. Normalizing flows have been increasingly used to define flexible and theoretically sound models for marked temporal point processes~\cite{intfree, ppflows}.

\section{Neural Temporal Point Process}
Buoyed by the predictive prowess of deep-learning models in modeling the dynamics of temporal sequences, modern MTPP models utilize a neural network with the probabilistic modeling ability of MTPP to enhance its predictive power~\cite{rmtpp,nhp,sahp,thp,fullyneural,gupta2021learning,intfree,karishma}. Specifically, they combine the continuous-time approach from the point process literature with modern deep learning approaches such as RNNs and transformers. Thus, these models can better capture the complex relationships between future events and historical events. The most popular approaches~\cite{rmtpp, nhp, sahp, thp} use different variants of neural networks to model the time- and mark distribution.

\xhdr{Recurrent Marked Temporal Point Process} RMTPP is the first-ever neural network-based MTPP model~\cite{rmtpp}. The underlying model of RMTPP is a two-step procedure that embeds the event sequence using a recurrent neural network (RNN) and then derives the formulation of CIF using this embedding. Specifically, given the sequence $\cm{S}_k = \{e_1, \cdots, e_k\}$, an RNN determines its vector representation denoted by $\bs{h}_k$. Later, RMTPP uses this representation of the sequence over an \textit{exponential} function to formulate the CIF. 
\begin{equation}
    \lambda^*(t) = \exp(\bs{w}_h \bs{h}_k + \bs{w}_{\Delta} \Delta_{t,k} + \bs{b}), 
\end{equation}
where, $\bs{w}_{\bullet}$ and $\bs{b}$ are trainable parameters. Due to this formulation, the modeling prowess of RMTPP is limited by the expressive power of the exponential function.

\xhdr{Neural Hawkes Process} NHP modified the LSTM architecture to model the continuous time of events in a sequence. Later, it uses the embedding from the LSTM to determine the CIF using a \textit{softplus} function~\cite{nhp}. Such a formulation is more expressive; however, it does not have a closed-form for the likelihood.

\xhdr{Fully Neural Point Process} FNP is a fully neural network-based intensity function for TPP. The underlying framework idea of their approach is to model the cumulative conditional intensity function, \ie, $-\int_{t_{i-1}}^{t} \lambda^*(\tau) d \tau$,  using a neural network. Later, the CIF can be obtained by differentiating this w.r.t time. This approach allows the model to compute the log-likelihood efficiently.

\xhdr{Point Process and Self-Attention} Transformer Hawkes process (THP)~\cite{thp} and self-attentive Hawkes Process~\cite{sahp} combine the transformer architecture to formulate a point process. In detail, these architectures obtain the embedding of the sequence using a transformer and then formulate the CIF using the obtained embeddings. 

\subsection{Intensity-free formulation of Temporal Point Process}
Here, we describe an intensity-free formulation for MTPP that estimates the temporal distribution of events using normalizing flows. MTPP models with a neural network-based intensity function have shown incredible prowess in learning the dynamics of CTES. However, these models face many constraints while sampling future events in a sequence. \citet{rmtpp} and~\citet{thp} are limited by the design choice for their intensity function,~\citet{nhp} requires approximating the integral using Monte Carlo and thus lacks closed-form sampling for future events. Lastly, ~\citet{fullyneural} lacks a defined formulation for a proper density function and uses an expensive sampling procedure. We summarize these drawbacks of existing MTPP models in Table~\ref{back_tab:mdl_details}. 

\begin{table}[t]
  \caption{Drawbacks and advantages of neural MTPP frameworks.}
  \vspace{-3mm}
  \centering
  \resizebox{\textwidth}{!}{
  \begin{tabular}{l|cccccc}
  \toprule
  \textbf{Model} & \textbf{RMTPP~\cite{rmtpp}} & \textbf{NHP~\cite{nhp}} & \textbf{FNP~\cite{fullyneural}} & \textbf{SAHP~\cite{sahp}} & \textbf{THP~\cite{thp}} & \textbf{Log-Normal~\cite{intfree}} \\ \hline
  Closed-form Likelihood & \checkmark & \text{\sffamily X} & \checkmark & \text{\sffamily X} & \text{\sffamily X} & \checkmark\\
  Robust $\lambda$ & \text{\sffamily X} & \checkmark & \checkmark & \checkmark & \checkmark & \checkmark \\
  Closed-form Sampling &  \checkmark & \text{\sffamily X} & \text{\sffamily X} & \text{\sffamily X} & \text{\sffamily X} & \checkmark\\
  \bottomrule
  \end{tabular}
  }
  \label{back_tab:mdl_details}
  \vspace{-3mm}
\end{table}

To overcome these drawbacks, \citet{intfree} propose a simple yet efficient \textit{intensity-free} formulation for modeling the inter-event arrival times in an MTPP. Specifically, they use temporal normalizing flows~\cite{shakir} over an RNN layer that performs on par with the state-of-the-art method. In particular, they use a log-normal distribution for inter-event arrival times:
\begin{equation}
t_{i}-t_{i-1} \sim \textsc{LogNormal} \left( \mu (\bs{h}_i), \sigma(\bs{h}_i) \right),
\end{equation}
where $\mu$ and $\sigma$ represent the mean and the variance of the log-normal distribution, respectively. The mean and variance are derived from the recurrent neural network layer with output $\bs{h}_{\bullet}$. Thus, the probability density function of the inter-arrival times will be:
\begin{equation}
    p\left( \Delta_t| \mu, \sigma \right) = \frac{1}{\sqrt{2\pi} \Delta_t \sigma} \exp \left( - \frac{(\log \Delta_t - \mu)^2}{2\sigma^2} \right)
\end{equation}
Such a log-normal distribution for inter-arrival times facilitates faster and closed sampling with stable convergence~\cite{intfree}, \ie, we can sample the inter-arrival times simply using a normal-distribution based sampling. Standard intensity-driven MTPP models rely on Ogata's thinning~\cite{ogata} or an inverse sampling~\cite{rl_utkarsh}, which operates using iterative acceptance-rejection protocol and, therefore, can be expensive. 

\section{Graph Convolution/Attention Networks}
In some applications, data cannot be represented in a Euclidean space and is, thus, represented as graphs with complex relationships and interdependency between nodes. Modeling this complex nature has led to several developments in neural graph networks that bridge the gap between deep learning and spectral graph theory~\cite{gcn,deferrard}. The most popular GNN models are graph-convolution networks (GCN)~\cite{gcn} and graph-attention networks (GAT)~\cite{gat}. Both these approaches work on the principle of \textit{neighborhood} aggregation; however, GCN explicitly assigns a non-parametric weight during the aggregation process and implicitly captures the weight via an end-to-end neural network architecture to better model the importance between neighborhood nodes. However, GAT uses a learnable attention weight for each node in the neighborhood~\cite{attention}.

\subsection{Graphs in Recommendation Systems}
Traditional graph embedding approaches focused on incorporating the node neighborhood proximity in a classical graph in their embedding learning process~\cite{node2vec,lbsn2vec}. In detail,~\citet{birank} adopts a label propagation mechanism to capture the inter-node influence and hence the collaborative filtering effect. Later, it determines the most probable purchases for a user via the items she has interacted with based on the structural similarity between the historical purchases and the new target item. However, the performance of these approaches is inferior to model-based CF methods since they do not optimize a recommendation-specific loss function. The recently proposed graph convolutional networks (GCNs)~\cite{gcn} have shown an incredible prowess for recommendation tasks in user-item graphs. The attention-based variant of GCNs, graph attention networks (GATs)~\cite{gat} are used for recommender systems in information networks~\cite{socialgcn,ngcf}, traffic networks~\cite{traffic, dcrnn} and social networks~\cite{pinsage,yfumob}. Furthermore, the heterogeneous nature of these information networks comprises multi-faceted influences that led to approaches with \textit{dual}-GCNs across both user and item domains~\cite{dualgcn,socialgcn}. However, these models have limited ability to learn highly heterogeneous data, \eg, a POI network with disparate weights, location-category as node feature, and varied sizes. Thus, limited research has been done on utilizing these models for spatial recommendations.

\section{Self-Attention}
Attention in deep learning is a widely used technique to get a weighted aggregation of different components of a model~\cite{attention}.~\citet{transformer} proposed an attention-based sequence-to-sequence method that achieved state-of-the-art performance in machine translation. Thus, there have been several applications of such models in domains including product recommendations~\cite{sasrec,tisasrec}, modeling spatial mobility preferences of users~\cite{geosasrec}, image generation~\cite{image_transformer}, \etc\ Here, we present a detailed description of the underlying sequence encoder-decoder module in~\citet{transformer}. Here, we provide an overview of the self-attention framework used in this thesis. 

A self-attention mechanism requires a fixed length input sequence, say $\bs{R} = [\bs{r}_1, \bs{r}_2, \cdots, \bs{r}_n]$, where $\bs{r}_{i} \in \mathbb{R}^{D}$ denotes the embedding of the event in the $i$-th position and $n$ denotes the sequence length. The process for obtaining these embeddings can vary as per the modeling problem. For \eg, in NLP tasks, the embeddings are representations of words in a sentence~\cite{bert,transformer}, and in recommender systems, they represent the items bought by a user~\cite{sasrec,tisasrec}.

Using the input embedding sequence, a self-attention model first injects a position encoding $\bs{r}_i \rightarrow \bs{r}_{i} + \bs{p}_{i}$, to every event embedding. Later, at every index, it calculates the weighted sum of all embeddings using three linear transformations as below:
\begin{equation}
    \overline{\bs{R}} = \mathrm{Attention}(\bs{R}\bs{W}^Q, \bs{R}\bs{W}^K, \bs{R}\bs{W}^V),
\end{equation}
where, $\bs{W}^Q, \bs{W}^K, \bs{W}^V \in \mathbb{R}^{D \times D}$ denote the projection matrices for \textit{queries}, \textit{keys}, and \textit{values} respectively. $\overline{\bs{R}}$ denotes the sequence of output embeddings and the function $\mathrm{Attention}(\cdot)$ is defined as:
\begin{equation}
    \mathrm{Attention}(\bs{Q}, \bs{K}, \bs{V}) = \mathrm{softmax}\left(\frac{\bs{Q}\bs{K}^{\top}}{\sqrt{D}} \right) \bs{V},
\end{equation}
In addition, it introduces a \textit{causality} between events in the sequence by forbidding all links between $\bs{Q}_i$ and $\bs{K}_j$ where $j > i$. 

To introduce a non-linearity into the present formulation, the self-attention procedure includes a point-wise feed-forward layer described below:
\begin{equation}
    \bs{V}_i = \mathrm{PFFN}(\overline{\bs{R}}_i) = \mathrm{ReLU} (\overline{\bs{R}}_i \bs{W}_1 + \bs{b}_1) \bs{W}_2 + \bs{b}_2,
\end{equation}
Lastly, based on the design choice, a sequence-to-sequence mechanism based on self-attention may include multiple stacked attention blocks, residual connections, layer normalization, and dropout between layers~\cite{transformer}. The final output embedding at index $i$, in this case, $\bs{V}_i$, represents a weighted aggregation of the history, \ie, all the events that have occurred before the current index. In recent years, neural MTPP models have used self-attention as the underlying mechanism to capture the dynamics of a sequence~\cite{thp,sahp,karishma,neuroseqret,proactive}.

Thus, in this chapter, we have a detailed overview of the techniques necessary for understanding the subsequent chapters of this thesis. In the following chapters, we describe the key technical contributions of this thesis. 

%% file: chapters/004_imtpp.tex
\newcommand{\amovies}{Movies\xspace}
\newcommand{\atoys}{Toys\xspace}
\newcommand{\taxi}{Taxi\xspace}
\newcommand{\ret}{Twitter\xspace}
\newcommand{\so}{SO\xspace}
\newcommand{\fq}{Foursquare\xspace}
\newcommand{\cel}{Celebrity\xspace}
\newcommand{\hth}{Health\xspace}
\newcommand{\mei}{PFPP\xspace}

\newcommand{\imtppobs}{$\textsc{IMTPP}_\mathcal{S}$}
\newcommand{\imtpplog}{$\textsc{IMTPP}_\mathcal{R}$}

\newcommand{\up}{\uppercase}
\newcommand{\PP}{{\text{Pr}}}
\newcommand{\set}[1]{\{  #1 \}}
\newcommand{\uk}[1]{\overline{#1}}
\newcommand{\lk}[1]{\underline{#1}}
\newcommand{\mis}{\epsilon}
\newcommand{\Ccal}{\mathcal{C}}

\newcommand{\pr}{{\text{prior}}}
\newcommand{\Sdata}{\mathcal{S}}
\newcommand{\Mdata}{\mathcal{M}}
\newcommand{\pmx}{\mathbb{P}_{\theta,x}}
\newcommand{\qmy}{\mathbb{Q}_{\phi,y}}
\newcommand{\prmx}{\mathbb{P}_{\pr,y}}
\newcommand{\prdt}{p_{\pr,\Delta}}
\newcommand{\qb}{\bm{q}}
\newcommand{\pdt}{p_{\theta,\Delta}}
\newcommand{\qdt}{q_{\phi,\Delta}}
\newcommand{\indicator}[1]{{\llbracket #1 \mathbb{R}bracket }}
\newcommand{\given}{\,|\,}
\newcommand*{\argmin}{\mathop{\mathrm{argmin}}}
\newcommand*{\argmax}{\mathop{\mathrm{argmax}}}
\newcommand{\nn}{\nonumber}
\renewcommand{\sb}{\bm{s}}
\newcommand{\mb}{\bm{m}}
\newcommand{\hb}{\bm{h}}
\newcommand{\vb}{\bm{v}}
\newcommand{\wb}{\bm{w}}
\newcommand{\ab}{\bm{a}}
\newcommand{\Wb}{\bm{W}}
\newcommand{\Ub}{\bm{U}}
\newcommand{\Vb}{\bm{V}}
\newcommand{\Gb}{\bm{G}}
\newcommand{\gb}{\bm{g}}
\newcommand{\bb}{\bm{b}}
\newcommand{\Qb}{\bm{Q}}
\newcommand{\cb}{\bm{c}}
\renewcommand{\indicator}[1]{{\llbracket #1 \rrbracket }}

\section{Introduction} \label{imtp_sec:intro}
Designing an accurate recommender system is conditioned on the availability of high-quality sequential data with no missing events. In this chapter, we study the problems associated with missing events in continuous time sequences and propose our solution to model and impute these missing events. 
In recent years, marked temporal point processes (MTPP) have shown an outstanding potential to characterize asynchronous events localized in continuous time. However, most of the MTPP models~\cite{Valera2014,wang2017human,neural_poisson,rmtpp,thp,sahp} --- with a few recent exceptions~\cite{shelton, mei_icml} --- have considered only the settings where the training data is completely observed or, in other words, there is no missing observation at all. While working with fully observed data is ideal for understanding any dynamical system, this is not possible in many practical scenarios. We may miss observing events due to constraints such as crawling restrictions by social media platforms; privacy restrictions (certain users may disallow collection of certain types of data); budgetary factors such as data collection for exit polls; or other practical factors, \eg, a patient may not be available at a certain time. This results in the poor predictive performance of MTPP models~\cite{rmtpp,thp,sahp} that skirt this issue.

Statistical analysis in the presence of missing data has been widely researched in literature in various contexts~\cite{rnn_miss, medical, traffic, miss_nips18}. \citet{little2019statistical} offer a comprehensive survey. It provides three models that capture missing data mechanisms in increasing order of complexity, \emph{viz.}, MCAR (missing completely at random), MAR (missing at random), and MNAR (missing not at random). Recently,~\citet{shelton} and \citet{mei_icml} proposed novel methods to impute missing events in continuous-time sequences via MTPP from the viewpoint of the MNAR mechanism. However, they focus on imputing missing data in between prior available observed events rather than predicting observed events in the face of missing events. Moreover, they deploy expensive learning and sampling mechanisms, which make them often intractable in practice, especially in the case of learning from a sequence of streaming events. For example,~\citet{shelton} applies an expensive MCMC sampling procedure to draw missing events between the observation pairs, which requires several simulations of the sampling procedure upon the arrival of a new sample. On the other hand, \citet{mei_icml} uses a bi-directional RNN, which re-generates all missing events by making a completely new pass over the backward RNN whenever one new observation arrives. As a consequence, it suffers from quadratic complexity with respect to the number of observed events. On the other hand, the proposal of \citet{shelton} depends on a predefined influence structure among the underlying events, which is available in linear multivariate parameterized point processes. In more complex point processes with neural architectures, such a structure is not explicitly defined, which further limits their applicability in real-world settings.
 
\subsection{Our Contribution}
In this chapter, we present our solution to overcome the above limitations via a novel modeling framework for point processes called \textbf{\imtpp} (\textbf{I}ntermittently-observed \textbf{M}arked \textbf{T}emporal \textbf{P}oint \textbf{P}rocesses)~\cite{imtpp}, which characterizes the dynamics of both observed and missing events as two coupled MTPP, conditioned on the history of previous events. In our setup, the generation of missing events depends both on the previously occurring missing events as well as the previously observed events. Therefore, they are MNAR ({missing not at random}), in the context of the literature on missing data~\cite{little2019statistical}. In contrast to the prior models~\cite{mei_icml,shelton}, \imtpp aims to learn the dynamics of both observed and missing events, rather than simply imputing missing events in between the known observed events, which is reflected in its superior predictive power over those existing models.

Precisely, \imtpp\ represents the missing events as latent random variables, which, together with the previously observed events, seed the generative processes of the subsequent observed and missing events. Then it deploys three generative models--- MTPP for observed events, prior MTPP for missing events, and posterior MTPP for missing events using recurrent neural networks (RNN) that capture the nonlinear influence of the past events. We also show that such a formulation can be easily extended to imputation tasks and still achieve significant performance gains over other models. \imtpp includes several technical innovations over other models that significantly boost its modeling and prediction accuracy. In detail, our contributions are:
\begin{compactitem}
\item In a notable departure from almost all existing MTPP models~\cite{rmtpp,mei_icml,de2016learning} which rely strongly on conditional intensity functions, we use a \emph{lognormal} distribution to sample the arrival times of the events. As suggested by~\citet{intfree}, such distribution allows efficient sampling as well as a more accurate prediction than the standard intensity function-based models.

\item The built-in RNNs in our model are designed to make \emph{forward} computations. Therefore, they incrementally update the dynamics upon the arrival of a new observation. Consequently, unlike the prior models, it does not require us to re-generate all the missing events in response to the arrival of an observation, which significantly boosts the efficiency of both learning and prediction as compared to both the previous approaches~\cite{mei_icml,shelton}.
\end{compactitem}

Our modeling framework allows us to train \imtpp\ using an efficient variational inference method, that maximizes the evidence lower bound (ELBO) of the likelihood of the observed events. Such a formulation highlights the connection of our model with the variational autoencoders (VAEs)~\cite{vrnn}. However, in sharp contrast to traditional VAEs, where the random noises or seeds often do not have immediate interpretations, our random variables bear concrete physical explanations, \ie, they are missing events, which renders our model more explainable than an off-the-shelf VAE. In addition, an extension to \imtpp, called \imtppp can identify the optimal positions of missing events in a sequence. Finally, we perform exhaustive experiments with eight diverse real-world datasets across different domains to show that \imtpp can model missing observations within a stream of observed events and enhance the predictive power of the original generative process for a full observation scenario.

\section{Related Work}\label{imtp_sec:relwork}
Our work is broadly related to the literature on (i) missing data models for discrete-time series; and (ii) missing data models for temporal point processes.

\subsection{Missing Data Models for Discrete-Time Series}
Our current work is also related to existing missing data models for discrete-time series, which do not necessarily consider MTPP. In principle, training sequential models in the presence of missing data is essential for robust predictions across a wide range of applications \eg, traffic networks~\cite{traffic}, modeling disease propagation~\cite{medical_2} and wearable sensor data~\cite{restful}. Motivated by these applications, there has been a significant effort in recent years in designing learning tools for sequence models with missing data~\cite{rnn_miss, medical, time_gan}. In particular, the proposal by \citet{rnn_miss} compensates for a missing event by applying a time decay factor to the previous hidden state in a GRU before calculating the new hidden state. \citet{medical} captures the effect of missing data by incorporating future information using bidirectional-RNNs. While these approaches do not provide explicit generative models of missing events, a few other models generate them by imputing them in between available observations. For example, \citet{brits} proposed a method of imputing missing events using a bi-directional RNN;~\citet{time_gan} employs a generative adversarial approach for generating missing events conditioned on the observed events. \citet{suggest_ijcai} and \citet{ealstm} are used for imputing in time-series. However, they cannot be used to sample \textit{marks} of missing events and, thus, cannot be extended to imputation in continuous-time event sequences. Thus, these models are complementary to our proposal as they do not work with temporal point processes.

\subsection{Missing Data Models for Temporal Point Process}
Very recently, there has been a growing interest in modeling MTPP in the presence of missing observations. However, they deploy expensive learning and sampling mechanisms on an apriori-known complete sequence of observations. More specifically, \citet{shelton} proposed a way of incorporating missing data by generating \textit{children} events for the observed events. They rely strongly on an expensive MCMC sampling procedure to draw missing events between the observation pairs. In order to adapt to such a protocol, we need to run the entire sampling routine several times whenever a new observation arrives. Such a method is extremely time-consuming and often intractable in practice. Moreover, they require an underlying multivariate parenthood structure that is not available in a complicated neural setting. Our work is closely related to the proposal by Mei \emph{et. al.}~\cite{mei_icml}. It employs two RNNs, in which the forward RNN--- initialized on $t=0$ -- models the observation sequence, and the backward RNN -- initialized on $t=T$ -- models the missing observations. To operate a backward RNN in an online setting, we need to pass the entire sequence of observations into it whenever a new sample arrives, which in turn makes it super expensive in practice. While re-running these methods after batch arrivals--- instead of re-running after every single arrival--- may appear as a compromised solution; however, that is ineffective in practice. Other approaches include the proposal by \citet{xu2017learning}, which proposes a training method for MTPP when the future and past events of a sequence window are censored; the work by~\citet{rasmussen_miss}, which assumes certain characteristics of missing data, and~\citet{miss_eq} is limited to spatial modeling.

\section{Problem Setup} \label{imtp_sec:psetup}
In this section, we first introduce the notations and then the setup of our problem of learning marked temporal point processes with observed and missing events over continuous time.

\subsection{Preliminaries and Notations}
We characterize an MTPP using the sequence of observed events $\Sdata_k=\{e_i=(x_i,t_i) | i \in[k] , t_i<t_{i+1}\}$, where the details of the notations is given in chapter~\ref{chapter:background}. As highlighted in Section~\ref{imtp_sec:intro}, there may be instances where an event has actually taken place but not recorded with the observed event sequence $\Sdata$. To this end, we introduce the \emph{MTPP for missing events}--- a latent MTPP--- which is characterized by a sequence of hidden events $\Mdata_r=\{\epsilon_j=(y_j,\tau_j)|j\in[r], \tau_j<\tau_{j+1}\}$ where $\tau_{j}\in\mathbb{R}^+$ and $y_j\in \Ccal$ are the times and the marks of the $j$-th missing events. Therefore, $\Mdata_r$ defines the set of first $r$ missing events. 
Moreover, we denote the inter-arrival times of the missing events as, $\Delta_{\tau,r} = \tau_{r}-\tau_{r-1}$. Note that $\tau_\bullet$, $y_\bullet$, $\Mdata_\bullet$ and $\Delta_{\tau,\bullet}$ for the MTPP of missing events share similar meanings with $t_\bullet$, $x_\bullet$, $\Sdata_\bullet$ and $\Delta_{t,\bullet}$ respectively for the MTPP of observed events. Here we further define two critical notations $\lk{k}$ and $\uk{k}$ as follows:
\begin{equation}
\label{imtp_eq:lk}\lk{k} =\argmin_r \set{\tau_r \given t_k < \tau_r < t_{k+1}},
\end{equation}
\begin{equation}
\label{imtp_eq:uk}\uk{k} =\argmax_r \set{\tau_r \given t_k < \tau_r < t_{k+1}}
\end{equation}
Here, $\lk{k}$ and $\uk{k}$ are the indices of the first and the last missing events respectively, among those which have arrived between $k$-th and $k+1$-th observed events. Figure~\ref{imtp_fig:illus} (a) illustrates our setup. In practice, the arrival times ($t$ and $\tau$) of both observed and missing events are continuous random variables, whereas the marks ($x$ and $y$) are discrete random variables. Therefore, following the state-of-the-art MTPP models~\cite{rmtpp, nhp}, we model a density function to draw the event timings and a probability mass function to draw marks.

\subsection{Overcoming Missing Events}
Our goal via \imtpp\ is to design an MTPP model which can generate the subsequent observed ($e_{k+1}$) and missing events ($\epsilon_{r+1}$) in a recursive manner, conditioned on the history of all events $\Sdata_k\cup \Mdata_r$ that have occurred thus far. Given the input sequence of observations $\Sdata_K$ consisting of the first $K$ observed events $\set{e_1,e_2,...,e_K}$, we first train our generative model and then recursively predict the next observed event $e_{K+1}$. Though \imtpp can also predict the missing events, we evaluate the predictive performance only on observed events since the missing events are not available in practice. We also evaluate the imputation performance of our model by predicting synthetically deleted events. Note that this setting is in contrast to the proposal of~\cite{mei_icml} that aims to impute the missing events based on the \emph{entire} observation sequence $\Sdata_K$ using a bi-directional RNN. Specifically, whenever one new observation arrives, it re-generates all missing events by making a completely new pass over the backward RNN. As a result, such an imputation method not only suffers from quadratic complexity with respect to the number of observed events, but it also has limited practicability, as future events are not available beyond the current timestamp in a streaming or online setting. Furthermore, their approach is tailored towards imputing missing events based on complete observations and is not well suited to predicting observed events in the face of missing observations. In contrast, our proposal is designed to generate {subsequent} observed and missing events in between previously observed events. Therefore, it does not require to re-generate all missing events whenever a new observation arrives, which allows it to enjoy a linear complexity with respect to the number of observed events and can be easily extended to online settings.

\section{Components of \imtpp} \label{imtp_sec:model}
\begin{figure*}[t]
\centering
\includegraphics[height=6cm]{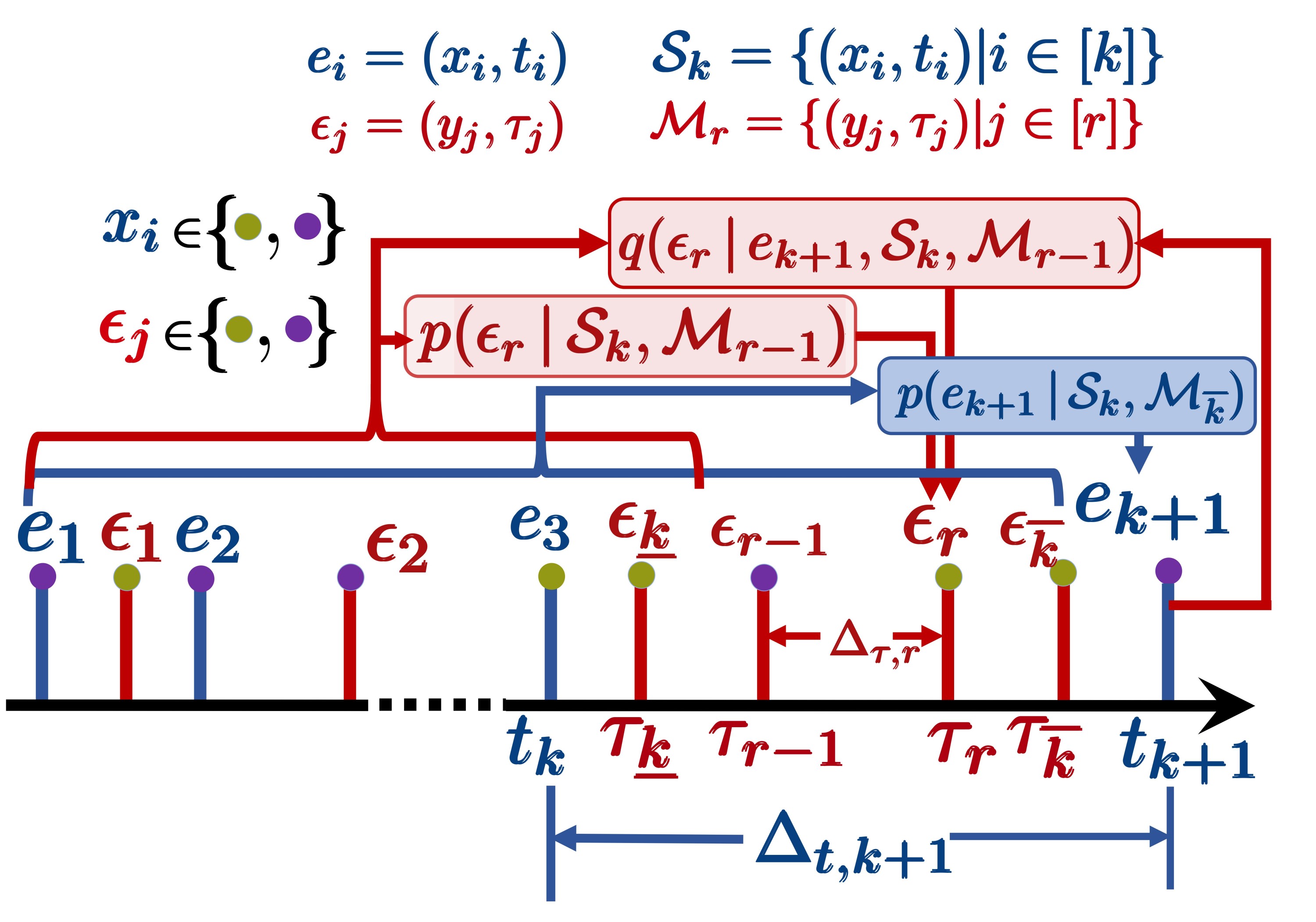}
\vspace{-2mm}
\caption{The overall neural architecture of \imtpp. The figure illustrates the notations, the observed, and missing point processes in \imtpp. The components concerning observed events and missing events are marked with blue and red, respectively. The figure also illustrates the generation process for events $e_{k+1}$ and $\mis_r$.}
\label{imtp_fig:overall}
\vspace{-2mm}
\end{figure*}

At the very outset, \imtpp, our proposed generative model, connects two stochastic processes -- one for the observed events, which samples the observed, and the other for the missing events -- based on the history of previously generated missing and observed events. Note, that the sequence of training events that are given as input to \imtpp consists of only the observed events. We model the missing event sequence through latent random variables, which, along with the previously observed events, drive a unified generative model for the complete (observed and missing) event sequence. The overall neural architecture of \imtpp, including the different processes for observed and missing events, is given in Figure~\ref{imtp_fig:overall}.

More specifically, given a stream of observed events denoted as $\Sdata_K =\set{e_1=(x_1,t_1), e_2=(x_2,t_2), \ldots, e_K=(x_K,t_K)}$, if we use the maximum likelihood principle to train \imtpp, then we should maximize the marginal log-likelihood of the observed stream of events, \ie,  $\log p(\Sdata_K)$. However, computation of $\log p(\Sdata_K)$ demands marginalization with respect to the set of latent missing events $\Mdata_{\uk{K-1}}$, which is typically intractable. Therefore, we resort to maximizing a variational lower bound or evidence lower bound (ELBO) of the log-likelihood of the observed stream of events $\Sdata_K$. Mathematically, we note that:
\vspace{-0.5cm}
\begin{align}
    & p(\Sdata_K)  =   \prod_{k=0} ^{K-1}\int_{\Mdata_{\uk{k}}}  p(e_{k+1} \given \Sdata_{k}, \Mdata_{\uk{k}})\, p(\Mdata_{\uk{k}})\, d\omega(\Mdata_{\uk{k}})\nn\\[-1.4ex]
 & = \mathbb{E}_{q(\Mdata_{\uk{K-1}} \given \Sdata_K)}\prod_{k=0} ^{K-1} \frac{p(e_{k+1}\given \Sdata_{k}, \Mdata_{\uk{k}}) \displaystyle \prod_{r=\lk{k}} ^{\uk{k}} p(\mis_r \given \Sdata_{k}, \Mdata_{r-1})}{ \displaystyle \prod_{r=\lk{k}} ^{\uk{k}} q(\mis_r \given e_{k+1}, \Sdata_{k}, \Mdata_{r-1}) }
\end{align}
where $\omega(\Mdata)$ is the measure of the set $\Mdata$, $q$ is an approximate posterior distribution that aims to interpolate missing events
$\mis_r$ within the interval $(t_k,t_{k+1})$, based on the knowledge of the next observed event $e_k$,
along with all previous events $\Sdata_k \cup \Mdata_{r-1}$, and $\lk{k}$, $\uk{k}$. Recall that $\lk{k}$ $(\uk{k})$ is the index $r$ of the first (last) missing event  $\mis_r$ among those which have arrived between $k$-th and $k+1$-th observed events, \ie, $\lk{k}=\argmin_r \set{\tau_r\given t_k < \tau_r < t_{k+1}}$ and $\uk{k}=\argmax_r \set{\tau_r\given t_k < \tau_r < t_{k+1}}$.
Next, by applying Jensen inequality\footnote{https://en.wikipedia.org/wiki/Jensen's\_inequality} over the likelihood, $\log p(\Sdata_K) $ is at-least:
\begin{equation}
\mathbb{E}_{q(\Mdata_{\uk{K-1} \given \Sdata_K} )} \sum_{k=0} ^{K-1} \log p(e_{k+1}\given \Sdata_{k}, \Mdata_{\uk{k}}) -\sum_{k=0} ^{K-1} \sum_{r=\lk{k}} ^{\uk{k}} \text{KL} \bigg[q(\mis_r \given e_{k+1}, \Sdata_{k}, \Mdata_{r-1}) || p(\mis_r \given \Sdata_{k}, \Mdata_{r-1}) \bigg],\label{imtp_eq:elbo}
\end{equation}
While the above inequality holds for any distribution $q$, the quality of this lower bound depends on the expressivity of $q$, which we would model using a deep recurrent neural network. Moreover, the above lower bound suggests that our model consists of the following components.

\begin{compactitem}
\item \textbf{MTPP for observed events.} The distribution $p(e_{k+1} \given \Sdata_{k}, \Mdata_{\uk{k}})$ models the MTPP for observed events, which generates the $(k+1)$-th event, $e_{k+1}$, based on the history of all $k$ observed events $\Sdata_k$ and all missing events $\Mdata_{\uk{k}}$ generated so far.

\item \textbf{Prior MTPP for missing events.} The distribution $p(\mis_{r} \given \Sdata_{k}, \Mdata_{{r-1}})$ is the prior model of the MTPP for missing events. It generates the $r$-th missing event $\mis_{r}$ after the observed event $e_k$, based on the prior information--- the history with all $k$ observed events $\Sdata_k$ and all missing events $\Mdata_{{r-1}}$ generated so far.

\item \textbf{Posterior MTPP for missing events.} Given the set of observed events represented by $\Sdata_{k+1}=\set{e_1,e_2, \ldots, e_{k+1}}$, the distribution $q(\mis_{r} \given e_{k+1},\Sdata_{k}, \Mdata_{{r-1}})$  generates the $r$-th missing event $\mis_{r}$, after the knowledge of the subsequent observed event $e_{k+1}$ is taken into account, along with information about all previously observed events $\Sdata_k$ and all missing events $\Mdata_{{r-1}}$ generated so far.
\end{compactitem}

\section{Architecture of \imtpp} \label{imtp_sec:detailed}
\begin{figure*}[t]
\centering
\begin{subfigure}{0.45\columnwidth}
  \centering
  \includegraphics[height=5cm]{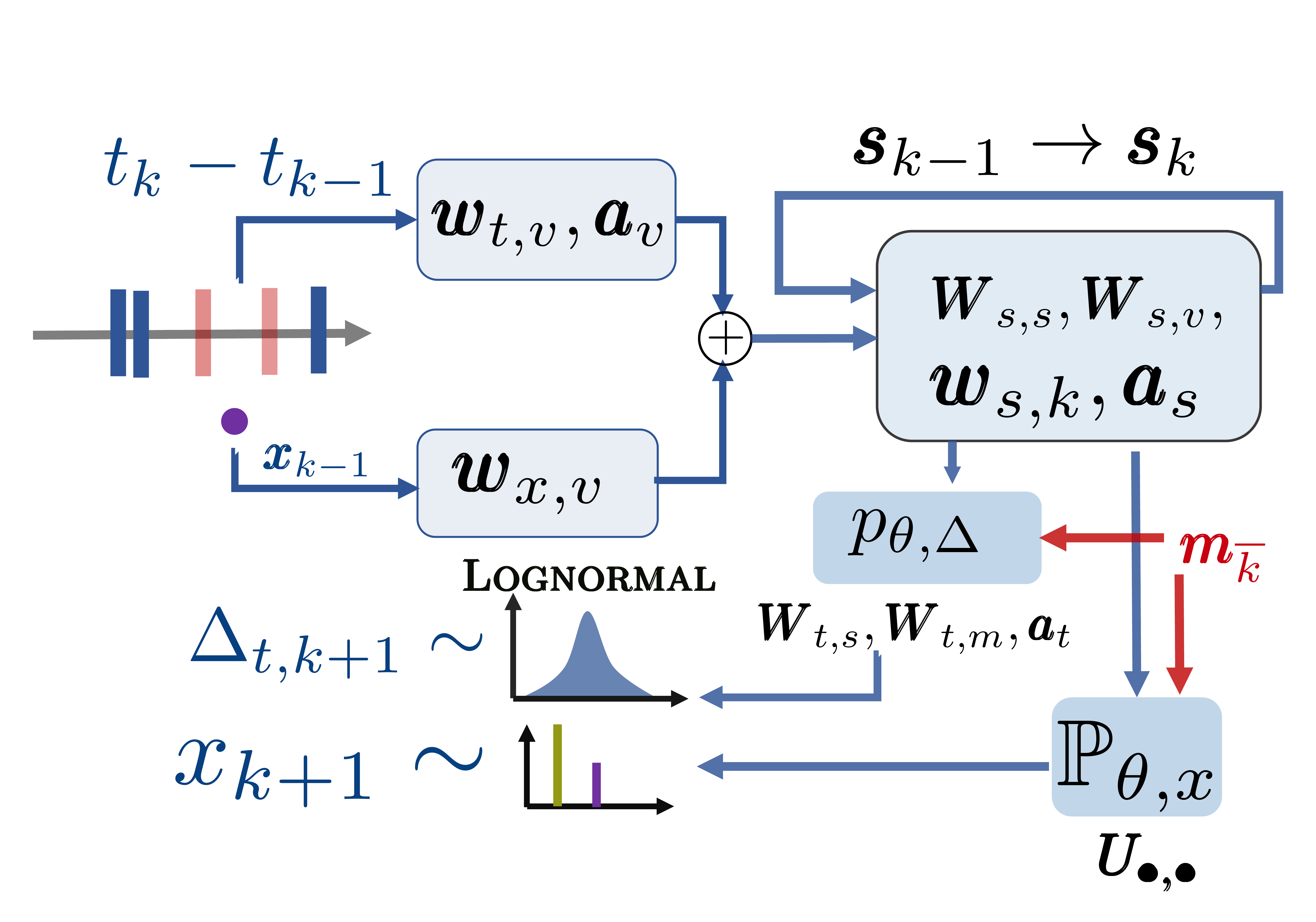}
  \caption{MTPP for observations $p_{\theta}$}
\end{subfigure}
\hspace{0.3cm}
\begin{subfigure}{0.45\columnwidth}
  \centering
  \includegraphics[height=5cm]{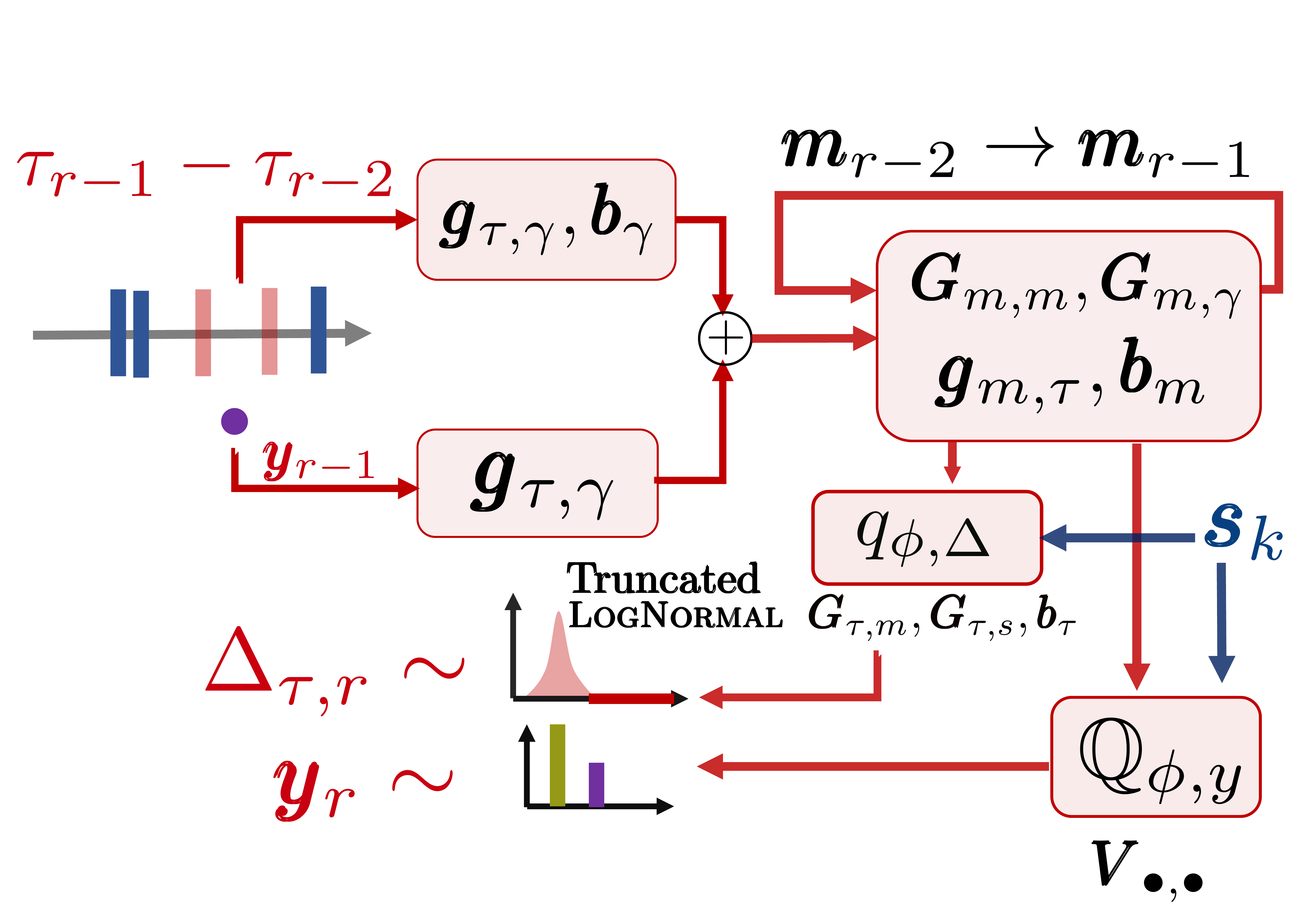}
  \caption{Posterior MTPP for missing events $q_\phi$}
\end{subfigure}
\vspace{-2mm}
\caption{Architecture of different processes in \imtpp. Panel (a) shows the neural architecture of the MTPP of observations $p_{\theta}$. Panel (b) shows the neural architecture of the posterior MTPP of missing events $q_{\phi}$. The information of $e_{k+1}$ is truncated to the lognormal distribution for missing data generation, whereas the lognormal distribution for observed is non-truncated.}
\vspace{-4mm}
\label{imtp_fig:illus}
\end{figure*}

We first present a high-level overview of deep neural network parameterization of different components of the \imtpp model and then describe component-wise architecture in detail. Finally, we briefly present the salient features of our proposal.

\subsection{High-level Overview}
We parameterize different components of \imtpp, introduced in the previous section using deep neural networks.
More specifically, we approximate the \emph{MTPP for observed events}, $p(e_{k+1} \given \Sdata_{k}, \Mdata_{\uk{k}})$ using $p_{\theta}$ and the \emph{posterior MTPP for missing events} $q(\mis_{r} \given e_{k+1},\Sdata_{k}, \Mdata_{{r-1}})$ using $q_{\phi}$, both implemented as neural networks with parameters $\theta$ and $\phi$ respectively. We set the \emph{prior MTPP for missing events} $p(\mis_{r} \given \Sdata_{k}, \Mdata_{{r-1}})$ as a known distribution $p_{\pr}$ using the history of all the events it is conditioned on. In this context, we design two recurrent neural networks (RNNs) which embed the history of observed events $\Sdata$ into the hidden vectors $\sb$ and the missing events $\Mdata$ into the hidden vector $\mb$, similar to several state-of-the art MTPP models~\cite{rmtpp,nhp,mei_icml}. In particular, the embeddings $\sb_k$ and $\mb_r$ encode the influence of the arrival time and the mark of the first $k$ observed events from $\Sdata_k$ and first $r$ missing events from $\Mdata_r$ respectively.
Therefore, we can represent the model for predicting the next observed event as: 
 \begin{align}
  p(e_{k+1} \given \Sdata_{k}, \Mdata_{\uk{k}}) = p_{\theta}(e_{k+1} \given \sb_{k}, \mb_{\uk{k}}).
 \end{align}
Following the above MTPP model for observed events, both the prior MTPP model and the posterior MTPP model for missing events offer similar conditioning with respect to $\sb_\bullet$ and $\mb_\bullet$. Identical to other MTPP models~\cite{rmtpp, nhp}, the RNN for the observed events updates $\sb_{k-1}$ to $\sb_{k}$ by incorporating the effect of $e_{k}$. Similarly, the RNN for the missing events updates $\mb_{r-1}$ to $\mb_{r}$ by taking into account the event $\mis_{r}$.

Each event has two components, its \textit{mark} and the \textit{arrival-time}, which are discrete and continuous random variables respectively. Therefore, we characterize the event distribution as a density function which is the product of the density function ($\pdt,\qdt,\prdt$) of the inter-arrival time and the probability distribution ($\pmx,\qmy,\prmx$) of the mark, \ie,
\begin{equation}
p_{\theta}(e_{k+1} = (x_{k+1},t_{k+1})\given \Sdata_{k}, \Mdata_{\uk{k}}) = \pmx(x_{k+1} \given \Delta_{t,k+1} , \sb_{k}, \mb_{\uk{k}}) \bm{\cdot} \pdt(\Delta_{t,k+1} \given \sb_{k}, \mb_{\uk{k}}), \label{imtp_eq:mt-1}
\end{equation}
\begin{equation}
q_{\phi}(\mis_{r}=(y_{r},\tau_r) \given e_{k+1},\Sdata_{k}, \Mdata_{{r-1}}) = \mathbb{Q}_{\phi,y}(y_{r} \given\Delta_{\tau,r} , e_{k+1},\sb_{k}, \mb_{{r-1}})\,\bm{\cdot}   \qdt(\Delta_{\tau,r} \given e_{k+1},\sb_{k}, \mb_{{r-1}}), \label{imtp_eq:mt-2}
\end{equation}
\begin{equation} 
p_{\pr}(\mis_{r}=(y_r,\tau_r) \given \Sdata_{k}, \Mdata_{{r-1}}) = \mathbb{P}_{\pr,y}(y_{r} \given \Delta_{\tau,r}  ,\sb_{k}, \mb_{{r-1}})\cdot p_{\pr,\tau}(\Delta_{\tau,r}  \given \sb_{k}, \mb_{{r-1}}), \label{imtp_eq:mt-3}
\end{equation}
where, as mentioned, the inter-arrival times $\Delta_{t,k}$ and $\Delta_{\tau,r}$ are given as $\Delta_{t,k}=t_{k}-t_{k-1}$ and $\Delta_r=\tau_r-\tau_{r-1}$. Moreover, $\pdt$, $\qdt$, and $\prdt$ denote the density of the inter-arrival times for the observed events, posterior density and the prior density of the inter-arrival times of the missing events, and $\pmx$, $\qmy$, and $\prmx$ denote the corresponding probability mass functions of the mark distributions. Figure~\ref{imtp_fig:illus} denotes the neural architecture of the MTPP for observed events and the posterior MTPP for missing events in \imtpp. The Prior MTPP for missing events has a similar architecture as standard MTPP models.

\subsection{Parameterization of $p_{\theta}$}
Given $k$ observed events and $r=\uk{k}$ missing events, the generative model $p_{\theta}$ samples the next event $e_{k+1}$ based on $\Sdata_k$ and $\Mdata_{r}$. To this aim, the underlying neural network takes the embedding vectors $\hb_{}$ and $\sb$ as input and provides the density $\pdt$ and $\pmx$ as output, which in turn are used to draw the event $e_{k+1}$. More specifically, we realize  $p_{\theta}$ in Eq.~\ref{imtp_eq:mt-1} as:
\begin{compactitem}
\item[(1)] \textbf{Input layer:} The first level is the input layer, which takes the last event as input and represents it through a suitable vector. In particular, upon arrival of $e_k$, it computes the corresponding vector $\vb _k$ as:
\begin{equation}
 \vb_{k}=\wb_{t,v} t_k+\wb_{x,v}x_k+\wb_{t,\Delta}(t_k-t_{k-1}) +\ab_v,
\end{equation}
where $\wb_{\bullet,\bullet}$ and $\ab_v$ are trainable parameters.

\item[(2)] \textbf{Hidden layer:} The next level is the hidden layer that embeds the sequence of observations into finite-dimensional vectors $\sb_\bullet$, computed using RNN. Such a layer takes $\vb_i$ as input and feeds it into an RNN  to update its hidden states in the following way.
\begin{equation}
 \sb_{k}=\tanh(\Wb_{s,s}   \sb_{k-1} + \Wb  _{s,v} \vb_{k} +  (t_k-t_{k-1}) \wb_{s,k}  + \ab_s),
\end{equation}
where $\Wb_{s,\bullet}$ and $\ab_s$ are trainable parameters. This hidden state $\sb_{k}$ can also be considered as a sufficient statistic of $\Sdata_{k}$, the sequence of the first $k$ observations.

\item[(3)] \textbf{Output layer:} The next level is the output layer which computes both $\pdt (\cdot) $ and $\pmx(\cdot)$ based on $\sb_k$ and $\mb_{\uk{k}}$.
%
%
To this end, we have the density of inter-arrival times as:
\begin{equation}
\pdt(\Delta_{t,k+1} \given \sb_{k}, \mb_{\uk{k}}) = \textsc{Lognormal}\left(\mu_{e}(\sb_k,\mb_{\uk{k}}), \sigma^2 _{e}(\sb_k,\mb_{\uk{k}})\right), \label{imtp_eq:llambda}
\end{equation}
with  $[\mu_{e}(\sb_k,\mb_{\uk{k}}), \sigma _{e}(\sb_k,\mb_{\uk{k}})] = \Wb_{t,s} ^\top \sb_{k} + \Wb_{t,m} ^\top \mb_{\uk{k}} +\ab_{t} $; and, the mark distribution as,
\begin{equation}
\pmx(x_{k+1}=x \given \Delta_{t,k+1},\sb_{k}, \mb_{\uk{k}}) =\frac{\exp(\Ub _{x,s} ^\top \sb_k + \Ub_{x,m}  ^\top \mb_{\uk{k}})}{\sum_{x'\in \Ccal}\exp(\Ub _{x',s} ^\top \sb_k + \Ub_ {x',m}  ^\top \mb_{\uk{k}})}, 
\end{equation}
The distributions are finally used to draw the inter-arrival time $\Delta_{t,k+1}$  and the mark $x_{k+1}$ for the event $e_{k+1}$. The sampled inter-arrival time $\Delta_{t,k+1}$ gives $ t_{k+1} = t_k +\Delta_{t,k}$. Here, the mark distribution is independent of $\Delta_{t,k+1}$.
\end{compactitem}
Finally, we note that $\theta=\set{\Wb_{\bullet,\bullet}, \wb_{\bullet,\bullet}, \Ub_{\bullet,\bullet},\ab_\bullet}$ are trainable parameters. We would like to highlight that, the proposed lognormal distribution of inter-arrival times $\Delta_{t,k}$ allows an easy re-parameterization trick--- $\textsc{Lognormal}(\mu_e,\sigma_e) = \exp(\mu_e + \sigma_e \cdot \textsc{Normal}(0,1))$---which mitigates variance of estimated parameters and facilitates faster training.
 
\subsection{Parameterization of $q _{\phi}$}
At the very outset, $q _{\phi}(\bullet\given e_k , \sb_k,\mb_{r-1} )$ (Eq.~\ref{imtp_eq:mt-2})  generates missing events that are likely to be omitted during the interval $(t_k,t_{k+1})$ after the knowledge of the subsequent observed event $e_{k+1}$ is taken into account. To ensure that missing events are generated within desired interval, $(t_k,t_{k+1})$, whenever an event is drawn with $\tau_r > t_{k+1}$, then $q _{\phi}(\bullet\given e_{k+1} , \sb_k,\mb_{r-1} )$ is set to zero and $\uk{k}$ is set to $r-1$. Otherwise, $\uk{k}$ is flagged as $\lk{k}$. Note that, $q _{\phi}(\bullet\given \sb_k,\mb_{r-1} )$ generates \emph{all} potential missing events in this interval. That said, it generates multiple events sequentially in one single run in contrast to the $p_{\theta}$. Similar to $p _{\theta}$, it has also a three level architecture.  

\begin{compactitem}    
\item[(1)] \textbf{Input layer:}
Given the subsequent observed event $t_{k+1}$ along with $\Sdata_k$ and $\mis_{r-1} = (y_{r-1},\tau_{r-1})$ arrives with $\tau_{r-1}< t_{k+1}$ or equivalently if $r-1 \neq \uk{k}$, then we first convert $\tau_{r-1}$ into a suitable representation as follows:
\begin{equation}
\bm{\gamma}_{r-1}=\gb_{\tau,\gamma} \tau_{r-1}+\gb_{y,\gamma}y_{r-1}+ \gb_{\Delta,\gamma}(\tau_{r-1}-\tau_{r-2})+\bb_\gamma,
\end{equation}
where $\gb_{\bullet,\bullet}$ and $\bb_\gamma$ are trainable parameters.
 
\item[(2)] \textbf{Hidden layer:} Similar to the hidden layer used in the $p_{\theta}$ model, the hidden layer here too embeds the sequence of missing events into finite-dimensional vectors $\mb_\bullet$, computed using RNN in a recurrent manner. Such a layer takes $\bm{\gamma}_{r-1}$ as input and feeds it into an RNN to update its hidden states in the following way:
\begin{equation}
\mb_{r-1} = \tanh \left(\Gb_{m,m} \mb_{r-2} + \Gb_{m,\gamma} \bm{\gamma}_{r-1} + (\tau_{r-1}-\tau_{r-2}) \bm{g}_{m,\tau} + \bb_m \right), \label{imtp_eq:lstmphi}
\end{equation}
where $\Gb_{\bullet,\bullet}, \bm{g}_{\bullet,\bullet}$  and $\bb_m$ are trainable parameters.

\item[(3)] \textbf{Output layer:}
The next level is the output layer which computes both $\qdt (\cdot) $ and $\qmy(\cdot)$ based on $\mb_r$ and $\sb_{k}$.
To compute these quantities, it takes five signals as input:
(i) the current update of the hidden state $\mb_r$ for the RNN in the previous layer,
(ii) the current update of the hidden state $\sb_{k}$ that embeds the history of observed events,
(iii) the timing of the last observed event, $t_{k}$,
(iv) the timing of the last missing event, $\tau_{r-1}$, and
(v) the timing of the next observation, $t_{k+1}$.
To this end, we have the density of inter-arrival times as:
\begin{align*}
\qdt(\Delta_{\tau,r} \given e_{k+1}, \sb_{k}, \mb_{r-1}) = \textsc{Lognormal}\left(\mu_{\mis}(\mb_{r-1},\sb_{k}), \sigma^2 _{\mis}(\mb_{r-1},\sb_{k})\right) \nn 
& \\ \odot \indicator{\tau_{r-1}+\Delta_{\tau,r}< t_{k+1}}, \label{imtp_eq:llambda-2}
\end{align*}
with  $[\mu_{\mis}(\mb_{r-1},\sb_{k}), \sigma  _{\mis}(\mb_{r-1},\sb_{k})] = \Gb_{\tau,m} ^\top \mb_{r-1} + \Gb_{\tau,s} ^\top \sb_{k} +\bb_{\tau} $; and, the mark distribution as,
\begin{equation}
\pmx(y_{r}=y \given \Delta_{\tau,r},e_{k+1}, \sb_{k}, \mb_{r-1}) =\frac{ \indicator{\tau_{r-1}+\Delta_{\tau,r}< t_{k+1}} \odot \exp(\Vb _{y,s} ^\top \sb_k + \Vb_{y,m}  ^\top \mb_{r-1})}{\sum_{y'\in \Ccal}  \exp(\Vb _{y',s} ^\top \sb_k + \Vb_{y',m}  ^\top \mb_{r-1})}, \label{imtp_eq:llambda-2x}
\end{equation}
Here, $\indicator{\cdot}$ denotes the indicator function of whether the sampled times of missing events are within the current observed time-interval. 

Hence, we have:
\begin{align*}
& \Delta_{\tau,r} \sim \qdt(\bullet \given e_{k+1}, \sb_{k}, \mb_{r-1})\nn\\
& \text{If } \Delta_{\tau,r} < t_{k+1}-\tau_{r-1}:\nn\\
&\qquad\qquad \tau_{r}   =\tau_j+\Delta \tau,\\
&\qquad \qquad y_r \sim \pmx(y_{r}=y \given \Delta_{\tau,r}, e_{k+1}, \sb_{k}, \mb_{r-1})\nn\\
&\qquad \qquad \uk{k}=\infty\ \texttt{(Allow more missing events)}\nn\\
& \text{Otherwise: }\nn\\
&\qquad\qquad \uk{k}=r-1.
\end{align*}
\end{compactitem}
Here, note that the mark distribution depends on $\Delta_{\tau, r}$. $\phi=\set{\Gb_{\bullet,\bullet}, \gb_{\bullet,\bullet}, \Vb_{\bullet,\bullet},\bb_\bullet}$ are trainable parameters. The  distribution in Eq.~\ref{imtp_eq:llambda-2x} ensure that given the first $k+1$ observations, $q _{\phi}$ generates the missing events only for $(t_{k},t_{k+1})$ and not for further subsequent intervals.  

\subsection{Prior MTPP model $p_{\pr}$}
We model the prior density (Eq.~\ref{imtp_eq:mt-3}) of the arrival times of the missing events as, 
\begin{equation}
\prdt(\Delta_{\tau,r} \given \sb_{k}, \mb_{r-1}) = \textsc{Lognormal}\left(\mu(\sb_k,\mb_{r-1}), \sigma^2 (\sb_k,\mb_{r-1})\right),
\end{equation}
with  $[\mu(\sb_k,\mb_{r-1}), \sigma^2 (\sb_k,\mb_{r-1}] =\qb_{\mu,m} ^\top \mb_{r-1} +\qb_{\mu,s} ^\top\sb_{k} +\cb $; and, the mark distribution of the missing events as,
\begin{equation}
\prmx(y_{r}=y \given   \Delta_{\tau,r}, \sb_{k}, \mb_{r-1}) = \frac{\exp(\Qb _{y,s} ^\top \sb_k + \Qb_{y,m}  ^\top \mb_{r-1})}{\sum_{y'\in \Ccal}  \exp(\Qb _{y',s} ^\top \sb_k + \Qb_{y',m}  ^\top \mb_{r-1})},
\end{equation}
All parameters $\Qb_{\bullet,\bullet}$, $\qb_{\bullet,\bullet}$ and $\cb$ are \textit{scaled} a-priori using a hyper-parameter $\overline{\mu}$. Thus, $\overline{\mu}$ determines the importance of the $p_{\pr}$ in the missing event sampling procedure of \imtpp. We specify the optimal value for $\overline{\mu}$ based on the prediction performance in the validation set.

\subsection{Training $\theta$ and $\phi$} \label{imtp_sec:opti}
Note that the trainable parameters for observed and posterior MTPPs are $\theta=\{\wb_{\bullet,\bullet}, \Wb_{\bullet,\bullet}, \ab_{\bullet}, \Ub_{\bullet,\bullet}\}$ and  $\phi=\{\gb_{\bullet,\bullet}, \Gb_{\bullet,\bullet}, \bb_{\bullet}, \Vb_{\bullet,\bullet}\}$ respectively. Given a history $\Sdata_K$ of observed events, we aim to learn $\theta$ and $\phi$ by maximizing ELBO, as defined in Eq.~\ref{imtp_eq:elbo}, \ie,
\begin{align}
 &\max_{\theta,\phi} \text{ELBO}(\theta,\phi).
\end{align}
We compute optimal parameters $\theta^*$ and $\phi^*$ that maximizes ELBO($\theta,\phi$) using stochastic gradient descent (SGD)~\cite{rumelhart1986learning}. More details regarding the hyper-parameter values are given in Section~\ref{imtp_sec:expts}.

\subsection{Optimal Position for Missing Events} \label{imtp_sec:imtppplusplus}
To better explain the missing event modeling procedure of \imtpp while simultaneously enhancing its practicability, we present a novel application of \imtppp, a novel variant that offers a trade-off between the number of missing events and the model scalability~\cite{imtppp}. In sharp contrast to the original problem setting of generating missing events between observed events, \imtppp is designed to impute a fixed number of events in a sequence. Specifically, given an input sequence and a user-determined parameter of the number of missing events to be imputed (denoted by $\overline{N}$), \imtppp determines the optimal time and mark of $\overline{N}$ events that when included with the observed MTPP achieve superior event prediction prowess. Note that these events may be missing at random positions that are not considered while training \imtppp. \imtppp achieves this by constraining the missing event sampling procedure of the posterior MTPP ($\qdt(\bullet)$) to limited iterations while simultaneously maximizing the likelihood of observed MTPP. Mathematically, it optimizes the following objective:
\begin{equation}
\max_{q_{\mathrm{imp}, \Delta}} \mathbb{E}_{q_{\mathrm{imp}, \Delta}} \sum_{k=0}^{K-1} \log p(e_{k+1}\given \Sdata_{k}, \Mdata_{\overline{N}}), \quad
\mathrm{where}\, \int_{0}^{T} q_{\mathrm{imp}, \Delta} dt = \overline{N}, \label{imtp_eq:plus}
\end{equation}
where $q_{\mathrm{imp}, \Delta}$ and $p(e_{k+1})$ denote the constrained posterior MTPP and the observed MTPP. However, determining the optimal position of missing events is a challenging task as while imputing events, the generator must consider the dynamics of future events in the sequence. Therefore, \imtppp includes a  two-step training procedure: (i) training observed and missing MTPP using the training set with unbounded missing events (as in Section \ref{imtp_sec:opti}); and then (ii) fine-tuning the parameters of the constrained posterior MTPP and observed MTPP by maximizing the objective in Eq.~\ref{imtp_eq:plus}. For the latter stage, we use the optimal positions of $\overline{N}$ missing events sampled from the posterior MTPP determined by their occurrence probabilities. Later, we assume these events represent all missing ($\Mdata_{\overline{N}}$), followed by a fine-tuning using Eq.~\ref{imtp_eq:plus}.

\subsection{Salient Features of \imtpp}
It is worth noting the similarity of our modeling and inference framework to variational autoencoders~\citep{vrnn}, with $q_{\phi}$ and $p_{\theta}$ playing the roles of encoder and decoder, respectively, while $p_{\pr}$ plays the role of the prior distribution of latent events. However, the random seeds in our model are not simply noise as they are interpreted in autoencoders. They can be concretely interpreted in \imtpp as missing events, making our model physically interpretable.

Secondly, note that the proposal of~\cite{mei_icml} aims to impute the missing events based on the entire observation sequence $\Sdata_K$, rather than to predict observed events in the face of missing events. For this purpose, it uses a bi-directional RNN and, whenever a new observation arrives, it re-generates all missing events by making a completely new pass over the backward RNN. As a consequence, such an imputation method suffers from quadratic complexity with respect to the number of observed events. 

In contrast, our proposal is designed to generate {subsequent} observed and missing events rather than imputing missing events in between observed events\footnote{\scriptsize However,  note that we also use the posterior distribution $q_{\phi}$ to impute missing events between already occurred events.}. To that aim, we only make forward computations, and therefore, it does not require us to re-generate all missing events whenever a new observation arrives, which makes it much more efficient than~\cite{mei_icml} in terms of both learning and prediction. Through our experiments, we also show the exceptionally time-effective operation of \imtpp over other missing-data models.

Finally, unlike most of the prior works~\cite{rmtpp,sahp,mei_icml,nhp,shelton,thp} we model our distribution for inter-arrival times using lognormal. Such a modeling procedure has major advantages over intensity-based models -- (i) scalable sampling during prediction as opposed to Ogata's thinning/inverse sampling; and (ii) efficient training via re-parametrization. Moreover, our generative procedure for missing events requires iterative sampling in the absence of new observed events and such an unsupervised procedure can largely benefit from the prowess of intensity-free models in forecasting future events in a sequence \cite{prathamesh}.

While~\citet{intfree} also uses model inter-arrival times using lognormal; they do not focus on predicting observations in the face of missing events. However, it is important to reiterate (see~\citet{intfree} for details) that this modeling choice offers significant advantages over intensity-based models in terms of providing ease of re-parameterization trick for efficient training, allowing a closed-form expression for expected arrival times and usability for supervised training as well.

\xhdr{Importance of \imtppp} On a broader level, \imtppp may be similar to \imtpp, however, they vary significantly. Specifically, the main distinctions are: (i) \imtppp offers higher practicability as it can be used for predicting future events and for imputing a fixed number of missing events; (ii) \imtpp cannot achieve the latter as it involves an unconstrained procedure for generating missing events; and (iii) \imtppp has an added feature to identify the optimal position of missing events in a sequence. Moreover, as the training procedure of \imtppp involves a pre-training step, the missing event generator has the knowledge of future events in a sequence. This is a sharp contrast to \imtpp which only involves forward temporal computations. To the best of our knowledge, \imtppp is the first-of-its-kind application of neural point process models that can solve several real-world problems, ranging from smooth learning curves to extending the sequence lengths.

\section{Experiments} \label{imtp_sec:expts}
In this section, we report a comprehensive empirical evaluation of \imtpp along with its comparisons with several state-of-the-art approaches. Our code uses Tensorflow\footnote{\scriptsize https://www.tensorflow.org/} v.1.13.1 and Tensorflow-Probability v0.6.0\footnote{\scriptsize https://www.tensorflow.org/probability}. Through these experiments, we aim to answer the following research questions.
\begin{compactitem}
\item[\textbf{RQ1}] What is the mark and time prediction performance of \imtpp in comparison to the state-of-the-art baselines? Where are the gains and losses?
\item[\textbf{RQ2}] How does \imtpp perform in presence of limited data?
\item[\textbf{RQ3}] How does the efficiency of \imtpp compare with the proposal of~\citet{mei_icml}?
\end{compactitem}

\subsection{Experimental Setup}
Here we present the details of all datasets, baselines, and the hyperparameter values used.
\xhdr{Datasets}
For our experiments, we use eight real datasets from different domains: Amazon movies (\amovies)~\cite{julian}, Amazon toys (\atoys)~\cite{julian}, NYC-Taxi (\taxi), \ret~\cite{retweet_data}, Stackoverflow (\so)~\cite{rmtpp}, \fq~\cite{lbsn2vec}, Celebrity~\cite{nagrani17}, and Health~\cite{ecg}. The statistics of all datasets are summarized in Table~\ref{imtp_tab:dset_details} and we describe them as follows:
\begin{compactitem}
 \item \textbf{Amazon Movies} \cite{julian}. For this dataset, we consider the reviews given to items under the category "Movies" on Amazon. For each item, we consider the time of the written review as the time of the event in the sequence and the rating (1 to 5) as the corresponding mark.

\item \textbf{Amazon Toys} \cite{julian}. Similar to Amazon Movies, but here we consider the reviews given to items under the category "Toys".

\item \textbf{NYC Taxi}\footnote{\scriptsize https://chriswhong.com/open-data/foil\_nyc\_taxi/}. Here, each sequence corresponds to a series of timestamped pick-up and drop-off events of a taxi in New York City, and location IDs are considered event marks.

\item \textbf{Twitter} \cite{retweet_data}. Similar to~\cite{nhp}, we group retweeting users into three classes based on their connectivity: an ordinary user (degree lower than the median), a popular user (degree lower than 95-percentile), and \textit{influencers} (degree higher than 95-percentile). Each stream of retweets is treated as a sequence of events with retweet time as the event time, and user class as the mark.

\item \textbf{Stack Overflow}. Similar to \cite{rmtpp}, we treat the badge awarded to a user on the \textit{stack overflow} forum as a mark. Thus we have each user corresponding to a sequence of events with \emph{times} corresponding to the time of mark affiliation. 

\item \textbf{Foursquare}. As a novel evaluation dataset, we use Foursquare (a location search and discovery app) crawls~\cite{lbsn2vec} to construct a collection of check-in sequences of different users from \emph{Japan}. Each user has a sequence with the mark corresponding to the \emph{type} of the check-in location (e.g. "Jazz Club") and the time as the timestamp of the check-in.

\item \textbf{Celebrity}~\cite{nagrani17}. In this dataset, we consider the series of frames extracted from YouTube videos of multiple celebrities as event sequences where event-time denotes the video-time, and the \emph{type} is decided upon the coordinates of the frame where the celebrity is located.

\item \textbf{Health}~\cite{ecg}. The dataset contains ECG records for patients suffering from heart-related problems. Since the length of the ECG record for a single patient can be up to a few million, we sample smaller individual sequences and consider each such sequence as independent with event type as the normalized change in the signal value and the time of recording as event time.
\end{compactitem}

\begin{table}[t!]
  \caption{Statistics of all real datasets used in this chapter.}
  \vspace{-3mm}
  \centering
  \resizebox{\textwidth}{!}{
  \begin{tabular}{l|cccccccc}
  \toprule
  \textbf{Dataset} & \textbf{Movies} & \textbf{Toys} & \textbf{Taxi} & \textbf{Twitter} & \textbf{SO} & \textbf{Foursquare} & \textbf{Celebrity} & \textbf{Health} \\ \hline
  Sequences $|\cm{D}|$ & 27747 & 14365 & 11000 & 22000 & 6103 & 2317 & 10000 & 10000\\ 
  Mean Length $\mathbb{E}{|{\cm{H}_T}|}$ & 48.27 & 35.30 & 15.79 & 108.84 & 72.48 & 145.53 & 120.8 & 297.3\\
  Event Types $\mathbb{E}{|\Ccal|}$ & 5 & 5 & 5 & 3 & 22 & 10 & 16 & 5\\
  \bottomrule
  \end{tabular}
  }
  \label{imtp_tab:dset_details}
  \vspace{-3mm}
\end{table}

\xhdr{Baselines} We compare \imtpp with the following state-of-the-art baselines for modeling continuous-time event sequences:
\begin{compactitem}
\item \textbf{HP}~\cite{hawkes}. A conventional Hawkes process or self-exciting multivariate point process model with an exponential kernel \ie, the past events raise the intensity of the next event of the same type.
\item \textbf{SMHP}~\cite{smhp}. A self-modulating Hawkes process wherein the intensity of the next event is not ever-increasing as in standard Hawkes but learned based on past events.
\item \textbf{RMTPP}~\cite{rmtpp}. A state-of-the-art neural point process that embeds sequence history using inter-event time differences and event marks using a recurrent neural network. 
\item \textbf{SAHP}~\cite{sahp}. A self-attention-based Hawkes process that learns the embedding for the temporal dynamics using a weighted aggregation of all historical events.
\item \textbf{THP}~\cite{thp}. The transformer Hawkes process extends the transformer model~\cite{transformer} to include time and mark influences between events to calculate the \textit{conditional} intensity function for the arrival of future events in the sequence.
\item \textbf{\mei}~\cite{mei_icml}. A particle filtering process for MTPP that learns the sequence dynamics using a bi-directional recurrent neural network.
\item \textbf{HPMD}~\cite{shelton}. Models the sequences using linear multivariate parameterized point processes and learns the inter-event influence using a predefined structure.
\end{compactitem}
We omit the comparisons with other MTPP models~\cite{fullyneural, intfree, xiao2017wasserstein, xiaointaaai, nhp} as they have already been outperformed by these approaches. Moreover, recent research~\cite{li2018learning} has shown that the performance of other RNN-based models such as~\citet{nhp} is comparable to RMTPP~\cite{rmtpp}.

\xhdr{Evaluation protocol} 
Given a stream of $N$ \emph{observed} events $\Sdata_N$, we split them into training $\Sdata_K$ and test set $\Sdata_N\backslash \Sdata_K$, where the training set (test set)
consists of first 80\% (last 20\%) events, \ie, $K=\lceil 0.8N\rceil$. We train \imtpp and the baselines on $\Sdata_K$ and then evaluate the trained models on the test set $\Sdata_N\backslash \Sdata_K$ in terms of (i) mean absolute error (MAE) of predicted times, and (ii) mark prediction accuracy (MPA). 
\begin{equation}
MAE = \frac{1}{|\Sdata_N\backslash \Sdata_K|}\sum_{e_i\in \Sdata_N\backslash \Sdata_K} \hspace{-1mm}\mathbb{E}[|t_i-\hat{t}_i|], \quad MPA = \frac{1}{|\Sdata_N\backslash \Sdata_K|}\sum_{e_i\in \Sdata_N\backslash \Sdata_K}\mathbb{P} (x_i=\hat{x}_i),
\end{equation}
Here $\hat{t_i}$ and $\hat{x_i}$ are the predicted time and mark the $i$-th event in the test set. Note that such predictions are made only on observed events in real datasets. For time prediction, given the varied temporal distribution across the datasets, we normalize event times across each dataset~\cite{rmtpp}. We report results and confidence intervals based on three independent runs.

\subsection{Implementation Details}
\xhdr{Parameter Settings}
For our experiments, we set $\dim(\vb_\bullet)=16$, and $\dim(\gamma_\bullet)=32$, where $\vb_\bullet$ and $\gamma_\bullet$ are the output of the first layers in $p^*_{\theta}$  and $q^*_{\phi}$ respectively; the sizes of hidden states as $\dim(\hb_{\bullet})=64$ and $\dim(\bm{z}_{\bullet})=128$; batch-size $B=64$. In addition, we set an $l_2$ regularizer over the parameters with regularizing coefficient of $0.001$.

\xhdr{System Configuration}
All our experiments were done on a server running Ubuntu 16.04. CPU: Intel(R) Xeon(R) Gold 5118 CPU @ 2.30GHz, RAM: 125GB and GPU: NVIDIA Tesla T4 16GB DDR6.

\xhdr{Baseline Details}\label{imtp_app:basline}
Since HP and SMHP~\citep{smhp} generate a sequence of events of a specified length from the weights learned over the training set, we generate $|N|$ sequences as per the data as $\mathcal{S} = \{s_1, s_2, \cdots s_N\}$ each with maximum sequence length. For evaluation, we consider the first $l_i$ set of events for each sequence $i$. For RMTPP, we set hidden dimension and BPTT is selected among $\{32, 64\}$ and $\{20, 50\}$ respectively. For THP, and SAHP, we set the number of attention heads as $2$, hidden key-matrix, and value-matrix dimensions are selected among $\{32, 64\}$. If applicable, for each model we use a dropout of $0.1$. For PFPP, we set $\gamma = 1$ and use a similar procedure to calculate the embedding dimension as in the THP. All other parameter values are the ones recommended by the authors of the corresponding models. 


\subsection{Event Prediction Performance} \label{imtp_sec:imtpp_rq2}
To address RQ1, we evaluate the event prediction ability of \imtpp. More specifically, we compare the performance of \imtpp with all the baselines introduced above across all six datasets. Tables~\ref{imtp_tab:main_mae} and~\ref{imtp_tab:main_mpa} summarizes the results, which sketches the comparative analysis in terms of mean absolute error (MAE) on time and mark prediction accuracy (MPA), respectively. From the results, we make the following observations:
\begin{compactitem}
\item \imtpp exhibits steady improvement over all the baselines in most of the datasets in the case of both time and mark prediction. However, for Stackoverflow and Foursquare datasets, THP outperforms all other models, including \imtpp, in terms of MPA.

\item RMTPP is the second-best performer in terms of MAE of time prediction almost in all datasets. In fact, in Stackoverflow (SO) dataset, it shares the lowest MAE together with \imtpp. However, there is no consistent second-best performer in terms of MPA. Notably, \mei and \imtpp, which take into account missing events, are the second-best performers for four datasets.

\item Both \mei~\cite{mei_icml} and HPMD~\cite{shelton} fare poorly with respect to \imtpp in terms of both MAE and MPA. This is because \mei focuses on imputing missing events based on complete observations and is not well suited to predict observed events in the face of missing observations. In fact, \mei\ does not offer a joint training mechanism for the MTPP for observed events and the imputation model. Rather it trains an imputation model based on the observation model learned a-priori. On the other hand, HPMD only assumes a linear Hawkes process with a known influence structure. Therefore it shows poor performance with respect to \imtpp. 
\end{compactitem}

\begin{table*}[t]
\footnotesize
\caption{Performance of all the methods in terms of mean absolute error across all datasets on the 20\% test set. Numbers with bold font (boxes) indicate the best (second best) performer. Results marked \textsuperscript{$\dagger$} are statistically significant (two-sided Fisher's test with $p \le 0.1$) over the best baseline.}
\vspace{-3mm}
\centering
\begin{tabular}{l|cccccccc}
\toprule
\textbf{Dataset} & \multicolumn{8}{c}{\textbf{Mean Absolute Error (MAE)}} \\ \hline 
 & \amovies & \atoys & \taxi & \ret & \so & \fq & \cel & \hth \\ \hline \hline
HP~\cite{hawkes} & 0.060 & 0.062 & 0.220 & 0.049 & 0.010 & 0.098 & 0.044 & 0.023\\
SMHP~\cite{smhp} & 0.062 & 0.061 & 0.213 & 0.051 & 0.008 & 0.091 & 0.043 & 0.024\\
RMTPP~\cite{rmtpp} & \fbox{0.053} & \fbox{0.048} & \fbox{0.128} & \fbox{0.040} & \textbf{0.005} & \fbox{0.047} & \fbox{0.036} & \fbox{0.021}\\
SAHP~\cite{sahp} & 0.072 & 0.073 & 0.174 & 0.081 & 0.017 & 0.108 & 0.051 & 0.027\\
THP~\cite{thp} & 0.068 & 0.057 & 0.193 & 0.047 & \fbox{0.006} & 0.052 & 0.040 & 0.026\\
\mei~\cite{mei_icml} & 0.058 & 0.055 & 0.181 & 0.042 & 0.007 & 0.076 & 0.039 & 0.022\\
HPMD~\cite{shelton} & 0.060 & 0.061 & 0.208 & 0.048 & 0.008 & 0.087 & 0.043 & 0.023\\
\imtpp & \textbf{0.049}\textsuperscript{$\dagger$} & \textbf{0.045}\textsuperscript{$\dagger$} & \textbf{0.108}\textsuperscript{$\dagger$} & \textbf{0.038}\textsuperscript{$\dagger$} & \textbf{0.005} & \textbf{0.041}\textsuperscript{$\dagger$} & \textbf{0.032}\textsuperscript{$\dagger$} & \textbf{0.019}\textsuperscript{$\dagger$}\\
\bottomrule
\end{tabular}
\label{imtp_tab:main_mae}
\vspace{3mm}
\caption{Performance of all the methods in terms of mark prediction accuracy (MPA). Numbers with bold font (boxes) indicate the best (second best) performer. Results marked \textsuperscript{$\dagger$} are statistically significant (two-sided Fisher's test with $p \le 0.1$) over the best baseline.}
\centering
\begin{tabular}{l|cccccccc}
\toprule
\textbf{Dataset} & \multicolumn{8}{c}{\textbf{Mark Prediction Accuracy (MPA)}} \\ \hline 
 & \amovies & \atoys & \taxi & \ret & \so & \fq & \cel & \hth \\ \hline \hline
HP~\cite{hawkes} & 0.482 & 0.685 & 0.894 & 0.531 & 0.418 & 0.523  & 0.229 & 0.405\\
SMHP~\cite{smhp} & 0.501 & 0.683 & 0.893 & 0.554 & 0.423 & 0.520 & 0.238  & 0.401\\
RMTPP~\cite{rmtpp} & 0.548 & 0.734 & 0.929 & \fbox{0.572} & 0.446 & 0.605 & 0.255 & 0.421\\
SAHP~\cite{sahp} & 0.458 & 0.602 & 0.863 & 0.461 & 0.343 & 0.459 & 0.227 & 0.353\\
THP~\cite{thp} & 0.537 & 0.724 & \fbox{0.931} & 0.526 & \textbf{0.458} & \textbf{0.624} & \fbox{0.268} & 0.425\\
\mei~\cite{mei_icml} & \fbox{0.559} & \fbox{0.738} & 0.925 & 0.569 & 0.437 & 0.582 & 0.256 & \fbox{0.427}\\
HPMD~\cite{shelton} & 0.513 & 0.688 & 0.907 & 0.558 & 0.439 & 0.531 & 0.247 & 0.409\\
\imtpp & \textbf{0.574}\textsuperscript{$\dagger$} & \textbf{0.746}\textsuperscript{$\dagger$} & \textbf{0.938}\textsuperscript{$\dagger$} & \textbf{0.577} & \fbox{0.451} & \fbox{0.612} & \textbf{0.273} & \textbf{0.438}\textsuperscript{$\dagger$}\\
\bottomrule
\end{tabular}
\label{imtp_tab:main_mpa}
\vspace{-2mm}
\end{table*}

\begin{figure}[t]
\centering
\begin{subfigure}{0.45\columnwidth}
  \centering
  \includegraphics[height=4cm]{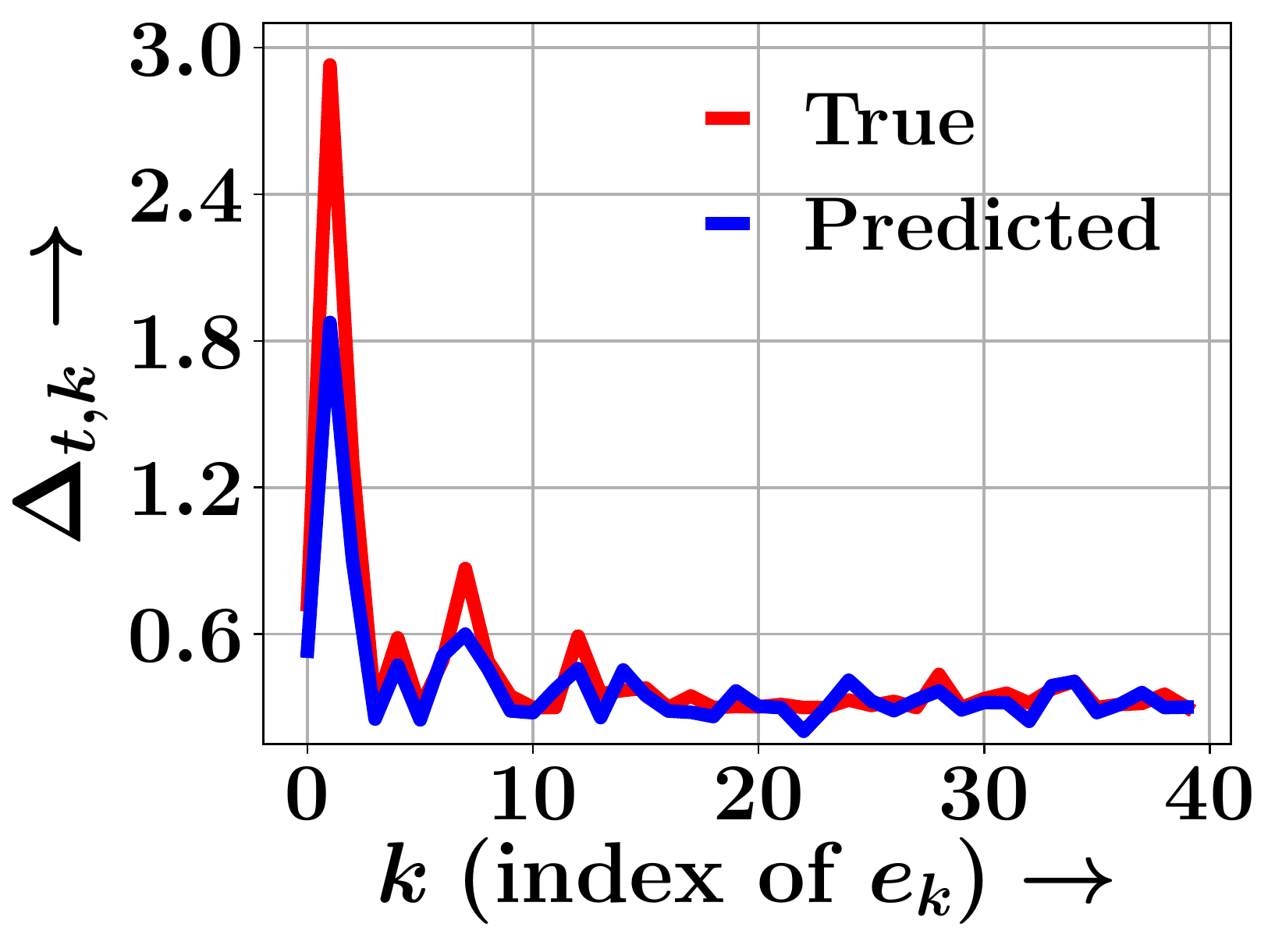}
  \caption{\amovies}
\end{subfigure}
\hspace{1cm}
\begin{subfigure}{0.45\columnwidth}
  \centering
  \includegraphics[height=4cm]{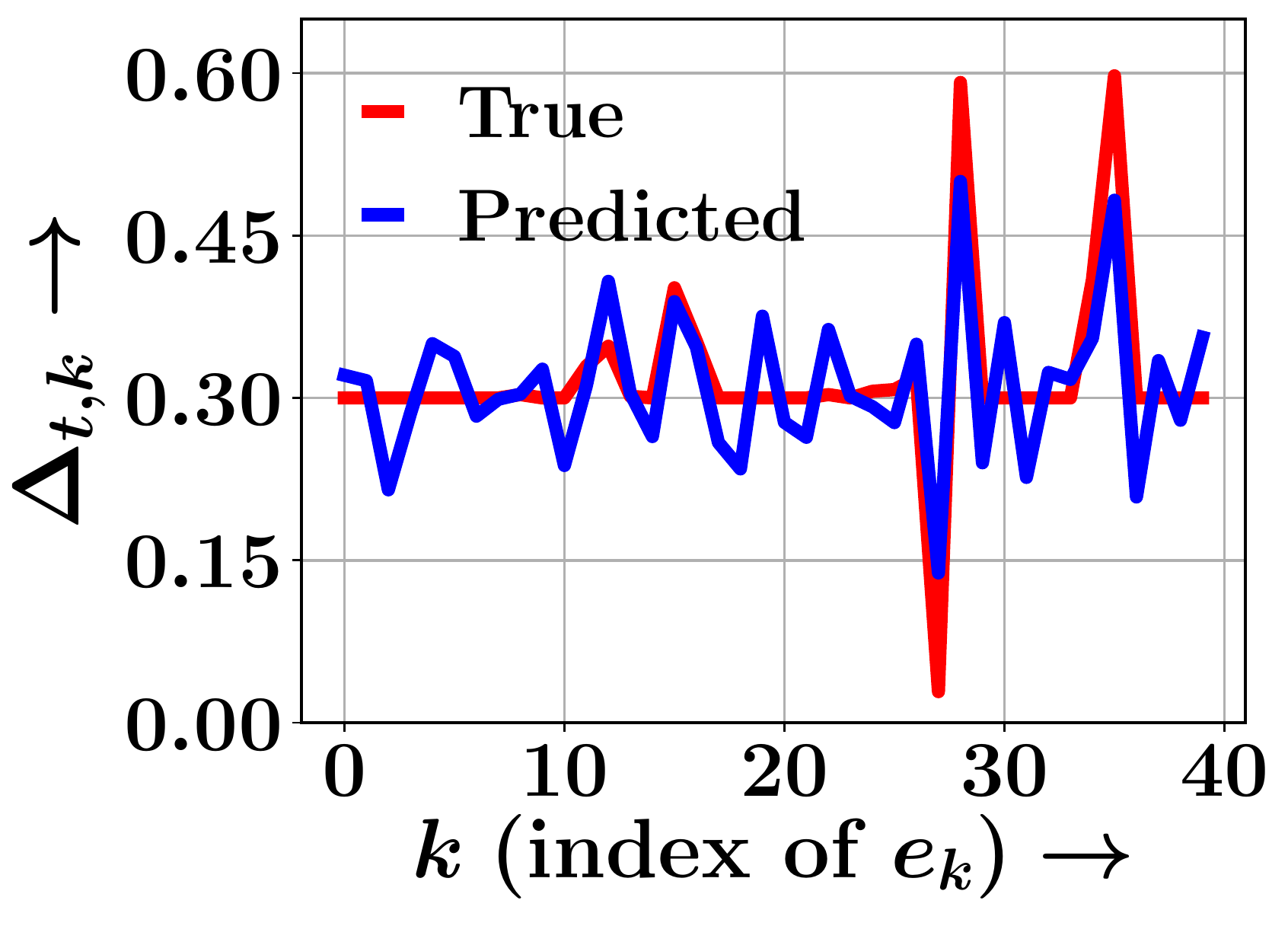}
  \caption{\atoys}
\end{subfigure}
\vspace{-2mm}
\caption{Real life examples of true and predicted inter-arrival times $\Delta_{t,k}$ of different events $e_k$, against $k$ for $k\in \set{k+1,\ldots,N}$. Panels (a) and (b) show the results for \amovies and \atoys datasets, respectively.}
\vspace{-2mm}
\label{imtp_fig:anecdote}
\end{figure}

\xhdr{Qualitative Analysis} In addition, we also perform a \emph{qualitative} analysis to identify if \imtpp can model the inter-event time-intervals in a sequence. Figure~\ref{imtp_fig:anecdote} provides some real-life event sequences taken from \amovies and \atoys datasets and the time-intervals predicted by \imtpp. The results qualitatively show that the predicted inter-arrival times closely match the true inter-arrival times. Moreover, the results also show that \imtpp can event efficiently model the large \textit{spikes} in inter-event time intervals.

\begin{figure}[t]
\centering
\begin{subfigure}{0.45\columnwidth}
  \centering
  \includegraphics[height=4cm]{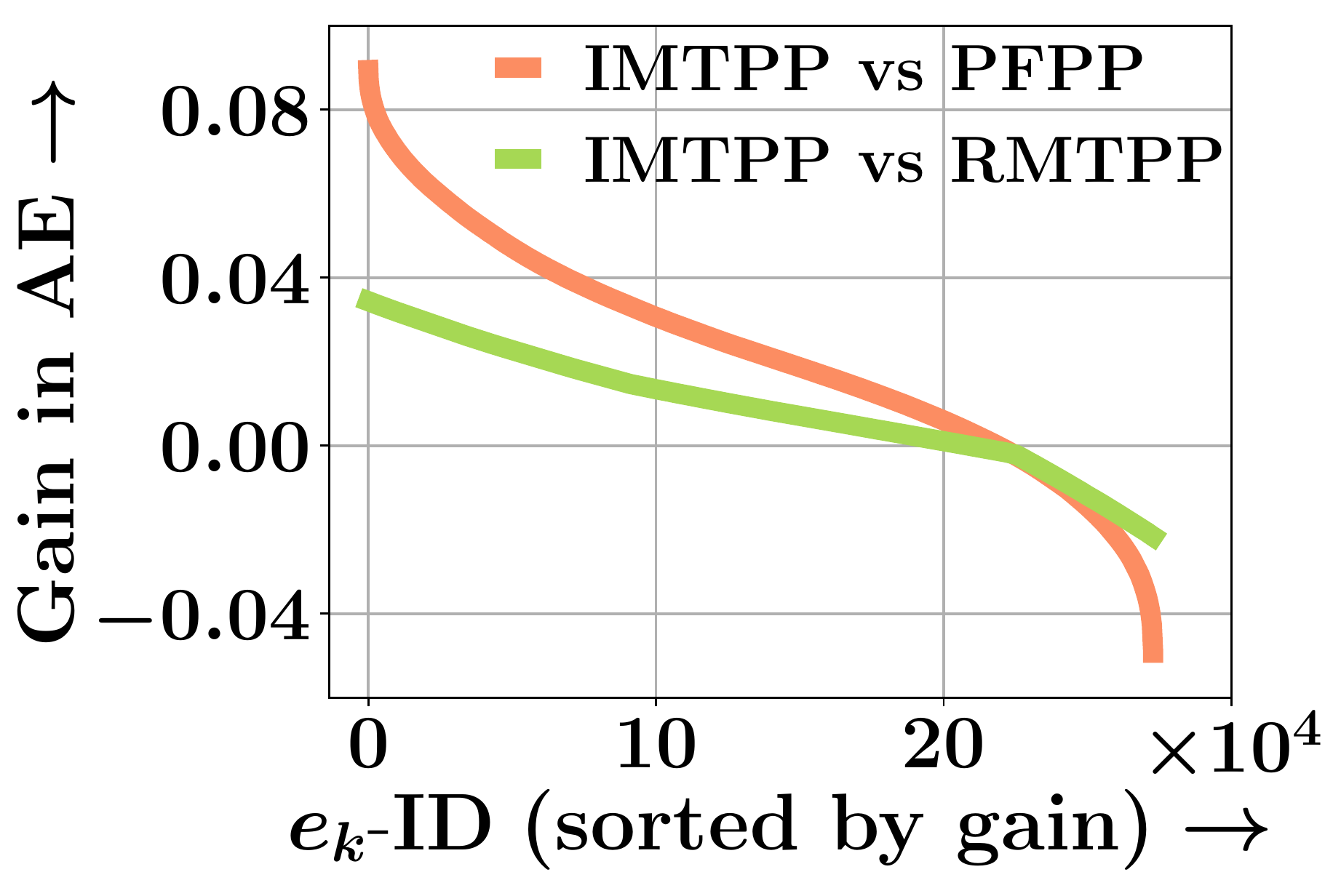}
  \caption{\amovies}
\end{subfigure}
\hspace{1cm}
\begin{subfigure}{0.45\columnwidth}
  \centering
  \includegraphics[height=4cm]{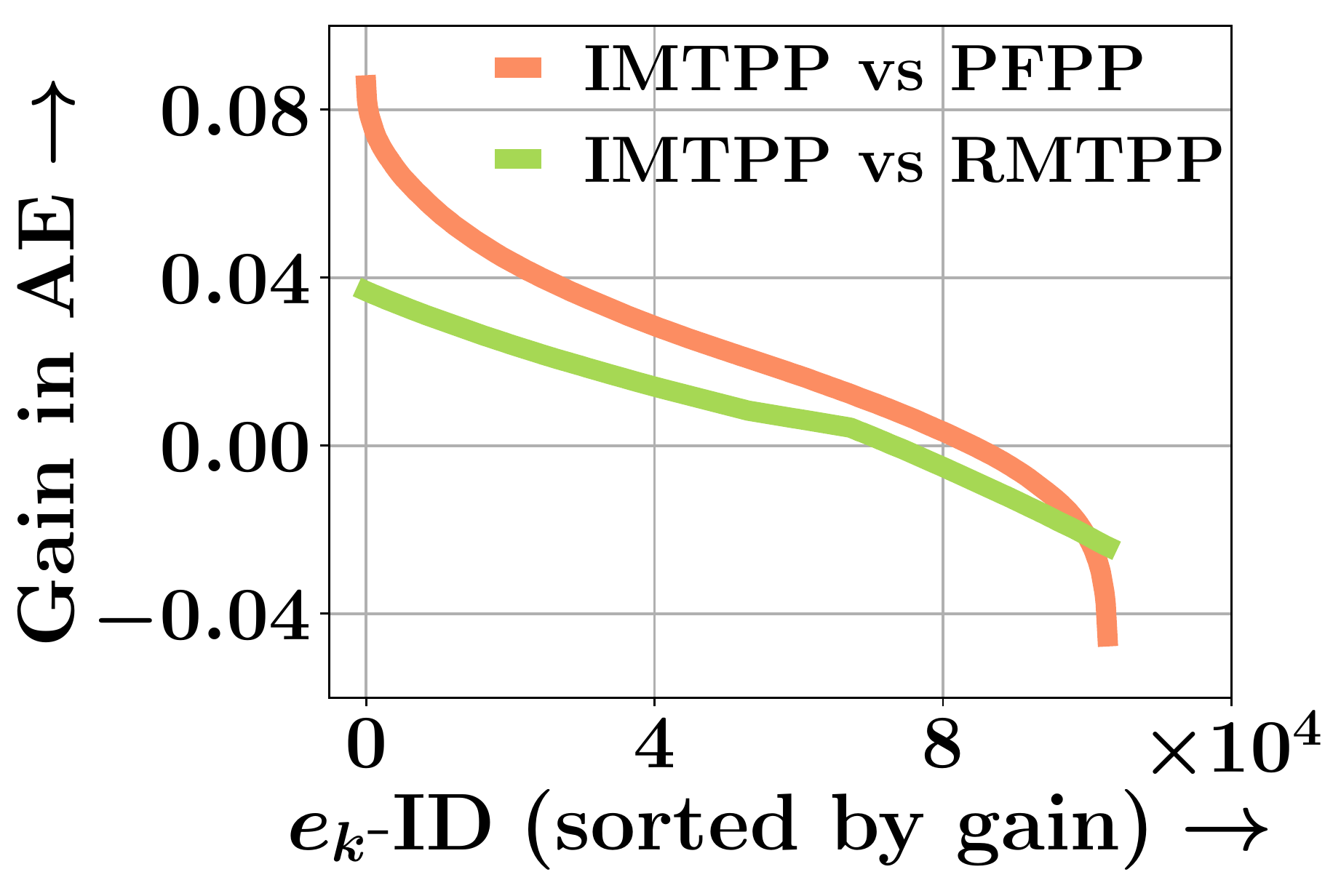}
  \caption{\atoys}
\end{subfigure}
\vspace{-2mm}
\caption{Performance gain in terms of $\text{AE}(\text{baseline})-\text{AE}(\imtpp)$--- the gain (above x-axis) or loss (below x-axis) of the average error per event $\mathbb{E}[|t_k-\hat{t_k}|]$ of \imtpp--- with respect to two competitive baselines: RMTPP and \mei. Events in the test set are sorted by decreasing gain of \imtpp along $x$-axis. Panels (a) and (b) show the results for \amovies and \atoys datasets, respectively.}
\vspace{-2mm}
\label{imtp_fig:drill-down}
\end{figure}
 
\xhdr{Drill-down Analysis}
Next, we provide a comparative analysis of the time prediction performance at the level of every event in the test set. To this end, for each observed event $e_i$ in the test set, we compute the gain (or loss) \imtpp achieves in terms of the time-prediction error per event $\mathbb{E}[|t_k-\hat{t_k}|]$, \ie, $\text{AE}(\text{baseline})-\text{AE}(\imtpp)$ for two competitive baselines, \eg, RMTPP and \mei for \amovies and \atoys datasets. Figure~\ref{imtp_fig:drill-down} summarizes the results, which shows that \imtpp outperforms the most competitive baseline \ie, RMTPP for more than 70\% events across both \amovies and \atoys datasets. It also shows that the performance gain of \imtpp over \mei is ever more prominent.

\begin{table*}[t]
\small
\centering
\caption{Mark prediction performance of Markov Chains and \imtpp across all datasets. We use MC of orders 1,2, and 3 and report results for the best-performing model.}
\vspace{-3mm}
\begin{tabular}{l|cccccccc}
\toprule
\textbf{Dataset} & \multicolumn{8}{c}{\textbf{Mark Prediction Accuracy (MPA)}} \\ \hline 
 & \amovies & \atoys & \taxi & \ret & \so & \fq & \cel & \hth\\ \hline \hline
MC & 0.542 & 0.702 & 0.829 & 0.548 & 0.443 & 0.575 & 0.249 & 0.416\\
\imtpp & \textbf{0.574} & \textbf{0.746} & \textbf{0.938} & \textbf{0.577} & \textbf{0.451} & \textbf{0.612} & \textbf{0.273} & \textbf{0.438}\\
\bottomrule
\end{tabular}
\vspace{-3mm}
\label{imtp_tab:markov}
\end{table*}

\xhdr{Performance Comparison with Markov Chains}
Previous research~\cite{rmtpp} has shown that the mark prediction performance of neural MTPP models is comparable to Markov Chains (MCs). Therefore, in addition to the mark prediction experiments with MTPP models in Table~\ref{imtp_tab:main_mpa}, we also report the results for the comparison between \imtpp and MCs of orders $1$ to $3$ in Table~\ref{imtp_tab:markov}. Note that we only report the results of the best-performing MC. The results show that the Markov models perform quite well in terms of mark prediction accuracy across all datasets. A careful investigation revealed that the datasets exhibit significant repetitive characteristics of marks in a small window. Thus for some datasets with large repetitions within a short history window, using a deep point process-based model is overkill. On the other hand, for NYC Taxi, the mobility distribution clearly shows long-term dependencies, thus severely hampering the performance of Markov Chains. In these cases, point process-based models show better performance by being able to model the inter-event complex dependencies more efficiently.

\begin{table*}[t!]
\small
\centering
\caption{Time prediction performance of \imtpp and its variants -- \imtppobs and \imtpplog in terms of MAE on the 20\% test set. Numbers with bold font indicate the best performer.}
\vspace{-3mm}
\begin{tabular}{l|cccccccc}
\toprule
\textbf{Dataset} & \multicolumn{8}{c}{\textbf{Mean Absolute Error (MAE)}} \\ \hline 
 & \amovies & \atoys & \taxi & \ret & \so & \fq & \cel & \hth\\ \hline \hline
\imtppobs & 0.054 & 0.047 & 0.115 & 0.042 & 0.005 & 0.044 & 0.034 & 0.021\\
\imtpplog & 0.056 & 0.051 & 0.120 & 0.041 & 0.006 & 0.043 & 0.037 & 0.023\\
\imtpp & \textbf{0.049} & \textbf{0.045} & \textbf{0.108} & \textbf{0.038} & \textbf{0.005} & \textbf{0.041} & \textbf{0.032} & \textbf{0.019}\\
\bottomrule
\end{tabular}
\label{imtp_tab:schur_mae}
\vspace{3mm}
\caption{Mark prediction performance of \imtpp and its variants in terms of MPA on the 20\% test set. Numbers with bold font indicate the best performer.}
\begin{tabular}{l|cccccccc}
\toprule
\textbf{Dataset} & \multicolumn{8}{c}{\textbf{Mark Prediction Accuracy (MPA)}} \\ \hline 
 & \amovies & \atoys & \taxi & \ret & \so & \fq & \cel & \hth \\ \hline \hline
\imtppobs & 0.569 & 0.742 & 0.929 & 0.574 & 0.450 & 0.603 & 0.267 & 0.425\\
\imtpplog & 0.563 & 0.724 & 0.927 & 0.568 & 0.449 & 0.598 & 0.261 & 0.433 \\
\imtpp & \textbf{0.574} & \textbf{0.746} & \textbf{0.938} & \textbf{0.577} & \textbf{0.451} & \textbf{0.612} & \textbf{0.273} & \textbf{0.438}\\
\bottomrule
\end{tabular}
\vspace{-3mm}
\label{imtp_tab:schur_mpa}
\end{table*}

\subsection{Ablation Study}
We also conduct an ablation study for two key contributions in \imtpp: missing event MTPP and the intensity-free modeling of time intervals. We denote \imtppobs as the variants of \imtpp without the missing MTPP and \imtpplog as the variant without the lognormal distribution for inter-event arrival times. More specifically, for \imtpplog we follow \cite{rmtpp} to determine an intensity function $\lambda^p_k$ for observed events using the output of the RNN, $\sb_{k}$.
\begin{equation}
\lambda^*_p (t_k) =\exp(\wb_{\lambda,s} \sb_{k} + \wb_{\lambda,m} \mb_{\uk{k}} + \wb_{\lambda,\Delta} (t_k-t_{k-1}) + \bb_{\lambda}),
\end{equation}
Later, we use the intensity function at a given timestamp to estimate the probability distribution of future events as:
\begin{equation}
p_{\theta, t}(t_{k+1}) = \lambda^*_p (t_k) \exp \left(- \int_{t_k}^{t} \lambda^*_p(\tau) \,d\tau \right),
\end{equation}
We report the performance of \imtpp and its variants in terms of MAE and MPA in Tables~\ref{imtp_tab:schur_mae} and~\ref{imtp_tab:schur_mpa} respectively. The results show that \imtpp outperforms \imtppobs and \imtpplog across all metrics. The performance gain of \imtpp over \imtppobs signifies the importance of including missing events for modeling event sequences. We also note that the performance gain of \imtpp over \imtpplog reinforces our modeling design of using an intensity-free model.

\begin{figure}[t]
\centering
\begin{subfigure}{0.45\columnwidth}
  \centering
  \includegraphics[height=4cm]{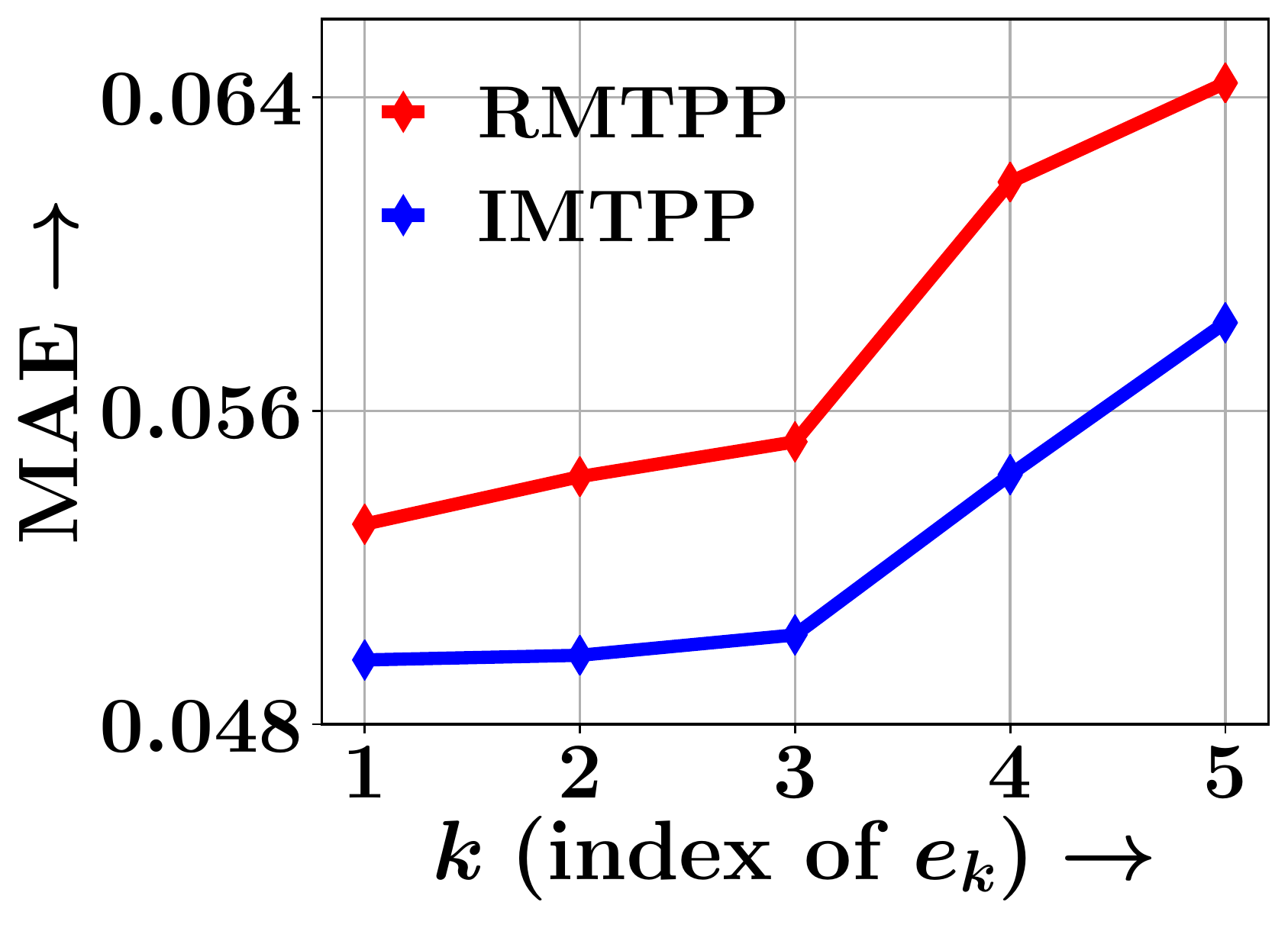}
  \caption{Movies, MAE}
\end{subfigure}
\hspace{1cm}
\begin{subfigure}{0.45\columnwidth}
  \centering
  \includegraphics[height=4cm]{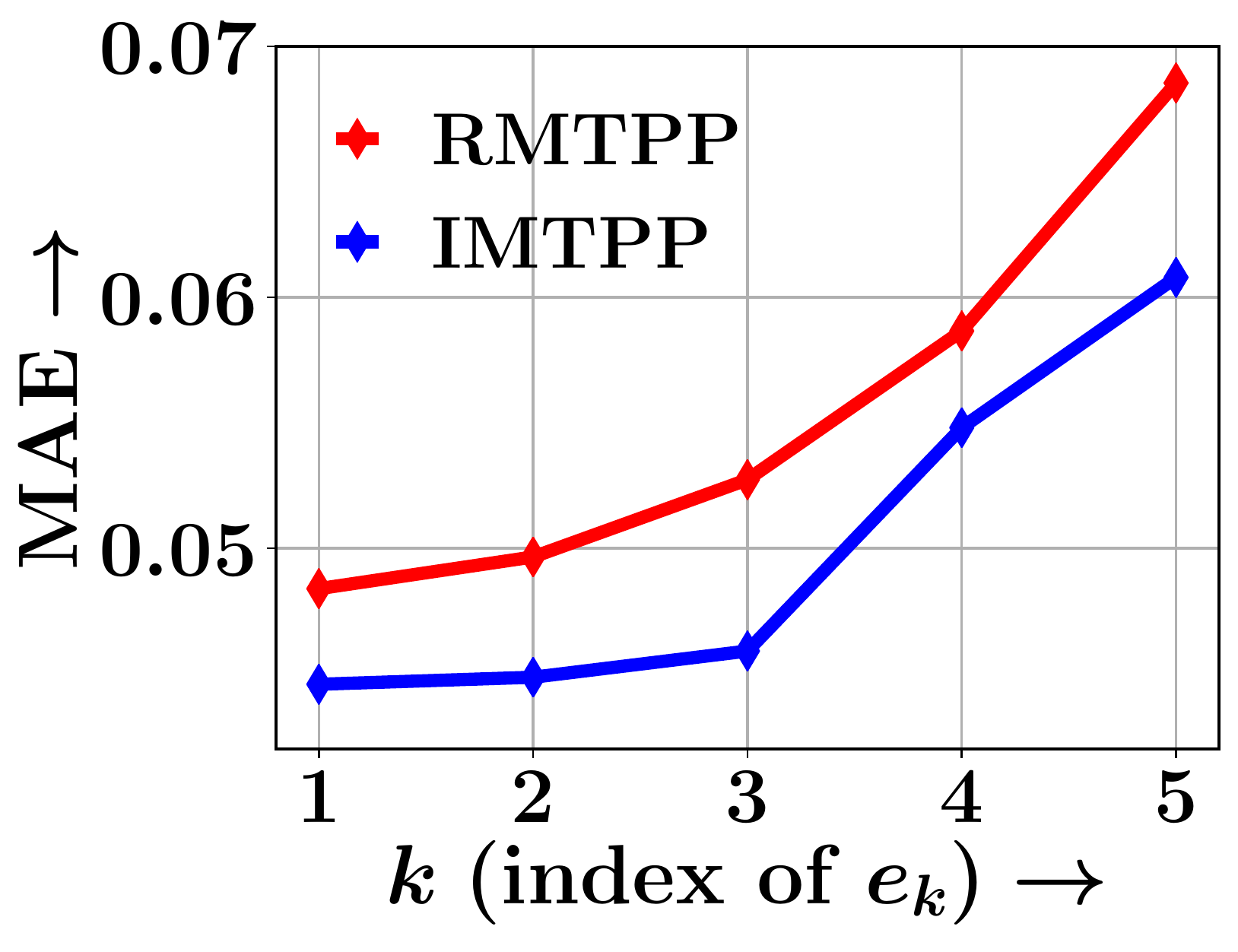}
  \caption{Toys, MAE}
\end{subfigure}
\begin{subfigure}{0.45\columnwidth}
  \centering
  \includegraphics[height=4cm]{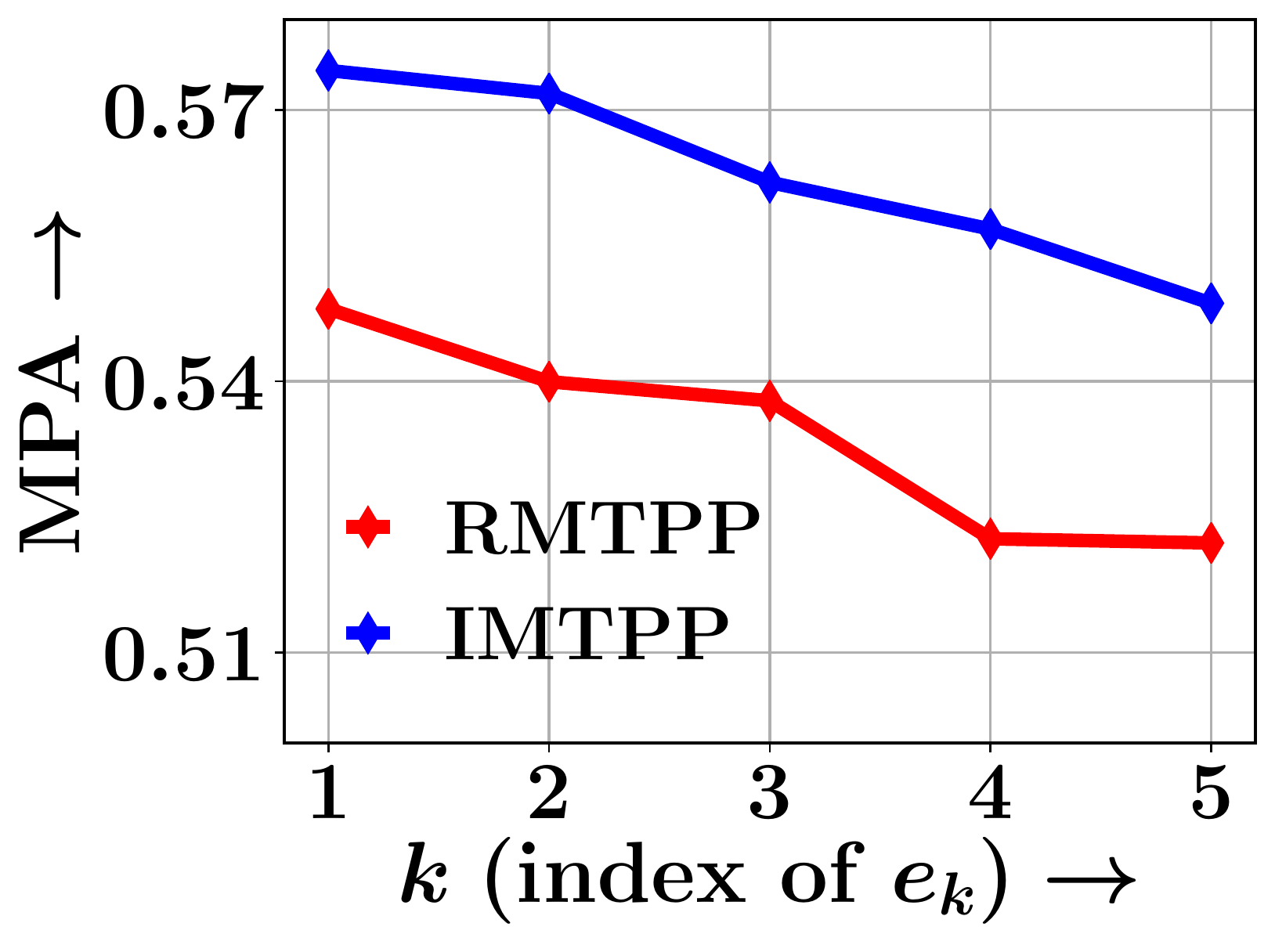}
  \caption{Movies, MPA}
\end{subfigure}
\hspace{1cm}
\begin{subfigure}{0.45\columnwidth}
  \centering
  \includegraphics[height=4cm]{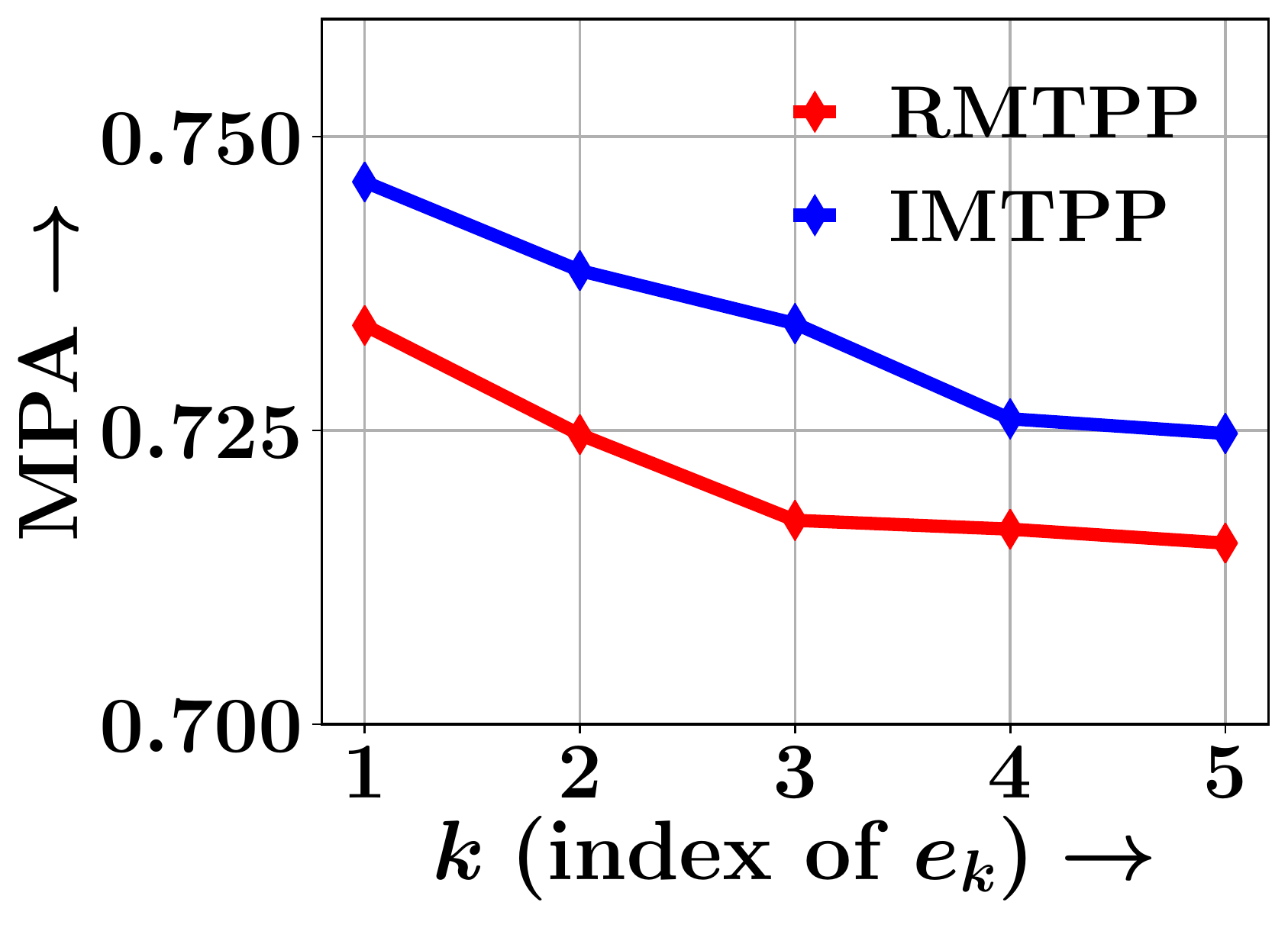}
  \caption{Toys, MPA}
\end{subfigure}
\vspace{-3mm}
\caption{Variation of the forecasting performance of \imtpp and RMTPP in terms of MAE and MPA at predicting next $i$-th event, against $i$ for \amovies and \atoys dataset. Panels (a--b) show the variation of MAE, while panels (c--d) show the variation of MPA. They show that as $n$ increases, the performance deteriorates for both the metrics and both datasets as the prediction task becomes more and more difficult.} 
\vspace{-3mm}
\label{imtp_fig:forecast}
\end{figure}

\subsection{Forecasting Future Events} \label{imtp_sec:imtpp_rq3}
To highlight the forecasting prowess of \imtpp against its competitors, we design a difficult event prediction task, where we predict the next $n$ events given only the current event as input. To do so, we keep sampling events using the trained model $p _{\hat{\theta}}$ and $q _{\hat{\phi}}$ till $n$-{th} prediction. Such an evaluation protocol effectively requires accurate inference of the missing data distribution since, unlike during the training phase, the future observations are not fed into the missing event model. To this end, we compare the forecasting performance of \imtpp against RMTPP, the most competitive baseline. Figure~\ref{imtp_fig:forecast} summarizes the results for \amovies and \atoys datasets, which shows that (i) the performances of all the algorithms deteriorate as $n$ increases and; (ii) \imtpp achieves ~5.5\% improvements in MPA and significantly better 10.12\% improvements in MAE than RMTPP across both datasets. The results further reinforce the ability of \imtpp to model the long-term distribution of events in a sequence.

\begin{figure}[t]
\centering
\begin{subfigure}{0.45\columnwidth}
  \centering
  \includegraphics[height=4cm]{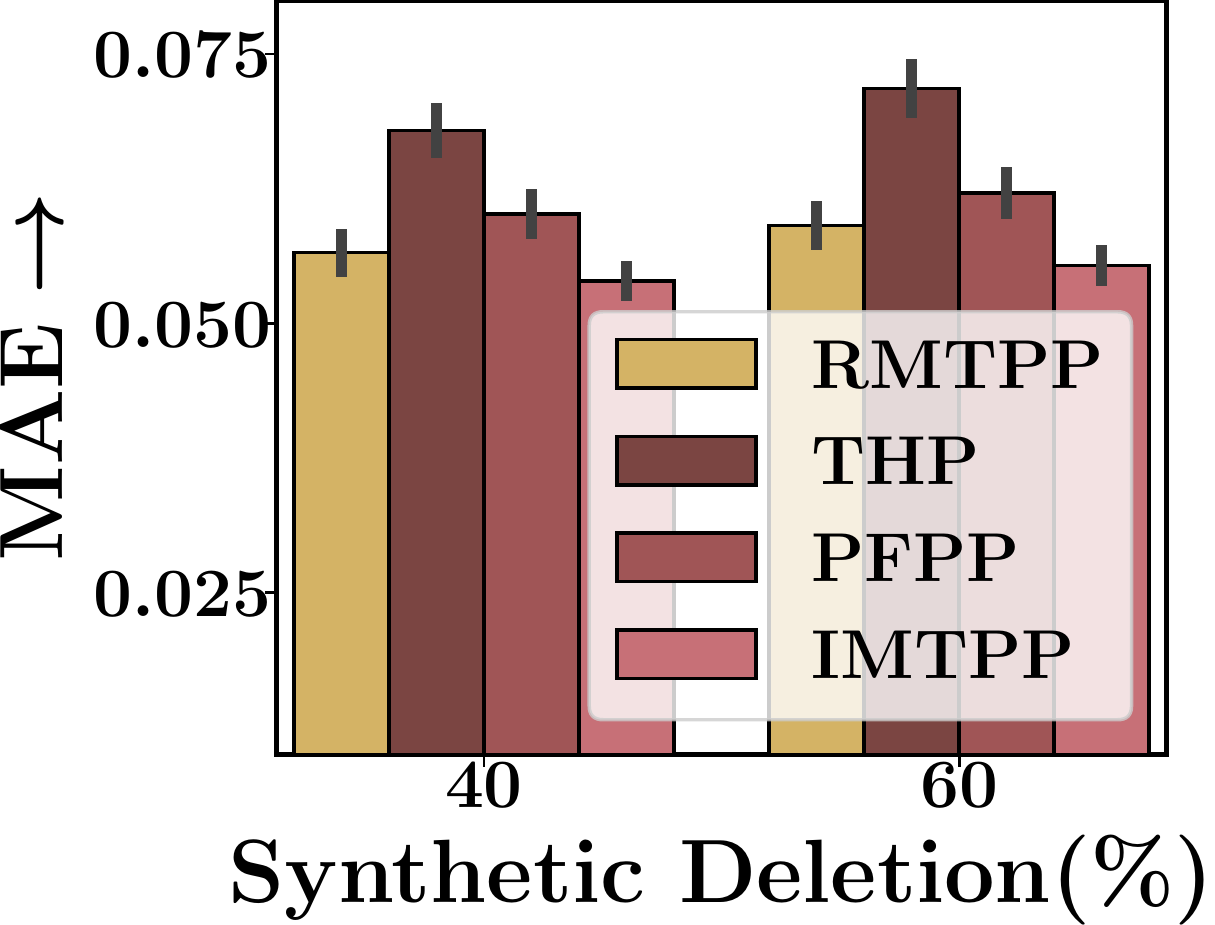}
  \caption{Movies, MAE}
\end{subfigure}
\hspace{1cm}
\begin{subfigure}{0.45\columnwidth}
  \centering
  \includegraphics[height=4cm]{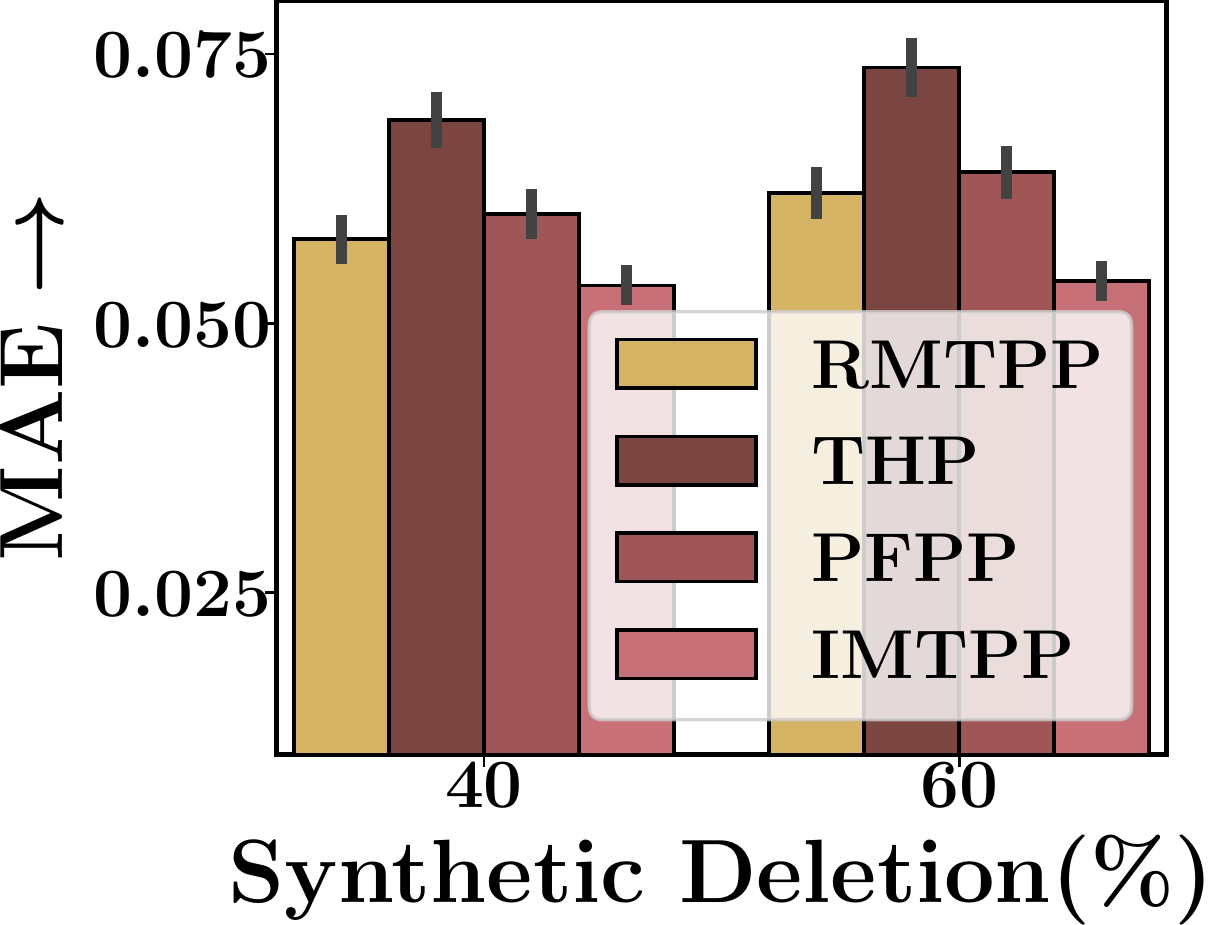}
  \caption{Toys, MAE}
\end{subfigure}
\begin{subfigure}{0.45\columnwidth}
  \centering
  \includegraphics[height=4cm]{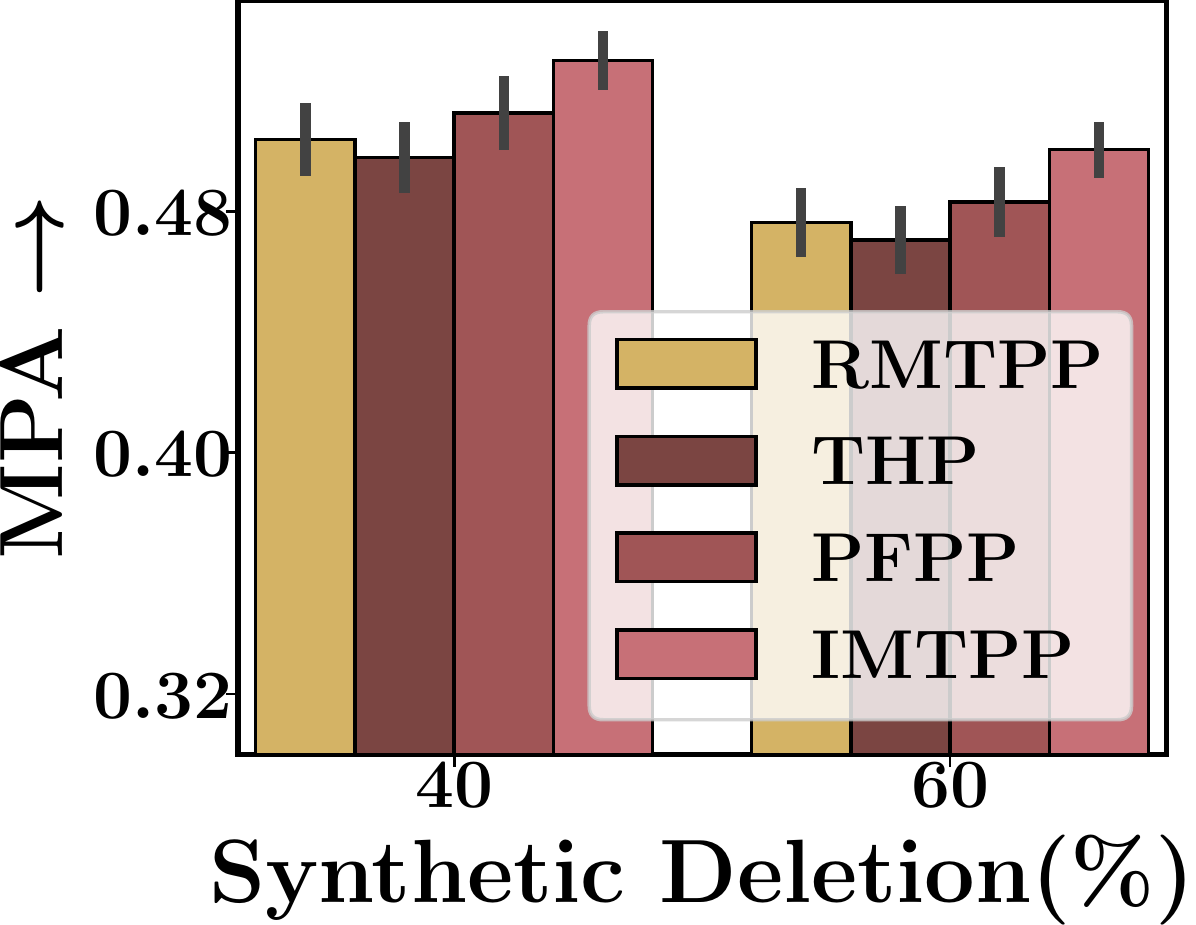}
  \caption{Toys, MPA}
\end{subfigure}
\hspace{1cm}
\begin{subfigure}{0.45\columnwidth}
  \centering
  \includegraphics[height=4cm]{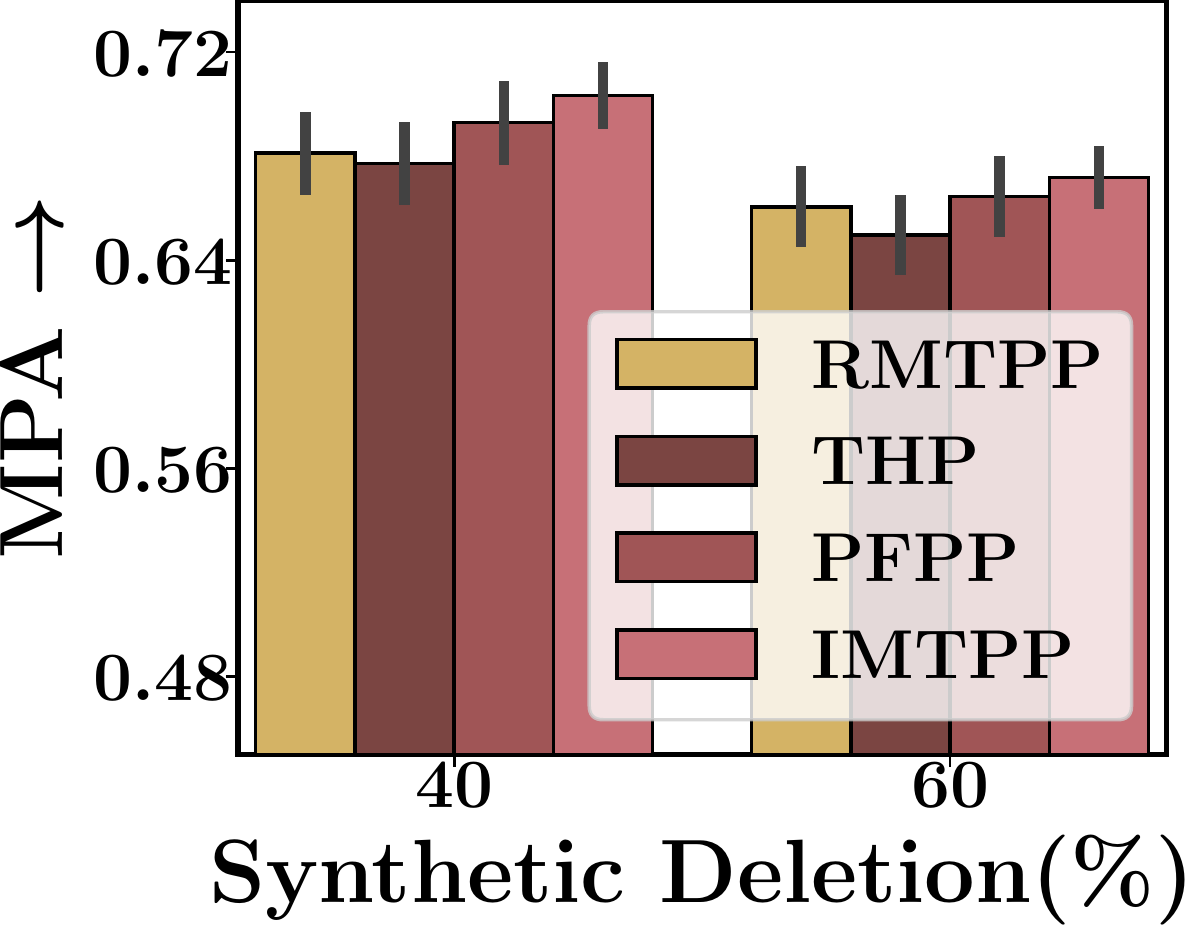}
  \caption{Movies, MPA}
\end{subfigure}
\vspace{-3mm}
\caption{Impact of missing observations on model performance for Movies and Toys dataset. We randomly delete 40\%  and 60\% events from the observed sequence and then train and test \imtpp and best performing baselines on the rest of the observed events. Panels (a--b) show the results for time prediction, while panels (c--d) show the results for mark prediction.} 
\vspace{-3mm}
\label{imtp_fig:missing}
\end{figure}

\subsection{Performance with Missing Data} \label{imtp_sec:miss_exp}
For RQ2 and to emphasize the applicability of \imtpp in the presence of missing data, we perform event prediction on sequences with limited training data. Specifically, we synthetically delete events from a sequence \ie, we randomly (via a normal distribution) delete $40\%$ (and $60\%$) of events from the original sequence and then train and test our model on the rest $60\%$ events($40\%$). Figure~\ref{imtp_fig:missing} summarizes the results across Movies and Toys datasets. From the results, we note that with synthetic data deletion, the performance improvement of \imtpp over best-performing baselines -- RMTPP, THP, and PFPP-- is significant even after 40\% of events are deleted. This is because \imtpp is trained to capture the missing events, and as a result, it can exploit the underlying setting with data deletion more effectively than the other models. Though this performance gains saturate with a further increase in missing data as the added noise in the datasets severely hampers the learning of both models. Interestingly, we note that RMTPP outperforms THP and PFPP even in situations with limited data.

\subsection{Scalability Analysis with \mei~\cite{mei_icml}} \label{imtp_sec:imtpp_rq4}
\subsubsection{With Complete Sequences} 
\begin{figure}[t]
\centering
\begin{subfigure}{0.45\columnwidth}
  \centering
  \includegraphics[height=4cm]{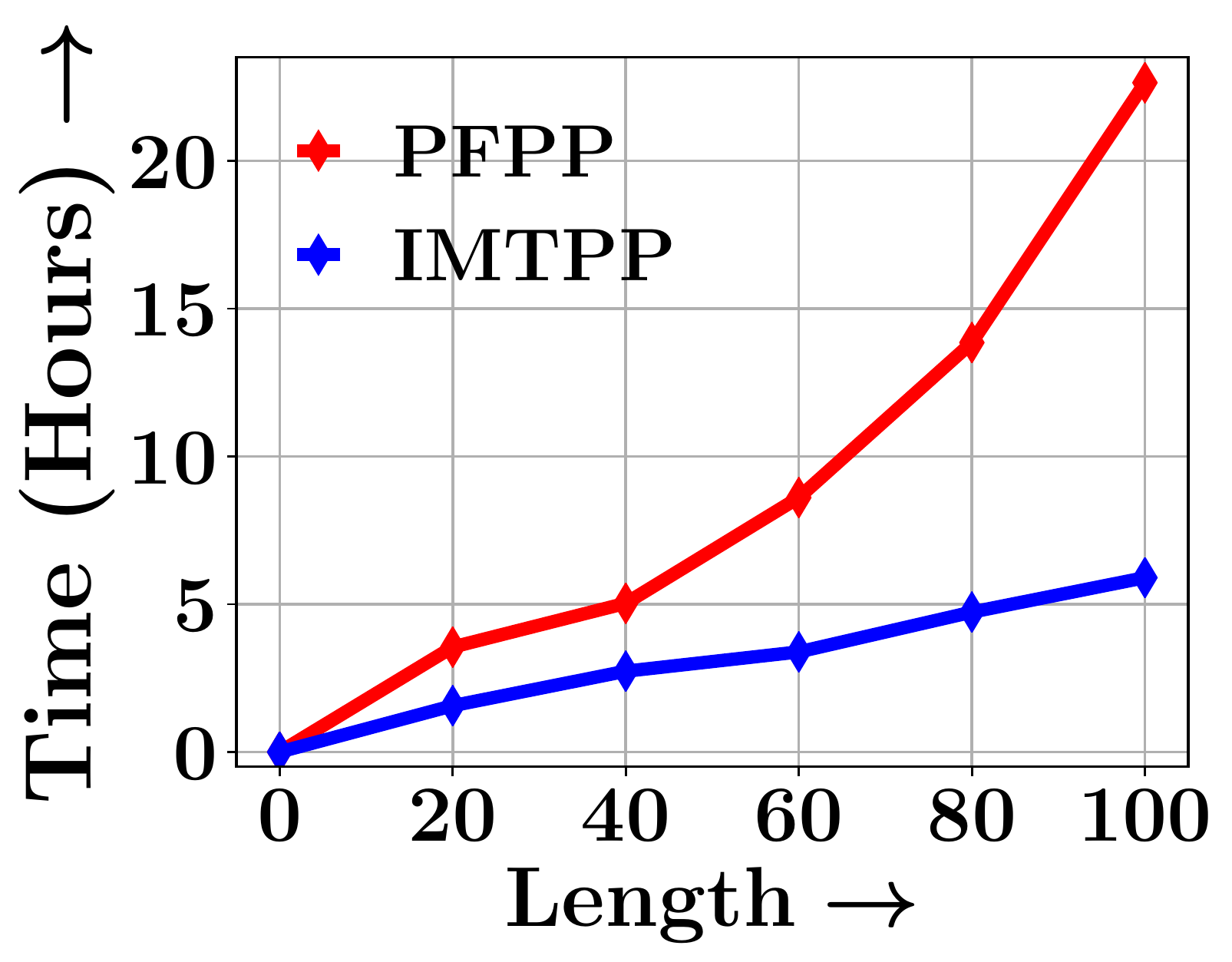}
  \caption{Time vs $|\Sdata_K|$}
\end{subfigure}
\begin{subfigure}{0.45\columnwidth}
  \centering
  \includegraphics[height=4cm]{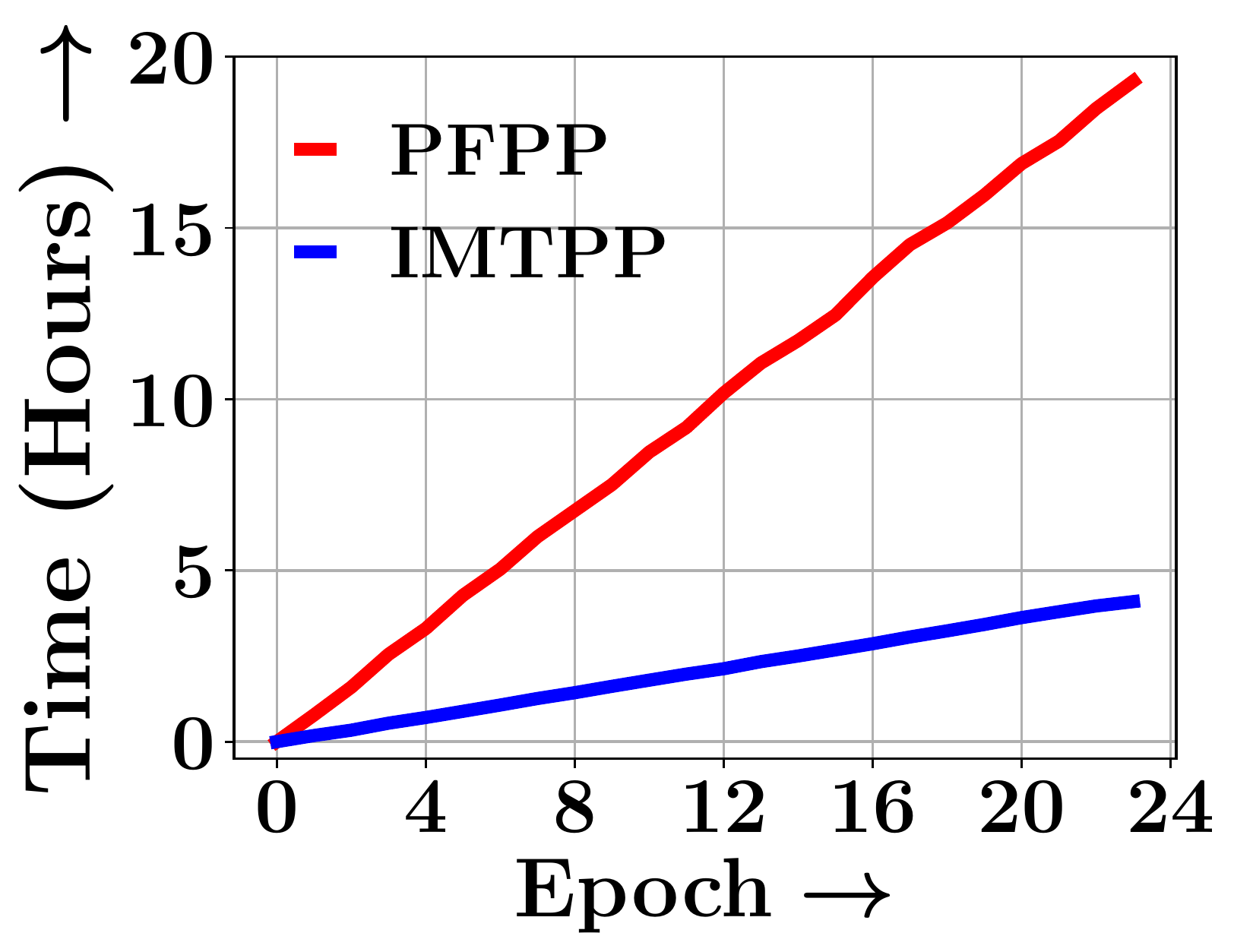}
  \caption{Time vs \# of epochs}
\end{subfigure}
\vspace{-3mm}
\caption{Runtime performance of PFPP and \imtpp for the largest dataset, \ie, Movies, with complete sequences. Panel (a) shows the time vs. length of the training sequence, and panel (b) shows the time vs. the number of epochs.} 
\label{imtp_fig:mei}
\vspace{-3mm}
\end{figure}

To highlight the time-effective learning ability of \imtpp, we compare the runtimes of \imtpp and \mei across no. of training epochs as well as the length of training sequence $|\Sdata_K|$. Figure~\ref{imtp_fig:mei} summarizes the results, which shows that \imtpp enjoys a better latency than \mei. In particular, we observe that the runtime of \mei increases quadratically with respect to $|\Sdata_K|$, whereas the runtime of \imtpp increases linearly. The quadratic complexity of \mei is due to the presence of a backward RNN, which requires a complete pass whenever a new event arrives. The larger runtimes of both models can be attributed to the massive size of \amovies dataset with $1.4$ million events.

\begin{table}[t!]
\small
\caption{Runtime comparison between \imtpp and \mei in a streaming setting. Here, DNF indicates that the code did not finish within 24:00hrs.}
\vspace{-3mm}
\centering
    \begin{tabular}{l|cccccccc}
  \toprule
  \textbf{Dataset} & \multicolumn{8}{c}{\textbf{Runtime (Hours)}} \\ \hline 
  & \amovies & \atoys & \taxi & \ret & \so & \fq & \cel & \hth\\ \hline \hline
  \imtpp & <2hr & <2hr & <1hr & <2hr & <1hr & <1hr & <2hr & <2hr \\
  \mei~\cite{mei_icml} & DNF & DNF & DNF & DNF & DNF & DNF & DNF & DNF\\
  \bottomrule
  \end{tabular}
  \label{imtp_tab:runtime}
  \vspace{-3mm}
\end{table} 

\subsubsection{Streaming-based Runtime}
Our setting differs significantly from \mei~\cite{mei_icml} as the latter requires the complete data distribution of missing and observed events. However, we evaluate if their model can be extended to our setting \ie, a \textit{streaming} setting wherein complete sequences are not available upfront rather, they arrive as we progress with time along a sequence. In a streaming setting, the model is trained as per the arrival of events, and the only way for \mei to be extended in this setting is to update the parameters at each arrival. This repetitive training is expensive and can withdraw the practicability of the model. To further assert our proposition, we evaluate the runtime of their model across the different datasets and compare it with \imtpp. We report the results for training across only a few epochs (10) in a streaming setting in Table~\ref{imtp_tab:runtime}. We note that across all datasets \mei fails to scale as expected. This delay in training could be attributed to the two-phase training of their model; (i) particle filtering, where they learn the underlying complete data distribution, and (ii) particle smoothing, in which they filter out the inadequate events generated during particle filtering. Secondly, since \mei requires complete data during training, an online setting would require repetitive parameter optimization on the arrival of each event. Thus, their model cannot be extended to such a scenario while maintaining its practicality. Thus, \imtpp is the singular practical approach for learning MTPPs in an online setting with intermittent observational data.

\subsection{Imputation Performance}
\begin{figure}[t]
\centering
\begin{subfigure}{0.24\columnwidth}
  \centering
  \includegraphics[height=3.5cm]{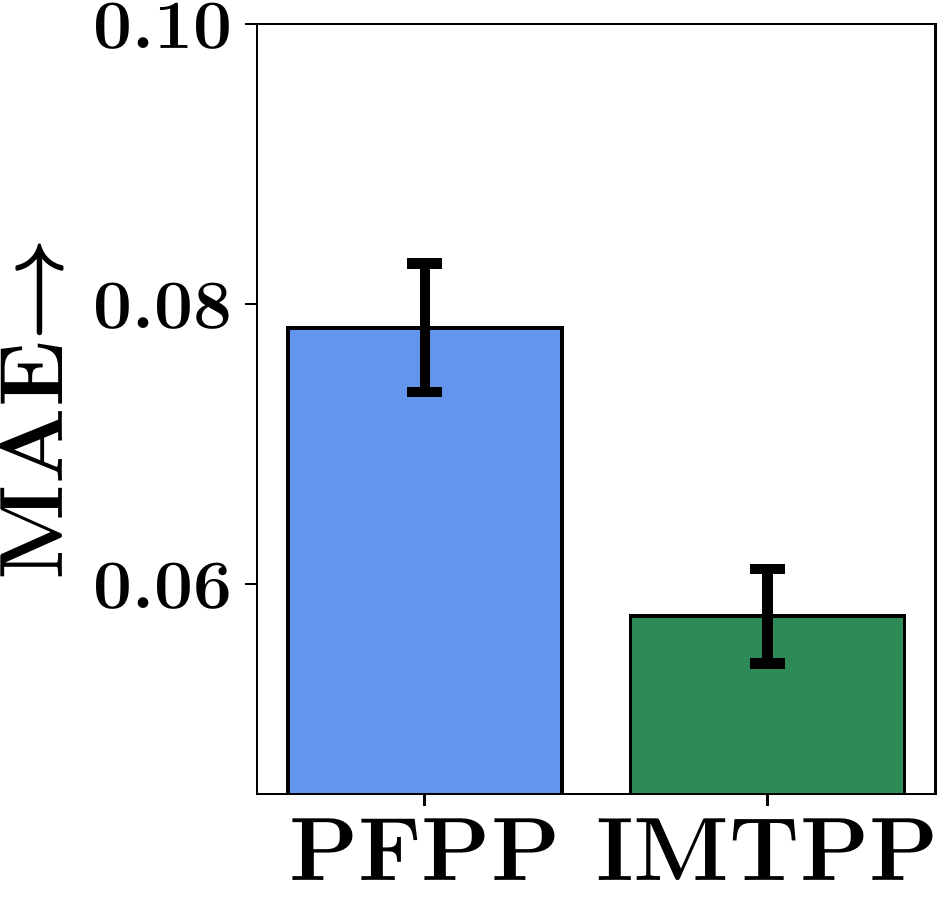}
  \caption{\amovies, MAE}
\end{subfigure}
\hfill
\begin{subfigure}{0.24\columnwidth}
  \centering
  \includegraphics[height=3.5cm]{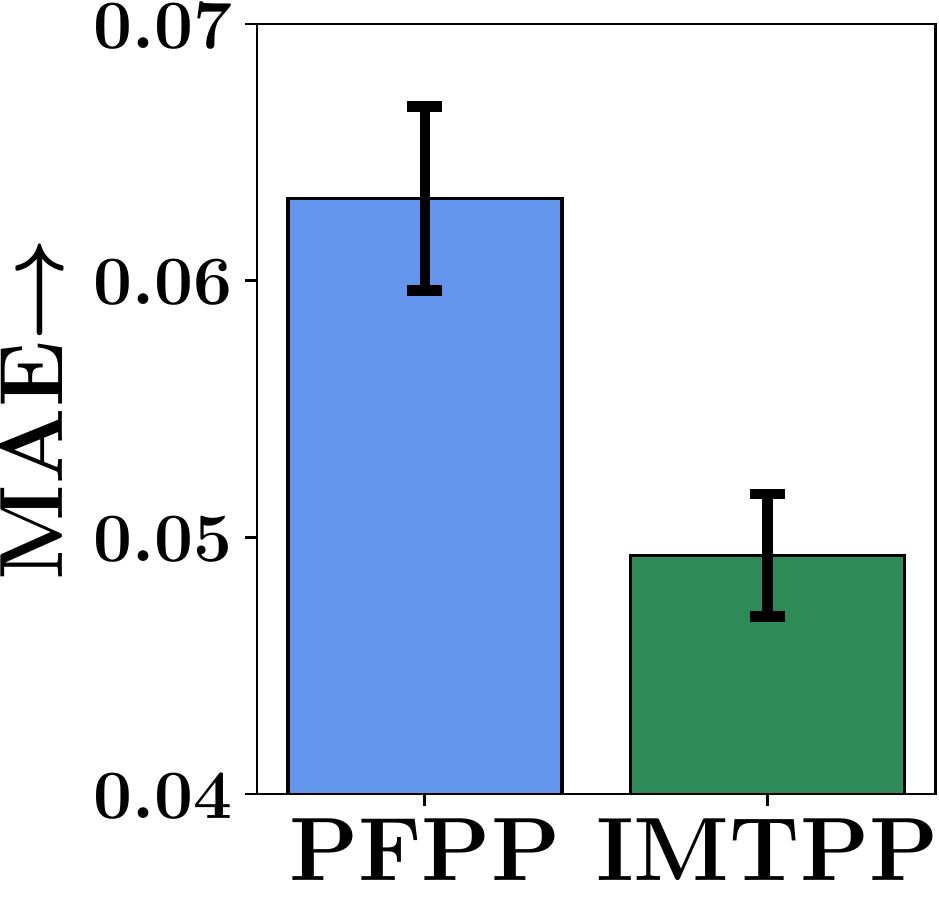}
  \caption{\atoys, MAE}
\end{subfigure}
\hfill
\begin{subfigure}{0.24\columnwidth}
  \centering
  \includegraphics[height=3.5cm]{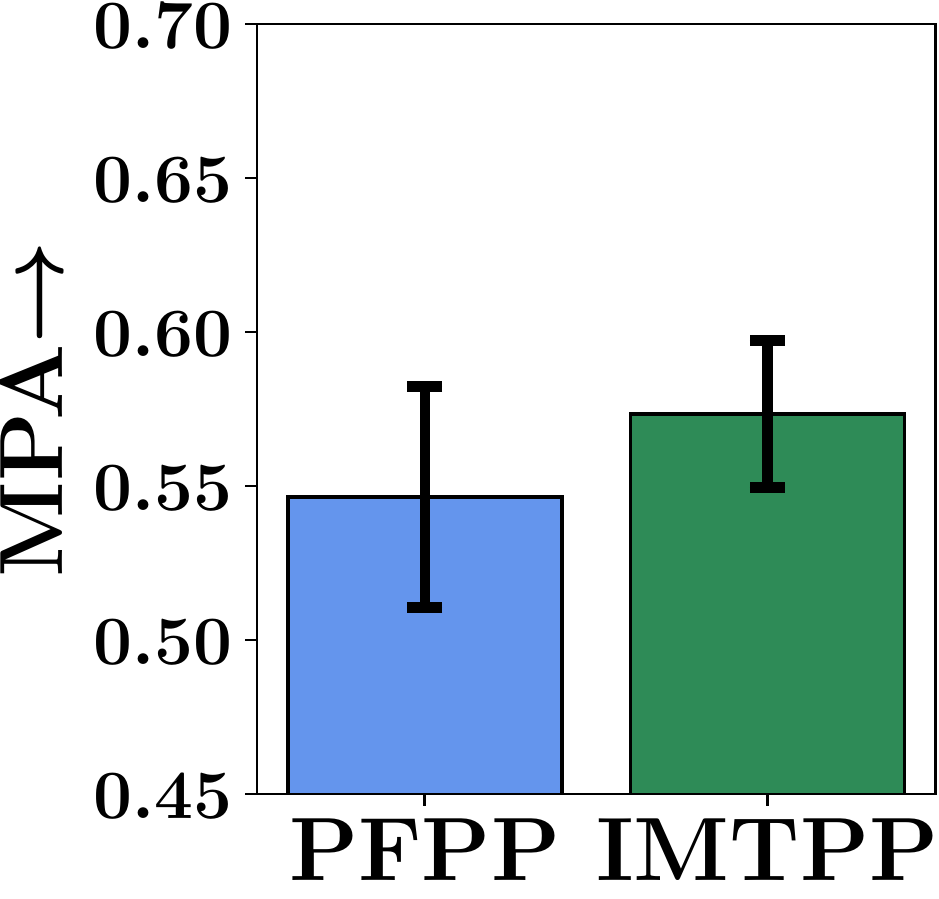}
  \caption{\amovies, MPA}
\end{subfigure}
\hfill
\begin{subfigure}{0.24\columnwidth}
  \centering
  \includegraphics[height=3.5cm]{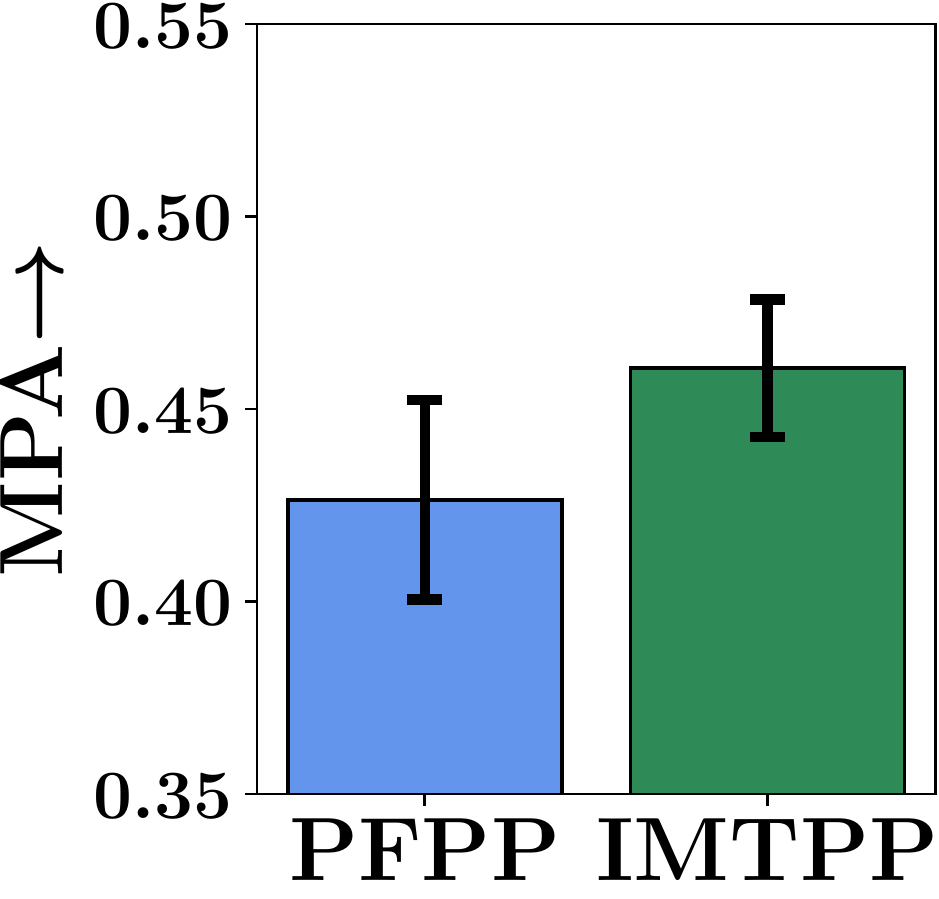}
  \caption{\atoys, MPA}
\end{subfigure}
\vspace{-3mm}
\caption{Missing event imputation performance of \imtpp and PFPP for Movies and Toys datasets. Panels (a--b) show the results for time prediction while panels (c--d) show the results for mark prediction.}
\label{imtp_fig:imp}
\vspace{-3mm}
\end{figure}

Here, we evaluate the ability of \imtpp and \mei to impute missing events in a sequence. Specifically, we evaluate the ability of both models to generate the missing events that were not present during training. Thus, for \amovies and \atoys datasets where we synthetically remove all the ratings between the first month and the third month for all the entities in both datasets. We evaluate across the imputed events for test sequences. One important thing to note is that \imtpp only takes into account the history, but \mei uses both, history and future events. We report the results across the Movies and Toys datasets in Figure~\ref{imtp_fig:imp}. To summarize, our results show that even with limited historical information, \imtpp outperforms \mei in time-prediction, whereas both models perform competitively for the mark prediction of missing events.

\begin{figure}[t]
\centering
\begin{subfigure}{0.45\columnwidth}
  \centering
  \includegraphics[height=4cm]{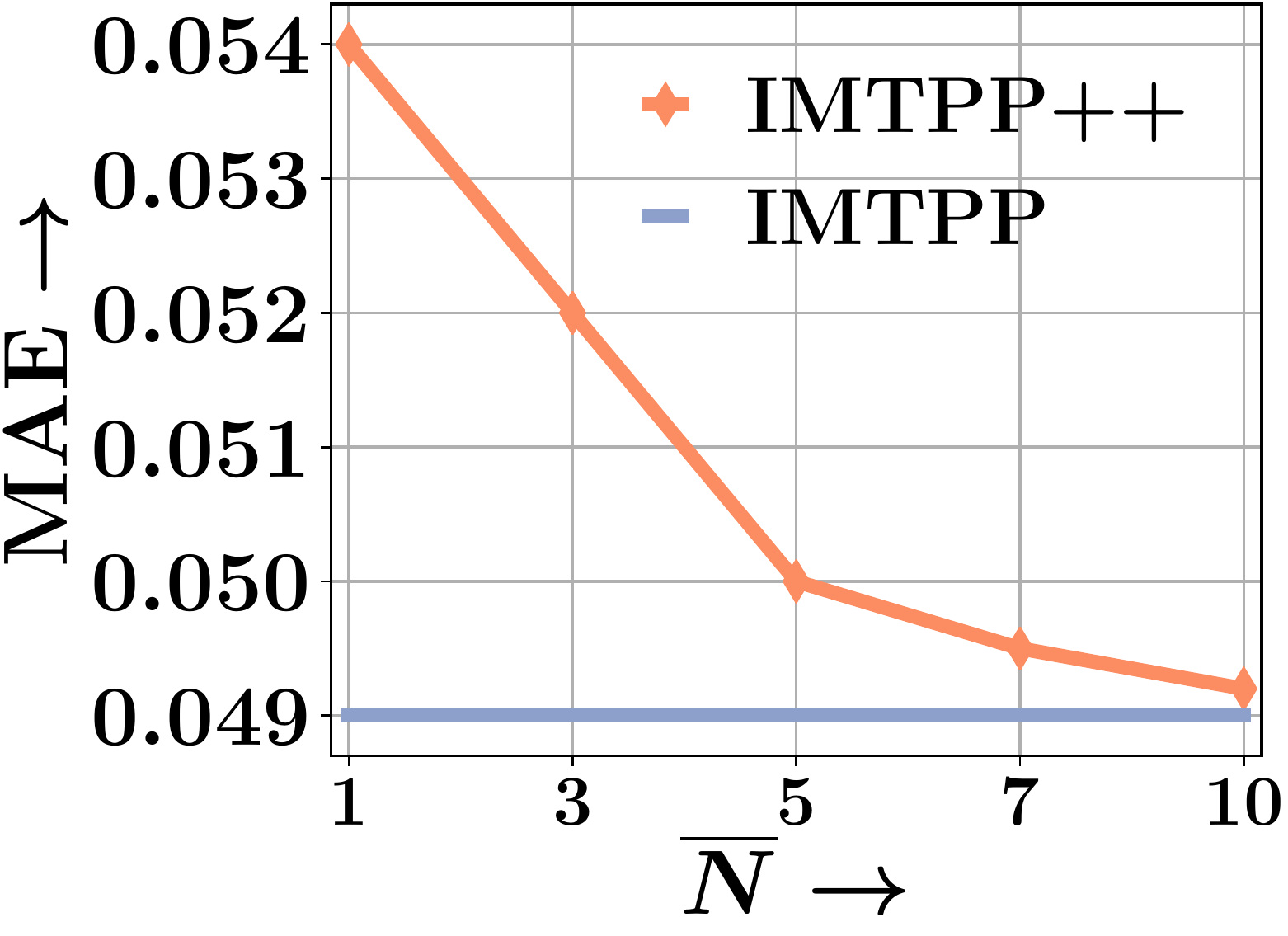}
  \caption{\amovies}
\end{subfigure}
\hspace{1cm}
\begin{subfigure}{0.45\columnwidth}
  \centering
  \includegraphics[height=4cm]{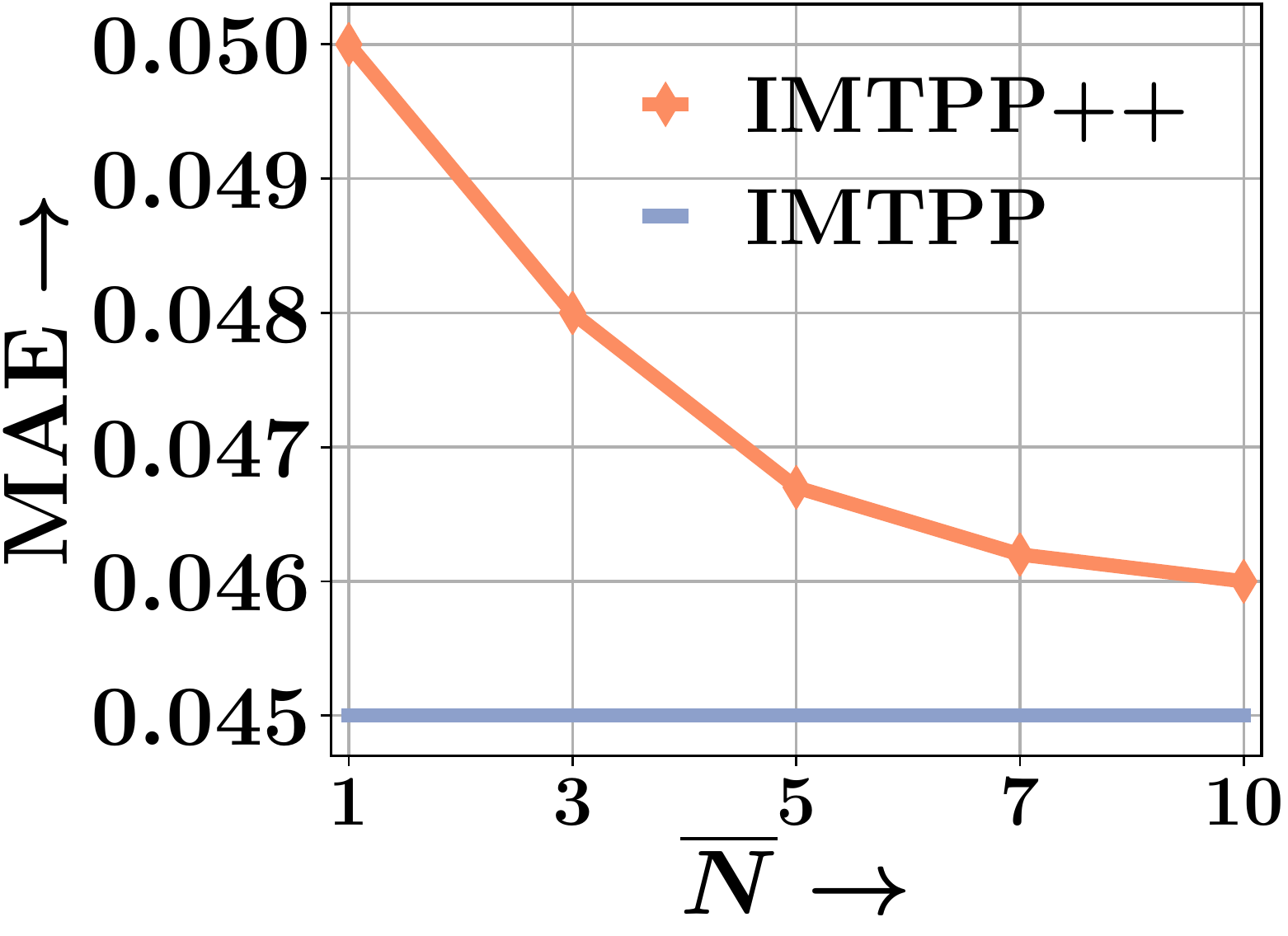}
  \caption{\atoys}
\end{subfigure}
\vspace{-2mm}
\caption{Predicting observed events using \imtppp across different values of $\overline{N}$. The results show that as we increase $\overline{N}$, the performance gap between \imtppp and \imtpp is decreased.}
\vspace{-2mm}
\label{imtp_fig:plus_pp}
\end{figure}

\begin{figure}[t]
\centering
\begin{subfigure}{0.45\columnwidth}
  \centering
  \includegraphics[height=4cm]{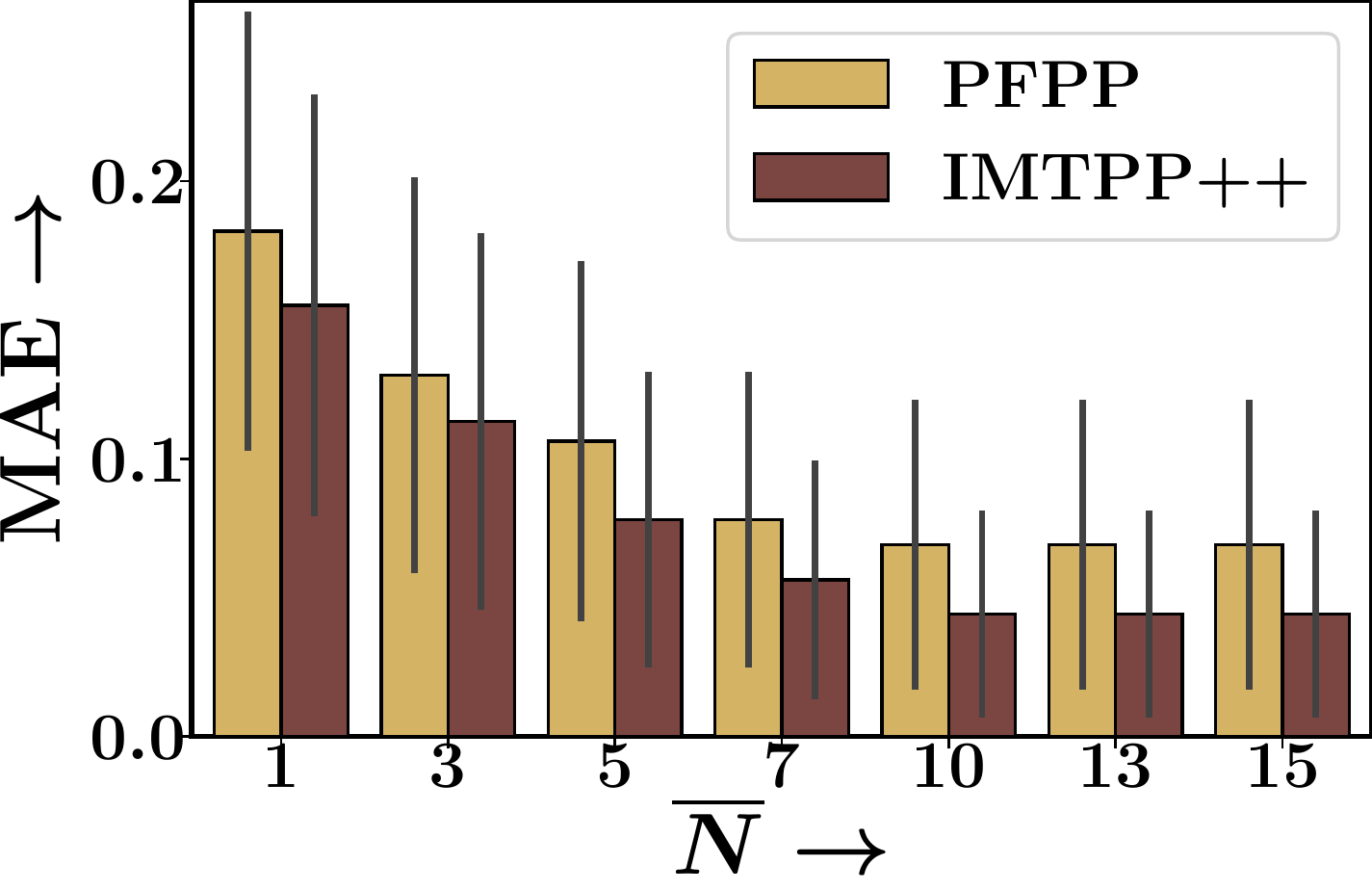}
  \caption{\amovies}
\end{subfigure}
\hspace{1cm}
\begin{subfigure}{0.45\columnwidth}
  \centering
  \includegraphics[height=4cm]{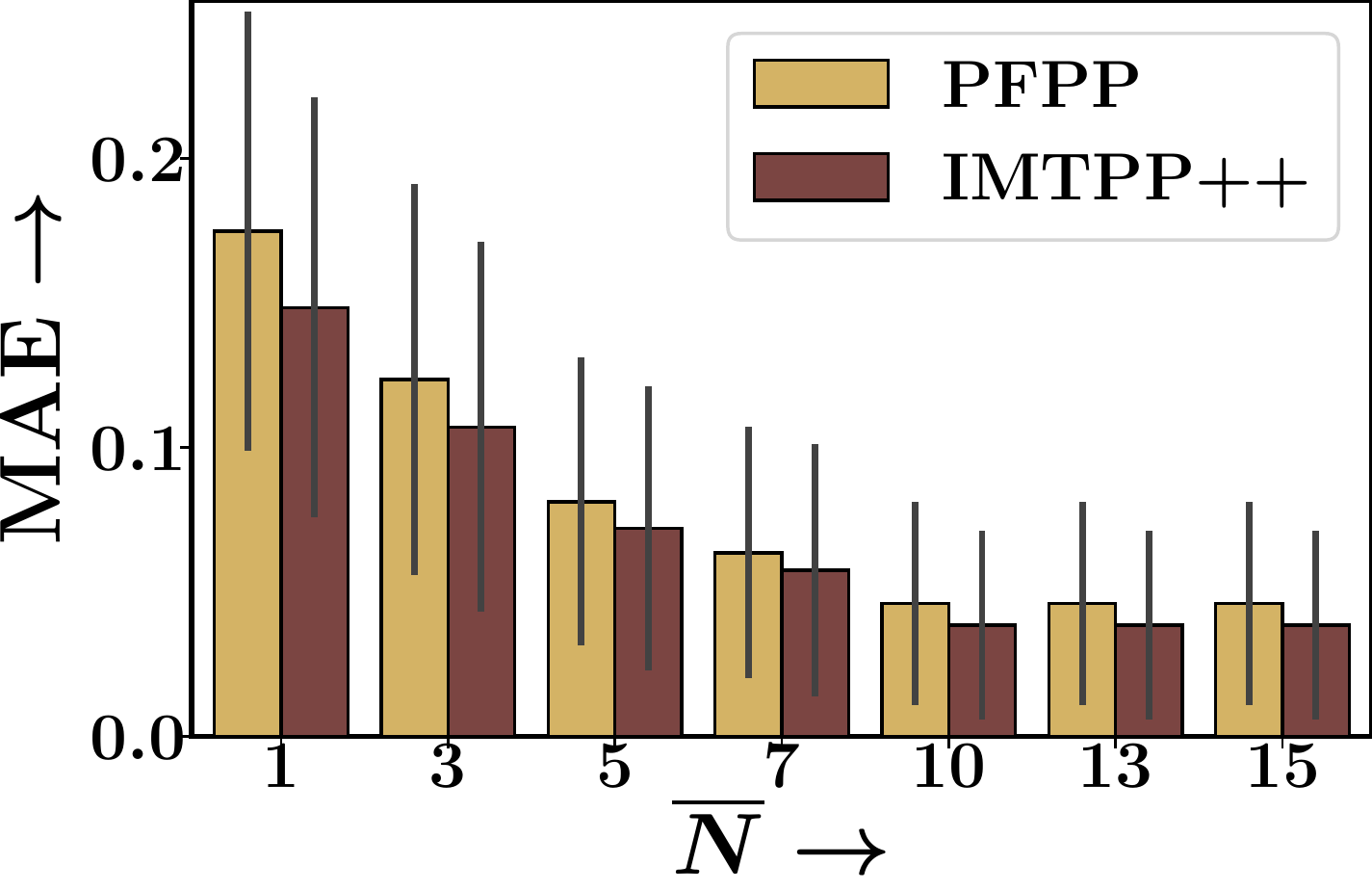}
  \caption{\atoys}
\end{subfigure}
\vspace{-2mm}
\caption{Compared the imputation performance of \imtppp and PFPP across different numbers of missing events. Note that this setting differs from Figure \ref{imtp_fig:imp}, as here the events are missing at random positions.}
\vspace{-2mm}
\label{imtp_fig:plus_pfpp}
\end{figure}

\subsection{Evaluating the Performance of \imtppp}
\xhdr{Observed Event Prediction using \imtppp} Here, we evaluate the ability of \imtppp to predict the observed events in a sequence. Specifically, we report the time prediction performance of \imtppp across a different number of permitted missing events ($\overline{N}$) and compare them with \imtpp \ie, with an unbounded number of missing events. Figure \ref{imtp_fig:plus_pp} summarizes our results which show that as we increase $\overline{N}$, the time prediction performance for \imtppp increases, and it narrows the performance gap with \imtpp. However, \imtpp still performs better than \imtppp. From the results, we conclude that \imtppp acts as a trade-off between the number of missing events and the prediction quality. This is a significant improvement over \imtpp as sampling missing events can be an expensive procedure. Moreover, as \imtppp involves fine-tuning over pre-trained \imtpp, it has an added advantage of fine-tuning at amounts of missing events. From our experiments, we found that fine-tuning \imtppp took less than 15 minutes across all values of $\overline{N}$. We also note that with small $\overline{N}$, \imtpp is comparable to RMTPP \ie, the best performer for time prediction.

\xhdr{Imputing Missing Events using \imtppp}
Our main contribution via \imtppp is to predict the missing events located randomly in a sequence. We evaluate this by performing an additional experiment using synthetic deletion. Specifically, we randomly sample $\overline{N}$ events from each sequence and tag them to be missing. Later, we evaluate the ability of \imtppp and PFPP in imputing these missing events. Figure \ref{imtp_fig:plus_pfpp} summarizes the results and we note that \imtppp easily outperforms PFPP across all values of $\overline{N}$. Moreover, we note that as we increase $\overline{N}$, the imputation performance becomes better. Naturally, it can be attributed to relatively \textit{lesser} variance in position of missing events with large $\overline{N}$. We reiterate that all confidence intervals are calculated using three independent runs.

\section{Conclusion} \label{imtp_sec:conc}
Modeling continuous-time events with irregular observations is a non-trivial task that requires learning the distribution of both -- observed and missing events. Standard MTPP models ignore this aspect and assume that the underlying data is complete with no missing events -- an ideal assumption that is not practicable in many settings. In order to solve these shortcomings, in \citet{imtpp}, we provide a method for incorporating missing events for training marked temporal point processes that simultaneously sample missing as well as observed events across continuous time. The proposed model \imtpp uses a coupled MTPP approach with its parameters optimized via variational inference. We further improve on \imtpp and propose \imtppp that has an added feature to identify the optimal position of missing events in a sequence. Experiments on several real datasets from diverse application domains show that our proposal outperforms other state-of-the-art approaches for predicting the dynamics of observed events. We also evaluate the ability of our models to impute synthetically deleted missing events within observed events. In this setting as well, our models outperform other alternatives along with better scalability and guaranteed convergence. Since including missing data, \textit{improves} over standard learning procedures, this observation opens avenues for further research that includes modeling or sampling missing data. 

%% file: chapters/005_axolotl.tex
\newcommand{\ours}{\textsc{Axo}\xspace}
\newcommand{\snorm}[1]{\big \lVert #1 \big \rVert}
\newcolumntype{a}{>{\columncolor{blue!5}}c}
\newcolumntype{b}{>{\columncolor{red!5}}c}


\section{Introduction}\label{axointro}
\begin{figure}[t!]
\centering
\begin{subfigure}[]{0.30\columnwidth}
\centering
\includegraphics[height=3.5cm]{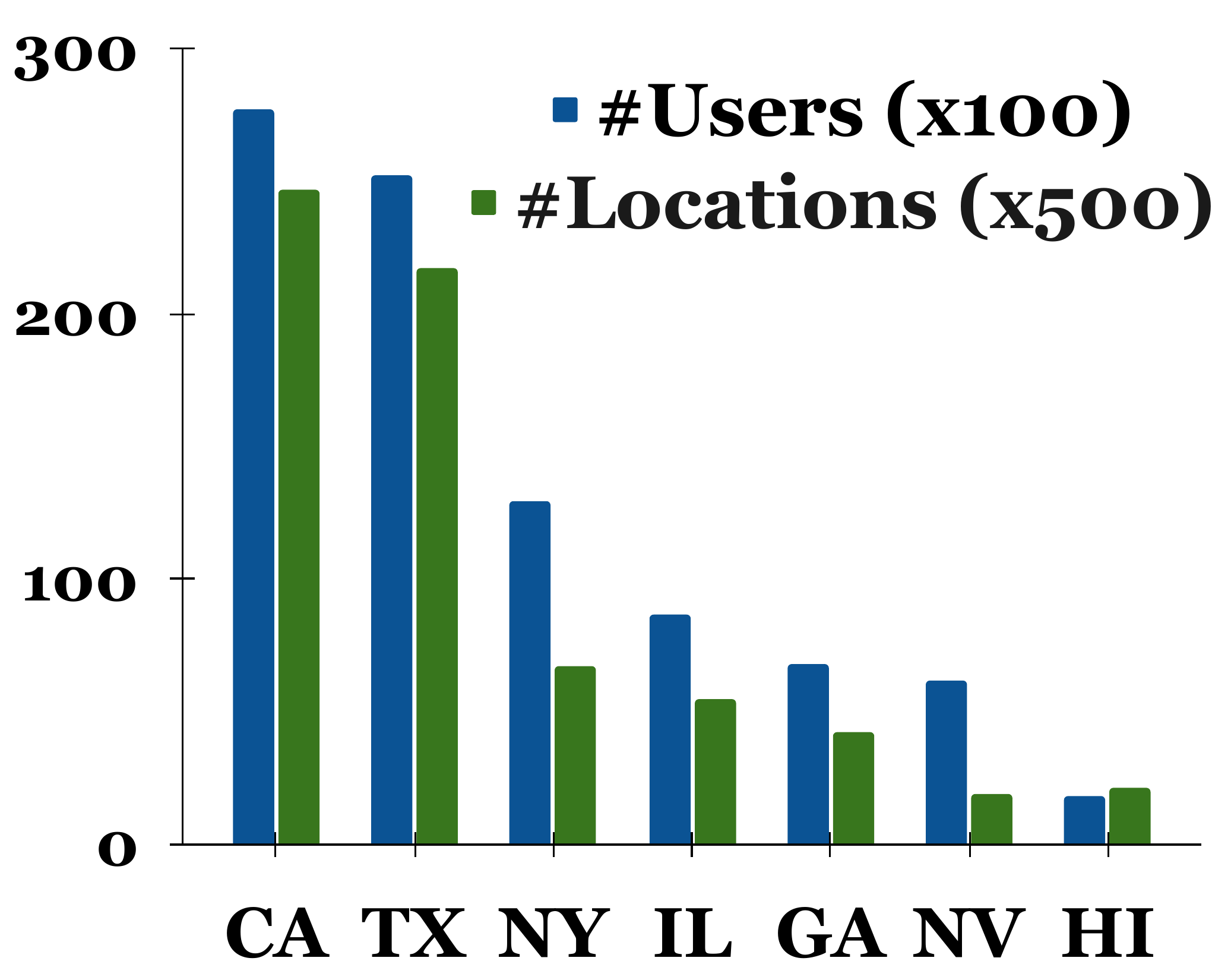}
\caption{State-wise Data}
\label{axofig:1a}
\end{subfigure}
\hfill
\begin{subfigure}[]{0.30\columnwidth}
\centering
\includegraphics[height=3.5cm]{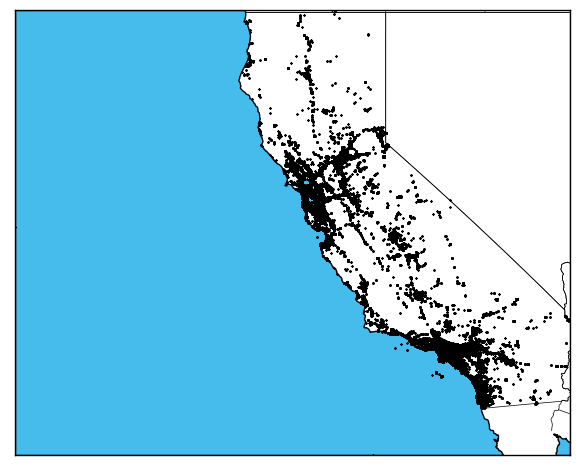}
\caption{California}
\label{axofig:1b}
\end{subfigure}
\hfill
\begin{subfigure}[]{0.30\columnwidth}
\centering
\includegraphics[height=3.5cm]{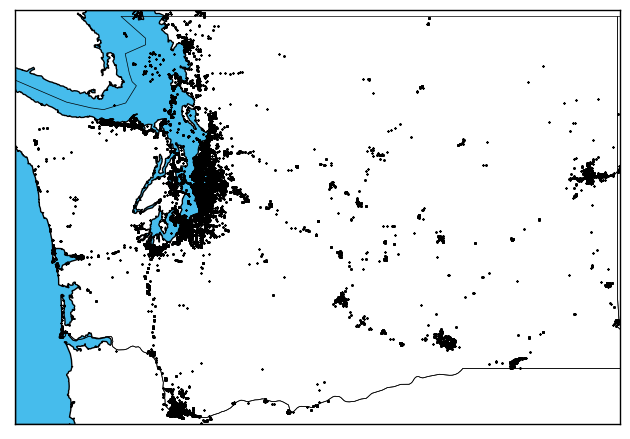}
\caption{Washington}
\label{axofig:1c}
\end{subfigure}
\caption{Skew in the volume of mobility data across different states in the US (Figure~\ref{axofig:1a}) and the large variation in region-specific density of mobility data between California and Washington in Figures ~\ref{axofig:1b} and ~\ref{axofig:1c} respectively (based on Gowalla dataset~\cite{scellato}).}
\label{axofig:distri}
\end{figure}

In this chapter, we present our solution to overcome the drawbacks of data scarcity for top-$k$ recommendations in spatial mobility networks. As POI (Points-of-Interest) gathering services such as Foursquare, Yelp, and Google Places are becoming widespread, there is significant research in extracting location preferences of users to predict their mobility behavior and recommend the next POIs that users are likely to visit~\cite{cara, locate, cheng2013you, deepmove, deepst, ruirui}. However, the quality of POI recommendations for users in regions where there is a severe scarcity of mobility data is much poorer in comparison to those from data-rich regions. This is a critical problem affecting the state-of-the-art approaches~\cite{regiontrans, www, similarwww}.
The situation is further exacerbated in recent times due to the advent of various restrictions on collecting personal data and growing awareness (in some geopolitical regions) about the need for personal privacy~\cite{privacy, privacy2}. It not only means that there is an overall reduction in the high-quality (useful) data\footnote{\scriptsize Nearly 80\% of the data generated by Foursquare users is discarded~\cite{fsq}.}, but, even more importantly, it introduces a \emph{high-skew} in the mobility data across different regions (see Figure~\ref{axofig:distri}) -- primarily due to varying views towards personal privacy across these regions.

Therefore, the current state-of-the-art methods struggle in low-data regions and the approaches that attempt to incorporate data from external sources suffer from the following limitations: 
\begin{inparaenum}[(i)]
\item limited to cold-start users from within a city~\cite{onlycity, onlycity2, ruiruiwww}, \item focused only on using traffic network ignoring the use of social network of users and location dynamics~\cite{regiontrans, www, deepst}, \item generate trajectories using a model learned on traffic-network images of source region~\cite{similarwww, privacy3}, thus vulnerable to recalibration on noisy images, \item operating only for users who are \emph{common across} locations~\cite{mamo, compare}, or \item adopting a limited level of transfer through domain-invariant features~\cite{adit} --like the spending capacity or the users' age, thus constrained by the feature unavailability in public datasets.  \end{inparaenum}
Unfortunately, none of these approaches, especially those based on visual data (\ie, traffic and location images), can be easily fine-tuned for a target POI network due to the varied spatial density, location category, and lack of user-specific features in POI datasets. 

\subsection{Our Contribution}
In this chapter, we present {\bfseries \axolotl}(\textbf{A}utomated \textit{cross} \textbf{Lo}cation-network \textbf{T}ransfer \textbf{L}earning), a novel meta learning-based approach for POI recommendation in limited-data regions while transferring model parameters learned at a data-rich region \emph{without any prerequisite of inter-region user overlap}~\cite{axolotl}. Specifically, we use a hierarchical multi-channel learning procedure with a novel meta-learning~\cite{maml, melu} extension for spatial mobility networks, called {\bf \textit{spatio-social} meta-learning (SSML)}, that learns the model parameters by jointly minimizing the region-specific social as well as location recommendation losses, and a cross-region transfer via \emph{clusters}~\cite{www, regiontrans} of user and locations with similar preferences by minimizing an alignment loss~\cite{what2tran, moreattn}, to achieve high performance even in extremely limited-data regions.
We represent the POI network of each region via a heterogeneous graph with users and locations as nodes and capture the user-location inter-dependence and their neighborhood structure via a \emph{twin graph attention} model~\cite{danser, dualgcn}.
The graph-attention model aggregates all four aspects of user and location influences~\cite{socialgcn, eljp} namely, (i) users' social neighborhood and their location affinity to construct user-specific and location-conditioned representations, and (ii) similarly for each location, its neighboring locations and associated user affinity to obtain a \emph{locality-specific} and \emph{user-conditioned} representations. We combine this multi-faceted information to determine the \emph{final} user and location representations that are used for POI recommendation. We highlight the region-size invariant performance of \axolotl by using regions with different spatial granularities, \ie, \textit{states} for Germany and US, respectively, and \textit{prefectures} for Japan. 

In summary, our key contributions via \axolotl are three-fold:
\begin{compactitem}
\item \textbf{Region-wise Transfer:} We address the problems associated with POI recommendation in limited data regions and propose \axolotl, a cross-region model transfer approach for POI recommendation that does not require common users and their traces across regions. It utilizes a novel \textit{spatio-social} meta-learning-based transfer and minimizes the divergence between user-location clusters with similar characteristics.

\item \textbf{User-Location Influences:} Our twin graph attention-based model combines user and location influences in heterogeneous mobility graphs. 
This is the first approach to combine these aspects for addressing the data scarcity problem with POI recommendations. \axolotl is robust to network size, trajectory spread, and check-in category variance making it more suitable for transfer across geographically distant regions (and even across different networks). 

\item \textbf{Detailed Empirical Evaluation:}  We conduct thorough experiments over 12 real-world points-of-interest datasets from the US, Japan, and Germany, at different region-wise granularity. They highlight the superior recommendation performance of \axolotl over state-of-the-art methods across all metrics. 
\end{compactitem}

\section{Related Work}
\label{axosec:relwork}
In this section, we introduce key related works. They mainly falls into the following categories: 1) Mobility Prediction; 2) Graph-based Recommendations; 3) Clustering in Spatial Datasets; and 4) Transfer Learning and Mobility.

\subsection{Mobility Prediction} 
Understanding the mobility dynamics of a user is widely studied using different data sources~\cite{zheng, cho}. Early efforts relied on taxi datasets to study individual trajectories~\cite{dcrnn, traffic}. However, these approaches are limited by the underlying datasets as it excludes two critical aspects of a mobility network; social friendships and location categories. The social network is used to model the influence dynamics across different users~\cite{locate, cara}, and the POI categories capture the different preferences of an individual~\cite{colab, cheng2013you}. We utilize user POI social networks for our model as these datasets provide both: a series of social dynamics for different users and location-specific interest patterns for a user. These are essential for tasks such as location-specific advertisements and personalized recommendations. Standard POI models that utilize an RNN~\cite{ganguly, cara, zheng, rnnlbsn, deepmove} or a temporal point process~\cite{imtpp, colab, reformd} are prone to making mistakes due to irregularities in the trajectories. These irregularities arise due to uneven data distributions, missing check-ins, and social links. Moreover, these approaches consider the check-in trajectory for each user as a sequence of events and thus have limited power to capture the user-location inter-dependence through their spatial neighborhood, \ie, the location-sensitive information that influences all neighborhood events. Some of the recent approaches~\cite{regiontrans, www}, harness the spatial characteristics by generating an image corresponding to each user trajectory and then utilizing a CNN as an underlying model. Such approaches based on visual data, and all CNN-based approaches, are limited by the image characteristics such as its resolution and scale of capture. Modern POI recommendation approaches such as \cite{lbsn2vec} model the spatial network as a graph and utilize a random-walk-based model, with~\cite{ijcailbsn} proposing a graph-based neural network-based model to incorporate structural information of the network. Unfortunately, none of these approaches are designed for mobility prediction in limited data regions.

\subsection{Graph based Recommendation}
Existing graph embedding approaches focus on incorporating the node neighborhood proximity in a classical graph in their embedding learning process~\cite{node2vec, lbsn2vec}. For \eg,~\cite{birank} adopts a label propagation mechanism to capture the inter-node influence and hence the collaborative filtering effect. Later, it determines the most probable purchases for a user via her interacted items based on the structural similarity between the historical purchases and the new target item. However, these approaches perform inferior to model-based CF methods since they do not optimize a recommendation-specific loss function.
The recently proposed graph convolutional networks (GCNs)~\cite{gcn} have shown significant promise for recommendation tasks in user-item graphs. The attention-based variant of GCNs, graph attention networks (GATs)~\cite{gat} are used for recommender systems in information networks~\cite{socialgcn, ngcf}, traffic networks~\cite{traffic, dcrnn} and social networks~\cite{pinsage, yfumob}. Furthermore, the heterogeneous nature of these information networks comprises of multi-faceted influences that led to approaches with \textit{dual}-GCNs across both user and item domains~\cite{dualgcn, socialgcn}. However, these models cannot be generalized for spatial graphs due to the disparate weights, location-category as node feature, and varied sizes.

\subsection{Clustering in Spatial Datasets}
Due to the disparate features in our spatial graph, identifying the optimal number of clusters for the source and target region is a challenging task. Thus, we highlight a few key related works for clustering POIs and users in a spatial graph. Standard community-detection algorithms for spatial datasets~\cite{colab, com1, com2, com3} are not suitable for grouping POIs as they ignore the graph structure, POI-specific features such as categories, geographical distances, and the order of check-ins in a user trajectory. 

Recent approaches~\cite{morris2019weisfeiler, sagpool, diffpool, abu2019mixhop, gunet} can automatically identify the number of clusters in a graph by capturing higher-order semantics between graph nodes, however, their application to graphs in spatial and mobility domains has certain challenges. In detail, (i) \citet{diffpool} can learn differentiable clusters for POIs and users for each region, however, these assignments are \textit{soft}, \ie, without definite boundaries between distant POI clusters and, moreover, have a quadratic storage complexity; (ii)~\citet{gunet} ignores the topology of the underlying spatial graph; (iii)~\citet{sagpool} can be extended to spatial graphs, however, has limited scalability due to its self-attention~\cite{transformer} based procedure; (iv)~\citet{morris2019weisfeiler} can incorporate higher-order structure in a POI graph using multi-dimensional Weisfeiler-Leman graph isomorphism; and (v)~\citet{abu2019mixhop} can learn inter-user and inter-POI relationships by mixing feature representations of neighbors at various distances. However, due to the presence of two types of graph nodes -- user and POI -- identifying higher-order relationships by solely considering POI or user nodes is challenging. In addition, ~\citet{spat} uses a differentiable grouping network to discover the latent dependencies in a spatial network but is limited to air-quality forecasting. 

\subsection{Transfer Learning and Mobility}
Transfer learning has long been addressed for tasks involving sparse data~\cite{what2tran, maml} with applications to recommender systems as well~\cite{manasi, melu, compare}. Transfer-based spatial applications deploy CNNs across regions and achieve significant improvements in limited data-settings~\cite{regiontrans, www}. However, these approaches are restricted to non-structural data and a graph-based approach has not been explored by the previous literature. \cite{ruiruiwww} extends meta-learning to enhance recommendations in a POI setting, but is limited to a specific region. Information transfer across graphs is not a trivial task~\cite{transgraph, kgmaml, adit}, and recent mobility models that incorporate graphs with meta-learning in~\cite{deepst, ltran} are either limited to traffic datasets and do not incorporate the social network or are limited to new trajectories~\cite{similarwww, privacy3}. From our experiments, we prove that a simple fine-tuning of the target data is susceptible to large cross-data variances, and thus, fine-tuning a generative model is not a trivial task in mobility-based networks.

\section{Problem Formulation} \label{axosec:problem}
We consider POI data for two regions, a \textit{source} and a \textit{target} denoted by $\ds{D}^{src}$ and $\ds{D}^{tgt}$. We denote the users and locations in source and target networks as $\cm{U}^{src}, \cm{P}^{src} \in \ds{D}^{src}$ and $\cm{U}^{tgt}, \cm{P}^{tgt} \in \ds{D}^{tgt}$ correspondingly with no common entries $\cm{U}^{src} \cap \cm{U}^{tgt} = \cm{P}^{src} \cap \cm{P}^{tgt} = \varnothing$. In other words, we do not need a common user between two regions to perform a cross-region mobility knowledge transfer. With a slight abuse of notation, the network for any region --either target or source-- is assumed to consist of users $|\cm{U}| = M$, locations $|\cm{P}| = N$ and an affinity matrix $\bs{R}=\{r\}_{M \times N}$. We populate entries in $\bs{R}$ as \textit{row-normalized} number of check-ins made by a user to a location (\ie, multiple check-ins mean higher value). This can be further weighed by the user-location ratings, if available. We denote a pair of users as $u_i, u_j$ and locations as $l_a, l_b$. We also assume that for each location $l_a$ we have one (or more) category label (such as Jazz Club, Cafe, \etc).

\begin{problem*}[\textbf{Target Region POI Recommendation}]
\label{axocheckin}
\textit{Given the mobility data of source and target regions, $\ds{D}^{src}$ and $\ds{D}^{tgt}$, our aim is to transfer the rich dynamics in the source region to improve POI recommendation for users in the target region. Specifically, maximize the following probability:
\begin{equation}
P^* = \arg \max \{ \mathbb{E}[r_{u_i, l_a}^{tgt}| \ds{D}^{src}, \ds{D}^{tgt} ] \}, \quad u_i \in \cm{U}^{tgt}, l_a \in \cm{P}^{tgt},
\end{equation}
where $\mathbb{E}[r^{tgt}_{u,l}]$ calculates the expectation of location $l_a$ in the target region being visited by the user $u_i$, thus $r_{u_i, l_a}^{tgt} \in \bs{R}^{tgt}$, given the mobility data of users from both source and target regions. Simultaneously for a region, our objective as personalized \emph{location} recommendation is to retrieve, for each user, a ranked list of candidate locations that are most likely to be visited by her based on the past check-ins available in the training set.}
\end{problem*}

\subsection{User-Location Graph Construction}
For each region, we construct a heterogeneous user-location graph $\cm{G}^{src}$ and $\cm{G}^{tgt}$ but we describe generically as $\cm{G}=\{\cm{U}\cup\cm{P}, \cm{E}\}$ with each user and location as a node. The disparate edges $\cm{E}_u, \cm{E}_l, \cm{E}_r \in \cm{E}$ determine the user-user, location-location and user-location relationships respectively. The structure of the graph is as follows: 
\begin{compactitem} 
\item An edge, $e_{u_i,u_j} \in \cm{E}_u$, between two users, $u_i$ and $u_j$ is denotes a social network friendship.
\item We form an edge, $e_{l_a, l_b} \in \cm{E}_l$, between two locations when any user has \emph{consecutive check-ins} between them -- \ie, a check in at $l_a$ (or $l_b$) followed immediately by $l_b$ (or $l_a$) with edge weight based on the \emph{geographical distance}~\cite{dcrnn, yfumob} between the two locations. Specifically, we use non-linear decay with distance as:
\begin{equation*}
w (e_{l_a, l_b}) = 
\begin{cases}
\exp \left(-\frac{d(l_a, l_b)}{\sigma^2}\right), & \mathrm{if~} d(l_a, l_b) \le \kappa, \\
0 & \mathrm{otherwise},
\end{cases}
\end{equation*}
where $d(l_a, l_b)$ is the \emph{haversine}~\cite{haversine} distance between the two locations.
\item A check-in by user $u_i$ at location $l_a$ results in a user-location edge, $e_{u_i,l_a} \in \cm{E}_r$.
\end{compactitem}

\noindent We also use the following notations to define different neighborhoods that will be used in our model description later: 
\begin{compactitem}
\item{$\mathcal{N}_{u_i} = \left\{u_k : e_{u_i,u_k} \in \mathcal{E}_u\right\}$: the social neighborhood of user $u_i$,}
\item{$\mathcal{N}_{l_a} = \left\{u_k : e_{u_k,l_a} \in \mathcal{E}_r\right\}$: the user neighborhood of location $l_a$,}
\item{$\mathcal{S}_{u_i} = \left\{l_k : e_{u_i, l_k} \in \mathcal{E}_r\right\}$: the location neighborhood of user $u_i$, and, finally,}
\item{$\mathcal{S}_{l_a} = \left\{l_k : e_{l_a, l_k} \in \mathcal{E}_l\right\}$: locations in the spatial vicinity of $l_a$.}
\end{compactitem}
Note that the graph $\mathcal{G}$ can also be enriched with all edges as weighted~\cite{ijcailbsn} conditioned on the availability of different features. However, we present a general framework that can be easily extended to such settings. 

\begin{table}[t!]
\small
\caption{\label{axotab:par} Summary of Notations Used.}
\vspace{-0.4cm}
\centering
\begin{tabular}{ll}
\toprule
\textbf{Notation} & \textbf{Description}\\
\midrule
$\ds{D}^{src}, \ds{D}^{tgt}$ & Datasets for source and target regions. \\
$\cm{U}^{src}, \cm{P}^{src}$ & Users and POIs for the source region. \\
$\cm{U}^{tgt}, \cm{P}^{tgt}$ & Users and POIs for the target region. \\
$\bs{R}$ & Normalized user to location preference matrix. \\
$\bs{U}_l, \bs{U}_s, \bs{U}_f$ & User's latent, location-based and final representations. \\
$\bs{L}_l, \bs{L}_s, \bs{L}_f$ & Location's latent, user-based and final representations. \\
$\Phi_{1\cdots4}$ & Graph Attention Networks in \axolotl. \\
$\psi_{1\cdots4}$ & MLP layers to calculate attention weights for $\Phi_{1\cdots4}$. \\
$\varphi_{1\cdots3}$ & MLPs for final user and location embeddings, and affinity prediction. \\
$\cm{L}_{p}, \cm{L}_{s}, \cm{L}_{c}$ & Affinity, social prediction and Cluster-based loss. \\
$K$ & Number of clusters in source and the target region. \\
$N_u$ & Number of target region-based updates for SSML. \\
$M_t$ & Iterations before cluster transfer. \\
$\bs{U}^{tgt}_c, \bs{L}^{tgt}_c$ & User and location cluster embeddings for target region. \\
$\bs{U}^{src}_c, \bs{L}^{src}_c$ & User and location cluster embeddings for source region. \\
\bottomrule
\end{tabular}
\end{table}

\section{\axolotl Framework}\label{axosec:framework}
In this section, we first describe in detail the basic model of \axolotl along with its training procedure. Then we present the key feature of \axolotl, {\it viz.}, its ability to transfer model parameters learned from a data-rich region to a data-scarce region. For a specific region, we embed all users, $\cm{U}$, and all locations, $\cm{P}$, through matrices $\bs{U}=\{\bs{u}_{i}\}_{M \times D}$ and $\bs{L}=\{\bs{l}_{a}\}_{N \times D}$ respectively with $D$ as the embedding dimension. A summary of all notations is given in Table~\ref{axotab:par}.

\subsection{Basic Model}
\begin{figure}[t] 
\centering
  \includegraphics[width=0.6\linewidth]{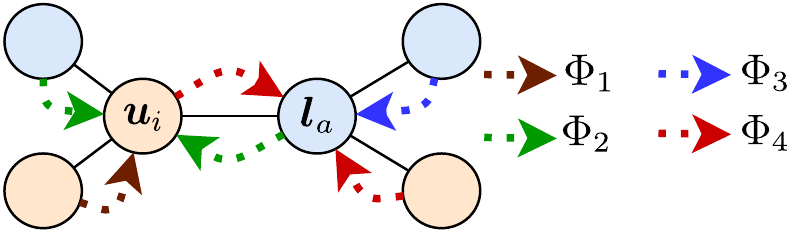}
  \vspace{-4mm}
  \caption{Different node and edge types in our graph model (users are in yellow and locations are in blue). The dashed arrows represent various influences and corresponding GATs ($\Phi_i, i=1,2,3,4$) in \axolotl.}
\label{axofig:social}
\vspace{-2mm}
\end{figure}

\begin{figure}[t!] 
\centering
  \includegraphics[height=7cm]{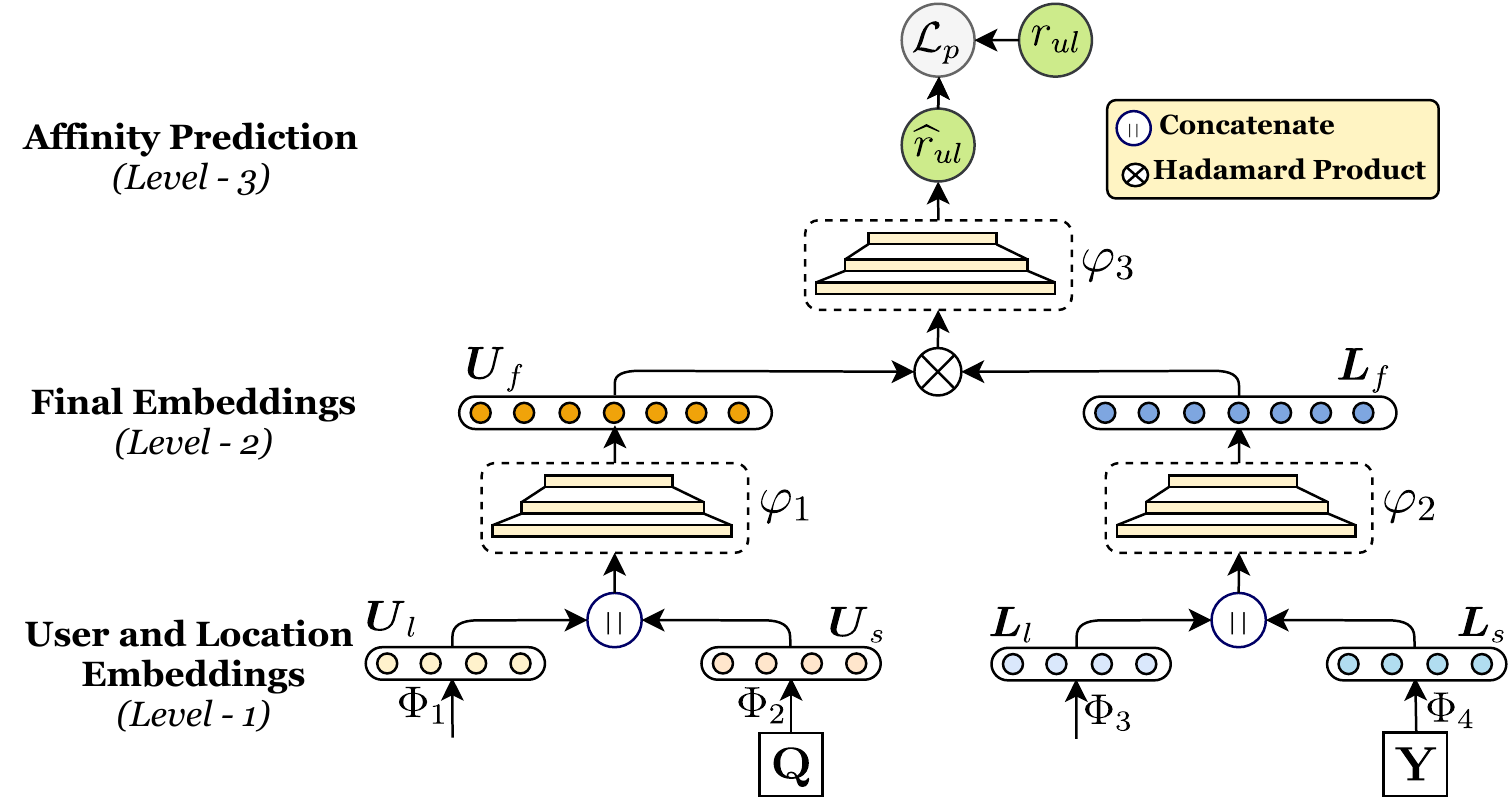}
\vspace{-4mm}  
  \caption{System architecture of \ours-basic with level-wise embedding computation and affinity prediction. User-latent ($\bs{U}_l$) and location conditioned ($\bs{U}_s$) embeddings are combined to a \textit{final} user embedding ($\bs{U}_f$), and similarly for location, $\bs{L}_l$ and $\bs{L}_s$ are combined to $\bs{L}_f$.}
\label{axofig:arch}
\vspace{-2mm}
\end{figure}

In the graph model of \axolotl, we capture the four aspects of influence propagation --namely, user-latent embeddings ($\bs{U}_l \in \mathbb{R}^{M \times D}$), location-conditioned user embeddings ($\bs{U}_s\in \mathbb{R}^{M \times D}$), location-latent embeddings ($\bs{L}_l \in \mathbb{R}^{N \times D}$), and user-conditioned location embeddings ($\bs{L}_s \in \mathbb{R}^{N \times D}$)-- illustrated in Figure ~\ref{axofig:social}. The basic model of \axolotl captures these four aspects using different graph attention networks, resulting in a twin-graph architecture as shown in  Figure~\ref{axofig:arch}. In the rest of this section, we first describe each of the graph attention components and the prediction model in the basic model of \axolotl. Subsequently, we delineate the information transfer component that operates over this basic model.

\noindent\textbf{GAT for User Latent Embedding($\Phi_1$):}
In each iteration, using the available user embeddings, $\bs{U}$, we aggregate each user's social neighborhood to obtain a new \emph{latent} representation of the user. 
We denote this embedding as $\bs{U}_l$ and calculate as follows:
\begin{equation} 
\bs{u}_{l, i} = \sigma  \bigg( \sum_{u_k \in \cm{N}_{u_i}} \bs{\alpha}^{\Phi_1}_{u_i, u_k}\left( \bm{W}_{\Phi_1} \bs{u}_k + \bm{b}_{\Phi_1} \right)  \bigg), \, \bs{u}_{l,i}\in \bs{U}_l,
\end{equation}
where $ u_i \in \cm{U}, \cm{N}_{u_i},\sigma, \bm{W}_{\Phi_1}$ and $\bm{b}_{\Phi_1}$ are the user node ${u}_i$, nodes in the social neighborhood of ${u}_i$, the activation function, the weight matrix and the bias vector respectively. $\bs{\alpha}^{\Phi_1}_{u_i, u_k}$ determines the attention weights between user embeddings $\bs{u}_k$ and $\bs{u}_i$  given by:
\begin{equation}
\label{axoeqn:alpha}
\bs{\alpha}^{\Phi_1}_{u_i, u_k} = \frac{\exp \big(\psi_1(\bs{u}_i, \bs{u}_k) \big)}{\sum_{u_j \in \cm{N}_{u_i}} \exp \big(\psi_1(\bs{u}_i, \bs{u}_j) \big)},
\end{equation}
where, $\psi_1(\bs{u}_i, \bs{u}_j) = \mathtt{LeakyReLU} (\bm{G}_{\Phi_1} \otimes(\bs{u}_i \mathbin\Vert \bs{u}_j))$ calculates the inter-user attention weights with learnable parameter $\bm{G}_{\Phi_1}$.

\noindent\textbf{GAT for Location-conditioned User Embeddings ($\Phi_2$):} To encapsulate the influence on a user based on her \cins as well as those by her social neighborhood, our embeddings must include location information for all check-ins made by different users in her social proximity. For this purpose, we first need to aggregate the location embeddings for every check-in made by a user ($u_i$) as her location-based embedding, $\bs{Q} = \{ \bs{q}_i \}_{M \times D}$. We note that the category of a check-in location is arguably the root cause for a user to visit the location, and thus, to capture the location-specific category and the user-category affinity in these embeddings, we populate $\bs{Q}$ using a max-pooling aggregator across each location embedding weighted by the probability of a category to be in a user's check-in locations. That is, 

\begin{equation}
\bs{q}_{i} = \mathtt{MaxPool} \bigg [ \sum_{l_k \in \cm{S}_{u_i}} \mathrm{p}^{u_i}_{(l_k)} \cdot \bs{l}_k \bigg], \quad \forall \bs{q}_{i} \in \bs{Q},
\end{equation}
where $\cm{S}_{u_i}$, $\mathrm{p}^{u_i}_{(l_k)}$ respectively denote the location neighborhood of $u_i$ and the probability of a POI-category to be present in the past check-ins of $u_i$. We calculate $\mathrm{p}^{u_i}_{(l_k)}$ as the fraction of check-ins made by the user to POIs of the specific category with the total number of her check-ins. Mathematically,
\begin{equation}
\mathrm{p}^{u_i}_{(l_k)} = \frac{\text{Number of check-ins by $u_i$ at POIs with category same as $l_k$}}{\text{Total number of check-ins by $u_i$}},
\end{equation}
We calculate these probabilities for every POI category and these values are user-specific. Moreover, the values of $\mathrm{p}^{u_i}_{(l_k)}$ can be considered as the explicit category preferences of a user $u_i$. Later, to get the influence of neighborhood locations on a user, we aggregate the location-based neighbor embeddings $\bs{Q}$ for each user (${u}_i$) based on her social network as:
\begin{equation}
\bs{u}_{s,i} = \sigma \bigg( \sum_{u_k \in \cm{N}_{u_i}} \bs{\alpha}^{\Phi_2}_{u_i, u_k}\left( \bm{W}_{\Phi_2} \bs{q}_k + \bm{b}_{\Phi_2}\right) \bigg), \, \bs{u}_{s, i}\in \bs{U}_s,
\end{equation}
where $\bs{\alpha}^{\Phi_2}_{u_i, u_k}$ is the attention weight for quantifying the influence a user has on another through its check-ins and is formulated using $\psi_2$ similar to $\psi_1$ (Eqn ~\ref{axoeqn:alpha}).
The resulting embedding $\bs{U}_s$ is the location-conditioned user embedding.

\noindent\textbf{GAT for Location Latent Embedding ($\Phi_3$):} Similar to $\Phi_1$, we aggregate the neighborhood of each location, $\bs{l}_a \in \bs{L}$, to get a \textit{latent} representation of each location. To factor the inter-location edge-weight in our embeddings, we sample locations from the neighborhood with probability proportional to $w(e_{l_a, l_b})$, \ie, the closer the locations higher their repetitive sampling. These sampled locations represent the vicinity of the check-in and we encapsulate them to get the latent location representation.
\begin{equation}
\bs{l}_{l, a} = \sigma \bigg( \sum_{l_k \in \cm{S}_{l_a}} \bs{\alpha}^{\Phi_3}_{l_a, l_k}\left( \bm{W}_{\Phi_3} \bs{l}_k + \bm{b}_{\Phi_3} \right) \bigg), \, \forall l_a \in \cm{P}, \bs{l}_{l,a}\in \bs{L}_l,
\end{equation}
where $\bs{\alpha}^{\Phi_3}_{l_a, l_k}$ is again formulated as in Equation \ref{axoeqn:alpha}, using $\psi_3(\bs{l}_a, \bs{l}_k)$.

\noindent\textbf{GAT for User-conditioned Location Embedding($\Phi_4$):} Similar to $\Phi_2$, we need to capture the influence of different users with check-ins nearby to the current location $l_a$. Thus we use a \emph{max-pool} aggregator to capture the user neighborhood of each location weighted by its affinity towards the location category. Through this, we aim to encapsulate the locality-specific user preferences, \ie, the \textit{counter}-influence of $\bs{Q}$ in $\Phi_2$, denoted as $\bs{Y} = \{\bs{y}_a\}_{N \times D}$. 
\begin{equation}
\bs{y}_{a} = \mathtt{MaxPool} \bigg[ \sum_{{u}_k \in \cm{N}_{l_a}} \mathrm{p}^{l_a}_{(u_k)} \cdot \bs{u}_k \bigg], \quad \forall \bs{y}_a \in \bs{Y}, 
\end{equation}
where, $\mathrm{p}^{l_a}_{(u_k)}$ denotes the category affinity of all users in the neighborhood $\cm{N}_{l_a}$ of a location $l_a$. We calculate $\mathrm{p}^{l_a}_{(u_k)}$ for each user as the fraction of check-ins of a user $u_k$ with the total check-ins for all users in $\cm{N}_{l_a}$ at the POIs with category same as $l_a$. Mathematically,
\begin{equation}
\mathrm{p}^{l_a}_{(u_k)} = \frac{\text{Number of check-ins by $u_k$ with category \textit{`cat'}}}{\text{Number of check-ins by users in $\cm{N}_{l_a}$ with category \textit{`cat'}}},
\end{equation}
where \textit{`cat'} denotes the category of POI $l_a$. Moreover, these probabilities are specific to each POI and can be interpreted as the affinity of nearby users towards the POI category. Similar to $\Phi_3$, we aggregate the location neighborhood using an edge-weight based sampling on $\bs{Y}$ to get user-conditioned location embedding $\bs{L}_s$ with parameters $\bm{W}_{\Phi_4}, \bm{b}_{\Phi_4}$ and $\bs{\alpha}^{\Phi_4}$ and $\psi_4(\bs{y}_a, \bs{y}_b)$. 

\subsection{Model Prediction}\label{axobasic_predict}
We combine the four representations of users and locations developed above using fully-connected layers, with a concatenated input of $\bs{U}_l$ and $\bs{U}_s$ for final user embedding $\bs{U}_f = \varphi_1\left(\bs{U}_l \mathbin\Vert \bs{U}_s\right)$; and similarly for locations $\bs{L}_l$ and $\bs{L}_s$ to obtain final location embedding $\bs{L}_f = \varphi_2\left(\bs{L}_l \mathbin\Vert \bs{L}_s\right)$. Finally, we estimate a user's affinity to check-in at a location by a Hadamard (element-wise) product between the corresponding representations. Formally, 
\begin{equation}
\widehat{r_{ul}} = \varphi_3\left(\bs{U}_f \, \otimes \, \bs{L}_f\right),
\end{equation}
where $\varphi_1(\cdot)$, $\varphi_2(\cdot)$ and $\varphi_3(\cdot)$ represent fully-connected neural layers. The parameters are optimized using a \textit{mean-squared} error that considers the difference between the user's predicted and the actual affinity towards a location, with $L_1$ regularization over the trainable parameters.
\begin{equation}
\cm{L}_{p} =  \sum_{(\forall u, l)} \snorm{\widehat{r_{ul}} - r_{ul}}^2 + \lambda_{p}\snorm{\Theta_{pred}}.
\end{equation}
$\Theta_{pred}$ refers to all the trainable parameters in \axolotl for a region-specific prediction including the weights for all attention networks. 

\subsection{\axolotl: Information Transfer}\label{axoinftran}
We now turn our attention to the central theme of \axolotl, namely, the training procedure for \axolotl along with its cluster-wise transfer approach. We reiterate that we do not expect any common users/POIs between source and target regions, and thus the only feasible way to transfer mobility knowledge using the trained model parameters and the user-POI embeddings. Specifically, there are two channels of learning for \axolotl, (i) Spatio-Social Meta-learning based optimization, and (ii) Region-wise cluster alignment loss.

\subsubsection{Spatio-Social Meta-Learning (SSML)}\label{axometa_part}
\begin{figure}[b!]
\centering
\begin{subfigure}[]{0.32\columnwidth}
\centering
\includegraphics[height=4cm]{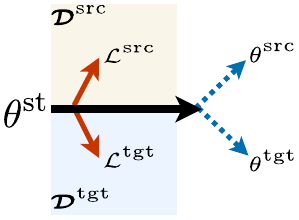}
\caption{MAML}
\label{axofig:a}
\end{subfigure}
\hfill
\begin{subfigure}[]{0.50\columnwidth}
\centering
\includegraphics[height=4cm]{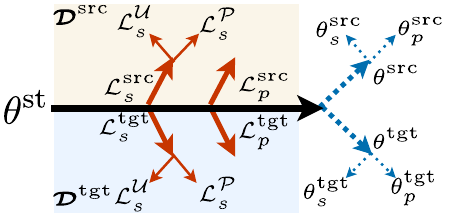}
\caption{SSML}
\label{axofig:b}
\end{subfigure}
\vspace{-4mm}
\caption{The difference between MAML(fig~\ref{axofig:a}) and our SSML (fig~\ref{axofig:b}) is that the latter optimizes parameters in a hierarchy, \ie, user and location parameters are updated for neighborhood prediction and then combined for POI recommendation. (Best viewed in color).}
\label{axofig:maml}
\end{figure}
Meta-learning has long been proposed to alleviate data scarcity problems in spatial datasets~\cite{deepst, www, ruiruiwww}. Specifically, in meta-learning, we aim to learn a joint parameter initialization for multiple tasks by simultaneously optimizing the prediction loss for each task. However, there is a high variance in data-quality between the regions $\ds{D}^{src}$ and $\ds{D}^{tgt}$ and thus a \textit{vanilla} meta-learning ---also called a model agnostic meta-learning (MAML)~\cite{maml}--- will not be sufficient as the target region is expected to have nodes with limited interactions. Such nodes, due to their low contribution to the loss function, may get neglected during the meta-procedure. We overcome this via a \textit{hierarchical} learning procedure that not only considers the location recommendation performance, but also the social neighborhood of all user and location nodes. We call the resulting learning procedure as \textbf{spatio-social meta-learning} (SSML) and the contrast between this approach and standard model agnostic meta-learning approach (MAML)~\cite{maml} is schematically given in Figure ~\ref{axofig:maml}. 

Specifically, we consider the two tasks of \begin{inparaenum}[(i)] \item optimizing the POI recommendation loss function $\cm{L}_{p}$ across both source and target regions, and, \item neighborhood prediction for each node in both the networks~\cite{hsml}.\end{inparaenum} We initialize the parameters for the recommender system with global initial values ($\theta^{st}$) shared across both source and target. Note that by $\theta^{st}$ we mean the parameters for all GATs ($\Phi_{1\cdots4}$) and prediction MLPs ($\varphi_{1\cdots3}$) and thus exclude region-specific, user and location embedding matrices, $\bs{U}^{src}, \bs{L}^{src}, \bs{U}^{tgt}$ and $\bs{L}^{tgt}$.
We describe them here:

\noindent \underline{\textit{Neighborhood Prediction}}: For any region target or source, consider a user $u_i$ and her neighbor $u_j \in \cm{N}_{u_i}$, we obtain the probability of them being connected on the social network as $\hat{v}_{u_i, u_j} = \bs{u}_i \cdot \bs{u}_j^T$ where $\bs{u_\bullet} \in \bs{U}_f$ represents the \textit{final} representations of a user. Thus, for all regions, $\ds{D}^{src}$ and $\ds{D}^{tgt}$, we optimize the following cross-entropy loss:
\begin{equation}
\cm{L}^{\cm{U}}_{s}(\ds{D}^{\bullet}) = -\sum_{u_i \in \cm{U}^{\bullet}} \sum_{\substack{u_j \in \cm{N}_{u_i}\\u'_j \notin \cm{N}_{u_i}}} \left[ \log \left ( \sigma(\hat{v}_{u_i, u_j}) \right) + \log \left (1 - \sigma(\hat{v}_{u_i, u'_j}) \right) \right],
\end{equation}
where, $\hat{v}_{u_i, u_j}, \hat{v}_{u_i, u'_j}, \sigma$ denote the estimated link probability between two users connected by their social networks, with a negatively sampled user $u'_j$, \ie, a user not in $\cm{N}_{u_i}$, and the \textit{sigmoid} function.
Similarly, we calculate the probability of a location $l_b$ being in the spatial neighborhood of a location $l_a$ as $\hat{v}_{l_a, l_b} = \bs{l}_a \cdot \bs{l}_b^T$ and denote the neighborhood loss as $\cm{L}^{\cm{P}}_{s}(\ds{D}^{\bullet})$. For the region as a whole, say target, the net neighborhood loss is defined as:
\begin{equation}
\cm{L}^{tgt}_{s} = \cm{L}^{\cm{U}}_{s}(\ds{D}^{tgt}) + \cm{L}^{\cm{P}}_{s}(\ds{D}^{tgt}),
\end{equation}
Similarly, for source regions, we denote social loss as $\cm{L}^{src}_{s}$.

\noindent \underline{\textit{POI Recommendation}}: Since \axolotl is designed for limited data regions, we purposely incline the meta-procedure towards improved target-region predictions. Specifically, we alter the meta-learning procedure by optimizing the parameters with the recommendation loss for the target region ($\cm{L}^{tgt}_p$) for a predefined number of updates ($N_{u}$) and then optimize for the source region prediction loss($\cm{L}^{src}_p$).
\begin{align}
\nonumber \mathrm{Target:} & \, \theta^{tgt}_{k+1} \leftarrow 
\begin{cases} 
  \theta^{tgt}_k - \omega_1 \nabla_{\theta^{tgt}_k}\,\cm{L}^{tgt}_{p}(f_{\theta^{tgt}_k}), & \forall \, 1 \le k \le N_u, \\
  \theta^{st} - \omega_1 \nabla_{\theta^{st}}\,\cm{L}^{tgt}_{p}(f_{\theta^{st}}), & \mathrm{otherwise}, 
\end{cases}\\
 \mathrm{Source:} & \, \theta^{src} \leftarrow \theta^{st} - \omega_1 \nabla_{\theta^{st}}\,\cm{L}^{src}_{p}(f_{\theta^{st}}),
\end{align}
where $\theta^{st}, \theta^{tgt}, \theta^{src}, \cm{L}^{tgt}_{p}, \cm{L}^{src}_{p}, f_\theta$ are the global model-independent parameters, parameters for target and source region, prediction loss for target and source, and \axolotl output respectively. 

\noindent \underline{\textit{Final Update:}} The final update to global parameters is done by: (i) optimizing the region-specific social loss, $\cm{L}^{tgt}_{s}$ and $\cm{L}^{src}_{s}$, and (ii) minimizing the POI recommendation loss $\cm{L}^{tgt}_{p}, \cm{L}^{src}_{p}$ for both source and target regions.
\begin{equation}
\theta^{st} \leftarrow \theta^{st} - \omega_2 \cdot \left[ \nabla_{\theta^{st}}\,\cm{L}^{src}_{p}(f_{\theta^{st}}) + \nabla_{\theta^{tgt}_{N_u}}\,\cm{L}^{tgt}_{p}(f_{\theta^{tgt}_{N_u}}) \right] - \omega_3 \cdot \left[ \nabla_{\theta^{st}}\,\cm{L}^{src}_{s}(f_{\theta^{st}}) + \nabla_{\theta^{st}}\,\cm{L}^{tgt}_{s}(f_{\theta^{st}}) \right],
\label{axoeqn:update}
\end{equation}
where, $\omega_2, \omega_3$ denote the region-wise learning rates. 

\begin{algorithm}[t]
\small
\DontPrintSemicolon
\KwIn{\\$\ds{D}^{src}$: Source-Region Training Data, $\ds{D}^{tgt}$: Target-Region Training Data \\
$M_t$: Epoch-based checkpoint, $\cm{C}$: Clustering function \\
}
\KwOut{$\theta^{tgt}$: Trained \axolotl Parameters for Target Region}
$\theta^{st}, \theta^{tgt}, \theta^{src} \leftarrow$ Randomly initialize all parameters\\
\SetKwBlock{Begin}{function}{end function}
  \While {$\mathtt{epoch} < \mathtt{Max\_Epoch}$}
  {
    Parameter update via target-region social prediction: $\theta^{tgt}_{0} \leftarrow \theta^{st} - \omega \nabla_{\theta^{st}}\,\cm{L}^{tgt}_{s}(f_{\theta^{st}})$\\
    Parameter update via source-region social prediction: $\theta^{src}_{0} \leftarrow \theta^{st} - \omega \nabla_{\theta^{st}}\,\cm{L}^{src}_{s}(f_{\theta^{st}})$\\
    
    Calculate the target-region recommendation loss: $\cm{L}^{tgt}_p \leftarrow \mathtt{PredictionLoss}(\ds{D}^{tgt})$\\
    Initial update-before iterations: $\theta^{tgt}_{1} \leftarrow \theta^{st} - \omega_1 \nabla_{\theta^{st}}\,\cm{L}^{tgt}_{p}(f_{\theta^{st}})$\\
    \For{$k < N_u$}
    { 
      Iterative updates: $\theta^{tgt}_{k+1} \leftarrow \theta^{tgt}_k - \omega_1 \nabla_{\theta^{tgt}_k}\,\cm{L}^{tgt}_{p}(g_{\theta^{tgt}_k})$\\
    }
    Calculate the source-region recommendation loss: $\cm{L}^{src}_p \leftarrow \mathtt{PredictionLoss}(\ds{D}^{src})$\\
    Update for source parameters $\theta^{src} \leftarrow \theta^{st} - \omega_1 \nabla_{\theta^{st}}\,\cm{L}^{src}_{p}(f_{\theta^{st}}(\cm{R}_t))$\\
    
    Joint update for global parameters: As in Eqn. \ref{axoeqn:update} \\
    \uIf{$\mathtt{epoch}\mod M_t$}
  {
    $\cm{L}_c \leftarrow \mathtt{ClusterLoss}(\cm{C}, \ds{D}^{src}, \ds{D}^{src})$\\
    Initialize cluster-based transfer $\theta^{src},\theta^{tgt} \leftarrow \displaystyle\min \cm{L}_c$
  }
  $\mathtt{epoch}++$\\
  }\label{axoendfor}
  Fine-tune for target region: $\theta^{tgt} \leftarrow \texttt{FineTune}(\cm{L}^{tgt}_{p})$ \\
  Return model parameters: \textbf{return} $\theta^{tgt}$
\caption{Training Algorithm for \axolotl}\label{axoaxoalgo}
\end{algorithm}

\subsubsection{Cluster Alignment Loss}\label{axocluster_loss}
Recent research~\cite{www, regiontrans} has shown that enforcing similar patterns across specific POI clusters between source and target domains, \eg\ from one university campus to another, facilitates better knowledge transfer. Unfortunately, obtaining the necessary semantic information to align similar clusters across regions that these techniques require is not always practical in large-scale settings. For a POI network, a rudimentary approach to identify clusters would be to traverse across the categories associated with each location -- which may be quite expensive to compute and will neglect the user dynamics as well. We avoid these approaches and use a \textit{lightweight} Euclidean-distance based \textit{k-means} clustering to identify a set of users and locations in source as well as target regions (separately) that have displayed \textit{similar} characteristics till the current iteration. Such a dynamic clustering mechanism over the contemporary GAT embeddings prevents the need for additional hand-crafting. In contrast to previous approaches~\cite{hsml, locate}, we utilize a \textit{hard}-assignment as unlike \textit{online} product purchases, the mobility of a user is bounded by geographical distance~\cite{cho}, and thus checkins to distant locations are very unlikely. We denote $\bs{U}^{tgt}_c$, $\bs{L}^{tgt}_c$, $\bs{U}^{src}_c$, $\bs{L}^{src}_c \in \mathbb{R}^{K\times D}$ and $\cm{C}$ as the cluster embedding matrices for target region-users, locations, source region-users, locations, and the clustering algorithm respectively. We calculate the cluster embedding by \textit{mean-pooling} the embeddings of elements in the cluster. For example, we calculate embeddings for user clusters in the target region as:
\begin{equation}
\bs{u}^{tgt}_{c} = \mathtt{MeanPool}\left[\bs{u}_i \cdot \Psi(u_i, c)\right]  \, \forall u_i \in \cm{U}^{tgt}, \bs{u}^{tgt}_{c} \in \bs{U}^{tgt}_{c},
\end{equation}
where, $\Psi(u_i, c)$ is the indicator function denoting whether user $u_i$ belongs to cluster $c$. Similarly we calculate $\bs{L}^{tgt}_c$, $\bs{U}^{src}_c$ and $\bs{L}^{src}_c$. For minimizing the divergence between the similar users and POIs across regions, instead of engineering an explicit alignment between clusters, we use an attention-based approach to identify and align clusters with similar patterns and minimize the corresponding \textit{weighted} $L_2$ loss~\cite{moreattn, what2tran}.
\begin{equation}
\cm{L}_c =  \sum_{\forall c^t_u \in \bs{U}^{tgt}_c} \sum_{\forall c^t_l \in \bs{L}^{tgt}_c} \snorm{\bs{c}^t_u - \sum_{c^s_u \in \bs{U}^{src}_c}\bs{\beta}^u_{c^t_u, c^s_u} \bs{c}^s_u}^2 + \, \snorm{\bs{c}^t_l - \sum_{c^s_l \in \bs{L}^{src}_c}\bs{\beta}^l_{c^t_l, c^s_l} \bs{c}^s_l}^2,
\end{equation}
where $\bs{\beta}^u, \bs{\beta}^l \in \mathbb{R}^{K \times K}$, are the attention matrices for user and location clusters, respectively. Each index in $\bs{\beta}^u, \bs{\beta}^l$ denotes the weight for a target and source cluster for users and location, respectively. A key modeling distinction between $\bs{\alpha}$ and $\bs{\beta}$ is that the latter includes \textit{self}-contribution of the node under consideration and in $\bs{\beta}$ we only aim to capture the contribution by the source-cluster on the particular target-cluster. Therefore $\bs{\beta}$ is calculated similarly as $\bs{\alpha}$ (Equation~\ref{axoeqn:alpha}) after restricting to only the inter-cluster interactions.

\subsubsection{Overcoming the Curse of Pre-Training}\label{axocurse}
Transfer learning, by definition, requires the transfer source to be \textit{pre}-trained, \ie, for the information propagation across clusters of users and locations, the set of weights for the source should be trained before initiating the transfer. In our setting, we do not extensively train the source parameters separately, as these parameters are jointly learned via meta-learning. This could be a severe bottleneck for the cluster-based transfer as it may lead to inaccurate information sharing across clusters as the source parameters are also simultaneously being learned. We reconcile these two by adopting a \textit{checkpoint}-based transfer approach by performing transfer based on the number of epochs for parameter optimization of the model. Specifically, we optimize the region-specific model parameters for $M_t$ epochs with each epoch across the entire source and target data. We \textit{checkpoint} this model state and consider the user-location cluster embeddings to initialize transfer by optimizing the all-region parameters through cluster-alignment loss, $\cm{L}_c$~\cite{what2tran}. Optimizing $\cm{L}_c$ updates the user and POI embedding by backpropagating the difference between similar source and target clusters. We minimize $\cm{L}_c$ via stochastic gradient descent(SGD)~\cite{rumelhart1986learning}. This checkpoint-based optimize-transfer cycle continues for fixed iterations, and then weights are later \textit{fine-tuned}~\cite{maml}. This learning procedure is described between line \#12 and line \#15 in Algorithm~\ref{axoaxoalgo}.

\subsubsection{Significance of using SSML with Cluster Loss} Here we highlight the importance of the two tasks in SSML -- neighborhood prediction and POI recommendation that are achieved by minimizing the loss functions $\cm{L}^{\cm{U}}_{s}$, $\cm{L}^{\cm{P}}_{s}$, and $\cm{L}_{p}$ respectively for each region. Particularly, the task of neighborhood prediction of each region is a combination of predicting the spatial neighbors of a POI (via $\cm{L}^{\cm{P}}_{s}$) and the social network of a user (via $\cm{L}^{\cm{U}}_{s}$). The former ensures that the embeddings of POIs located within a small geographical area can capture the latent features of the particular area~\cite{locate, www}. Such a feat is not achievable by minimizing the difference between POIs in a common cluster, as the clusters are determined explicitly from these embeddings, whereas the spatial graph is constructed using the distance between POIs, and thus is a better estimate of neighborhoods within a region. Similarly, minimizing $\cm{L}^{\cm{U}}_{s}$ ensures that user embeddings capture the flow of POI-preferences between socially connected users~\cite{ngcf} that cannot be captured via an embedding-based clustering. Thus, the task of predicting the neighborhood of a node can lead to better POI recommendations for users closer to a locality.

\section{Experiments}
\label{axosec:experiments}
We perform check-in recommendations in the test data to evaluate \axolotl across three geo-tagged activity streams from different countries. With our experiments, we aim to answer the following research questions:
\begin{compactitem}
\item[\textbf{RQ1}] Can \axolotl outperform state-of-the art baselines for location recommendation in sparse regions?
\item[\textbf{RQ2}] What are the contributions of different modules in \axolotl? 
\item[\textbf{RQ3}] How are the weights in \axolotl transferred across regions? 
\end{compactitem}
All our models are implemented in Tensorflow on a server running Ubuntu 16.04. CPU: Intel(R) Xeon(R) Gold 5118 CPU @ 2.30GHz, RAM: 125GB, and GPU: NVIDIA V100 GPU.

\subsection{Experimental Settings}
\textbf{Dataset Description.}
For our experiments, we combine POI data from two popular datasets, Gowalla~\cite{scellato} and Foursquare~\cite{lbsn2vec}, across 12 different regions of varied granularities from the United States(US), Japan(JP), and Germany(DE). For each country, we construct 4 datasets: one with large check-in data and three with limited data. We adopt a commonly followed data cleaning procedure~\cite{cara, locate} ---for \emph{source datasets}, we filter out locations with less than 10 check-ins, users with less than 10 check-ins and less than 5 connections. For {target datasets}, these thresholds are set at 5, 5, and 2 respectively. Higher criteria are used for source datasets to minimize the effects of noisy data during transfer. 
The statistics of the twelve datasets is given in Table \ref{axotab:data} with each acronym denoting the following region: (i) CA: California(US), (ii) WA: Washington(US), (iii) MA: Massachusetts(US), (iv) OH: Ohio(US), (v) TY: Tokyo(JP), (vi) HY: Hyogo(JP), (vii) KY: Kyoto(JP), (viii) AI: Aizu(JP) (ix) NR: North-Rhine Westphalia(DE), (x) BW: Baden-W\"urttemberg(DE), (xi) BE: Berlin(DE), and (xii) BV: Bavaria(DE). We consider CA, TY, and NR as the source regions and WA, NY, MA and KY, HY, AI, and BV, BW, and BE as the corresponding target regions.

\noindent \textbf{Evaluation Protocol:}
For each region, we consider the first 70\% data, based on the time of check-in as training, 10\% as validation, and the rest as test data for both Gowalla and Foursquare. For each region, we use the training data to get a list of the top-k most probable check-in locations for each user and compare it with ground-truth check-ins in the test. Note that there is no user or location overlap between source and target regions and for each user we only recommend check-ins located in the specific region. For evaluation, we use: \textit{Precision@k} and \textit{NDCG@k}, with $k = 1, 5, 10$, and report confidence intervals based on three independent runs.

\noindent \textbf{Parameter Settings:} For all experiments, we adopt a three layer architecture for MLPs with dimensions $\varphi_1, \varphi_2 =\{32\rightarrow32 \rightarrow D \}$ and $\varphi_3= \{32\rightarrow32\rightarrow1\}$. Other variations for the MLP had insignificant differences. We keep $M_t = \{4, 6\}, N_u = \{4, 8\}, K = \{20, 50, 100\}, \lambda_p = 0.01$, $\kappa = 50km, D=16$ and batch-size in $\{16, 32\}$. Unless otherwise mentioned, we use these parameters in all our experiments. For meta- and cluster-based transfer, we set $\omega_1, \omega_2, \omega_3 = 0.001$ and learning-rate as $0.01$, as recommended for training a meta-learning algorithm ~\cite{maml}.

\begin{table*}[t!]
\caption{Statistics of datasets used in this chapter. The source region columns are highlighted, followed by target regions. The datasets are further partitioned based on the country of origin (US, Japan, and Germany).}
\vspace{-2mm}
\label{axotab:data}
\centering
\resizebox{\textwidth}{!}{
\begin{tabular}{l|accc|accc|accc}
\toprule
\textbf{Property} & \textbf{CA} & \textbf{WA} & \textbf{MA} & \textbf{OH} & \textbf{TY} & \textbf{KY} & \textbf{AI} & \textbf{HY} & \textbf{NR} & \textbf{BV} & \textbf{BW} & \textbf{BE}\\
\hline
\#Users ($|\cm{U}|$) & 3518 & 1959 & 1623 & 1322 & 6361 & 1445 & 2059 & 1215 & 1877 & 923 & 682 & 1015\\
\#Locations ($|\cm{P}|$) & 42125 & 16758 & 10585 & 8509 & 11905 & 2055 & 4561 & 2255 & 13049 & 7290 & 8381 & 5493\\
\#User-User Edges ($|\mathcal{E}_u|$) & 26167 & 5039 & 3702 & 3457 & 32540 & 2973 & 5723 & 2124 &  5663 & 2833 & 2370 & 3122\\
\#Location-Location Edges ($|\mathcal{E}_l|$) & 250360 & 94400 & 46225 & 34182 & 146021 & 15237 & 37470 & 14055 & 66086 & 31823 & 40253 & 29308\\
\#User-Location Edges ($|\mathcal{E}_r|$) & 310350 & 114897 & 52097 & 37431 & 189990 & 17573 & 40503 & 17036 & 75212 & 37673 & 50889 & 31915\\
\bottomrule
\end{tabular}
}
\vspace{-0.3cm}
\end{table*}

\subsection{Methods} \label{axobaselines}
We compare \axolotl with the state-of-the-art methods based on their architectures below:
\begin{compactenum}[(1)]
\item{\bf Methods based on Random Walks:}
\begin{asparadesc}
\item [Node2Vec~\cite{node2vec}] Popular random-walks-based embedding approach that uses parameterized breadth- and depth-first search to capture representations.
\item [Lbsn2Vec~\cite{lbsn2vec}] State-of-the-art random-walk-based POI recommendation, it uses a random-walk-with-stay scheme to jointly sample user check-ins and social relationships to learn node embeddings.
\end{asparadesc}
\item \textbf{Graph-based POI Recommendation}
\begin{asparadesc}
\item [Reline~\cite{reline}] State-of-the-art multi-graph based POI recommendation algorithm. Traverses across location and user graphs to generate individual embeddings.
\end{asparadesc}

\item \textbf{Methods based on Matrix Factorization:}
\begin{asparadesc}
\item [GMF~\cite{gmf}] Standard matrix factorization, which is optimized using a personalized prediction loss for users.
\item [NMF~\cite{ncf}] Collaborative filtering-based model that applies MLPs above the concatenation of user and item embeddings to capture their interactions.
\end{asparadesc}

\item \textbf{Methods based on Graph Neural Networks:}
\begin{asparadesc}
\item [NGCF~\cite{ngcf}] State-of-the-art graph neural network recommendation framework that encodes the collaborative signal with connectivity in the user-item bipartite graph.
\item [DANSER~\cite{danser}] Uses \textit{dual} graph-attention networks across the item and user networks and predicts using a reinforcement policy-based algorithm. 
\end{asparadesc}
\item \textbf{Methods using Transfer Learning:}
\begin{asparadesc}
\item [MDNN~\cite{homanga}] An MLP based meta-learning model that performs \textit{global} as well as \textit{local} updates together.
\item [MCSM~\cite{manasi}] A neural architecture with parameters learned through an optimization-based meta-learning.
\item [MeLU~\cite{melu}] State-of-the-art \textit{meta}-learning based recommendation system. Estimates user preferences in a data-limited query set by using a data-rich support set across the concatenation of user and item representations.
\item [MAMO~\cite{mamo}] Modifies MAML~\cite{maml} to incorporate a heterogeneous information network by item content information and memory-based mechanism.
\item [PGN~\cite{pretrain}] A state-of-the-art meta-learning procedure for pre-training neural graph models to better capture the user and item embeddings. Specifically, it includes a three-step procedure -- a neighborhood sampler, a GNN-based aggregator, and meta-learning-based updates. For our experiments, we apply PGN over graph attention networks~\cite{gat}.
\end{asparadesc}
\end{compactenum}
As mentioned in Section \ref{axointro}, other techniques either collectively learn parameters across \textit{common} users in both domains\cite{cmf, compare} or either utilize to meta-path based approach~\cite{mhin}, and thus are not suitable for our setting. Furthermore, to demonstrate the drawbacks of traditional transfer learning, we report results for the following variants of \axolotl:
\begin{compactitem}[(1)]
\item [\textbf{\ours-f:}] We train \ours-basic on the source data and fine-tune the weights for the target data as a standard transfer-learning setting.
\item [\textbf{\ours-m:}] For this variant \ours-basic model is trained on both regions using only the proposed spatio-social meta-learning, \ie, without cluster-based optimization.
\end{compactitem}
The main contribution including both SSML and cluster-based transfer is termed \axolotl. 

\noindent \textbf{Baseline Implementations:} Here, we present the implementation details for each of our baselines. Specifically, for region-specific models, we follow a standard practice of optimizing their parameters on the training set of the target region and then predicting for users in the corresponding test set. For MDNN and MCSM, train the parameters on both regions using their standard meta-learning procedure. For MeLU, we modify the training protocol to perform \textit{global-updates} using the user-POI pairs for the target regions and \textit{local-updates} using the source-region parameters. For more details, please refer to Section 3.2 in~\cite{melu}. A similar training procedure is followed for MAMO. Lastly, for PGN we pre-train the model parameters on the source-region check-ins and then fine-tune on the target region as per their three-step optimizing procedure.

\begin{table*}[t!]
\caption{\label{axotab:main} Performance comparison between state-of-the-art baselines, \axolotl and its variants (\ours-f and \ours-m). The first column represents the \textit{source} and the corresponding \textit{target} regions. The grouping is done based on baseline details in Section \ref{axobaselines}. Numbers with bold font indicate the best-performing model. All results marked $\dagger$ are statistically significant (\ie, two-sided Fisher's test with $p \le 0.1$) over the best baseline.}
\vspace{-2mm}
\centering
\resizebox{\textwidth}{!}{
\begin{tabular}{c|c|cc|c|cc|cc|ccccc|aaa}
\toprule
$\bs{\mathcal{D}}^{src} \rightarrow \bs{\mathcal{D}}^{tgt}$ & \textbf{Metric} & \textbf{N2V} & \textbf{L2V} & \textbf{Reline} & \textbf{GMF} & \textbf{NMF} & \textbf{Danser} & \textbf{NGCF} & \textbf{MDNN} & \textbf{MCSM} & \textbf{MeLU} & \textbf{MAMO} & \textbf{PGN} & \textbf{Axo-f} & \textbf{Axo-m} & \textbf{Axolotl}\\ \hline

\multirow{5}{*}{\textbf{CA} $\rightarrow$ \textbf{WA}}
 & Prec@1 & 0.3053 & 0.3493 & 0.3912 & 0.3709 & 0.4306 & 0.4792 & 0.4716 & 0.4639 & 0.4287 & 0.5094 & 0.4783 & 0.5038 & 0.4605 & 0.5529 & \textbf{0.5537}$\dagger$\\
 & Prec@5 & 0.5618 & 0.6042 & 0.6248 & 0.6096 & 0.6630 & 0.7124 & 0.6983 & 0.6733 & 0.6694 & 0.7041 & 0.6974 & 0.7067 & 0.6793 & 0.7714 & \textbf{0.7736}$\dagger$\\ 
 & Prec@10 & 0.5652 & 0.5718 & 0.6198 & 0.5982 & 0.6537 & 0.7012 & 0.6981 & 0.6787 & 0.6436 & 0.7003 & 0.6827 & 0.6941 & 0.6832 & 0.7431 & \textbf{0.7489}$\dagger$\\ 
 \cline{2-17}
 & NDCG@5 & 0.3804 & 0.4533 & 0.5386 & 0.5228 & 0.5688 & 0.5809 & 0.5646 & 0.5743 & 0.5680 & 0.6064 & 0.5814 & 0.5932 & 0.5788 & 0.6473 & \textbf{0.6503}$\dagger$\\ 
 & NDCG@10 & 0.4372 & 0.4976 & 0.6067 & 0.5884 & 0.6461 & 0.6710 & 0.6587 & 0.6511 & 0.6449 & 0.6721 & 0.6687 & 0.6783 & 0.6572 & 0.7199 & \textbf{0.7285}$\dagger$\\ 
\hline

 \multirow{5}{*}{\textbf{CA} $\rightarrow$ \textbf{OH}}
 & Prec@1 & 0.4453 & 0.4787 & 0.5239 & 0.4863 & 0.5055 & 0.5437 & 0.5386 & 0.5035 & 0.5019 & 0.5492 & 0.5238 & 0.5391 & 0.5394 & 0.6083 & \textbf{0.6102}$\dagger$\\
 & Prec@5 & 0.5618 & 0.5694 & 0.6022 & 0.5745 & 0.6152 & 0.6767 & 0.6643 & 0.6283 & 0.6198 & 0.6759 & 0.6693 & 0.6774 & 0.6480 & 0.7118 & \textbf{0.7134}$\dagger$\\ 
 & Prec@10 & 0.5651 & 0.5703 & 0.6173 & 0.5802 & 0.6247 & 0.6759 & 0.6681 & 0.6382 & 0.6233 & 0.6659 & 0.6628 & 0.6648 & 0.6531 & 0.7023 & \textbf{0.7042}$\dagger$\\ 
 \cline{2-17}
 & NDCG@5 & 0.4504 & 0.4576 & 0.4973 & 0.4581 & 0.4850 & 0.5017 & 0.5132 & 0.5271 & 0.5136 & 0.5498 & 0.5231 & 0.5370 & 0.5344 & 0.6088 & \textbf{0.6110}$\dagger$\\ 
 & NDCG@10 & 0.5070 & 0.50934 & 0.5658 & 0.5781 & 0.5840 & 0.6129 & 0.6115 & 0.6049 & 0.5934 & 0.6180 & 0.5890 & 0.6134 & 0.6222 & 0.6654 & \textbf{0.6690}$\dagger$\\
\hline

 \multirow{5}{*}{\textbf{CA} $\rightarrow$ \textbf{MA}}
 & Prec@1 & 0.3964 & 0.3900 & 0.4239 & 0.4092 & 0.4476 & 0.4886 & 0.4761 & 0.4462 & 0.4581 & 0.4889 & 0.4731 & 0.4807 & 0.4539 & \textbf{0.5813}$\dagger$ & 0.5807\\
 & Prec@5 & 0.5618 & 0.5737 & 0.6282 & 0.5816 & 0.6273 & 0.6873 & 0.6791 & 0.6432 & 0.6400 & 0.6808 & 0.6814 & 0.6973 & 0.6400 & 0.7349 & \textbf{0.7373}$\dagger$\\ 
 & Prec@10 & 0.5651 & 0.5899 & 0.6215 & 0.6007 & 0.6266 & 0.6798 & 0.6620 & 0.6419 & 0.6302 & 0.6649 & 0.6587 & 0.6798 & 0.6545 & 0.7316 & \textbf{0.7317}$\dagger$\\
 \cline{2-17}
 & NDCG@5 & 0.4206 & 0.4491 & 0.5243 & 0.4860 & 0.5034 & 0.5614 & 0.5587 & 0.5219 & 0.5187 & 0.5759 & 0.5602 & 0.5683 & 0.5342 & 0.6392 & \textbf{0.6404}$\dagger$\\
 & NDCG@10 & 0.4785 & 0.5084 & 0.5528 & 0.5382 & 0.5476 & 0.6186 & 0.6037 & 0.5764 & 0.5622 & 0.6294 & 0.6140 & 0.6381 & 0.5785 & 0.6930 & \textbf{0.6941}$\dagger$\\
\hline
 
 \multirow{5}{*}{\textbf{TY} $\rightarrow$ \textbf{AI}}
 & Prec@1 & 0.4703 & 0.5407 & 0.6130 & 0.5811 & 0.6026 & 0.5930 & 0.6200 & 0.5960 & 0.6018 & 0.6462 & 0.6137 & 0.6390 & 0.6625 & 0.7190 & \textbf{0.7218}$\dagger$\\
 & Prec@5 & 0.5411 & 0.6008 & 0.6119 & 0.6044 & 0.6289 & 0.6519 & 0.6579 & 0.5996 & 0.6433 & 0.6610 & 0.6413 & 0.6583 & 0.6389 & 0.7216 & \textbf{0.7231}$\dagger$\\
 & Prec@10 & 0.5926 & 0.6122 & 0.6208 & 0.6135 & 0.6333 & 0.6540 & 0.6569 & 0.6152 & 0.6172 & 0.6430 & 0.6217 & 0.6535 & 0.6522 & \textbf{0.7090}$\dagger$ & 0.7017\\
 \cline{2-17}
 & NDCG@5 & 0.4489 & 0.5013 & 0.5139 & 0.5032 & 0.5390 & 0.5346 & 0.5958 & 0.5304 & 0.5799 & 0.6174 & 0.5689 & 0.6003 & 0.5850 & 0.6581 & \textbf{0.6631}$\dagger$\\
 & NDCG@10 & 0.5562 & 0.5796 & 0.5927 & 0.5818 & 0.6128 & 0.6145 & 0.6485 & 0.5802 & 0.6004 & 0.6298 & 0.5910 & 0.6344 & 0.6299 & \textbf{0.7082}$\dagger$ & 0.7052\\  
\hline

 \multirow{5}{*}{\textbf{TY} $\rightarrow$ \textbf{KY}}
 & Prec@1 & 0.4317 & 0.4812 & 0.5226 & 0.4926 & 0.5837 & 0.5719 & 0.5914 & 0.5518 & 0.5867 & 0.6173 & 0.5874 & 0.6074 & 0.6209 & 0.6847 & \textbf{0.6910}$\dagger$\\
 & Prec@5 & 0.5811 & 0.6002 & 0.6044 & 0.5869 & 0.6372 & 0.6579 & 0.6567 & 0.6213 & 0.6014 & 0.6521 & 0.6041 & 0.6794 & 0.6348 & 0.7011 & \textbf{0.7018}$\dagger$\\
 & Prec@10 & 0.5926 & 0.6098 & 0.6123 & 0.5995 & 0.6243 & 0.6555 & 0.6569 & 0.6135 & 0.6005 & 0.6695 & 0.6117 & 0.6741 & 0.6287 & 0.7019 & \textbf{0.7093}$\dagger$\\
 \cline{2-17}
 & NDCG@5 & 0.4789 & 0.5024 & 0.5016 & 0.5120 & 0.5878 & 0.5758 & 0.6291 & 0.6024 & 0.5818 & 0.6892 & 0.6499 & 0.6742 & 0.6458 & 0.7390 & \textbf{0.7454}$\dagger$\\
 & NDCG@10 & 0.5562 & 0.5752 & 0.5807 & 0.5621 & 0.6271 & 0.6485 & 0.6500 & 0.6318 & 0.6044 & 0.7192 & 0.6786 & 0.6983 & 0.6857 & 0.7720 & \textbf{0.7737}$\dagger$\\
\hline

 \multirow{5}{*}{\textbf{TY} $\rightarrow$ \textbf{HY}}
 & Prec@1 & 0.5208 & 0.5719 & 0.5832 & 0.5582 & 0.6030 & 0.6139 & 0.6083 & 0.5990 & 0.5752 & 0.6146 & 0.5824 & 0.6108 & 0.5854 & \textbf{0.6793}$\dagger$ & 0.6781\\
 & Prec@5 & 0.5411 & 0.5996 & 0.5938 & 0.6053 & 0.6018 & 0.6527 & 0.6719 & 0.6419 & 0.6322 & 0.6685 & 0.6459 & 0.6814 & 0.6484 & \textbf{0.7172}$\dagger$ & 0.7134\\
 & Prec@10 & 0.5726 & 0.6072 & 0.6059 & 0.6205 & 0.6138 & 0.6541 & 0.6569 & 0.6348 & 0.6206 & 0.6406 & 0.6283 & 0.6746 & 0.6268 & 0.6994 & \textbf{0.7019}$\dagger$\\ 
 \cline{2-17}
 & NDCG@5 & 0.4891 & 0.4959 & 0.4888 & 0.5316 & 0.6183 & 0.5724 & 0.6195 & 0.6092 & 0.6151 & 0.6480 & 0.6204 & 0.6359 & 0.5968 & 0.7043 & \textbf{0.7068}$\dagger$\\
 & NDCG@10 & 0.5262 & 0.5736 & 0.5681 & 0.6047 & 0.6256 & 0.6461 & 0.6485 & 0.6380 & 0.6353 & 0.6793 & 0.6433 & 0.6684 & 0.6225 & 0.7298 & \textbf{0.7352}$\dagger$\\ 
\hline

\multirow{5}{*}{\textbf{NR} $\rightarrow$ \textbf{BE}}
 & Prec@1 & 0.4624 & 0.4721 & 0.4689 & 0.4627 & 0.4830 & 0.5211 & 0.5354 & 0.5277 & 0.5101 & 0.5471 & 0.5371 & 0.5439 & 0.5392& 0.5972 & \textbf{0.5981}$\dagger$\\
 & Prec@5 & 0.4680 & 0.4981 & 0.5783 & 0.5436 & 0.5836 & 0.6430 & 0.6214 & 0.6023 & 0.6148 & 0.6471 & 0.6252 & 0.6572 & 0.6392 & \textbf{0.6900}$\dagger$ & 0.6889\\ 
 & Prec@10 & 0.5254 & 0.5634 & 0.6092 & 0.5999 & 0.6124 & 0.6444 & 0.6250 & 0.6162 & 0.6029 & 0.6391 & 0.6281 & 0.6492 & 0.6215 & 0.6742 & \textbf{0.6749}$\dagger$\\ 
 \cline{2-17}
 & NDCG@5 & 0.4998 & 0.5077 & 0.5315 & 0.5189 & 0.5416 & 0.5920 & 0.6018 & 0.6297 & 0.6291 & 0.6603 & 0.6198 & 0.6569 & 0.6631 & 0.7370 & \textbf{0.7376}$\dagger$\\ 
 & NDCG@10 & 0.5532 & 0.5674 & 0.6285 & 0.5779 & 0.6248 & 0.6826 & 0.6604 & 0.6633 & 0.6516 & 0.6889 & 0.6749 & 0.6842 & 0.6811 & 0.7734 & \textbf{0.7741}$\dagger$\\ 
\hline

 \multirow{5}{*}{\textbf{NR} $\rightarrow$ \textbf{BV}}
 & Prec@1 & 0.3284 & 0.3794 & 0.4024 & 0.3878 & 0.4115 & 0.4588 & 0.4308 & 0.4465 & 0.4209 & 0.4896 & 0.4482 & 0.5163 & 0.4745 & 0.5562 & \textbf{0.5593}$\dagger$\\
 & Prec@5 & 0.4380 & 0.4559 & 0.5005 & 0.4843 & 0.4917 & 0.5649 & 0.5318 & 0.5250 & 0.5075 & 0.5657 & 0.5318 & 0.5739 & 0.5736 & 0.6296 & \textbf{0.6323}$\dagger$\\ 
 & Prec@10 & 0.4860 & 0.5062 & 0.5324 & 0.5150 & 0.5456 & 0.5882 & 0.5675 & 0.5744 & 0.5700 & 0.5809 & 0.5624 & 0.6044 & 0.5648 & \textbf{0.6357}$\dagger$ & 0.6355\\ 
 \cline{2-17}
 & NDCG@5 & 0.4378 & 0.4543 & 0.4713 & 0.4571 & 0.4868 & 0.5239 & 0.5243 & 0.5303 & 0.5271 & 0.5694 & 0.5473 & 0.5789 & 0.5700 & 0.6183 & \textbf{0.6226}$\dagger$\\ 
 & NDCG@10 & 0.4772 & 0.4886 & 0.5768 & 0.5131 & 0.5847 & 0.6043 & 0.5982 & 0.5906 & 0.6019 & 0.6232 & 0.5980 & 0.6267 & 0.6191 & 0.6708 & \textbf{0.6739}$\dagger$\\  
\hline
 
 \multirow{5}{*}{\textbf{NR} $\rightarrow$ \textbf{BW}}
 & Prec@1 & 0.3584 & 0.3858 & 0.4172 & 0.3966 & 0.4263 & 0.4767 & 0.4594 & 0.4520 & 0.4363 & 0.4773 & 0.4587 & 0.4758 & 0.4689 & \textbf{0.5066}$\dagger$ & 0.5062\\
 & Prec@5 & 0.4861 & 0.5197 & 0.5416 & 0.5282 & 0.5579 & 0.5670 & 0.5683 & 0.5564 & 0.5423 & 0.5849 & 0.5670 & 0.5913 & 0.5740 & 0.6198 & \textbf{0.6201}$\dagger$\\
 & Prec@10 & 0.4860 & 0.5487 & 0.5693 & 0.5354 & 0.5293 & 0.5878 & 0.5849 & 0.5694 & 0.5459 & 0.5805 & 0.5718 & 0.5866 & 0.5712 & \textbf{0.6306}$\dagger$ & 0.6297\\ 
 \cline{2-17}
 & NDCG@5 & 0.4102 & 0.4972 & 0.5122 & 0.4898 & 0.5167 & 0.5750 & 0.5731 & 0.5460 & 0.5281 & 0.5623 & 0.5372 & 0.5683 & 0.5563 & 0.6048 & \textbf{0.6075}$\dagger$\\ 
 & NDCG@10 & 0.4772 & 0.5532 & 0.5783 & 0.5552 & 0.5862 & 0.6217 & 0.6101 & 0.5962 & 0.5931 & 0.6183 & 0.5849 & 0.6283 & 0.6000 & 0.6598 & \textbf{0.6614}$\dagger$\\
\bottomrule
\end{tabular}
}
\vspace{-4mm}
\end{table*}

\subsection{Performance Comparison}
To address RQ1, we report on the performance of location recommendation of different methods across all our target datasets in Table~\ref{axotab:main}. From these results, we make the following key observations:
\begin{compactitem}
\item \axolotl, and its variant \ours-m that employ meta-learning, consistently yield the best performance on all the datasets. In particular, the complete \axolotl improves over the strongest baselines by 5-18\% across the metrics. These results signify the importance of meta-learning with external data to design solutions for limited-data regions.

\item \ours-f does not perform on par with other approaches. This observation further cements the advantage of a joint-learning over traditional fine-tuning. The performance gain by \axolotl over \ours-m highlights the importance of minimizing the divergence between the embeddings across the two regions.

\item Among meta-learning-based models, we note that MeLU~\cite{melu} and PGN~\cite{pretrain} perform better than other baseline models, however, they are easily outperformed by \axolotl. We also note that though \axolotl and PGN are graph-based meta-learning models, the performance difference can be attributed to the ability of \axolotl to include node features. Specifically, PGN only leverages the graph structure and cannot thus incorporate any heterogeneous auxiliary information about the entities, such as POI category and distances, whereas \axolotl captures all features of a spatial network. 

\item The characteristic of MeLU~\cite{melu} to include samples from the data-rich network into its meta-learning-based procedure leads to significant improvements over other baselines even with its MLP-based architecture. These improvements are more noteworthy for smaller datasets like Bavaria (BV). However, with the inclusion of graph attention networks, \axolotl captures the complex user-location dynamics better than MeLU.

\item Danser~\cite{danser} and NGCF~\cite{ngcf} perform comparable to meta-learning based baselines in some datasets -- \eg\, MA and WA. This is due to a sufficiently moderate dataset size to fuel their \textit{dual} graph neural networks. Danser~\cite{danser} particularly incorporates a reinforcement learning-based policy optimization which particularly leads to better modeling power albeit at the cost of more computation. However, for extremely limited-data regions and Precision@1 predictions, the input-data size alone is not sufficient to accurately train all parameters.

\item Despite Reline~\cite{reline} being the state-of-the-art multi-graph-based model for location recommendation, other methods that incorporate complex structures using dual-GCNs or meta-learning are able to easily outperform it, even under sparse data conditions.
\end{compactitem}

\noindent To sum up, our empirical analysis suggests the following: \begin{inparaenum}[(i)] \item the state-of-the-art models, including fine-tuning based transfer approaches, are not suitable for location recommendation in a limited-data region, \item \axolotl is a powerful recommender system not only for mobility networks with limited-data, but also in general, and \item for data-scarce regions, forcing embeddings to adapt with similar source-region clusters has significant performance gains. \end{inparaenum}

\subsection{Ablation Study}\label{axoablation}
\begin{figure}[t]
\centering
\begin{subfigure}{0.3\columnwidth}
  \centering
  \includegraphics[height=3.5cm]{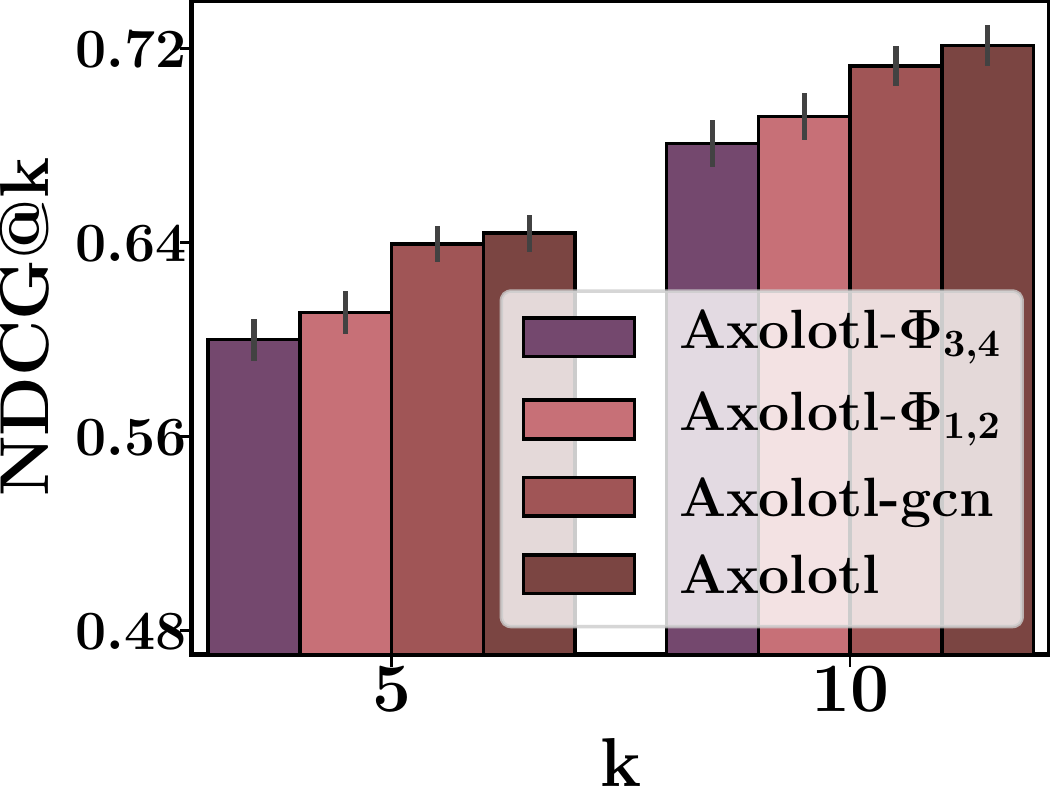}
  \caption{Washington (US)}
\end{subfigure}
\hfill
\begin{subfigure}{0.3\columnwidth}
  \centering
  \includegraphics[height=3.5cm]{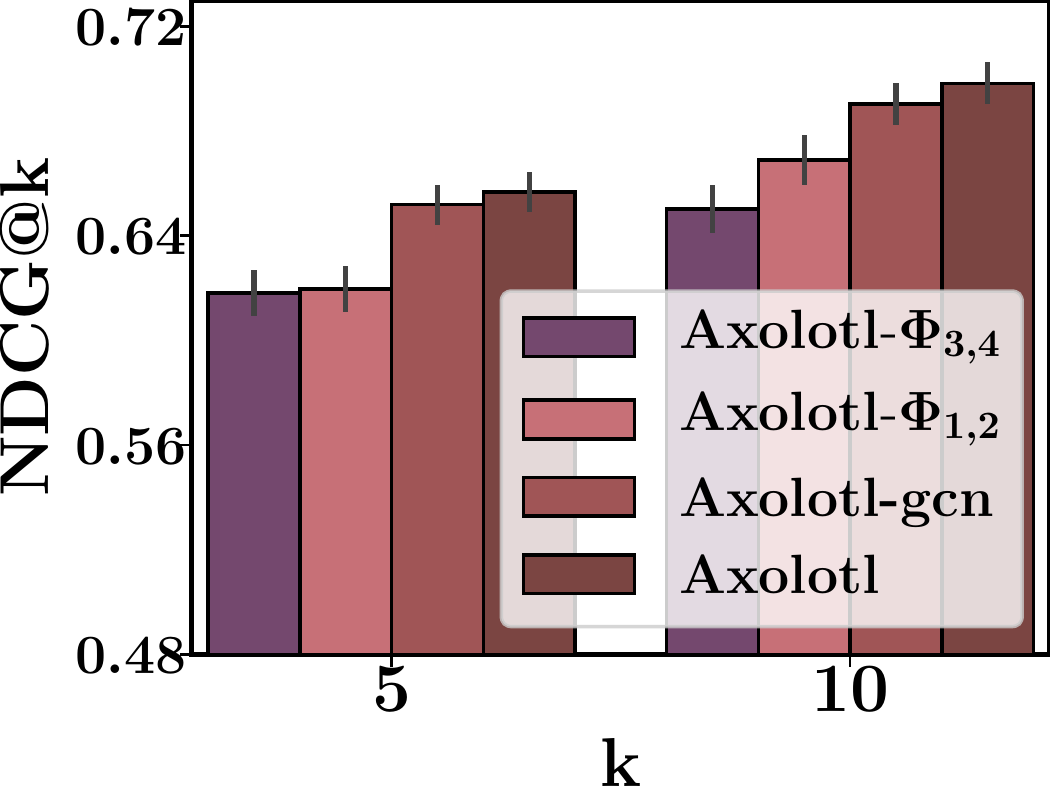}
  \caption{Aizu (JP)}
\end{subfigure}
\hfill
\begin{subfigure}{0.3\columnwidth}
  \centering
  \includegraphics[height=3.5cm]{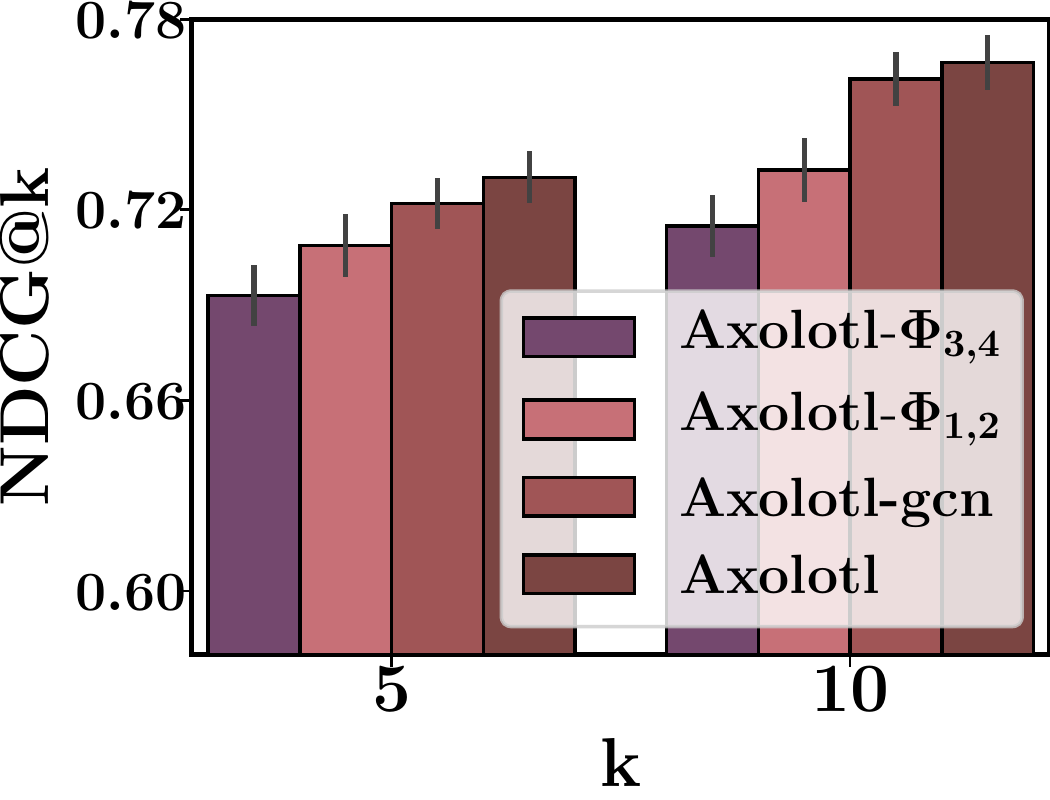}
  \caption{Berlin (DE)}
\end{subfigure}
\vspace{-3mm}
\caption{Ablation Study across different graph architectures possible in \axolotl along with a GCN variant.}
\label{axofig:ablation}
\vspace{-2mm}
\end{figure}

To address RQ2, we conduct an ablation study for two key contributions in \axolotl: user-GATs ($\Phi_1$ and $\Phi_2$) and location-GATs ($\Phi_3$ and $\Phi_4$). For estimating the contribution of user-GATs, we use meta-learning and alignment loss-based training only for $\Phi_1$ and $\Phi_2$. We denote this variant as \axolotl-$\Phi_{1,2}$. Similarly, for location-GATs, we use \axolotl-$\Phi_{3,4}$. We also include a GCN~\cite{gcn} based implementation of \axolotl denoted as \axolotl-gcn. From the results in Figure \ref{axofig:ablation}, we observe that \axolotl with joint training of user and location GATs has better prediction performance than \axolotl-$\Phi_{1,2}$ and \axolotl-$\Phi_{3,4}$. Interestingly, transferring across users leads to better prediction performance than transferring across locations. This could be attributed to a larger difference in the number of locations between source and target regions in comparison to the number of users. \axolotl-gcn has significant performance improvements over the preceding approaches, with \axolotl further having modest improvements due to weighted neighborhood aggregation.

\begin{figure}[t]
\centering
\begin{subfigure}{0.3\columnwidth}
  \centering
  \includegraphics[height=3.5cm]{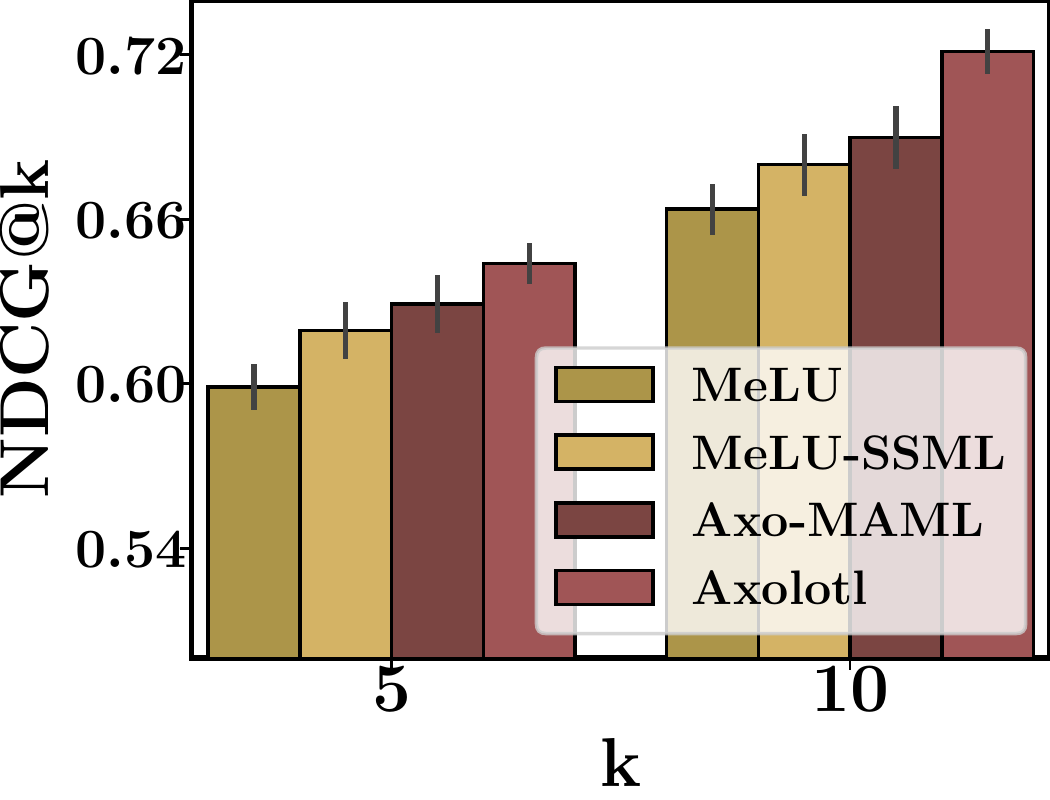}
  \caption{Washington (US)}
\end{subfigure}
\hfill
\begin{subfigure}{0.3\columnwidth}
  \centering
  \includegraphics[height=3.5cm]{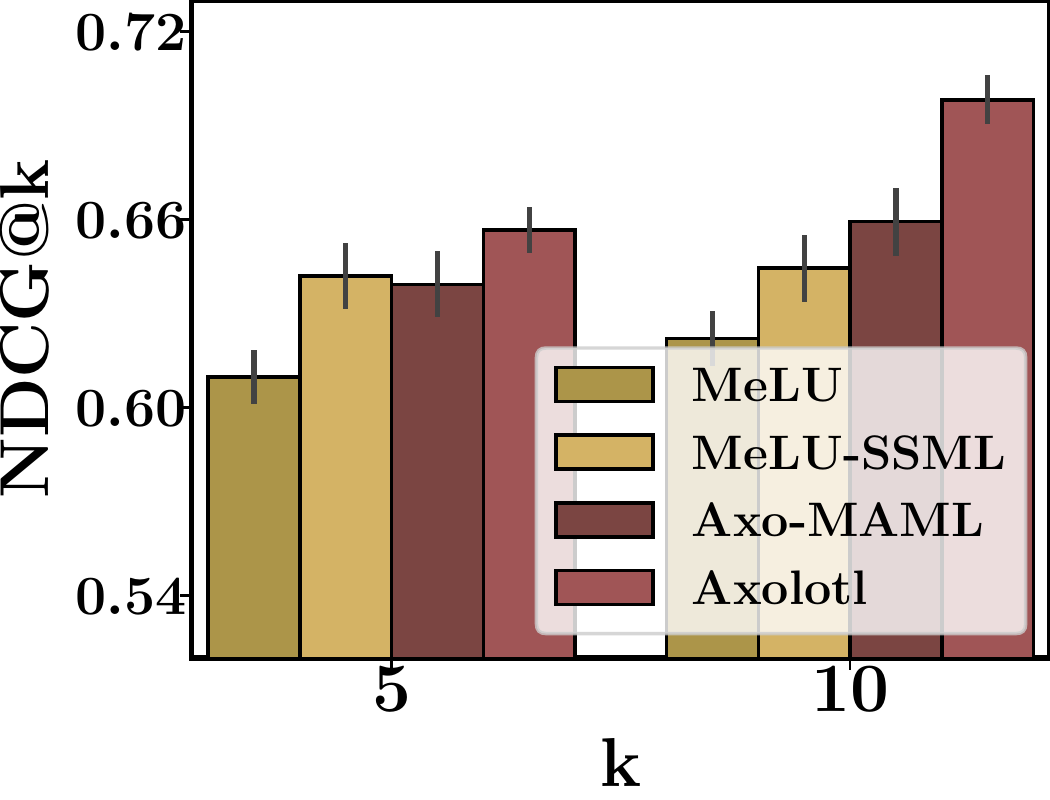}
  \caption{Aizu (JP)}
\end{subfigure}
\hfill
\begin{subfigure}{0.3\columnwidth}
  \centering
  \includegraphics[height=3.5cm]{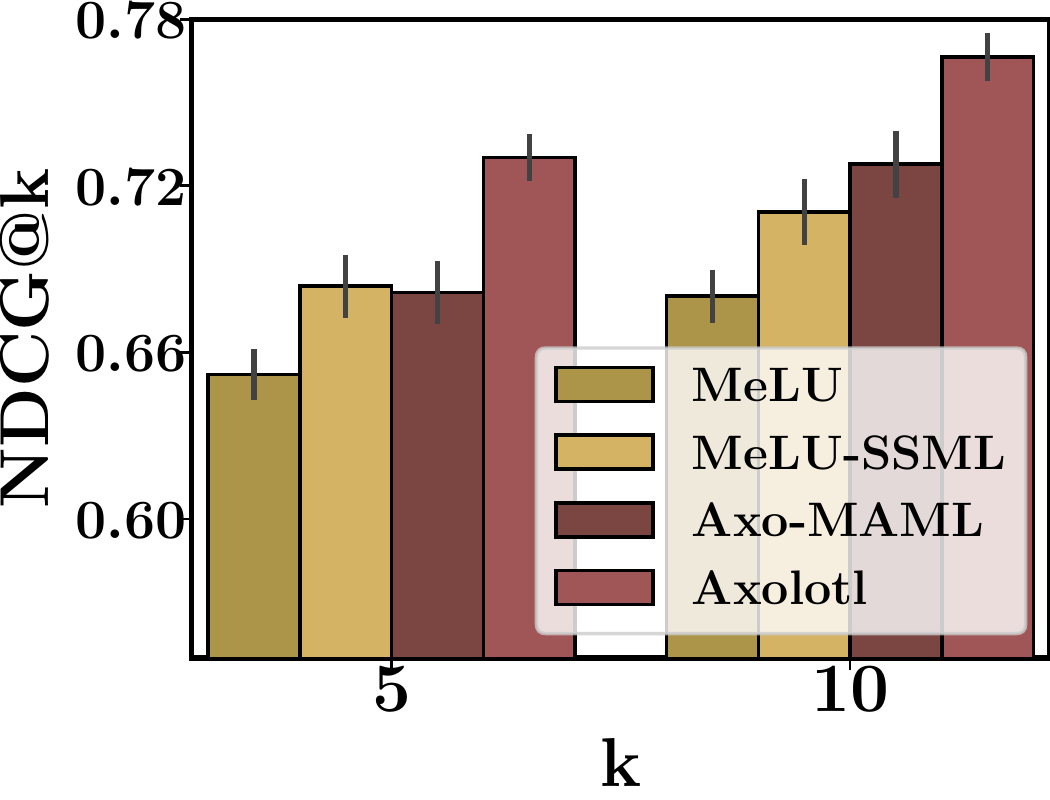}
  \caption{Berlin (DE)}
\end{subfigure}
\vspace{-3mm}
\caption{Contribution of novel spatio-social meta-learning in \axolotl and the corresponding gains over MeLU.}
\label{axofig:pnp}
\end{figure}

\noindent \textbf{Contribution of SSML:} To further assert the importance of our SSML (Section~\ref{axometa_part}), we compare the state-of-the-art MAML-based model, viz., MeLU, with our proposed learning procedure, and also include results after training \ours-basic with MAML. From the results given in Figure \ref{axofig:pnp}, we note that MeLU, when trained with SSML, easily outperforms its MAML-based counterpart. It demonstrates not only the effectiveness of the proposed learning method but also its versatility to be incorporated with other baselines. This claim is substantiated further by the poorer performance of \ours-basic with MAML over the complete \axolotl model. 

\subsection{Transfer of Weights across Regions}\label{axotran_weight}
\begin{figure}[t]
\centering
\begin{subfigure}{0.3\columnwidth}
  \centering
  \includegraphics[height=3.5cm]{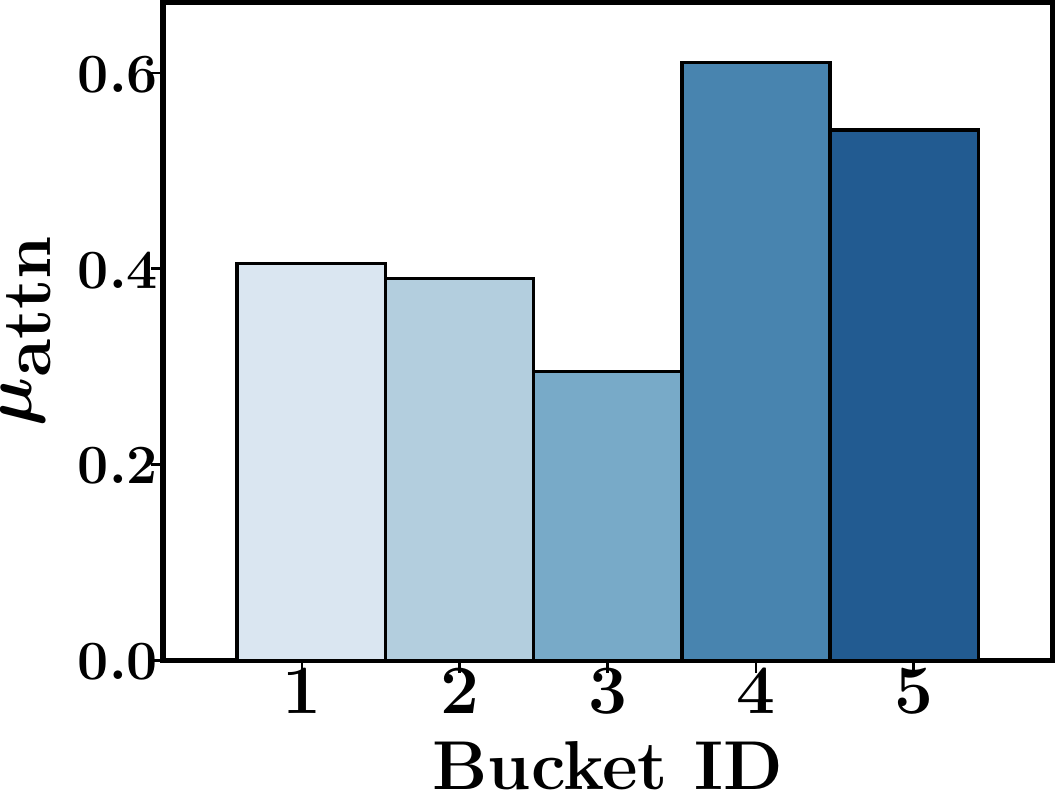}
  \caption{Washington (US)}
\end{subfigure}
\hfill
\begin{subfigure}{0.3\columnwidth}
  \centering
  \includegraphics[height=3.5cm]{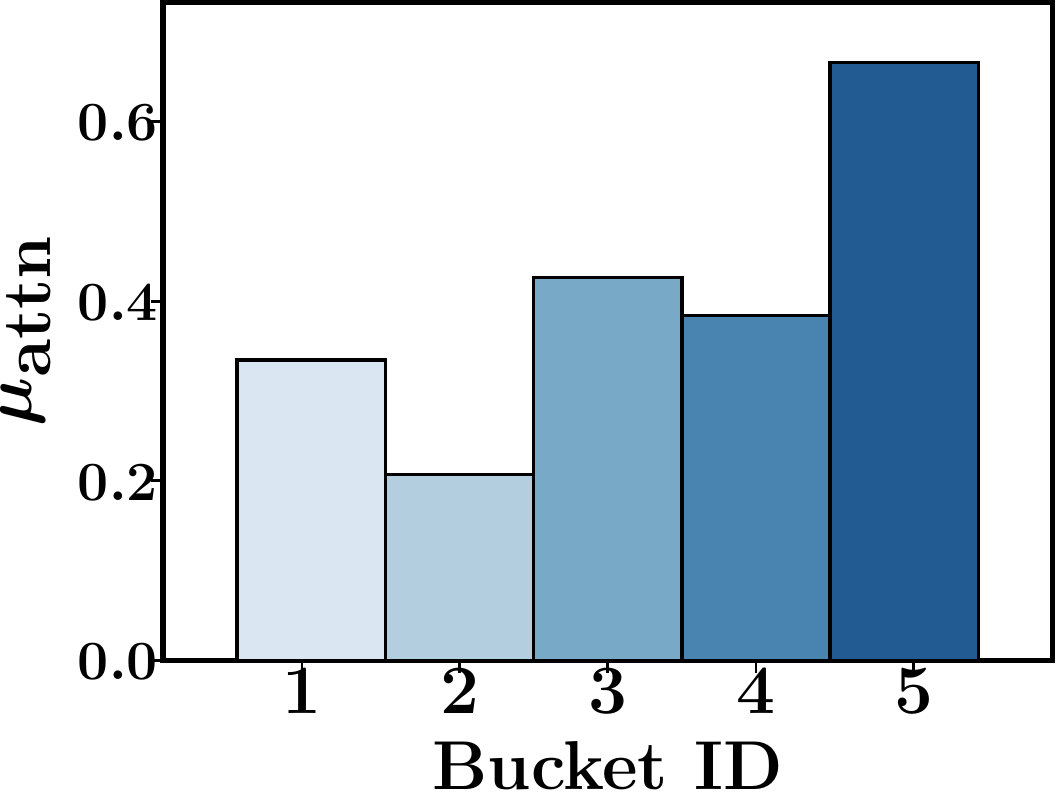}
  \caption{Aizu (JP)}
\end{subfigure}
\hfill
\begin{subfigure}{0.3\columnwidth}
  \centering
  \includegraphics[height=3.5cm]{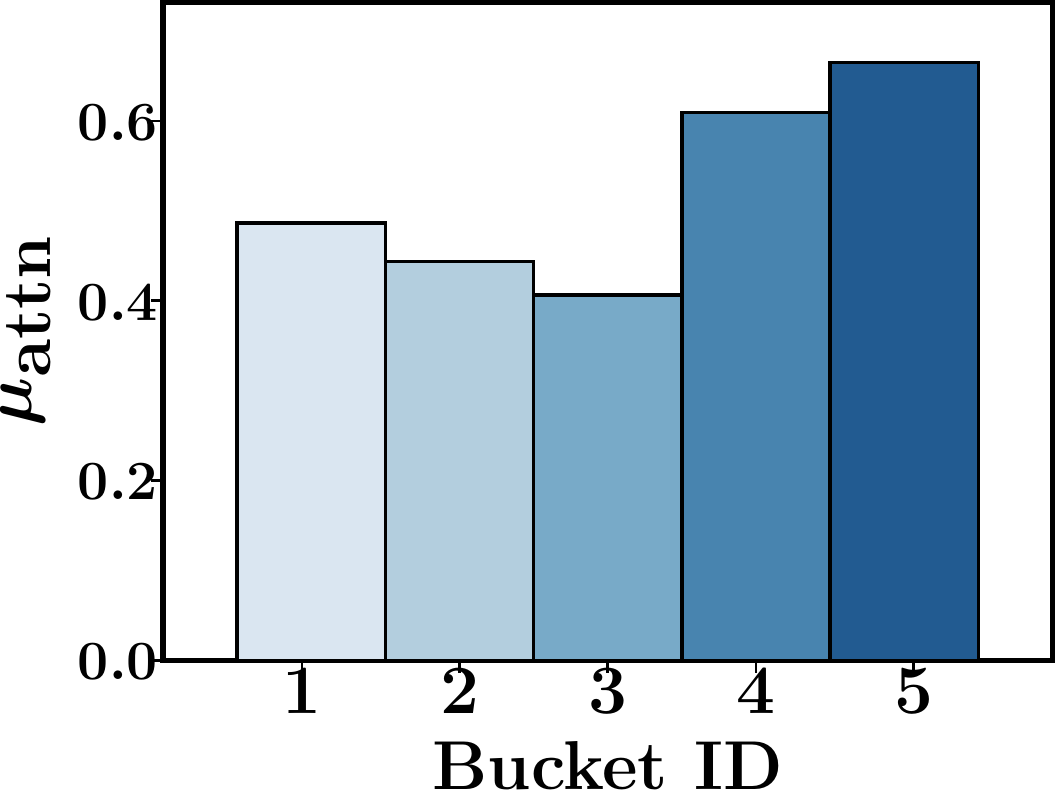}
  \caption{Berlin (DE)}
\end{subfigure}
\vspace{-3mm}
\caption{Bucket-wise Mean Attention Transfer Weights between Source and Target regions.}
\label{axofig:tran_weights}
\vspace{-2mm}
\end{figure}

Another important contribution we make is the cross-region transfer via cluster-based alignment loss. To address RQ3, we show that \axolotl encapsulates the cluster-wise analogy by plotting the attention-weights corresponding to the similarity between location clusters ($\bm{L}^{tgt}_c$ and $\bm{L}^{src}_c$). We quantify the similarity using \textit{Damerau-Levenshtein} distance~\cite{dldistance} across category distribution for all clusters in source and target regions. Later, we group them into \textit{five} equal buckets as per their similarity score, \ie, bucket-5 will have clusters with higher similarity as compared to other buckets. Figure \ref{axofig:tran_weights} shows the mean attention value across each bucket for all datasets. We observe that \axolotl is able to capture the increase in inter-cluster similarity by increasing its attention weights. This feature is more prominent for Aizu and comparable for Berlin.

\section{Conclusion}
\label{axosec:concl}
In conclusion, we developed a novel architecture called \axolotl that incorporates mobility data from other regions to design a location recommendation system for data-scarce regions. We also propose a novel procedure called the spatio-social meta-learning approach that captures the regional mobility patterns as well as the graph structure. We address the problems associated with an extremely data-scarce region and devise a suitable cluster-based alignment loss that enforces similar embeddings for communities of users and locations with similar dynamics. Experiments over diverse mobility datasets revealed that \axolotl is able to significantly improve over the state-of-the-art baselines for POI recommendation in limited data regions and even performs considerably better across datasets and data-rich source regions. 

%% file: chapters/006_reformd.tex
\newcommand{\nflow}{\mathcal{L}\mathcal{N}}
\newcommand{\org}{\text{src}}
\newcommand{\tgt}{\text{tgt}}


\section{Introduction}\label{reforintro}
There exists a high variability in mobility data volumes across different regions, which deteriorates the performance of spatial recommender systems that rely on region-specific data. In addition, the existing techniques to overcome spatial data-scarcity are limited by their inability to model the sequentially in POI visits by a user~\cite{axolotl, regiontrans, www, similarwww}. Addressing this problem is necessary as recent research has shown that accurate advertisements on Points-of-Interest (POI) networks, such as Foursquare and Instagram, can achieve up to 25 times the return-on-investment~\cite{fsqstats}. Consequently, predicting the time-evolving mobility of users, \ie, \emph{where} and \emph{when}, is of utmost importance to power systems relying on spatial data. 

In this chapter, we present our solution to overcome the drawbacks of data scarcity for sequential POI recommendations. Current approaches~\cite{locate, cho, lbsn3} overlook the temporal aspect of a recommender system as it involves modeling continuous-time check-in sequences -- which is challenging with standard neural architectures~\cite{rmtpp, thp, nhp}. The problem is further aggravated by the variation in volumes of mobility data across regions due to the growing awareness of personal data privacy~\cite{privacy, privacy2}. As highlighted in Chapter~\ref{chapter:axolotl}, there exists a high variability in mobility data volumes across different regions, which deteriorates the performance of spatial recommender systems that rely on region-specific data. Consequently, this scarcity of data affects the performance of all neural models.

In recent years, neural marked temporal point processes (MTPP) models have outperformed other neural architectures for characterizing asynchronous events localized in continuous time with applications including healthcare~\cite{rizoiu_sir}, finance~\cite{sahp, bacry}, and social networks~\cite{imtpp, nhp, thp}. These models have also been used in spatial networks and for predicting user mobility patterns. However, these models are either: (i) limited to predicting the time of user-location interactions, \ie, the time when the user will \cin next and do not predict the precise location~\cite{chandan}, (ii) restricted to \emph{one} dataset without a foreseeable way to easily utilize external information and disregard the opportunity to reuse trained parameters from external datasets~\cite{colab,deepjmt}. Thus, none of these approaches can be used for designing spatial mobility prediction models for limited data regions.

\subsection{Our Contribution}
We present \reformd(\textbf{Re}usable \textbf{F}lows f\textbf{or} \textbf{M}obility \textbf{D}ata), a novel transfer learning framework that learns spatial and temporal distribution of check-ins using normalizing flows(NFs) on a check-in-rich \emph{source} region and transfers them for efficient prediction in a check-in-scarce \emph{target} region~\cite{reformd}. Specifically, we consider the series of check-ins made by a user as her check-in sequence and model these sequences for all users from a region using a neural MTPP and learn the inter-check-in time interval and spatial-distance distributions as two independent NFs~\cite{flowbook, shakir}. To make the learned spatial and temporal NFs \emph{invariant} of the underlying region, we restrict our model to learn the distribution of inter-check-in time intervals and spatial distance. These features are unaffected by the network characteristics that vary across regions -- POI categories and user affinities towards these POIs. Therefore, these NFs can be easily extended for prediction in other mobility regions. The ability of NFs to provide faster sampling and closed-form training for continuous-time event sequences~\cite{intfree} makes them a perfect medium to transfer mobility information. Moreover, for transferring across regions, we cluster the check-in sequences of each region, with each cluster containing check-in sequences with \emph{similar} spatial and temporal check-in patterns and only transfer the parameters across these clusters. In summary, the key contributions we make via \reformd are three-fold:
\begin{compactitem}
\item We propose \reformd, a transfer-learning model for predicting mobility dynamics in check-in-scarce datasets by incorporating mobility parameters trained on a check-in-rich region.
\item We present a novel NF-based transfer over the MTPP that not only enables a faster sampling of time and distance features of next check-in but also achieves high performance even with limited fine-tuning on the target region.
\item Finally, we empirically show that \reformd outperforms the state-of-the-art models by up to 20\% and 23\% for check-in-category and time prediction and can easily be extended to product recommendation datasets.
\end{compactitem}

\section{Related Works}\label{reforrw}
In this section, we introduce key related works. It mainly falls into the following categories: 1) Sequential POI Recommendation; and 2) Temporal Point Processes with Spatial Data.

\subsection{Sequential POI Recommendation}
Recent sequential POI prediction models consider the check-in trajectory for each user as a sequence of events and utilize an RNN-based learning~\cite{cheng2013you, cara, lbsn3} with some variants that incorporate the spatial features as well~\cite{rnnlbsn, deepmove}. Another approach \cite{tribe} is a generic model for predicting user trajectories as well as the next product recommendation. Recent approaches for check-in time prediction are limited to a single dataset~\cite{deepjmt, chandan, colab}. They also model event times as random variables rather than sequential flows and, thus, cannot be used for transfer across regions.

\subsection{Marked Temporal Point Processes with Spatial Data}
MTPPs have emerged as a powerful tool to model asynchronous events localized in continuous time~\cite{daley2007introduction, hawkes}, which have a wide variety of applications, for \eg,  information diffusion, disease modeling, finance, etc. Driven by these motivations, in recent years, there has been a surge of works on TPPs~\cite{rizoiu_sir, imtpp, farajtabar2017fake}. Modeling the event sequences via a neural network led to further developments, including neural Hawkes process~\cite{nhp} and several other neural models of TPPs~\cite{xiaointaaai, xiao2017wasserstein, fullyneural}, but cannot incorporate heterogeneous features as in spatial networks. Moreover, recent works that deploy MTPP for predicting user mobility patterns are either: (i) limited to predicting the time of user-location interactions rather than actual locations~\cite{chandan}, (ii) restricted to \emph{one} dataset without a foreseeable way to easily utilize external information~\cite{colab}, or (iii) disregard the opportunity to reuse trained parameters from external datasets by jointly embedding the check-in and time distributions~\cite{deepjmt}. Thus, none of these approaches can be used for designing mobility prediction models for limited data regions. The approach most similar to our model is~\cite{intfree} that learns the inter-event time intervals using NFs, but ignores the spatial dynamics and is limited to a single data source.

\section{Problem Setup}\label{reforproblem}
We consider the mobility records for two regions with non-overlapping locations and users, \textit{source} and \textit{target} as $\ds{D}^{\org}$ and $\ds{D}^{\tgt}$ respectively. These notations are consistent with Chapter~\ref{chapter:axolotl}. For any region, we represent a user trajectory as a sequence of check-ins represented by $\cm{S}_k=\{e_i=(c_i,t_i, d_i) | i \in[k] , t_i<t_{i+1}, d_i<d_{i+1}\}$, where $t_i\in\mathbb{R}^+$ is the check-in time, $d_i\in\mathbb{R}^+$ is the total distance traveled, and $c_i\in \mathcal{C}$ is a discrete category of the $i$-th check-in with $\cm{C}$ as the set of all categories, and $\cm{S}_k$ denotes the first $k$ check-ins. We represent the inter-check-in times and distances as, $\Delta_{t,k} = t_{k}-t_{k-1}$ and $\Delta_{d,k} = d_{k}-d_{k-1}$ respectively and model their distribution using NFs. Our goal is to capture these region invariant dynamics in the source region for mobility prediction in the target region, \ie given the check-in sequence for the target region, $\cm{S}^{\tgt}_K$ and the MTPP trained on source region, we aim to predict the time and category of the \textit{next} check-in, $e^{\tgt}_{K+1}$.

\section{Model Description}\label{reformodel}
We divide the working of \reformd into two parts: (i) the neural MTPP to capture mobility dynamics specific to a region, and (ii) the transfer of NFs trained on the source region to the target region. The underlying framework of \reformd is given in Figure~\ref{reforfig:model}.

\subsection{Region-Specific MTPP}
We model the check-in sequences using an MTPP that we build on a recurrent neural network (RNN). The RNN is used to obtain time-conditioned vector representation of sequences, as in~\cite{rmtpp, nhp, fullyneural}. Later, via these embeddings, we estimate the mark distribution and inter-event time and space densities using a three-stage architecture:
\begin{compactitem}
\item[(1)] \textbf{Input stage:}
In this stage, we represent the incoming check-in at index $k$, $e_k$ using a suitable vector embedding, $\bs{v}_k$ as:
\begin{equation}
\bs{v}_k = \bs{w}_{c} c_k + \bs{w}_{t} \Delta_{t,k} + \bs{w}_{d}\Delta_{d,k} + \bs{b}_v,
\end{equation}
where $\bs{w}_{\bullet}, \bs{b}_{\bullet}$ are trainable parameters,  $\bs{v}_k$ denotes the vector embedding for check-in $e_k$, and $c_k$ denotes the category of the $k$-th \cin respectively. Here, $\Delta_{t,k} = t_{k}-t_{k-1}$ and $\Delta_{d,k} = d_{k}-d_{k-1}$ denote the difference between \cins in terms of time and distance respectively. 

\item[(2)] \textbf{Update stage:} In this stage, we update the hidden state representation of the RNN to include the current check-in $e_k$ as:
\begin{equation}
\bs{s}_k = \tanh(\bs{G}_{s} \bs{s}_{k-1} + \bs{G}_{v} \bs{v}_{k} + \bs{g}_{t} \Delta_{t,k} + \bs{g}_{d} \Delta_{d,k} + \bs{b}_s),
\end{equation}
where $\bs{G}_{\bullet}, \bs{g}_{\bullet}, \bs{b}_{\bullet}$ are trainable parameters and $\bs{s}_k$ denotes the RNN hidden state, \ie, a \textit{cumulative} embedding for all previous check-ins till the current time $t_k$.

\begin{figure}[t]
\centering
\includegraphics[height=5cm]{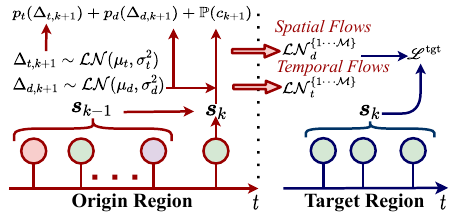}
\vspace{-0.4cm}
\caption{\label{reforfig:model} Architecture of \reformd with flow-based transfer between source region (red) and target region (blue).}
\vspace{-0.2cm}
\end{figure}

\item[(3)] \textbf{Output stage:}
Given the trajectory embedding $\bs{s}_k$, we predict the \textit{next} check-in time and the check-in category. Unlike~\cite{rmtpp, nhp} that learn the time distribution using an \textit{intensity}-based formulation using the RNN hidden state, we model the \textit{density} of arrival times using a \textit{lognormal}~\cite{intfree} flow denoted as $p_t(\Delta_{t, k+1})$ conditioned on $s_k$. More details are given in Chapter~\ref{chapter:background}.
\begin{equation}
p_t(\Delta_{t, k+1} | \bs{s}_k) = \texttt{LogNormal} \big(\mu_t(\bs{s}_k), \sigma^2_t(\bs{s}_k)\big),
\label{reforeqn:flow}
\end{equation}
with $[\mu_t(\bs{s}_k), \sigma^2_t(\bs{s}_k)] = [\bs{W}_{\mu} \bs{s}_k + \bs{\mu}_t, \bs{W}_{\sigma^2} \bs{s}_k + \bs{\sigma^2}_t]$ denote the \textit{mean} and \textit{variance} of the time distribution. Such a formulation reduces model complexity and facilitates faster training and sampling in a closed-form~\cite{intfree}.

To predict the time of the next check-in, we sample the probable time difference between the current and the next check-in as $\Delta_{t, k+1} \thicksim \nflow_t\big(\mu_t(\bs{s}_k), \sigma^2_t(\bs{s}_k)\big)$, where $\nflow_t$ denotes the learned lognormal parameters. The time of the next check-in is the sum of the sampled time difference and the \textit{current} check-in time, $\widehat{t_{k+1}} = t_k + \Delta_{t, k+1}$. Similar to the temporal flow, we also model the inter-check-in density of spatial distances using a lognormal denoted as $p_d(\Delta_{d, k+1} | \bs{s}_k)$. We interpret this distribution as the \textit{spatial} flow for a region.

\end{compactitem}

The inter-location spatial distance plays a crucial role in determining the next POI~\cite{cho, cara}. Unlike time, the distances between two check-in locations are \textit{unchanged} throughout the data. Previous approaches~\cite{rmtpp, nhp} ignore these spatial features and rely solely on the past check-in-categories. This affects the prediction accuracy as the travel distance to the location is a critical element for determining POI recommendations.
Moreover, in a sequential setting, the distance that the user will travel for her next check-in is not known. Our MTPPs, being generative models, and spatial flows overcome this drawback as we can sample the probable travel distance for the next check-in from the spatial flow as $\Delta_{d, k+1} \thicksim \nflow_d\big(\mu_d(\bs{s}_k), \sigma^2_d(\bs{s}_k)\big)$. Then, for predicting the next check-in, we use the sampled distance $\Delta_{d, k+1}$ and RNN hidden state $\bs{s}_k$ via attention-weighted embedding~\cite{attention}.
\begin{equation}
\bs{s}^*_k = \bs{s}_k + \alpha \cdot \bs{w}_f \Delta_{d, k+1},
\label{reforeqn:fusion}
\end{equation}
where $\alpha, \bs{w}_f$ denote the attention weight, a trainable parameter, and $\bs{s}^*_k$ denotes the \textit{updated} hidden state. We then predict the next check-in category as:
\begin{equation}
\mathbb{P}(c_{k+1} = c| \bs{s}^*_k) = \frac{\exp (\bs{V}_{s, c} \bs{s}^*_k + \bs{b}_{s, c})}{\sum_{\forall c' \in \cm{C}} \exp (\bs{V}_{s, c'} \bs{s}^*_k + \bs{b}_{s, c'})},
\end{equation}
where $\bs{V}_{s, \bullet}, \bs{b}_{s, \bullet}$ are trainable parameters and $\bullet$ denotes the entry corresponding to a category. $\mathbb{P}(c_{k+1} = c| \bs{s}^*_k)$ denotes the probability of next check-in being of category $c$ with $c \in \cm{C}$.

\noindent \textbf{Optimization:}
Given the set of all sequences $\cm{S}$ for a region $\ds{D}$, we maximize the joint likelihood for the next check-in, the lognormal density distribution of spatial and temporal normalizing flows.
\begin{equation}
\mathscr{L} = \sum_{\forall \cm{S}}\sum_{k = 1}^{|\cm{S}|} \log \big( \mathbb{P}(c_{k+1}|\bs{s}^*_k) \cdot p_t(\Delta_{t, k+1} | \bs{s}_k) \cdot p_d(\Delta_{d, k+1} | \bs{s}_k) \big ).
\label{reforeqn:likelihood}
\end{equation}
where $\mathscr{L}$ denotes the joint likelihood, which we represent as the sum of the likelihoods for all user sequences. We learn the parameters of \reformd using Adam~\cite{adam} optimizer.

\subsection{Flow-based Transfer}
For transferring the mobility parameters across the regions, we follow the standard transfer learning procedure of training exclusively on the \textit{source} region and then fine-tuning for the target region~\cite{transfer,ltran}. However, the affinity of a user towards a POI evolves with time~\cite{cho, locate}. For example, a POI with frequent user check-ins during the summer season might not be an attractive option in winter. We include these insights by training multiple independent normalizing flows, each for spatial and temporal densities. Specifically, we cluster the check-in sequences in the source region into $\cm{M}$ perfectly equal clusters based on the \textit{median} of the occurrence times for all the check-ins. Later, for each cluster of sequences, we train spatial and temporal flows independently. Here, our net likelihood changes to include the sum of all $\cm{M}$ likelihoods, $\mathscr{L} = \sum_{i = 1}^{\cm{M}} \mathscr{L}_i$, where $\mathscr{L}_i$ is the joint likelihood for trajectories in clusters $\cm{M}$.

\noindent As in the source region, we divide the user trajectories in the target region as well into $\cm{M}$ clusters, and for trajectories in target-cluster $m^{\tgt}_i$, we attentively factor the spatial and temporal flows corresponding to source-cluster $m^{\org}_i$. Mathematically, for the temporal flows in the target region, our density of arrival times changes to:
\begin{equation}
[\mu_t(\bs{s}_k), \sigma^2_t(\bs{s}_k)]^{\tgt} = [\bs{W}_{\mu} \bs{s}_k + \bs{b}_{\mu} + \phi_t \bs{\mu}^{\org}_t, \bs{W}_{\sigma} \bs{s}_k + \bs{b}_{\sigma} + \phi_t \bs{\sigma}^{\org}_t],
\end{equation}
where $\bs{s}_k, \phi_t, \bs{\mu}^{\org}_t, \bs{\sigma}^{\org}_t$ denotes the hidden state representation for the target region, attention parameter for temporal flow and the learned flow parameters of \textit{mean} and \textit{variance} for cluster $m_i$ in the source region. Similarly, our spatial flows for the target region include the source flow parameters with an attention parameter $\phi_d$. For faster convergence, we share $\phi_t$ and $\phi_d$ across all $\cm{M}$. Other model components are the same as in the source region and we maximize the joint likelihood for the target region as in Equation~\ref{reforeqn:likelihood}.

We highlight that our choice to divide the sequences based on \textit{median} of user trajectories rather than the individual check-in locations is driven by the following technical point: in the latter case, the \emph{net flow} --be it spatial or temporal-- would be the sum of lognormal flows for each set. Such a formulation is undesirable since the result is neither \emph{closed} nor does it remain a lognormal~\cite{logsum}, thus requiring involved techniques to approximate them~\cite{barouch1986sums}, which we would like to explore in future work. However, with the current formulation, we can learn the parameters of different flows independently.

\section{Evaluation}\label{reforeval}
In this section, we conduct an empirical evaluation of \reformd. Specifically, we address the following research questions.
\begin{compactitem}
\item[\textbf{RQ1}] Can \reformd outperform state-of-the-art baselines for time and check-in prediction?
\item[\textbf{RQ2}] What is the advantage of transferring via normalizing flows?
\item[\textbf{RQ3}] Can we extend \reformd for non-spatial datasets?
\end{compactitem}
For evaluating mobility prediction, we consider six POI datasets from the US and Japan~\cite{lbsn2vec}. All our models are implemented in Tensorflow on an NVIDIA Tesla V100 GPU.

\begin{table}[t!]
\caption{Statistics of datasets used in this chapter. The source region columns are highlighted, followed by target regions, and are partitioned based on the country of source (the US and Japan).}
\label{refortab:data}
\vspace{-0.4cm}
\centering
\small
\begin{tabular}{l|accc|accc}
\toprule
\textbf{Property} & \textbf{NY} & \textbf{MI} & \textbf{NE} & \textbf{VI} & \textbf{TY} & \textbf{AI} & \textbf{CH} & \textbf{SA}\\
\hline
\#Users or \#Sequences ($|\cm{S}|$) & 25.6k & 6.7k & 4.1k & 6.5k & 32.1k & 10.9k & 7.5k & 11.4k\\
\#Categories ($|\cm{C}|$) & 403 & 364 & 311 & 357 & 376 & 319 & 286 & 289\\
Mean Length ($\mu_{|\cm{S}|}$) & 57.17 & 66.21 & 48.56 & 56.33 & 61.72 & 56.60 & 63.60 & 53.08\\
\bottomrule
\end{tabular}
\end{table}

\subsection{Experimental Settings}
\xhdr{Dataset Description:}
We use POI data from Foursquare~\cite{lbsn2vec} in the United States(US) and Japan(JP). For each country, we construct four datasets: one with large check-in data and three with limited data. The statistics of all datasets are given in Table \ref{refortab:data} with each acronym denoting the following region: (i) NY: New York(US), (ii) MI: Michigan(US), (iii) NV: Nevada(US), (iv) VI: Virginia(US), (v) TY: Tokyo(JP), (vi) CH: Chiba(JP), (vii) SA: Saitama(JP) and (viii) AI: Aichi(JP). We consider NY and TY as the source regions and MI, NV, VI, CH, SA, and AI as the corresponding target regions. For each region, we consider the time of check-in and category as event time and mark and normalize the times based on the minimum and maximum event times. We set the embedding and RNN hidden dimension to $64$ and $\cm{M} = 3$ for all our experiments. 

\xhdr{Evaluation Protocol:}
We split each stream of, say $N$ check-ins $\cm{S}_N$ into training and test set, where the training set (test set) consists of the first 80\% (last 20\%) check-in. We evaluate models using standard metrics~\cite{rmtpp} of (i) mean absolute error (MAE) of predicted and actual check-in times, $\frac{1}{|\cm{S}|}\sum_{e_i\in \cm{S}}[|t_i-\widehat{t}_i|]$ and (ii) mark (check-in category) prediction accuracy (MPA), \ie, $\frac{1}{|\cm{S}|}\sum_{e_i\in \cm{S}} \#(c_i=\widehat{c}_i)$. Here $\widehat{t_i}$ and $\widehat{c_i}$ are the predicted time and category of the $i$-th check-in. Moreover, the clustering of sequences into different sets is done based solely on the training data, and using these thresholds, we assign clusters to sequences in the test data.

\xhdr{Baseline Implementation Details}
We use similar baseline model implementations as in Chapter~\ref{chapter:imtpp}. 
For RMTPP, we set hidden dimension, and the back-propagation through time parameter (BPTT) is selected among $\{32, 64\}$ and $\{20, 50\}$, respectively. For THP, and SAHP, we set the number of attention heads as $2$, hidden, key-matrix and value-matrix dimensions are selected among $\{32, 64\}$. If applicable, for each model, we use a dropout of $0.1$. For PFPP, we set $\gamma = 1$ and use a similar procedure to calculate the embedding dimension as in the THP. All other parameter values are the ones recommended by the respective authors.

\vspace{-0.3cm}
\subsection{Baselines}
We compare the performance of \reformd with the following state-of-the-art methods:
\begin{compactitem}
\item \textbf{NHP~\cite{nhp}}: Models an MTPP using continuous-time LSTMs for capturing the temporal evolution of sequences.
\item \textbf{RMTPP~\cite{rmtpp}}: A recurrent neural network that models time differences to learn a representation of the past events.
\item \textbf{SAHP~\cite{sahp}}: A self-attention model to learn the temporal dynamics using an aggregation of historical events.
\item \textbf{THP~\cite{thp}}: Extends the transformer model~\cite{transformer} to include the \textit{conditional} intensity of event arrival and the inter-mark influences.
\end{compactitem}
We omit comparison with other continuous-time models~\cite{fullyneural, intfree, xiao2017wasserstein, xiaointaaai, hawkes} as they have already been outperformed by these approaches.

\begin{table}[t]
\caption{Performance of all the methods in terms of mean absolute error (MAE) on the 20\% test set. Numbers with bold font (boxes) indicate the best (second best) performer. Results marked \textsuperscript{$\dagger$} are statistically significant (two-sided Fisher's test with $p \le 0.1$) over the best baseline.}
\vspace{-3mm}
\centering
\small
\begin{tabular}{l|cccccc}
\toprule
\textbf{Dataset} & \multicolumn{6}{c}{\textbf{Mean Absolute Error (MAE)}} \\ \hline
$\ds{D}^{\org} \rightarrow \ds{D}^{\tgt}$ & NY $\rightarrow$ MI & NY $\rightarrow$ NE & NY $\rightarrow$ VI & TY $\rightarrow$ AI & TY $\rightarrow$ CH & TY $\rightarrow$ SA \\ \hline \hline
NHP~\cite{nhp} & 0.0920 & 0.1710 & 0.1482 & 0.1146 & 0.1217 & 0.1288 \\
RMTPP~\cite{rmtpp} & \underline{0.0817} & \underline{0.1581} & \underline{0.1360} & \underline{0.1058} & \underline{0.1162} & \underline{0.1205} \\
SAHP~\cite{sahp} & 0.1132 & 0.1958 & 0.1705 & 0.1671 & 0.1696 & 0.1574\\
THP~\cite{thp} & 0.0983 & 0.1735 & 0.1652 & 0.1445 & 0.1426 & 0.1468\\
\reformd & \textbf{0.0672}\textsuperscript{$\dagger$} & \textbf{0.1317}\textsuperscript{$\dagger$} & \textbf{0.1089}\textsuperscript{$\dagger$} & \textbf{0.0858}\textsuperscript{$\dagger$} & \textbf{0.0897}\textsuperscript{$\dagger$} & \textbf{0.0973}\textsuperscript{$\dagger$}\\\hline
$\Delta$ (\%) & 17.74 & 16.69 & 19.92 & 18.90 & 22.80 & 19.25\\
\bottomrule
\end{tabular}
\label{refortab:main_mae}
\vspace{3mm}
\caption{Performance in terms of mark prediction accuracy (MPA) on the 20\% test set. Numbers with bold font (boxes) indicate the best (second best) performer. Results marked \textsuperscript{$\dagger$} are statistically significant (two-sided Fisher's test with $p \le 0.1$) over the best baseline.}
\centering
\small
\begin{tabular}{l|cccccc}
\toprule
\textbf{Dataset} & \multicolumn{6}{c}{\textbf{Mark Prediction Accuracy (MPA)}} \\ \hline
$\ds{D}^{\org} \rightarrow \ds{D}^{\tgt}$ & NY $\rightarrow$ MI & NY $\rightarrow$ NE & NY $\rightarrow$ VI & TY $\rightarrow$ AI & TY $\rightarrow$ CH & TY $\rightarrow$ SA \\ \hline \hline
NHP~\cite{nhp} & 0.1745 & 0.1672 & 0.1348 & 0.2162 & 0.4073 & 0.3820 \\
RMTPP~\cite{rmtpp} & 0.1761 & \underline{0.1684} & 0.1377 & 0.2293 & \underline{0.4250} & 0.4036 \\
SAHP~\cite{sahp} & 0.1587 & 0.1529 & 0.1303 & 0.1968 & 0.3864 & 0.3943 \\
THP~\cite{thp} & \underline{0.1793} & 0.1545 & \underline{0.1493} & \underline{0.2361} & 0.4229 & \underline{0.4057} \\
\reformd & \textbf{0.2159}\textsuperscript{$\dagger$} & \textbf{0.1868}\textsuperscript{$\dagger$} & \textbf{0.1631}\textsuperscript{$\dagger$} & \textbf{0.2588}\textsuperscript{$\dagger$} & \textbf{0.4474} & \textbf{0.4208}\\
\hline
$\Delta$ (\%) & 20.41 & 10.92 & 9.24 & 9.61 & 5.27 & 3.72 \\
\bottomrule
\end{tabular}
\label{refortab:main_mpa}
\vspace{-2mm}
\end{table}

\subsection{Prediction Performance}
Here we address RQ1, \ie, the ability of \reformd to model the dynamics of a check-in sequence. Tables~\ref{refortab:main_mae} and~\ref{refortab:main_mpa} summarize the results, which sketches the comparative analysis in terms of mean absolute error (MAE) on time and mark prediction accuracy (MPA) of the next \cin, respectively. Through the results, we make the following observations:
\begin{compactitem}
\item \reformd consistently yields the best performance on all the datasets. In particular, it improves over the strongest baselines by 10\% and 19\% for category and time prediction respectively. These results indicate the importance of spatial and temporal flow-based transfer from external data for prediction in limited-data regions. RMTPP~\cite{rmtpp} is the second-best performer in terms of MAE of time prediction almost for all the datasets.


\item  As per the results, we also establish that the enhancement in terms of performance gain due to transfer learning is more significant for mark prediction rather than time prediction. This is due to the consistent performance gains over all baselines in MPA; however, for MAE, the performance gap narrows. 
\end{compactitem}

We also highlight that there exists a performance gap between NHP and RMTPP. Recent research~\cite{li2018learning} has shown that these models show similar performances. However, these performances can be obtained after extensive parameter tuning for NHP, which we consider beyond the scope of this work.

\begin{figure}[t]
\centering
\begin{subfigure}[b]{0.45\columnwidth}
\centering
\includegraphics[height=4cm]{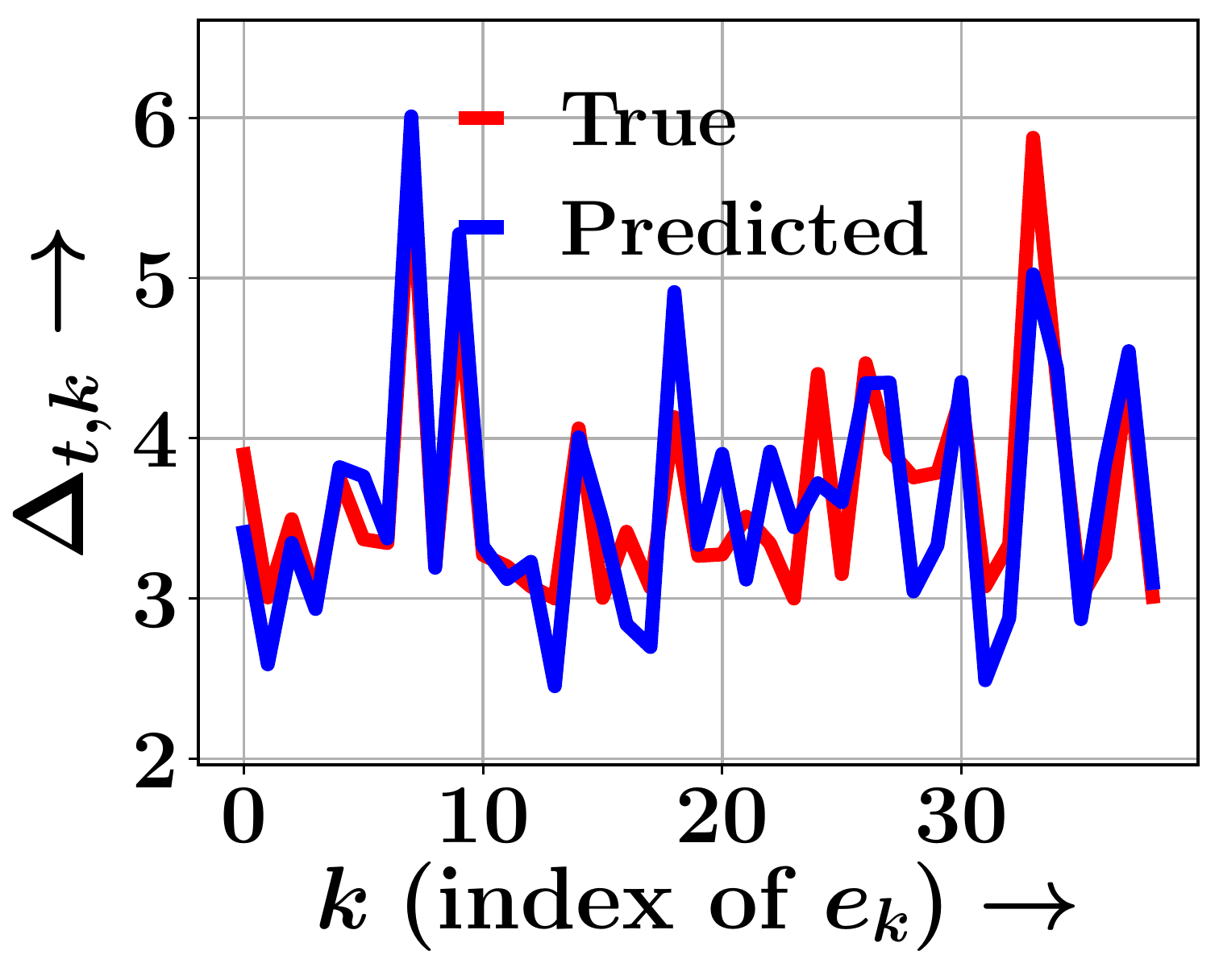}
\caption{Virginia}
\end{subfigure}
\hspace{1cm}
\begin{subfigure}[b]{0.45\columnwidth}
\centering
\includegraphics[height=4cm]{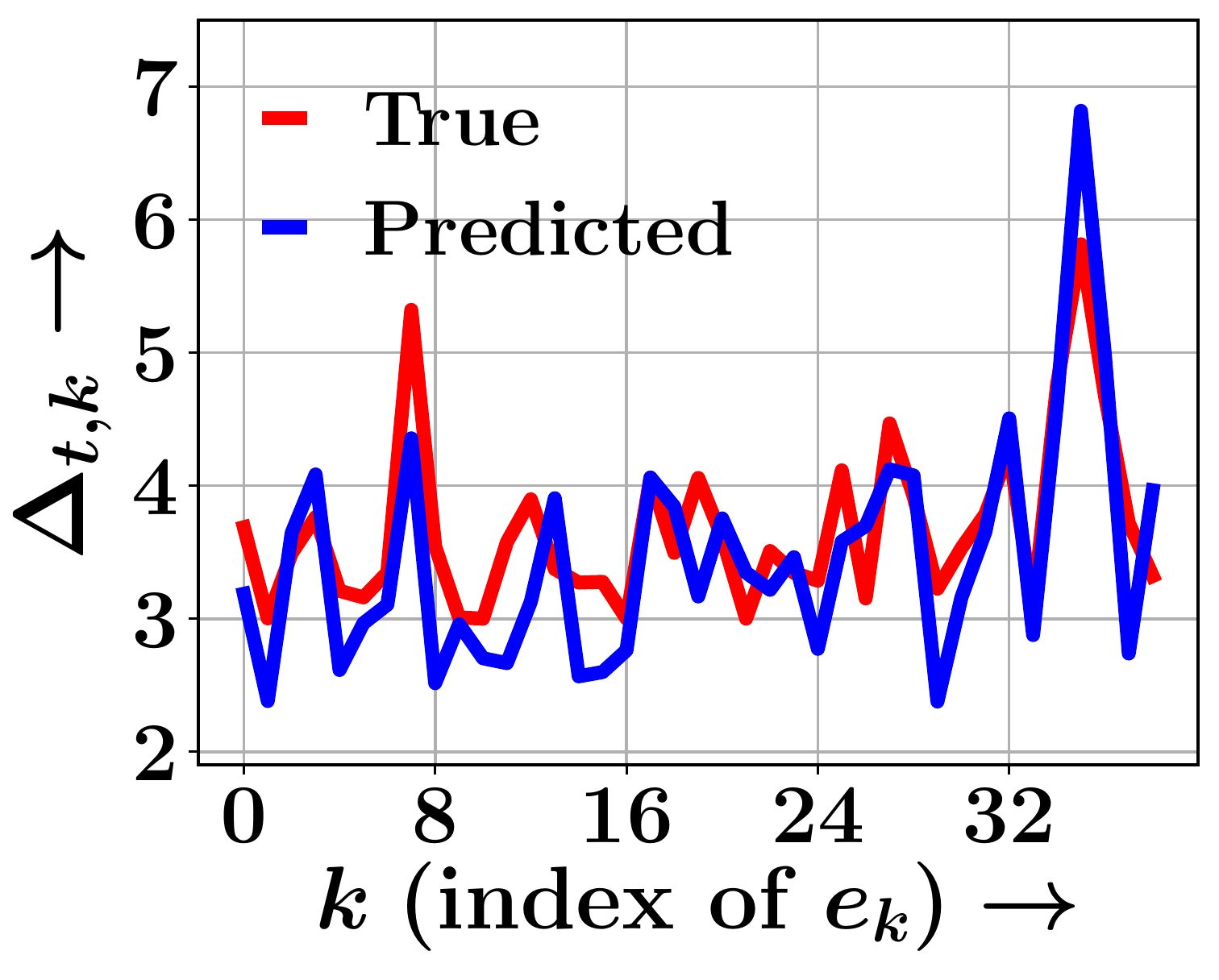}
\caption{Aichi}
\end{subfigure}
\vspace{-4mm}
\caption{\label{reforfig:qualitative} Real life \textit{true} and \textit{predicted} inter-arrival times $\Delta_{t,k}$ of different events $e_k$ for (a) Virginia and (b) Aichi.}
\end{figure}

\begin{table}[t]
\caption{Prediction performance of all the methods for product recommendation in Amazon datasets. Results marked \textsuperscript{$\dagger$} are statistically significant as in Tables ~\ref{refortab:main_mae} and ~\ref{refortab:main_mpa}.}
\vspace{-0.4cm}
\small
\centering
\begin{tabular}{l|cc|cc}
\toprule
& \multicolumn{2}{c|}{\textbf{Mark Prediction Accuracy}} & \multicolumn{2}{c}{\textbf{Mean Absolute Error}} \\ \hline
$\ds{D}^{\org} \rightarrow \ds{D}^{\tgt}$ & DM $\rightarrow$ AP & DM $\rightarrow$ BY & DM $\rightarrow$ AP & DM $\rightarrow$ BY\\ \hline
NHP~\cite{nhp}& 0.8773 & 0.5711 & 0.0903 & 0.1795\\
RMTPP~\cite{rmtpp} & 0.8975 & 0.5530 & \underline{0.0884} & \underline{0.1758}\\
SAHP~\cite{sahp} & 0.8931 & 0.5517 & 0.1439 & 0.2214\\
THP~\cite{thp} & \underline{0.9084} & \underline{0.5879} & 0.1253 & 0.2035\\
\reformd & \textbf{0.9129} & \textbf{0.6035} & \textbf{0.0756}\textsuperscript{$\dagger$} & \textbf{0.1564}\textsuperscript{$\dagger$} \\ \hline
$\Delta$ (\%) & 0.49 & 2.65 & 14.47 & 11.03\\
\bottomrule
\end{tabular}
\label{refortab:item}
\end{table}

\subsection{Qualitative Analysis} 
We also perform a qualitative analysis to demonstrate how \reformd is able to model the check-in time distribution. For this, we plot the actual inter-check-in time differences and the difference time predicted by \reformd in Figure~\ref{reforfig:qualitative} for Virginia and Aichi datasets. From the results, we note that the predicted inter-arrival times closely match with the true inter-arrival times, and \reformd is even able to capture large time differences (peaks). This reinforces the ability of \reformd to learn the temporal dynamics of a check-in sequence. Moreover, it demonstrates our claim that using external data can enhance recommendation performance. For other datasets, we noted that they displayed similar trends. 

\textbf{Runtime:} Here, we report on the run-time of \reformd to verify its applicability in real-world settings. Specifically, for all datasets, the times for training on the source and, later on, target regions are within 3 hours, thus within the range for practical deployment. In addition to shorter training times that are feasible for deployment, we further highlight that these values are mainly due to the inefficient CPU-based batch sampling. With GPU-based sampling alternatives, this run-time can be significantly improved, which we consider for future work.

\begin{figure}[t]
\centering
\begin{subfigure}[b]{0.45\columnwidth}
\centering
\includegraphics[height=4cm]{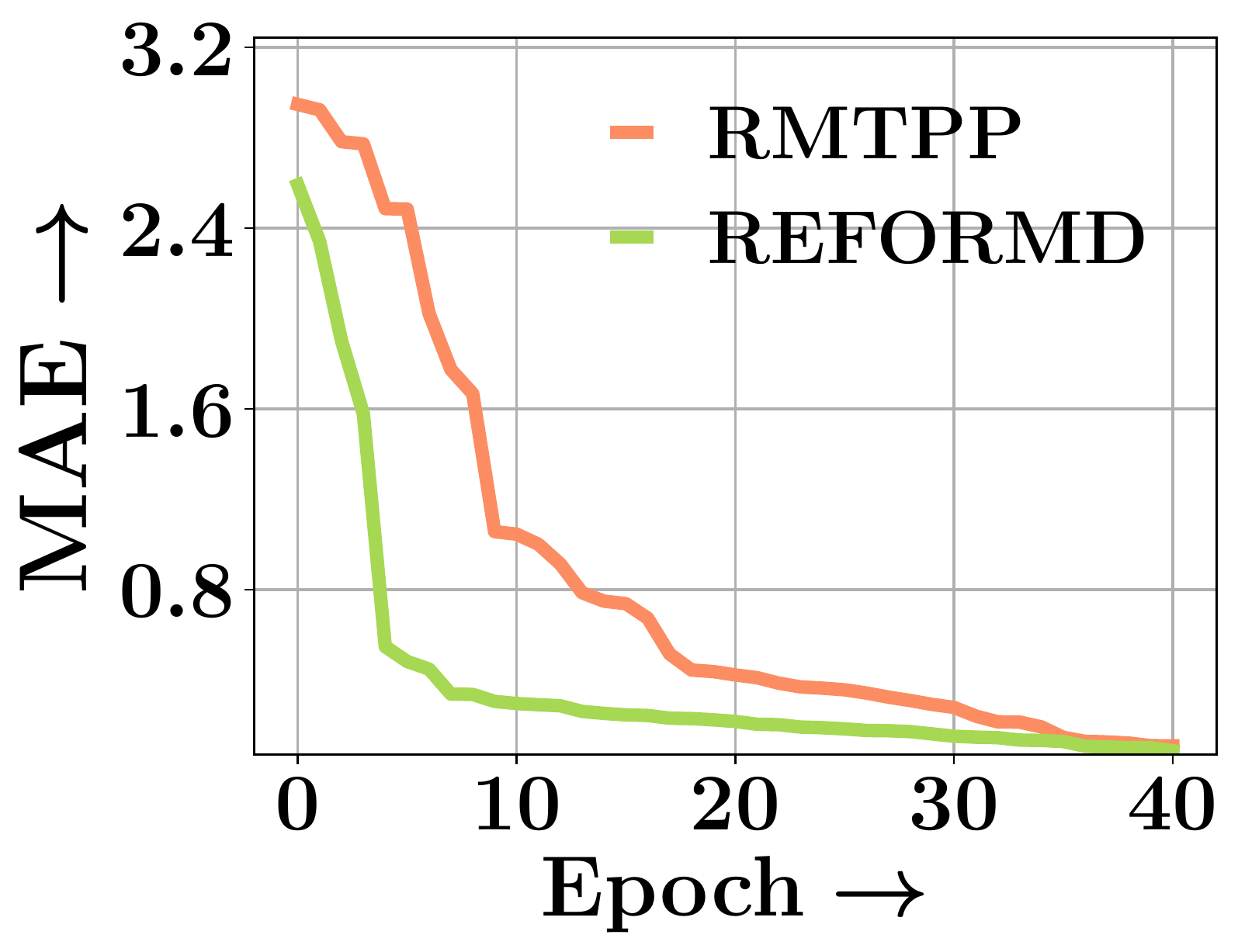}
\caption{Virginia}
\end{subfigure}
\hspace{1cm}
\begin{subfigure}[b]{0.45\columnwidth}
\centering
\includegraphics[height=4cm]{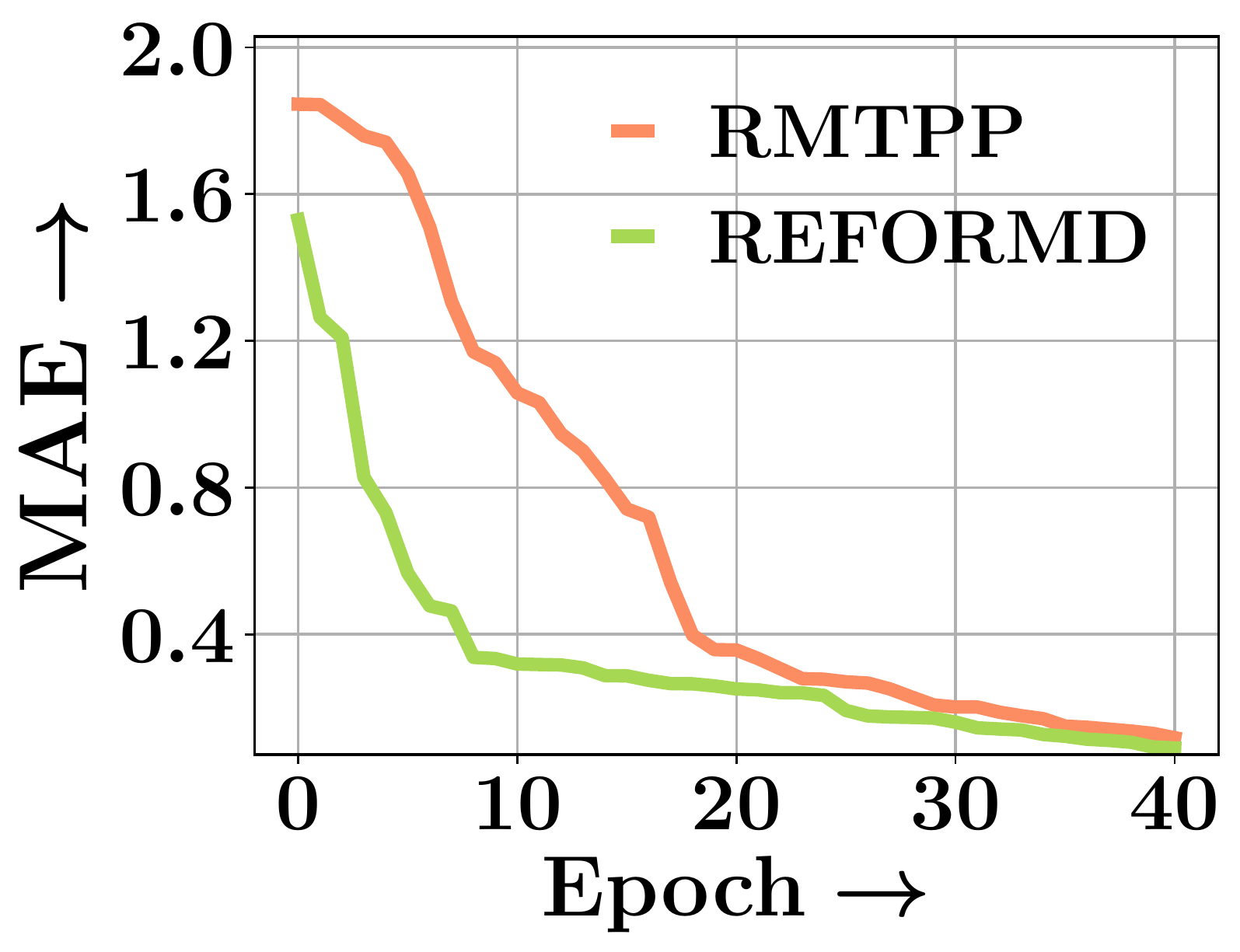}
\caption{Aichi}
\end{subfigure}
\vspace{-4mm}
\caption{\label{reforfig:epochs} Training curves of \reformd and RMTPP for time prediction with \textit{best} MAE for (a) Virginia (VI) and (b) Aichi (AI).}
\vspace{-4mm}
\end{figure}

\subsection{Advantages of Transfer}
To address RQ2, we report that \reformd outperforms other baselines and also exhibits a key feature of \textit{transfer} learning, \ie, quick parameter learning~\cite{transfer}. Exhibiting this property is necessary for transfer-learning models as it highlights the effect of source-region data and whether a transfer procedure was necessary. We highlight this characteristic by plotting the time prediction error (MAE) corresponding to the epochs trained on the target region for \reformd and the best time prediction model, \ie, RMTPP. Figure~\ref{reforfig:epochs} summarizes the results for Virginia and Aichi. From the results, we note that \reformd exhibits faster convergence than RMTPP for both datasets. More specifically, the flow-based transfer procedure of \reformd can outperform most baselines even with a fine-tuning of a few epochs. Thus, validating the effectiveness of the transfer-learning method. Moreover, the results also highlight the stable learning procedure of \reformd. Specifically, even after using the tuned parameters on a new region. \reformd displays consistently improving performance with an increase in epochs. 

\subsection{Product Recommendation}
To address RQ3, we further evaluate the performance of \reformd in product recommendation, \ie, without spatial coordinates. Consequently, we use purchase records for three item categories from Amazon~\cite{julian}, namely Digital Music (DM), Appliances (AP), and Beauty (BY). For each item, we use the user reviews as the events in a sequence with the time of the written review as the event time and the rating (1 to 5) as the corresponding mark. As in this case, we do not have a spatial density function $p_d(\Delta_{d, k+1})$, we change the \textit{fusion} equation \ref{reforeqn:fusion} to include the predicted time of next purchase as:
\begin{equation}
\bs{s}^*_k = \bs{s}_k + \alpha \cdot \bs{w}_f \Delta_{t, k+1},
\end{equation}
We consider Digital Music($|\cm{S}| = 12k$) as source and Appliances($|\cm{S}| = 7k$) and Beauty($|\cm{S}| = 6k$) as target. From the results in Table ~\ref{refortab:item} we make the following observations:
\begin{compactitem}
\item Even in the absence of spatial flows, \reformd outperforms other baselines across all metrics.
\item For product recommendations as well, we see a similar trend as in spatial datasets with RMTPP outperforming other baseline models in terms of time prediction.
\item Lastly, we note that THP performs competitively with \reformd for mark prediction.
\end{compactitem}
The results show that the performance gain due to \reformd is more significant for POI recommendation than product recommendation. Thus, highlighting the scope for improvement with transfer-learning methods for the product recommendations, which is beyond the current formulation of \reformd. 

\section{Conclusion}
In conclusion, we developed a novel architecture, called \reformd, a novel method for sequential POI recommendation that incorporates mobility data from other regions to design a location recommendation system for data-scarce regions. Specifically, it transfers mobility knowledge across regions by sharing the spatial and temporal NFs for continuous-time check-in prediction. Experiments over diverse mobility datasets revealed that \reformd is able to significantly improve over the state-of-the-art baselines for POI recommendation in limited data regions and even performs better than other models for product recommendation. 

%% file: chapters/007_revamp.tex
\newcommand{\tel}{Shanghai-Telecom\xspace}
\newcommand{\tdk}{TalkingData\xspace}
\newcommand{\cat}[1]{`\textit{#1}'}
\newcommand{\sap}[1]{smartphone-app}

\section{Introduction}
The rapid advancements in the smartphone industry and ubiquitous internet access have led to exponential growth in the number of available users and internet-based applications. These smartphones have become increasingly prevalent across the entire human population, with up to 345 million units sold in the first quarter of 2021\footnote{\scriptsize https://www.canalys.com/newsroom/canalys-worldwide-smartphone-market-Q1-2021 (Accessed October 2022)}. Consequently, the online footprint of a user spans multiple applications with an average smartphone owner accessing 10 applications (or apps) in a day and 30 apps in each month\footnote{\scriptsize https://buildfire.com/app-statistics/ (Accessed October 2022)}. These footprints can be perceived as the digitized nature of the user's proclivity in different domains. Recent research~\cite{revisit, revisit2} has shown that the online web activity of a user exhibits \textit{re-visitation} patterns, \ie, a user is likely to visit certain apps repetitively with similar time intervals between corresponding visits. \citet{remob} and~\citet{wwwmob} have shown that these online re-visitation patterns are analogous to their spatial mobility preferences, \ie, the current geographical location can influence the web-browsing activities of a user. Moreover, such cross-domain information of app preferences of a user can be collected without using any personally identifiable information (PII), and thus, maintain the privacy of a user~\cite{silkroad, cross, li}. Therefore, to enhance the performance of a points-of-interest (POI) recommendation system, it is crucial to model the app re-visitation users along with their location preferences. 

\subsection{Limitations of Prior Works}
\begin{figure}[t]
\centering
  \includegraphics[height=5cm]{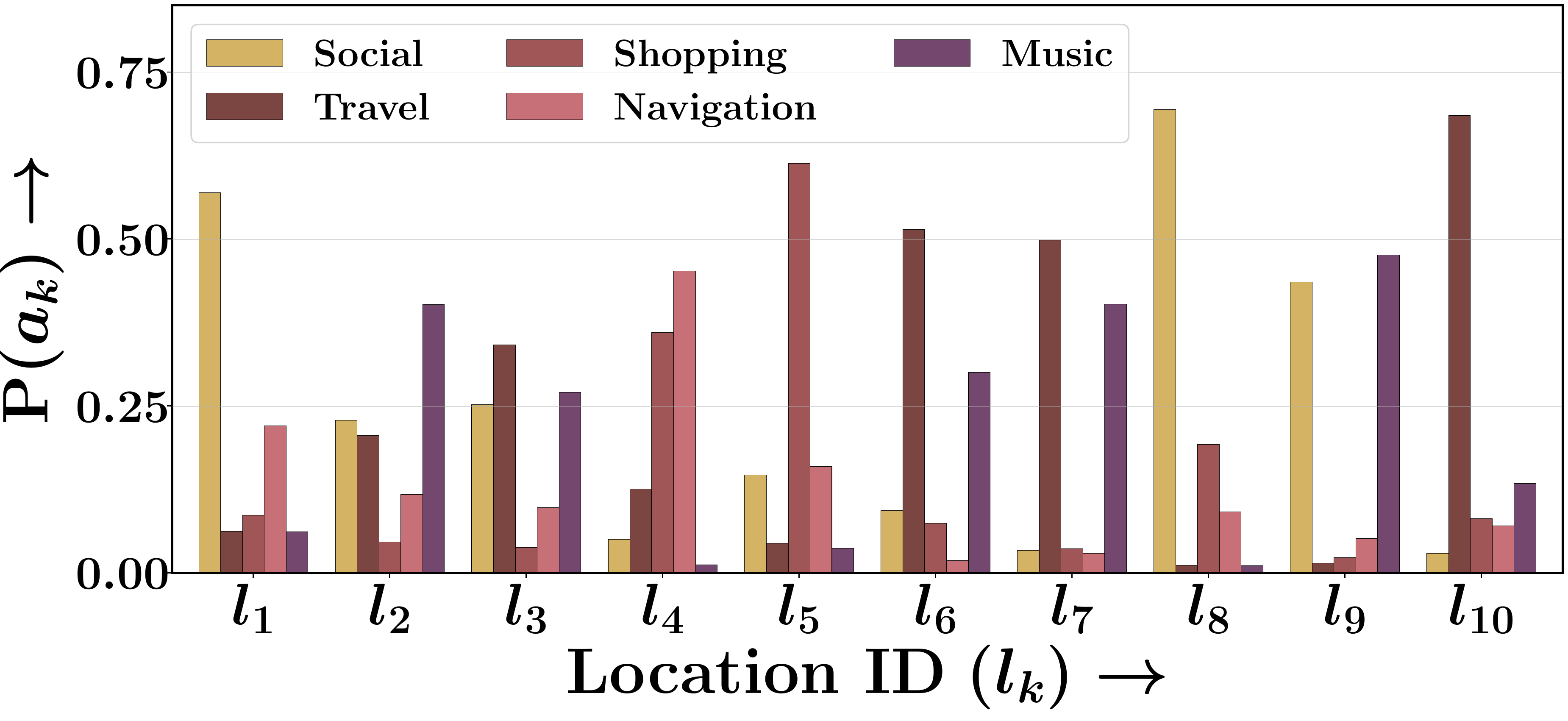}
  \vspace{-5mm}
\caption{The probability of a smartphone-app category --  among \cat{Social}, \cat{Travel}, \cat{Shopping}, \cat{Navigation}, and \cat{Music} -- to be used at ten most popular locations from \tel dataset. The plot indicates that the smartphone app usage depends on the \cin location.}
\vspace{-4mm}
\label{revfig:mdl_tel_cat}
\end{figure}

Modern POI recommendation approaches~\cite{deepmove, axolotl, lbsn2vec} utilize the standard features specific to a user and a POI -- a social network, geo-coordinates, and the category classifications of POIs -- to learn the mobility patterns of a user. The situation has been exacerbated in recent times due to the advent of restrictions on personal data collection and a growing awareness (in some geopolitical regions) about the need for personal privacy~\cite{privacy, privacy2}. Moreover, current approaches overlook two crucial aspects of urban computing -- the exponential growth of online platforms and the widespread use of smartphones. Undeniably, everyone carries and simultaneously uses their smartphones wherever they go. To highlight the importance of the relationship between POI and the apps being used, in Figure~\ref{revfig:mdl_tel_cat}, we plot the category of the app used by all users at the ten most popular locations from our \tel dataset~\cite{li}. The plot shows that the \cin locations can influence a user to visit apps of certain categories more than other apps. We note that this influence of a POI over the category of the app is applicable to multiple users. 

The correlation between spatial mobility and smartphone use is essential to address the problems related to user demographics~\cite{carat, f2f2}, trajectory analysis~\cite{coarse}, app recommendation~\cite{appusage2vec}, and to identify hotspots for network operators~\cite{wwwmob}. However, utilizing smartphone usage for sequential POI recommendations is not addressed in the past literature. The approaches most similar to our work are by~\citet{reapp2} and~\citet{f2f}. \citet{reapp2} utilizes a Dirichlet process to determine the next user location but it completely disregards the user's privacy, \ie, requires precise geo-coordinates. Lastly,~\citet{f2f} is limited to the cold-start recommendation.

\begin{figure}[t]
\centering
  \includegraphics[height=5cm]{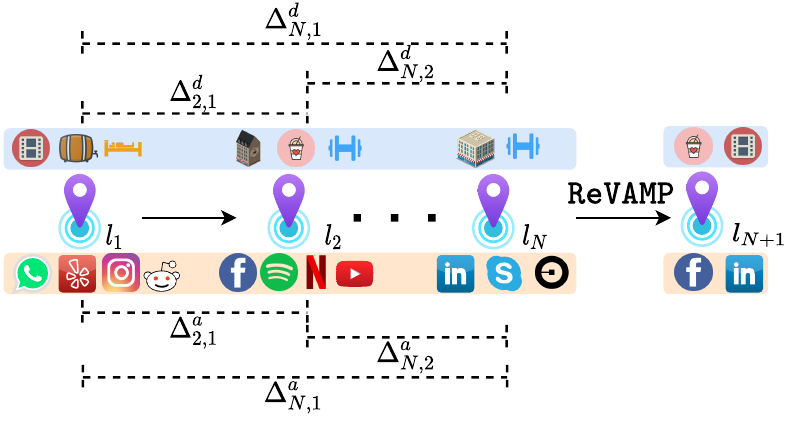}
\vspace{-6mm}
\caption{\revamp learns the dynamics of a \cin sequence via inter-\cin time, app, and POI variations. Here, $\Delta^a_{\bullet, \bullet}$ and $\Delta^d_{\bullet, \bullet}$ denote the smartphone-app and POI-based differences respectively.}
\vspace{-4mm}
\label{revfig:mdl_intro}
\end{figure}

\subsection{Our Contribution}
In this chapter, we present \textbf{\revamp}(\textbf{Re}lative position \textbf{V}ector for \textbf{A}pp-based \textbf{M}obility \textbf{P}rediction), a sequential POI recommendation model that learns the location and app affinities of smartphone users while simultaneously maintaining their privacy needs~\cite{revamp}. Specifically, we consider each \cin as an event involving a smartphone activity and the physical presence at a POI and \revamp models the correlation between the smartphone-app preferences and the spatial mobility preferences of a user. Parallelly, to preserve the privacy restrictions, it solely utilizes two aspects of urban mobility: (a) the types of smartphone apps used during a \cin and (b) the category of the \cin location. Thus, the proposed approach is not privy to any kind of identifiable information (PII) related features such as the precise smartphone app being accessed, \eg, \cat{Facebook}, \cat{Amazon}, \etc, the accurate geo-location, inter-\cin distance, or the social network of a user. Buoyed by the success of self-attention~\cite{transformer} models in sequence modeling, \revamp encodes the dynamic \cin preferences in the user trajectory as a weighted aggregation of all past \cins. Moreover, to better capture the evolving POI and app preferences, it models the variation between each \cin in the sequence using \textit{absolute} and \textit{relative} positional encodings~\cite{reltran, postran}. Specifically, we embed three properties associated with each \cin\ -- the smartphone app category, POI-category, and the time of \cin -- and model the temporal evolution as the inter-\cin embedding differences independently. Figure~\ref{revfig:mdl_intro} demonstrates how \revamp embeds and adaptively learns the inter-\cin dynamics between the app and POI categories to determine the next \cin location for a user. Moreover, \revamp grants the flexibility to predict the category of the most likely smartphone app to be accessed and the POI category at the next \cin. Predicting the app preferences of a user has limitless applications, ranging from smartphone app recommendation and bandwidth modeling by cellular network providers~\cite{appusage2vec, f2f, f2f2}. To summarize, the key contributions we make via \revamp are three-fold:
\begin{compactitem}
\item We propose a self-attention-based approach, called \revamp, to learn the POI preferences of a user via the coarse-grained smartphone usage logs. \revamp returns a ranked list of candidate POIs and the most likely app and POI category for the next \cin.
\item We preserve the privacy needs of a user by learning a personalized sequence encoding for every user. In detail, we force our model to learn the evolving spatial preferences using the variations between each \cin in the sequence based on app category, POI category, and time of the \cin. Thus, our approach is not privy to accurate geo-locations and social networks. 
\item Exhaustive experiments over two large-scale datasets from China show that \revamp outperforms other state-of-the-art methods for sequential POI recommendation, next app, and location-category prediction tasks. Moreover, we perform a detailed analysis of each component of \revamp and a convergence analysis.
\end{compactitem}

\section{Related Work} \label{revsec:relwork}
In this section, we highlight some relevant works to our work. It mainly falls into -- smartphone and mobility, sequential recommendation, and positional encodings for self-attention.

\subsection{Modeling Smartphone and Mobility}
Understanding the mobility dynamics of a user has wide applications ranging from location-sensitive advertisements, social community of user, and disease propagation~\cite{coarse, cho, colab}. Traditional mobility prediction models utilized function-based learning for spatial preferences but were highly susceptible to irregular events in the user trajectory~\cite{locate, cheng2013you}. Therefore, modern approaches~\cite{deepmove, cara, advlbsn} utilize a neural network to model the complex user-POI relationships, geographical features, travel distances, and category distribution. These approaches consider the user trajectory as a \cin sequence and train their model parameters by capturing the influences across different sequences. Other approaches~\cite{lbsn3, imtpp, reformd} include the continuous-time contexts for modeling the time-evolving preferences of a user. However, prior research has shown that users exhibit \textit{re-visitation} patterns on their web activities~\cite{revisit, revisit2} and these re-visitation patterns resonate with the mobility preferences of a user~\cite{reapp, reapp2}. As per the permissions given by a user to an app, leading corporations, such as Foursquare, utilize smartphone activities to better understand the likes and dislikes of a user to give better POI recommendations~\cite{use2}. The correlation between spatial mobility and smartphone use is essential to address the problems related to user demographics~\cite{carat, f2f2}, trajectory analysis~\cite{coarse}, app recommendation~\cite{appusage2vec}, and to identify hotspots for network operators~\cite{wwwmob}. However, utilizing smartphone usage for sequential POI recommendations is not addressed in the past literature. The works most similar to our work are by \citet{reapp2} and \citet{f2f}. \citet{reapp2} utilizes a Dirichlet process to determine the next user's location, but it completely disregards the user's privacy, \ie, requires precise geo-coordinates, and~\citet{f2f} is limited to \textit{cold-start} POI recommendation rather than sequential recommendations. 

\subsection{Sequential Recommendation}
Standard collaborative filtering (CF) and matrix factorization (MF) based recommendation approaches~\cite{ncf, f2f2} return a list of most likely items that a user will purchase in the future. However, these approaches ignore the temporal context associated with the preferences, \ie, it evolves with time. The task of a sequential recommender system is to continuously model the user-item interactions in the past purchases (or \cins) and predict future interactions. Traditional sequence modeling approaches such as personalized Markov chains~\cite{fpmc} combine matrix factorization with inter-item influences to determine the time-evolving user preferences. However, it has limited expressivity and cannot model complex functions. Neural models such as GRU4Rec~\cite{gru4rec} utilize a recurrent neural network (RNN) to embed the time-conditioned user preferences which led to multiple developments like GRU4Rec+~\cite{gru4recplus}. Recent research has shown that including attention~\cite{attention} within the RNN architecture achieved better prediction performances than standard RNN models even in the case of POI recommendations~\cite{attnloc1, rnnlbsn, attnloc4}. However, all these approaches were outperformed by the self-attention-based sequential recommendation models~\cite{sasrec, tisasrec}. In detail, the underlying model of \citet{sasrec} is a transformer architecture~\cite{transformer} that embeds user preferences using a weighted aggregation of all past user-item interactions. However, due to largely the heterogeneous nature of data in spatial datasets, \eg, POI category, geographical distance, \etc, extending such models for sequential POI recommendation is a challenging task.

\subsection{Relative Positional Encodings and Self-Attention}
The self-attention models are oblivious to the position of events in the sequence, and thus, the original proposal to capture the order of events used fixed function-based encodings~\cite{transformer}. However, recent research on positional encodings~\cite{reltran, postran} has shown that modeling the position as a relative pairwise function between all events in a sequence, in addition to the fixed-function encodings, achieves significant improvements over the standard method. Thus, such relative encodings have been used in a wide range of applications -- primarily for determining the relative word order in natural language tasks~\cite{tacl, rel_nlp3} and image order in computer vision problems~\cite{rel_images, rel_images2}. Such relative encodings have also been incorporated in item-based recommender systems~\cite{tisasrec} through \textit{time-interval} based inter-event relevance and in POI recommendation~\cite{geosasrec} through geographical distance-based variances. However, the former approach cannot be extended to model the heterogeneous nature of smartphone mobility data, and the latter requires precise geographical coordinates. Moreover, including the app-category and POI-category-based relevance is challenging because these are context-dependent, \ie, two categories such as \cat{Burger Joint} and \cat{Sushi Restaurant} differ in terms of the semantic meaning of the category term. Such differences are not explicit and must be learned via natural language embeddings.

\section{Problem Formulation} \label{revsec:psetup}
We consider a setting with a set of users as $\cm{U}$ and a set of locations (or POIs), $\cm{P}$. We embed each POI using a $D$ dimensional vector and denote the embedding matrix as $\bs{L} \in \mathbb{R}^{|\cm{P}| \times D}$. We represent the mobile trajectory of a user $u_i$ as a sequence of \cins, $e^{u_i}_k \in \cm{E}^{u_i}$, with each \cin comprising of the smartphone app and the POI details. For a better understanding of our model, let us consider a toy sequence with five \cins to POIs with categories, -- \cat{Bar}, \cat{Cafe}, \cat{Burger-Joint}, \cat{Cafe}, and \cat{Sushi Restaurant}, while using smartphone apps categories -- \cat{Social}, \cat{Shopping}, \cat{Game}, \cat{Social}, and \cat{Travel}, respectively. Thus, for this example, \revamp will use the details of the first four \cins to predict the last \cin.

\begin{definition}[\textsf{Check-ins}]
\label{checkin}
\textit{We define a \cin as a timestamped activity of a user with her smartphone and location details. Specifically, we represent the $k$-th \cin in $\cm{E}$ as $e_k = \left\{l_k, t_k, \cm{A}_k, \cm{S}_k \right\}$ where $l_k$ and $t_k$ denote the POI and \cin time respectively. Here, $\cm{A}_k$ denotes the categories set of the smartphone app accessed by a user, and $\cm{S}_k$ denotes the set of POI categories.}
\end{definition}

With a slight abuse of notation, we denote a \cin sequence as $\cm{E}$ and the set of all app- and location categories till a $k$-th \cin as $\cm{A}^*_k = \bigcup_{i=1}^{k} \cm{A}_k$ and $\cm{S}^*_k = \bigcup_{i=1}^{k} \cm{S}_k$ respectively. Now, we formally define the problem of sequential POI recommendations. For our example, $\cm{A}$ will consist of \cat{Shopping}, \cat{Game}, \cat{Social}, and \cat{Travel}, while $\cm{S}$ will include \cat{Bar}, \cat{Burger-Joint}, \cat{Cafe}, and \cat{Sushi Restaurant} respectively.

\begin{problem*}[\textbf{Personalized Sequential Recommendation}]
\label{problem}
\textit{Using the user's past \cin records consisting of app and POI categories, we aim to get a ranked list of the most likely locations the user is expected to visit in her next \cin. Specifically, we learn the time-evolving variation in smartphone and physical mobility to estimate her future preference towards different locations in her vicinity}. 
\end{problem*}
Mathematically, given the first $k$ \cins in a sequence as $\cm{E}_k$, we aim to identify the set of candidate POI for the next \cin, \ie, $e_{k+1}$, conditioned on the app- and location-categories of all \cins in the history. Specifically, we aim to maximize the following probability:
\begin{equation}
\mathbb{P}^* = \arg\max_{\Theta} \{ \mathbb{E}[e_{k+1}| \cm{E}_{k}, \cm{A}^*_k, \cm{S}^*_k]\}
\end{equation}
where $\mathbb{E}[e_{k+1}]$ calculates the expectation of $e_{k+1}$ being in the sequence of the user, $\cm{E}_{k}$ given the past \cins of a user. Here, $\Theta$ denotes the \revamp model parameters.

\begin{figure}[t]
\centering
  \includegraphics[height=5cm]{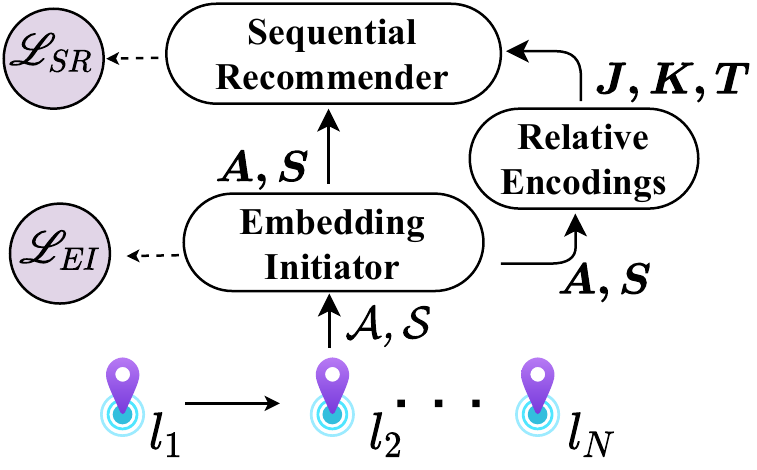}
  \vspace{-2mm}
\caption{Overview of the neural architecture of \revamp.}
\label{revfig:overall}
\end{figure}

\begin{figure}[t]
\centering
\hspace{-2cm}
\begin{subfigure}{0.4\columnwidth}
  \centering
  \includegraphics[height=4cm]{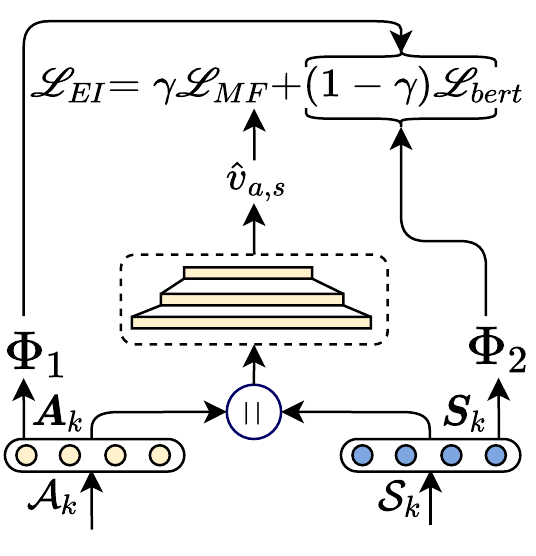}
  \caption{Embedding Initiator (EI)}
\end{subfigure}
\hspace{-1cm}
\begin{subfigure}{0.55\columnwidth}
  \centering
  \includegraphics[height=4cm]{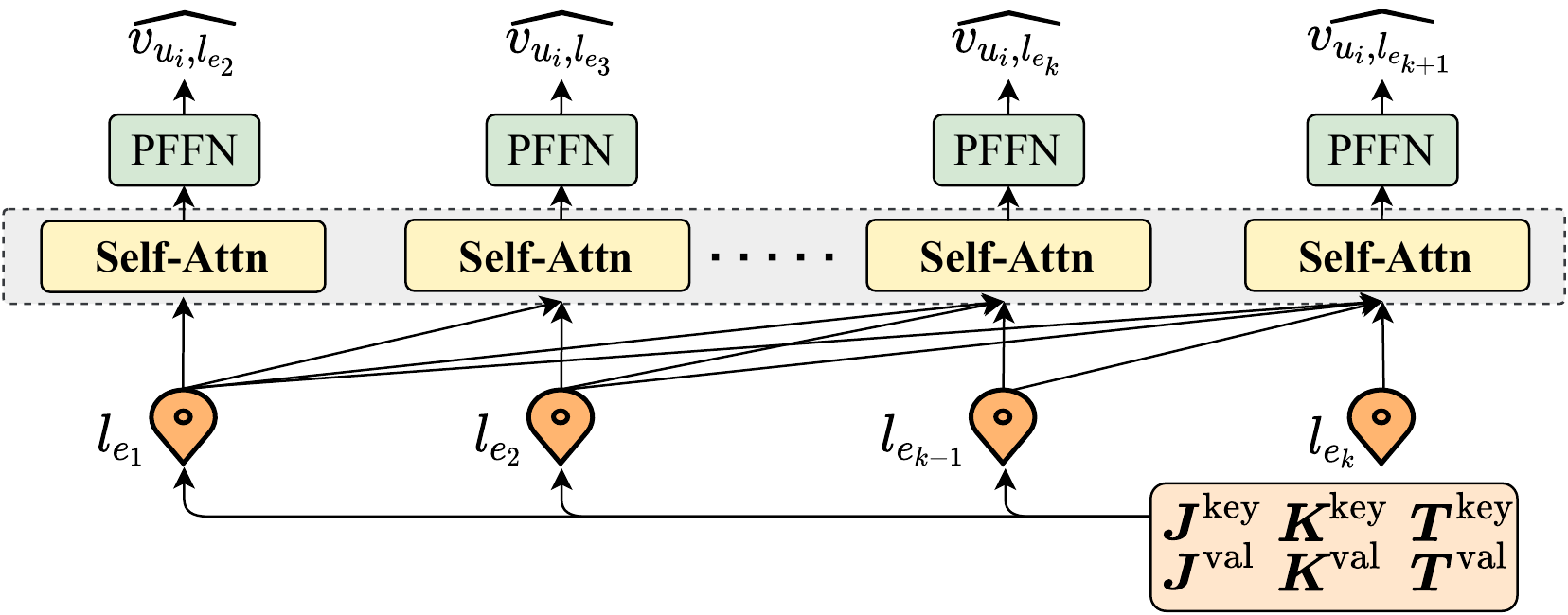}
  \caption{Sequential Recommender (SR)}
\end{subfigure}
\vspace{-3mm}
\caption{Architecture of different components in \revamp.
Panel (a) illustrates the setup of the Embedding Initiator (EI) that learns the category representations.
Panel (b) shows the self-attention architecture in Sequential Recommender (SR).
Note that the input to the self-attention is an aggregation of all past events and relative positional encodings.}
\label{revfig:model}
\end{figure}

\section{\revamp Framework}\label{revsec:model}
In this section, we first present a high-level overview of the deep neural network architecture of \revamp and then describe the component-wise architecture in detail.

\subsection{High-level Overview}\label{revsec:hlevel}
\revamp comprises two components -- (i) an embedding initiator (EI) and (ii) a sequential recommender (SR). Figure~\ref{revfig:overall} shows the overall architecture of \revamp with different components, and the schematic diagrams for both of them are given in Figure~\ref{revfig:model}. The workflow of \revamp includes three steps: (i) determining the embeddings of all POI and app categories using the EI module; (ii) calculating the relative positional encodings in terms of app category, POI category, and time of \cin, and determining embedding matrices for each; and (iii) using the category embeddings from EI module and the newly derived relative embeddings to determine the mobility preferences of a user via the SR module.

As we model the differences between the category of POI and smartphone apps across \cins, we must capture the semantic meaning associated with each category, for \eg, the difference between a \cat{Sushi restaurant} and a \cat{Cafe}. Accordingly, EI takes the \cin sequence of a user as input, learns the representations of all smartphone app- and POI-categories, and calculates the evolving user preferences as the variation between \cins.
\begin{equation}
\bs{A}, \bs{S} = G_{\mathrm{EI}}(\cm{E}_k, \cm{A}^*_k, \cm{S}^*_k),
\end{equation}
where $\bs{A}, \bs{S}$ denote the learned embeddings for app and POI categories, respectively, and $G_{\mathrm{EI}}(\bullet)$ denotes the Embedding Initiator. Moreover, \revamp works by modeling the variations between different \cins in a sequence. Specifically, it learns how the mobility preference of a user has evolved based on the difference in the current and past \cins. Capturing and feeding these differences to our self-attention model is a challenging task as we denote each \cin via POI and app category embeddings. Therefore, we derive these differences and simultaneously embed them to be fed into a self-attention model. These variations are used to assign relative positional encodings to the \cins in the stacked self-attention architecture in SR.
\begin{equation}
\bs{J}, \bs{K}, \bs{T} = f_{\mathrm{RE}}(\cm{E}_k, \bs{A}, \bs{S}),
\end{equation}
where $\bs{J}, \bs{K}, \bs{T}$ are the relative positional encodings for app categories, POI categories, and time respectively. Here, $f_{\mathrm{RE}}(\bullet)$ denotes the function to calculate these relative encodings. Note that these encodings are \textit{personalized}, \ie, they are calculated independently for each user. SR  then combines these relative encodings with absolute positional encodings to model the sequential POI preference of a user. Through this, we aim to get a ranked list of the most probable candidate POIs for the next \cin of a user. 
\begin{equation}
\widehat{l_{k+1}} = G_{\mathrm{SR}}(\cm{E}_k, \bs{J}, \bs{K}, \bs{T}),
\end{equation}
where $\widehat{l_{k+1}}$ is the candidate POI for the $k+1$-th \cin of a user and $G_{\mathrm{SR}}(\bullet)$ denotes the sequential recommender. Figure \ref{revfig:model} shows a schematic diagram of \revamp architecture. The training process of \revamp is divided into two steps -- train the category embeddings using EI and then use them for sequential recommendation in the SR section. More details are given in Section \ref{revsec:train}.

\xhdr{Preserving the user's privacy via \revamp} Here, we highlight the privacy-conscious nature of our underlying framework. In detail, the existing techniques that model the relationship between the app usage and the physical mobility of a user utilize the precise geo-coordinates, the precise apps used, and the details of all background apps~\cite{reapp, li, f2f}. Thus, these approaches have two major drawbacks: (i) they are a serious violation of the privacy of a user; and (ii) collecting accurate data of this granularity makes the problem highly synthetic in nature. Therefore, in \revamp, we do not incorporate any of this information that can compromise the privacy of a user. Specifically, we only use the category of the visited POI and the category of the only active app. Therefore, in our setting, it is difficult to identify individual users based on these coarse-grained records. Moreover, such cross-app data can be collected while simultaneously preserving the user's privacy~\cite{cross}.

\begin{table}[t!]
\small
\caption{\label{revtab:par} Summary of Notations Used.}
\vspace{-2mm}
\centering
\small
\begin{tabular}{l|l}
\toprule
\textbf{Notation} & \textbf{Description}\\
\hline
$\cm{U}, \cm{P}$ & Set of all user and locations\\
$e_k \in \cm{E}$ & $k$-th \cin in the sequence\\
$\cm{A}, \cm{S}$ & Set of all smartphone and POI categories\\
$\bs{A}, \bs{S}$ & Smartphone and POI category embeddings\\
$D$ & Embedding Dimension \\
$\bs{J}, \bs{K}, \bs{T}$ & App, POI, and time based relative encodings\\
$\bs{P}^{\mathrm{key}}, \bs{P}^{\mathrm{val}}$ & Absolute positional encodings\\
$\bs{J}^{\mathrm{key}}, \bs{K}^{\mathrm{key}}, \bs{T}^{\mathrm{key}}$ & Key matrices for relative embeddings\\
$\bs{J}^{\mathrm{val}}, \bs{K}^{\mathrm{val}}, \bs{T}^{\mathrm{val}}$ & Value matrices for relative embeddings\\
$\mathscr{L}_{\mathrm{MF}}, \mathscr{L}_{\mathrm{Bert}}$ & Trajectory and BERT-based loss\\
$\mathscr{L}_{\mathrm{Rec}}$ & POI recommendation loss for SR\\
$\mathscr{L}_{\mathrm{App}}, \mathscr{L}_{\mathrm{POI}}$ & Smartphone and POI category loss\\
\bottomrule
\end{tabular}
\vspace{-0.4mm}
\end{table}

\subsection{Embedding Initiator (EI)} \label{revsec:eim}
An important ability of \revamp is that it learns the mobility preferences conditioned only on the categories of smartphone apps and POI rather than the exact location coordinates and app preferences. Learning from such coarse data is a challenging task and training with \textit{randomly-initialized} embeddings may not capture the category semantics. For \eg, if \cat{Burger-Joints} and \cat{Asian-Restaurants} are frequently visited, then a training process with random initialization will lead to similar the trained embeddings. Therefore, our category embeddings must simultaneously capture the user preferences towards each category and the category semantics via pre-trained word embeddings. We highlight this through an example -- a user checks a smartphone app of category \cat{Social} frequently at two separate locations, say \cat{Cafe} and \cat{Sushi Restaurant}, then the category embeddings should capture the POI influence that persuaded a user to use apps of a similar category(\cat{Social} in this case) as well as the semantic difference between a coffee joint and an Asian restaurant. Therefore, we use a two-channel training procedure, wherein we use pre-trained embeddings to extract the semantic meaning of all app and location categories and learn user preferences towards these categories via a lightweight matrix factorization. Specifically, given a \cin sequence $e_k \in \cm{E}$ we follow a four-layer architecture:
\begin{compactitem}
\item[(1)] \textbf{Input Layer.} We initially embed the app and location categories, $\cm{A}^*$ and $\cm{S}^*$, as $\bs{A} \in \mathbb{R}^{|\cm{A}^*| \times D}$ and $\bs{S} \in \mathbb{R}^{|\cm{S}^*| \times D}$ respectively. Each row $\bs{a}_i \in \bs{A}$ represents a $D$-dimension representation of a smartphone app category. Similarly, $\bs{s}_i \in \bs{S}$ is a representation for a POI category.

\item[(2)] \textbf{MF Layer.} To learn the interaction between the app and POI categories, we follow a lightweight collaborative filtering approach, wherein we concatenate the entries in $\bs{A}$ and $\bs{S}$ that appear together in a \cin $e_k \in \cm{E}$. Specifically, we concatenate the app and POI category embeddings for a \cin and then use a feed-forward network. 
\begin{equation}
\widehat{v}_{a_i, s_j} = \mathrm{ReLU} \left( \bs{w}_{v}(\bs{a}_i || \bs{s}_j) + \bs{b}_v \right), 
\end{equation}
where $\widehat{v}_{a_i, s_j}$  denotes the probability of an app of category $a_i$ to be accessed at a POI of category $s_j$, $||$ denotes the concatenation operator, and $\bs{w}_{\bullet}, \bs{b}_{\bullet}$ are trainable parameters. We train our embeddings via a cross-entropy loss:
\begin{equation}
\mathscr{L}_{\mathrm{MF}} = -\sum_{k=1}^{|\cm{E}|} \sum_{\substack{a_i \in \cm{A}_k \\ s_j \in \cm{S}_k}} \bigg [ \log \big( \sigma(\widehat{v}_{a_i, s_j}) \big) + \log \big( 1 - \sigma(\widehat{v}_{a_i, s'_j}) \big)
+ \log \big( 1 - \sigma(\widehat{v}_{a'_i, s_j}) \big) \bigg ],
\end{equation}
where $\widehat{v}_{\bullet, \bullet}$ denotes the estimated access probability (i) $\widehat{v}_{a_i, s_j}$ between a \textit{true} app- and location-category, \ie, $a_i \in \cm{A}_k, s_j \in \cm{S}_k$, (ii) $\widehat{v}_{a'_i, s_j}$ for a negatively sampled app-category with a true location-category, \ie, $a_i \notin \cm{A}_k, s_j \in \cm{S}_k$, and (iii) $\widehat{v}_{a_i, s'_i}$ for a negatively sampled location-category with a true app-category, \ie, $a_i \in \cm{A}_k, s_j \notin \cm{S}_k$.

\item[(3)] \textbf{BERT Layer.}
To capture the real-world semantics of a category, we use a pre-trained BERT~\cite{bert} model with over 110M parameters. Specifically, we extract the embeddings for each smartphone app and POI category from the pre-trained model. Later, we maximize the similarity between these embeddings and our category representations, $\bs{A}$ and $\bs{S}$, by optimizing a mean squared loss.
\begin{equation}
\mathscr{L}_{\mathrm{Bert}} = \frac{1}{|\cm{E}|} \sum_{k=1}^{|\cm{E}|} \sum_{\substack{a_i \in \cm{A}_k \\ s_j \in \cm{S}_k}} \left [||\bs{a}_i - \Phi_1(a_i)||^2 + || \bs{s}_i - \Phi_2(s_i)||^2 \right],
\end{equation}
where, $\bs{a}_i \in \bs{A}$ and $\bs{s}_j \in \bs{S}$ are our trainable embedding for categories $a_i$ and $s_j$ respectively, and $\Phi_{\bullet}$ denotes a two-step function that extracts pre-trained embeddings for all categories and uses a feed-forward network to normalize the embedding dimension to $D$. Specifically,
\begin{equation}
\Phi_1(a_i) = \mathrm{ReLU} \big(\bs{w}_1 \cdot \cm{B}(a_i) + \bs{b}_1 \big),
\end{equation}
\begin{equation}
\Phi_2(s_i) = \mathrm{ReLU} \big(\bs{w}_2 \cdot \cm{B}(s_i) + \bs{b}_2\big),
\end{equation}
where $\cm{B}$ denotes the set of all pre-trained embeddings, $\cm{B}(a_i)$ and $\cm{B}(s_i)$ denote the extracted app and location category embedding, and $\bs{w}_{\bullet}, \bs{b}_{\bullet}$ are trainable parameters.

\item[(4)] \textbf{Optimization.}
We train our embeddings using a two-channel learning procedure consisting of app-location interaction loss, $\mathscr{L}_{\mathrm{MF}}$, and pre-trained embedding loss, $\mathscr{L}_{\mathrm{Bert}}$, by optimizing a \textit{weighted} joint loss.
\begin{equation}
\mathscr{L}_{\mathrm{EI}} \,=\, \gamma\mathscr{L}_{\mathrm{MF}} + (1 - \gamma) \mathscr{L}_{\mathrm{Bert}},
\end{equation}
where $\gamma$ denotes a scaling parameter. Later, we use $\bs{A}$ and $\bs{S}$ to identify the inter-\cin differences and model the POI preferences of a user.
\end{compactitem}

\subsection{Relative or Inter-\cin Variations}
Buoyed by the efficacy of relative encodings for self-attention models~\cite{reltran, postran}, \revamp captures the evolving preferences of a user as relative encodings based on three inter-\cin differences: (i) Smartphone App-based dynamics, (ii) Location category distribution, and (iii) Time-based evolution across the event sequence.

\xhdr{Smartphone App-based Variation}
Recent research~\cite{postran,reltran} has shown that users' preferences towards smartphone apps are influenced by their geo-locations and other POI-based semantics. Seemingly, it is more likely for a user to be active on a multiplayer game at a social joint rather than at her workplace. We quantify the differences in the app preferences of a user via the differences in the embeddings of the smartphone-app category being used at a \cin. However, in our datasets, every smartphone app is associated with at least one category, \ie, an app can belong to multiple categories, for \eg, Amazon belongs to only one category of \cat{Retail}, but PUBG (a popular mobile game) may belong to categories \cat{Game} and \cat{Action Game}. Calculating the variation based on different embedding is a challenging task. Therefore, we first calculate a ``net app-category'' embedding to denote the representation of all categories an app belongs to. Specifically, for each \cin $e_k$, we calculate the \textit{net} app-category as a mean of all category embeddings.
\begin{equation}
\bs{\mu}^a_k = \frac{1}{|\cm{A}_k|} \sum_{a_i \in \cm{A}_k}\bs{a}_i,
\end{equation}
where $\bs{\mu}^a_k, a_i \in \cm{A}_k, \bs{a}_i \in \bs{A}$ represent the net app-category embedding for a \cin $e_k$, the app-category used in the \cin and the corresponding embedding learned in the EI (see Section~\ref{revsec:eim}). Such an embedding allows us to simplify the input given to the self-attention mechanism in our recommender system. Following~\cite{reltran}, we use these embeddings to calculate a inter-\cin variance matrix $\bs{J} \in \mathbb{W}^{|\cm{E}| \times |\cm{E}|}$ for each \cin sequence. Specifically, the $i$-th row in matrix $\bs{J}$ denotes the difference between the mean app-category embedding of \cin $e_i$ with all other \cins in the sequence and is calculated as:
\begin{equation}
\bs{J}_{i,j} = \Bigg\lfloor \frac{ f_{\mathrm{cos}} (\bs{\mu}^a_i, \bs{\mu}^a_j) - \min_f(\cm{E})}{ \max_f(\cm{E}) - \min_f(\cm{E})} \cdot I_a \Bigg\rfloor,
\label{eq:J}
\end{equation}
where $f_{\mathrm{cos}}(\bullet, \bullet), \min_f(\cm{E}), \max_f(\cm{E})$ denote the function for normalized cosine-distance, the minimum and maximum cosine distance between the mean category embedding for any two \cins in a sequence. We use $I_a$ as a clipping constant and a \textit{floor} operator to discretize the entries in $\bs{J}$. Such discretization makes it convenient to extract positional encodings for the self-attention model in SR. 

\xhdr{POI-based Variation} We derive the inter-\cin differences between POI categories using a similar procedure for app-based differences. However,  POI can belong to multiple categories. Therefore, similar to our procedure for calculating app-based variations, we calculate a \textit{net} POI category embedding, $\bs{\mu}^l_k$ for each \cin as $\bs{\mu}^l_k = \frac{1}{|\cm{S}_k|} \sum_{s_i \in \cm{S}_k}\bs{s}_i$. Later, as in Eqn~\ref{eq:J}, we calculate the POI-based inter-\cin variance matrix $\bs{K} \in \mathbb{W}^{|\cm{E}| \times |\cm{E}|}$ using a clipping constant $I_l$. Here, the $i$-th row in matrix $\bs{K}$ denotes the difference between the mean POI-category embedding of \cin $e_i$ with all other \cins in the sequence.

\xhdr{Time-based Variation} Ostensibly, there may be irregularities in the smartphone app usage of a user. For \eg, a user browsing \cat{Amazon} may receive a message \cat{Twitter} that she immediately checks and then later continues her shopping on Amazon. Notably, the \cat{Amazon} app did not influence the user to access \cat{Twitter} and vice-versa, as such a change between apps was coincidental. To model these nuances in \revamp we use the time interval between accessing different smartphone apps. Specifically, similar to app- and POI-category based inter-\cin differences, we derive a \textit{time-based} variations matrix, $\bs{T} \in \mathbb{W}^{|\cm{E}| \times |\cm{E}|}$, using the absolute time-difference between each \cin.
\begin{equation}
\bs{T}_{i,j} = \Bigg\lfloor \frac{|t_i - t_j|}{t_{\min}} \cdot I_t \Bigg\rfloor,
\end{equation}
where $t_i, t_j, t_{\min},$ and $I_t$ denote the time of \cin $e_i$ and $e_j$, minimum time-interval between \cins of a user and the normalizing constant for time respectively. 

\subsection{Sequential Recommender (SR)}
In this section, we elaborate on the sequential recommendation procedure of \revamp that is responsible for modeling the app and POI preferences of a user and then recommending a candidate POI for the next \cin. Specifically, it uses a self-attention architecture consisting of five layers:
\begin{compactitem}
\item[(1)] \textbf{Input Layer.} The SR model takes the \cin sequence of a user ($\cm{E}$), relative app, POI, and time encodings, ($\bs{K}, \bs{J}$, and $\bs{T}$ respectively), and the \textit{mean} app and location category representations ($\bs{\mu}^a_{\bullet}, \bs{\mu}^l_{\bullet}$) as input to the self-attention model. Since the self-attention models require a fixed input sequence, we limit our training to a fixed number of \cins, \ie, we consider the $N$ most recent \cins in $\cm{E}$ for training our model, and if the number of \cins is lesser than $N$, we repeatedly add a \textit{[pad]} vector for the initial \cins within the sequence. 

\item[(2)] \textbf{Embedding Retrieval Layer.} 
Since the self-attention models are oblivious of the position of each \cin in the sequence, we use a trainable positional embedding for each \cin~\cite{sasrec, tisasrec}. Specifically, we initialize two distinct vectors denoted by $\bs{P}^{\mathrm{key}} \in \mathbb{R}^{N \times D}$ and $\bs{P}^{\mathrm{val}} \in \mathbb{R}^{N \times D}$ where the $i$-th rows, $\bs{p}^{\mathrm{key}}_i$ and $\bs{p}^{\mathrm{val}}_i$, denote the positional encoding for the \cin $e_i$ in the sequence. 
Similarly, we embed the relative positional matrices $\bs{K}$, $\bs{J}$, and $\bs{T}$ into encoding matrices $\bs{J}^{\mathrm{key}}, \bs{J}^{\mathrm{val}} \in \mathbb{R}^{N \times N \times D}$, $\bs{K}^{\mathrm{key}}, \bs{K}^{\mathrm{val}} \in \mathbb{R}^{N \times N \times D}$, and $\bs{T}^{\mathrm{key}}, \bs{T}^{\mathrm{val}} \in \mathbb{R}^{N \times N \times D}$ respectively.
\begin{equation}
\bs{K}^{\mathrm{key}} = \begin{bmatrix} \bs{k}^{\mathrm{key}}_{1,1} & \cdots & \bs{k}^{\mathrm{key}}_{1,N}\\ \vdots & \vdots & \vdots \\ \bs{k}^{\mathrm{key}}_{N,1} & \cdots & \bs{k}^{\mathrm{key}}_{N,N} \end{bmatrix}, \quad \bs{K}^{\mathrm{val}} = \begin{bmatrix} \bs{k}^{\mathrm{val}}_{1,1} & \cdots & \bs{k}^{\mathrm{val}}_{1,N}\\ \vdots & \vdots & \vdots \\ \bs{k}^{\mathrm{val}}_{N,1} & \cdots & \bs{k}^{\mathrm{val}}_{N,N} \end{bmatrix},
\label{revrel_emb}
\end{equation}
We use two separate matrices to avoid any further linear transformations~\cite{reltran}. Each entry in $\bs{K}^{\mathrm{key}}$ and $\bs{K}^{\mathrm{val}}$ denotes a $D$ dimensional vector representation of corresponding value in  in $\bs{K}$. We follow a similar procedure to initialize $\bs{J}^{\mathrm{key}}, \bs{J}^{\mathrm{val}}, \bs{T}^{\mathrm{key}}$ and $\bs{T}^{\mathrm{val}}$ for $\bs{J}$ and $\bs{T}$ respectively.

\item[(3)] \textbf{Self-Attention Layer.}
Given the \cin sequence of a user, the self-attention architecture learns the sequential preference of a user towards POIs. Specifically, for an input sequence consisting of POI embeddings of locations visited by a user, $\bs{L}^{\cm{E}} = (\bs{l}_{e_1}, \bs{l}_{e_2}, \cdots \bs{l}_{e_N})$ where $l_{e_i} \in e_i$ and $\bs{l}_{e_i} \in \bs{L}$ are the location visited in \cin $e_i$ the POI embedding for $l_{e_i}$ respectively, we compute a new sequence $\bs{Z} = (\bs{z}_1, \bs{z}_2, \cdots \bs{z}_N)$, where $\bs{z}_{\bullet} \in \mathbb{R}^{D}$. Each output embedding is calculated as a weighted aggregation of embeddings of all the POIs visited in the past. 
\begin{equation}
\bs{z}_i = \sum_{j=1}^{N} \bs{\alpha}_{i,j} \left( \bs{w}_{v, j}\bs{l}_{e_j} + \overline{\bs{\mu}}_j + \bs{p}^{\mathrm{val}}_j + \bs{j}^{\mathrm{val}}_{i,j} + \bs{k}^{\mathrm{val}}_{i,j} + \bs{t}^{\mathrm{val}}_{i,j} \right),
\label{reveq:tv}
\end{equation}
where $\bs{l}_{e_j}$ is the POI embedding, $\overline{\bs{\mu}}_j = \bs{\mu}^a_j + \bs{\mu}^l_j$ is the sum of the smartphone app and POI category mean embeddings, and $\bs{w}_{v, j}$ is a trainable parameter. The attention weights $\bs{\alpha}_{\bullet, \bullet}$ are calculated using a soft-max over other input embeddings as:
\begin{equation}
\bs{\alpha}_{i,j} = \frac{\exp \big(x_{i,j}\big)}{\sum_{k=1}^{N} \exp \big( x_{i,k}\big)},
\label{reveq:tsoft}
\end{equation}
where $x_{i,j}$ denotes the compatibility between two \cins -- $e_i$ and $e_j$ -- and is computed using both -- relative- as well as absolute-positional encodings.
\begin{equation}
x_{i,j} = \frac{\bs{w}_{q, i}\bs{l}_{e_i} \left( \bs{w}_{k, j}\bs{l}_{e_j} + \bs{p}^{\mathrm{key}}_j + \bs{j}^{\mathrm{key}}_{i,j} + \bs{k}^{\mathrm{key}}_{i,j} + \bs{t}^{\mathrm{key}}_{i,j} \right )^{\top}}{\sqrt{D}},
\label{reveq:tqk}
\end{equation}
where $\bs{w}_{q, \bullet}, \bs{w}_{k, \bullet}$ and $D$ denote the input \textit{query} projection, \textit{key} projection, and the embedding dimension respectively. We use the denominator as a scaling factor to control the dot-product gradients. As our task is to recommend a candidate POI for future \cins and should
only consider the first $k$ \cins to predict the $(k + 1)$-th \cin, we introduce a \textit{causality} over the input sequence. Specifically, we modify the
procedure to attention in Eqn.(\ref{reveq:tqk}) and remove all links between the future \cins and the current \cin.

\item[(4)] \textbf{Point-wise Layer.} As the self-attention lacks any non-linearity, we apply a feed-forward layer with two linear transformations with ReLU activation. 
\begin{equation}
\mathrm{PFFN}(\bs{z}_k) = \mathrm{ReLU} \left ( \bs{z}_k \bs{w}_{p,1} + \bs{b}_{1} \right ) \bs{w}_{p,2} + \bs{b}_2, 
\end{equation}
where $\bs{w}_{p, \bullet}, \bs{b}_{\bullet}$ are trainable layer parameters.
The combination of a self-attention layer and the point-wise layer is referred to as a self-attention \textit{block} and stacking self-attention blocks gives the model more flexibility to learn complicated dynamics~\cite{transformer}. Thus, we stack $M_b$ such blocks, and to stabilize the learning process, we add a residual connection between each such block.
\begin{equation}
\bs{z}^{(r)}_k = \bs{z}^{(r-1)}_k + \mathrm{PFFN} \big( f_{ln} (\bs{z}^{(r-1)}_k) \big),
\end{equation}
where $1 \le r \le M_b, f_{ln}(\bullet)$ denote the level of the current self-attention block and \textit{layer-normalization} function respectively. The latter is used to further accelerate the training of self-attention and is defined as follows:
\begin{equation}
f_{ln}(\bs{z}_k) = \beta \odot \frac{\bs{z}_k - \mu_z}{\sqrt{\sigma_z^2 + \epsilon}} + \gamma,
\end{equation}
where $\odot, \mu_z, \sigma_z, \beta, \gamma, \epsilon$ denote the element-wise product, mean of all the input embeddings, the variance of all input embeddings, learned scaling factor, bias term, and the Laplace smoothing constant respectively.

\item[(5)] \textbf{Prediction Layer.} A crucial distinction between \revamp and the standard self-attention model is that \revamp not only predicts the candidate POIs for the next \cin, but also the category of the smartphone app and the POI category to be used in the next \cin. Here, we describe the prediction procedure for each of them.

\textit{\underline{POI Recommendation:}} We predict the next POI to be visited by a user in the \cin sequence using a matrix-factorization~\cite{ncf} based approach between the transformer output $\bs{Z}^{(M_b)} = (z_1, z_2, \cdots z_k)$ and the embeddings of the POIs visited by the user, $(\bs{l}_{e_2}, \bs{l}_{e_{3}}, \cdots \bs{l}_{e_{k+1}})$
\begin{equation}
\widehat{v_{u_i, l_{e_k}}} = \bs{z}_{k-1} \bs{l}_{e_{k}}^{\top},
\end{equation}
where $\widehat{v_{u_i, l_{e_k}}}$ is the calculated probability of user, $u_i$, to visit the POI, $l_{e_k}$, for her next \cin. We learn the model parameters by minimizing the following cross-entropy loss.
\begin{equation}
\mathscr{L}_{\mathrm{Rec}} = -\sum_{u_i \in \cm{U}} \sum_{k=1}^{N} \left[ \log \left ( \sigma(\widehat{v_{u_i, l_{e_k}}} \right) + \log \left (1 - \sigma(\widehat{v_{u_i, l'_{e_k}}}) \right) \right] + \lambda || \Theta ||_F^2,
\label{reveq:cross}
\end{equation}
where $\widehat{v_{u_i, l'_{e_k}}}$ denotes the \cin probability for a negatively sampled POI, \ie, a randomly sampled location that will not be visited by a user. $\lambda, \sigma, \Theta$ denote regularization parameter, \textit{sigmoid} function, and the trainable parameters respectively.

\textit{\underline{{Predicting App Categories:}}} Predicting the next smartphone app to be accessed by a user has numerous applications ranging from smartphone system optimization, resource management in mobile operating systems, and battery optimization~\cite{appusage2vec, carat}. Therefore, to predict the category of the next app to be used, we follow a matrix-factorization approach to calculate the relationship between the user preference embedding, $\bs{z}_k$, and the mean of smartphone app embeddings for the next \cin. 
\begin{equation}
\widehat{q_{u_i, \cm{A}_k}} = \bs{z}_{k-1} {\bs{\mu}^a_{k}}^{\top},
\end{equation}
where $\widehat{q_{u_i, \cm{A}_k}}, \bs{\mu}^a_{k}$ denote the usage probability of apps of categories in $\cm{A}_k$ and the mean embedding for all apps used in \cin $e_k$. Later, we minimize a cross-entropy loss with negatively sampled apps, \ie, apps that were not used by the user, denoted as $\mathscr{L}_{\mathrm{App}}$.

\textit{\underline{{Predicting Location Categories:}}} As in app-category prediction, we calculate the preference towards a POI-category using the mean of POI category embedding $\bs{\mu}^l_k$ and learn the parameters by optimizing a similar \textit{cross-entropy} loss denoted as $\mathscr{L}_{\mathrm{POI}}$.
The net loss for sequential recommendation is a weighted combination of POI recommendation loss, app-category loss, and location-category loss.
\begin{equation}
\mathscr{L}_{\mathrm{SR}} = \mathscr{L}_{\mathrm{Rec}} + \kappa (\mathscr{L}_{\mathrm{App}}+ \mathscr{L}_{\mathrm{POI}}),
\end{equation}
Here, $\kappa$ is a tunable hyper-parameter for determining the contribution of category prediction losses. All the parameters of \revamp, including the weight matrices, relative-position weights, and embeddings are learned using an Adam optimizer\cite{adam}.
\end{compactitem}

\subsection{\revamp: Training} \label{revsec:train}
As mentioned in Section~\ref{revsec:hlevel}, \revamp involves a two-step training procedure. Specifically, it consists of the following steps: \begin{inparaenum}[(i)] \item learning the app and POI category embeddings using the embedding initiator(EI) and \item training the self-attention model in SR to recommend candidate POI to the user\end{inparaenum}. In detail, we first train the parameters of EI by minimizing the $\mathscr{L}_{EI}$ loss for multiple epochs and later use the trained category embeddings in SR and recommend candidate POI by minimizing the recommendation loss, $\mathscr{L}_{SR}$.

We highlight that a \textit{joint} training of both EI and SR is not suitable in the presence of relative positional encodings, as they are conditioned on the category embeddings learned in EI. Therefore, during joint training, an update in the category embedding will make the trained parameters of SR across the previous epochs unsuitable for prediction in the future. Moreover, these encodings are calculated \textit{relatively}, \ie, conditioned on the embedding of other categories in a sequence, and thus any change in the category embedding will affect the category embeddings. 

\section{Experiments}\label{revsec:expts}
In this section, we report a comprehensive empirical evaluation of \revamp and compare it with other state-of-the-art approaches. We evaluate the POI recommendation performance of \revamp using two real-world datasets from China. These datasets vary significantly in terms of data sparsity, the no. of app categories, and POI categories. With our experiments, we aim to answer the following research questions:
\begin{compactitem}
\item[\textbf{RQ1}] How does \revamp compare to cutting-edge models for sequential POI recommendation? What are the gains and losses?
\item[\textbf{RQ2}] What is the contribution of relative positional encodings?
\item[\textbf{RQ3}] What is the scalability of \revamp and the stability of the learning procedure?
\end{compactitem}
All our algorithms are implemented in Tensorflow on a server with Ubuntu 16.04. CPU: Intel (R) Xeon (R) Gold 5118 processor at 2.30GHz, RAM: 125GB, GPU: NVIDIA V100 32GB.

\begin{table}[t]
\small
\caption{Statistics of all datasets used in this chapter.}
\vspace{-2mm}
\centering
\begin{tabular}{c|ccccc}
\toprule
\textbf{Dataset} & $|\cm{U}|$ & $|\cm{P}|$ & $|\cm{E}|$ & $|\cm{A}|$ & $|\cm{S}|$ \\ \hline
\tel & 869 & 32680 & 3668184 & 20 & 17 \\
\tdk & 14544 & 37113 & 438570 & 30 & 366 \\ \bottomrule
\end{tabular}
\label{revtab:data}
\vspace{-2mm}
\end{table}

\subsection{Experimental Setup}
\begin{figure}[t!]
\centering
\begin{subfigure}{0.45\columnwidth}
  \centering
{\includegraphics[height=4cm]{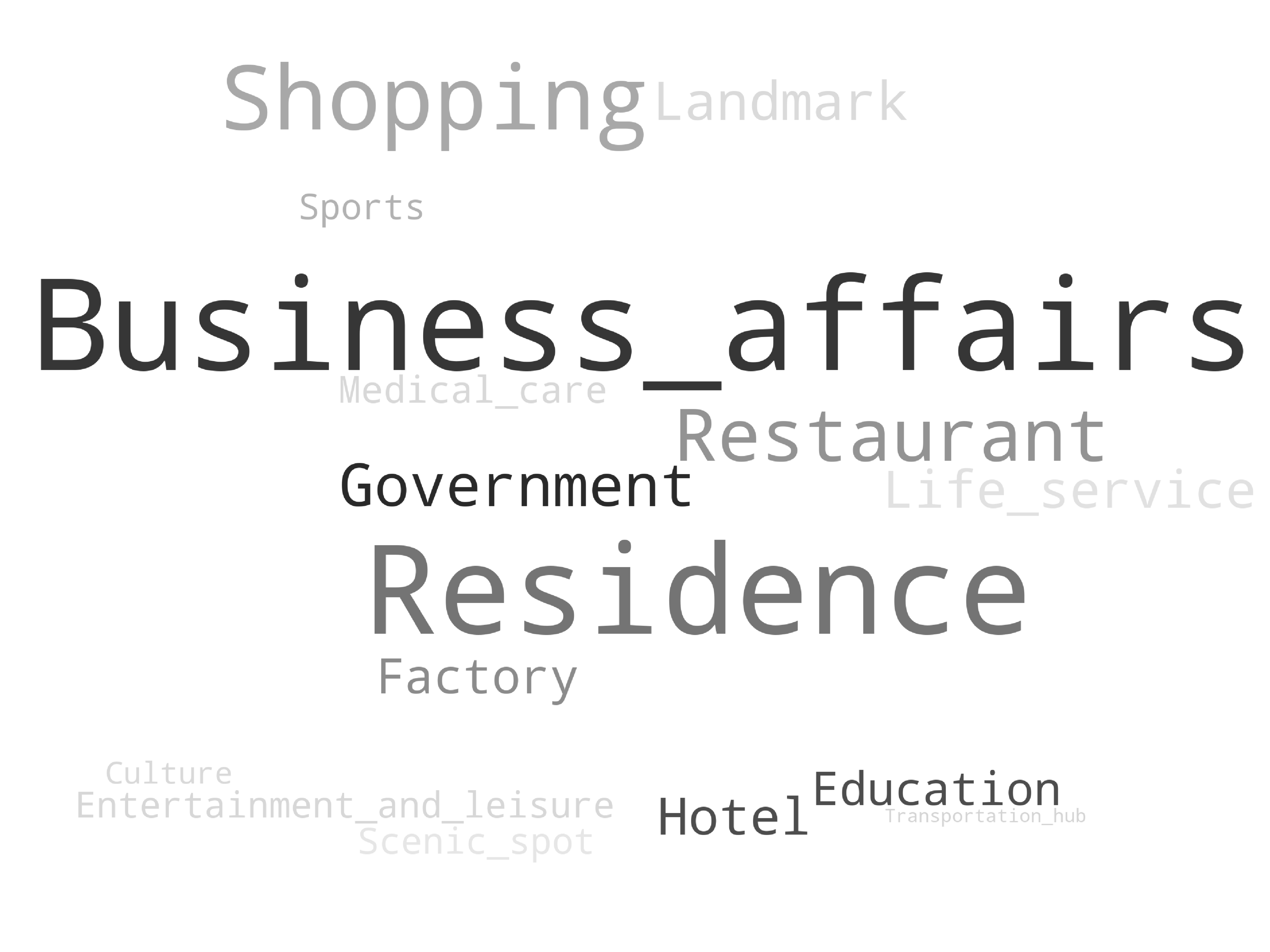}}
\caption{\tel}
\end{subfigure}
\hfill
\begin{subfigure}{0.45\columnwidth}
  \centering
{\includegraphics[height=4cm]{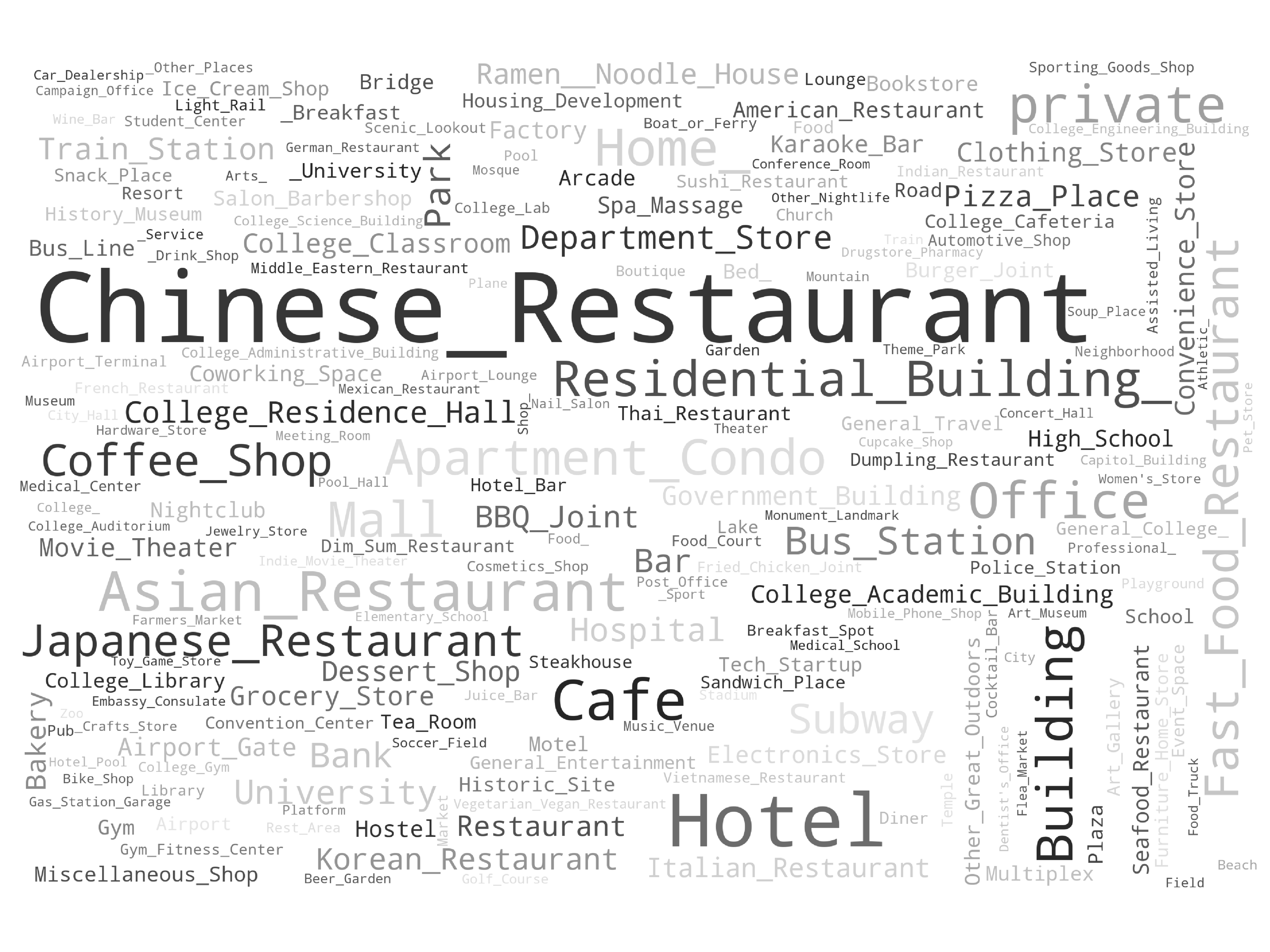}}
\caption{\tdk}
\end{subfigure}
\vspace{-2mm}
\caption{Word-cloud for POI Categories for both \tel and \tdk datasets. The larger the font size indicates a larger frequency of location of the category.}
\vspace{-2mm}
\label{revfig:wordcloud}
\end{figure}

\xhdr{Dataset Description}
As our goal is to recommend POIs to a user based on her smartphone usage, the mobility datasets used in our experiments must contain the user trajectory data, \ie, geographical coordinates, time of a \cin, as well as the smartphone-usage statistics -- applications used across different locations, the categories of different apps based on online app-stores, \etc\ Therefore we consider two popular large-scale datasets -- \textit{\tel} and \textit{\tdk} and their statistics are given in Table~\ref{revtab:data}. Moreover, we highlight the high variance between the category semantics of both datasets by plotting the location category word clouds in Figure~\ref{revfig:wordcloud}. The variance across the datasets is due to the different sources used for extracting location categories -- the at-hand location categories for \tel and Foursquare-based categories for \tdk.
\begin{compactenum}
\item[(1)] \textbf{\tel:}
This smartphone usage and the physical-mobility dataset was collected by a major network operator in China~\cite{li}. The trajectories were collected from Shanghai in April 2016. It contains the details of a user's physical mobility and time-and-geo-stamped smartphone app usage records. More specifically, for each user, we have the timestamped records of the smartphone apps being used and the different cellular-network base stations to which the smartphone was connected during the data collection procedure. For the region covered by each cellular network base station, we also have the details of the internal POIs and their corresponding categories. For our experiments, we consider each \textit{user$\rightarrow$base-station} entry as a \cin and all the apps and their categories associated with that \cin as the events in the sequence $\cm{E}$. We adopt a commonly followed data cleaning procedure~\cite{cara, locate} and filter out users and POI with less the five \cins.

\item[(2)] \textbf{\tdk:}
A large-scale public app-usage dataset that was released by TalkingData\footnote{\scriptsize www.talkingdata.com/ (Accessed October 2022)}, a leading data intelligence solution provider based in China. The original dataset released by the company~\cite{tk} consists of location- and timestamped records of smartphone app usage and physical trajectories of a user. However, in this dataset, we lack the categories associated with each POIs. We overcome this by extracting location categories and geo-coordinates from publicly available \cin records~\cite{lbsn2vec} for users in Foursquare -- a leading social mobility network, and mapping each \cin location in Foursquare to a location in the TalkingData based on geographical coordinates. For our experiments using this dataset, we restrict our \cin records to only the locations situated in mainland China. As in the \tel dataset, here as well we filter out the users and POI with lesser than five \cins. 
\end{compactenum}

\xhdr{Evaluation Metric}
We evaluate \revamp and the other sequential recommendation baselines using a widely used \textit{leave-one-out} evaluation, \ie, next \cin prediction task. Specifically, for each user, we consider the last \cin of the trajectory sequence as the test \cin, the second last \cin for validation, and all preceding events as the training set~\cite{sasrec, tisasrec}. For a fair evaluation, we also follow a common testing strategy wherein we pair each ground truth \cin in the test set with $100$ randomly sampled negative events, \ie, the \cins not associated to the user~\cite{ncf, sasrec}. Therefore, the task becomes to rank the negative \cins with the ground truth \cin. To evaluate the effectiveness of all approaches, we use Hits@$k$ and NDCG@$k$, with $k \in \{1, 5, 10\}$, and report the confidence intervals based on five independent runs.

\xhdr{Parameter Settings}
For all results in Section~\ref{revrq1} and~\ref{revrq2}, we set $N = 200$ and $N=100$ for \tel and \tdk respectively. We set $I_a = I_l = I_t = 64$, $D=64$, and $\lambda=0.002$, We search the batch-size in $\{128, 256\}$, the no of attention-heads in $\{1, 2, 4, 8\}$, $\kappa, \gamma $ are searched in $\{ 0.2, 0.5, 0.8\}$, and the dropout probability is set to $0.2$. 

\xhdr{Baselines} 
We compare \revamp with the state-of-the-art methods based on their architectures below:
\begin{compactenum}[(1)]
\item{\bf Standard Recommendation Systems.}
\begin{asparadesc}
\item [FPMC~\cite{fpmc}] FPMC utilizes a combination of factorized first-order Markov chains and matrix factorization for recommendation and encapsulates a user's evolving long-term preferences as well as short-term purchase-to-purchase transitions.
\item [TransRec~\cite{transrec}] A first-order sequential recommendation model that captures the evolving item-to-item preferences of a user through a translation vector.
\end{asparadesc}
\item{\bf POI Recommendation Systems.}
\begin{asparadesc}
\item [STGN~\cite{attnloc4}] Uses a modified LSTM network that captures the spatial and temporal dynamic user preferences between successive \cins using spatio-temporal gates. Hence, it requires the exact location coordinates as input to the model.
\end{asparadesc}
\item{\bf Smartphone App-based.}
\begin{asparadesc}
\item [AUM~\cite{reapp2}] Models the user's mobility as well as app-usage dynamics using a Dirichlet process to predict the next successive \cin locations.
\end{asparadesc}
\item{\bf Recurrent and Convolutional Neural Network.}
\begin{asparadesc}
\item [GRU4Rec+~\cite{gru4recplus}] A RNN-based approach that models the user action sequences for a session-based recommendation. It is an improved version of GRU4Rec~\cite{gru4rec} with changes in the loss function and the sampling techniques.
\item [Caser~\cite{caser}] A state-of-the-art CNN-based sequential recommendation method that applies convolution operations on the $N$-most recent item embeddings to capture the higher-order Markov chains.
\end{asparadesc}
\item{\bf Self-Attention.}
\begin{asparadesc}
\item [Bert4Rec~\cite{bert4rec}] A bi-directional self-attention~\cite{bert} based sequential recommendation model that learns user preferences using a Cloze-task loss function, \ie, predicts the artificially \textit{masks} events form a sequence.
\item [SASRec~\cite{sasrec}] A self-attention~\cite{transformer} based sequential recommendation method that attentively captures the contribution of each product towards a user's preference embedding.
\item [TiSRec~\cite{tisasrec}] A recently proposed enhancement to SASRec that uses \textit{relative}-position embeddings using the difference in the time of consecutive purchases made by the user.
\end{asparadesc}
\end{compactenum}
We omit comparisons across other approaches for sequential recommendations, such as GRU4Rec~\cite{gru4rec} and MARank~\cite{marank}, as they have already been outperformed by the current baselines. We calculate the confidence intervals based on the results obtained after three independent runs.

\begin{table}[t]
\centering
\caption{\label{revtab:tel_results} Next \cin recommendation performance of \revamp and state-of-the-art baselines for the \tel dataset. Here, we exclude a comparison with STGN~\cite{attnloc4} as it requires the precise geographical coordinates for \cin locations, which we lack in the \tel dataset. All results are statistically significant (\ie, two-sided Fisher's test with $p \le 0.1$) over the best baseline. Numbers with bold font (boxes) indicate the best (second best) performer.}
\vspace{-3mm}
\small
\begin{tabular}{c|ccccc}
\toprule
\textbf{Baselines} & \textbf{NDCG/Hits@1} & \textbf{NDCG@5} & \textbf{NDCG@10} & \textbf{Hits@5} & \textbf{Hits@10} \\ \hline \hline
FPMC~\cite{fpmc} & 0.5906 & 0.6021 & 0.6402 & 0.6162 & 0.6481 \\
TransRec~\cite{transrec} & 0.5437 & 0.5803 & 0.6055 & 0.5839 & 0.6081 \\
GRU4Rec+~\cite{gru4recplus} & 0.6291 & 0.6432 & 0.6796 & 0.6443 & 0.6867 \\
Caser~\cite{caser} & 0.6418 & 0.6472 & 0.6782 & 0.6507 & 0.6991 \\
AUM~\cite{reapp2} & 0.5718 & 0.6089 & 0.6358 & 0.6097 & 0.6433 \\
Bert4Rec~\cite{bert4rec} & 0.7031 & 0.7346 & 0.7442 & 0.7188 & 0.7301 \\
SASRec~\cite{sasrec} & 0.7279 & 0.7530 & \fbox{0.7562} & 0.7583 & 0.7648 \\
TiSRec~\cite{tisasrec} & \fbox{0.7284} & \fbox{0.7542} & 0.7558 & \fbox{0.7618} & \fbox{0.7663} \\ \midrule
ReVAMP & \textbf{0.7865} & \textbf{0.8021} & \textbf{0.8186} & \textbf{0.8203} & \textbf{0.8340} \\ \bottomrule
\end{tabular}
\vspace{3mm}
\caption{\label{revtab:tdk_results} Next \cin performance comparison between \revamp and state-of-the-art baselines for the \tdk dataset. All results are statistically significant over the best baseline as in Table \ref{revtab:tel_results}.}
\begin{tabular}{c|ccccc}
\toprule
\textbf{Baselines} & \textbf{NDCG/Hits@1} & \textbf{NDCG@5} & \textbf{NDCG@10} & \textbf{Hits@5} & \textbf{Hits@10} \\ \hline \hline
FPMC~\cite{fpmc} & 0.7224 & 0.7362 & 0.7704 & 0.7408 & 0.7892 \\ 
TransRec~\cite{transrec} & 0.6872 & 0.6892 & 0.7691 & 0.6902 & 0.7784 \\ 
GRU4Rec+~\cite{gru4recplus} & 0.7319 & 0.7654 & 0.7913 & 0.7703 & 0.7962 \\ 
Caser~\cite{caser} & 0.7321 & 0.7802 & 0.8079 & 0.8157 & 0.8482 \\ 
STGN~~\cite{attnloc4} & 0.6694 & 0.7981 & 0.8132 & 0.8224 & 0.8549 \\ 
AUM~\cite{reapp2} & 0.7184 & 0.7395 & 0.7646 & 0.7782 & 0.8179 \\ 
Bert4Rec~\cite{bert4rec} & 0.7728 & 0.8247 & 0.8281 & 0.8614 & 0.8743 \\ 
SASRec~\cite{sasrec} & 0.8295 & \fbox{0.8621} & 0.8680 & \fbox{0.9027} & \fbox{0.9108} \\ 
TiSRec~\cite{tisasrec} & \fbox{0.8307} & 0.8619 & \fbox{0.8693} & 0.8998 & 0.9014 \\ \midrule
ReVAMP & \textbf{0.8793} & \textbf{0.9324} & \textbf{0.9371} & \textbf{0.9492} & \textbf{0.9594} \\ \bottomrule
\end{tabular}
\vspace{-5mm}
\end{table}

\subsection{Performance Comparison} \label{revrq1}
In this section, we address RQ1 and report the location recommendation performance of different methods across both mobility datasets. The results for \tel and \tdk datasets are given in Table~\ref{revtab:tel_results} and Table~\ref{revtab:tdk_results}, respectively. From these results, we make the following observations.
\begin{compactitem}[$\bullet$]
\item \revamp consistently outperforms all other baselines for sequential mobility prediction across both datasets. The superior performance signifies the importance of including the smartphone usage pattern of a user to determine her mobility preferences. We also note that the performance gains over other self-attention-based models -- Bert4Rec~\cite{bert4rec}, SASRec~\cite{sasrec}, and TiSRec~\cite{tisasrec} further reinforce our claim that including \textit{relative} positional encodings based on the smartphone, spatial and temporal characteristics can better learn the mobility preferences.

\item We also note that the self-attention-based models, such as Bert4Rec, SASRec, TiSRec, and \revamp consistently yield the best performance on all the datasets and easily outperform CNN and RNN-based models, namely Caser~\cite{caser} and GRU4Rec+~\cite{gru4recplus}. This further signifies the unequaled proficiency of the transformer~\cite{transformer} architecture to capture the evolution of user preferences across her trajectory sequence. More importantly, it outperforms the state-of-the-art location recommendation model STGN~\cite{attnloc4} that uses the additional information of precise geographical coordinates of each POI location.

\item \revamp also outperforms the other smartphone-activity-based approach, AUM~\cite{reapp2} by up to 34\% across different metrics. 

\item We also note that neural baselines such as Caser~\cite{caser}, GRU4Rec+~\cite{gru4recplus} achieve better results as compared to FPMC~\cite{fpmc} and TransRec~\cite{transrec}. It asserts the utmost importance of designing modern recommender systems using neural architectures. Moreover, GRU4Rec+ achieves a similar performance compared to Caser. 
\end{compactitem}

\noindent To sum up, our empirical analysis suggests the following: \begin{inparaenum}[(i)] \item the state-of-the-art models, including self-attention and standard neural models, are not suitable for modeling mobile-user trajectories, and \item \revamp achieves better recommendation performance as it captures the mobility dynamics as well as the smartphone-activity of a user. \end{inparaenum}

\begin{figure}[t]
\centering
\begin{subfigure}{0.45\columnwidth}
  \centering
{\includegraphics[height=4cm]{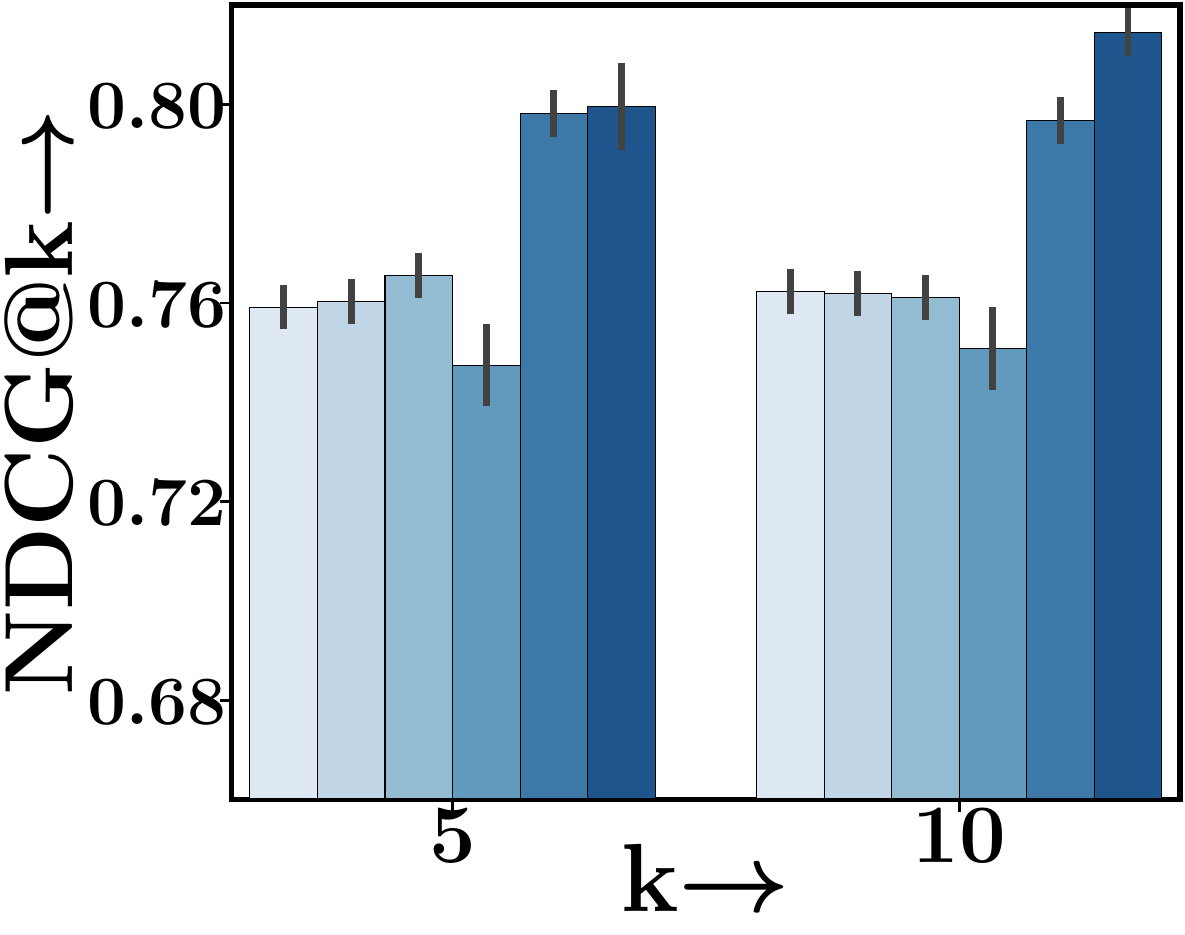}}
\caption{\tel}
\end{subfigure}
\hfill
\begin{subfigure}{0.45\columnwidth}
  \centering
{\includegraphics[height=4cm]{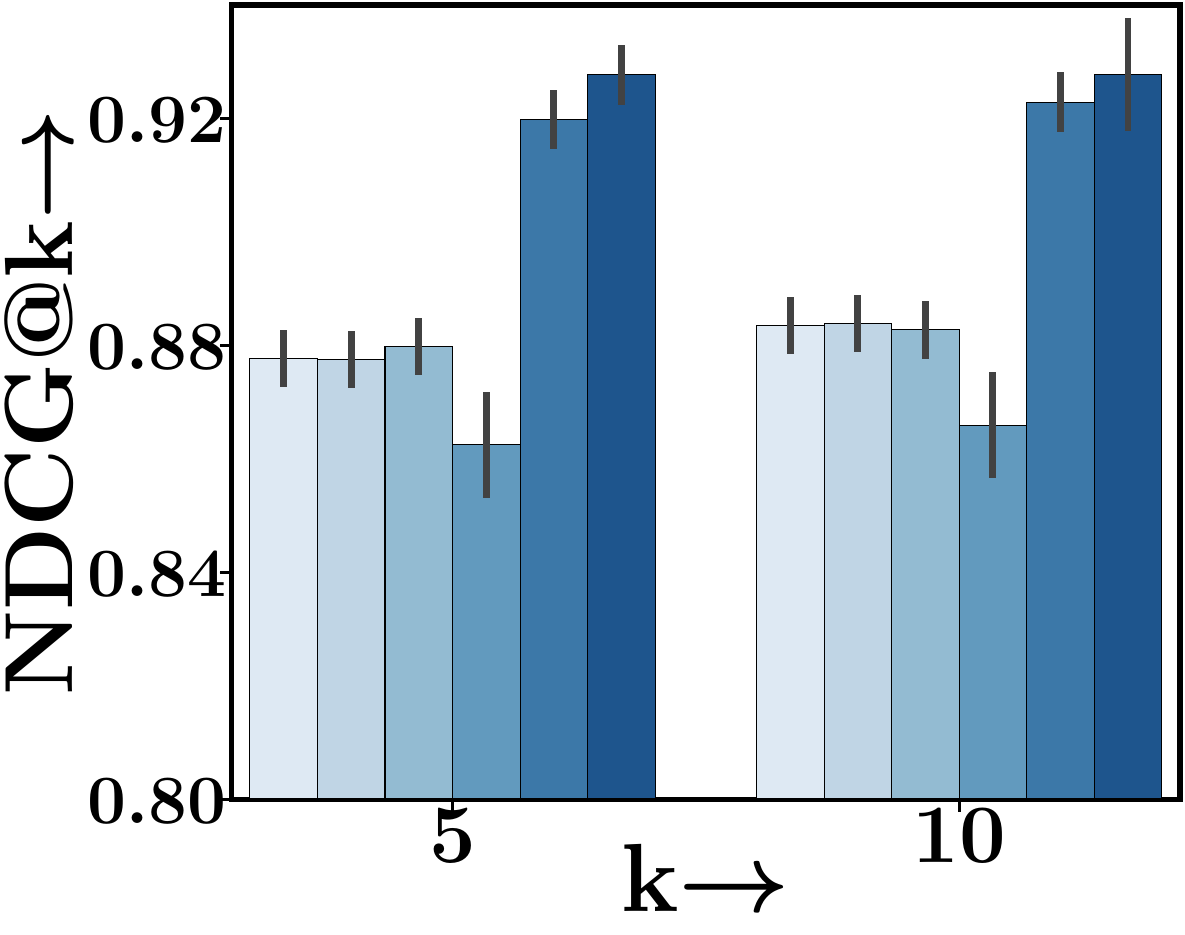}}
\caption{\tdk}
\end{subfigure}
{\includegraphics[width=0.6\linewidth]{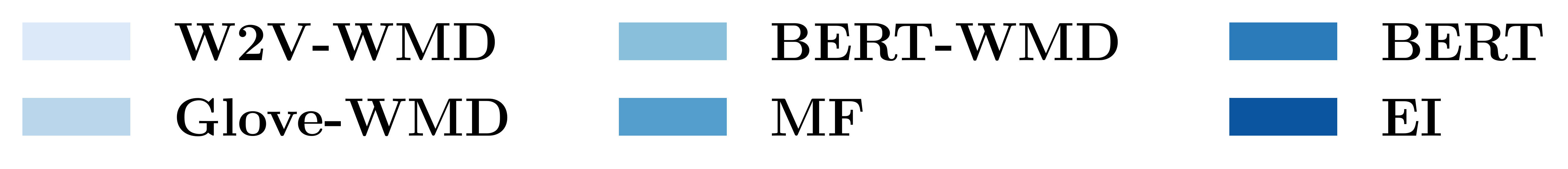}}
\vspace{-2mm}
\caption{POI recommendation performance of \revamp with different methods for obtaining the relative positional encodings, \ie, the inter-\cin differences between app- and location embeddings. Here, the time-based representations are kept consistent across all the models.}
\vspace{-2mm}
\label{revfig:init}
\end{figure}

\subsection{Ablation Study} \label{revrq3}
To address RQ2, we perform an ablation study to estimate the efficacy of different components in the \revamp architecture. More specifically, we aim to calculate the contribution of (i) the embedding initiator and (ii) relative positional embeddings.

\xhdr{Analysis of Embedding Initiator}
We reiterate that EI, defined in Section \ref{revsec:eim}, is used to learn the semantic meaning of each app- and POI category as well as the influence between these embeddings in a mobility sequence. We accomplish this via a joint loss that consists of -- minimizing the divergence between the category vector and the pre-trained BERT\cite{bert} vectors and a collaborative-filtering (CF) loss. These trained embeddings are later used to learn the inter-\cin differences through relative positional encodings. To emphasize its importance, we compare the prediction performances of \revamp with different procedures to learn category embeddings and thus the relative embeddings. Specifically, we consider: \begin{inparaenum}[(i)] \item word-movers-distance(WMD)~\cite{wmd} between the word2vec~\cite{w2v} representations of each category, \item WMD on Glove~\cite{glove} based representations, \item WMD based on BERT~\cite{bert} initialized vectors, \item simple collaborative filtering based parameter training, \item using pre-trained BERT, and (vi) the proposed EI model\end{inparaenum}. From the results in Figure~\ref{revfig:init}, we note that our proposed EI achieves the best prediction performance compared to other approaches. We also note that standard pre-trained BERT vectors outperform other WMD-based approaches.

\xhdr{Relative Positional Encodings}
Relative positional embeddings are a crucial element in our model. We calculate the performance gains due to the different relative encodings -- app-, time- and location-based by estimating the recommendation performance of the following approaches: \begin{inparaenum}[(i)] SASRec~\cite{sasrec}; \item TiSRec~\cite{tisasrec}; \item  \revamp with time-based relative positional encoding called \revamp-t; \item  \revamp with app-based encodings, denoted as \revamp-a; \item  \revamp with location-based encodings, denoted as \revamp-l; and \item the complete \revamp model with all relative encodings\end{inparaenum}.

\begin{figure}[t]
\centering
\begin{subfigure}{0.45\columnwidth}
  \centering
{\includegraphics[height=4cm]{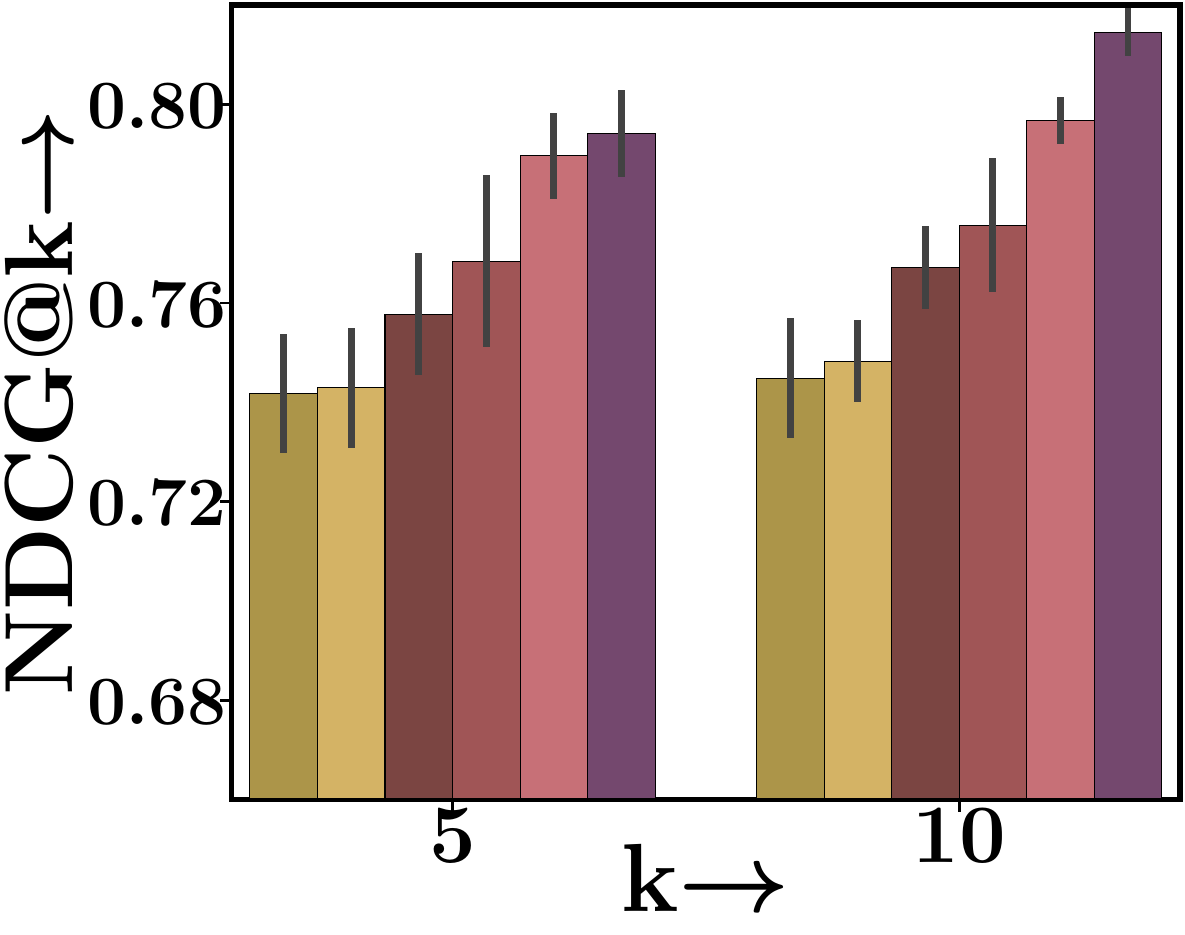}}
\caption{\tel}
\end{subfigure}
\hfill
\begin{subfigure}{0.45\columnwidth}
  \centering
{\includegraphics[height=4cm]{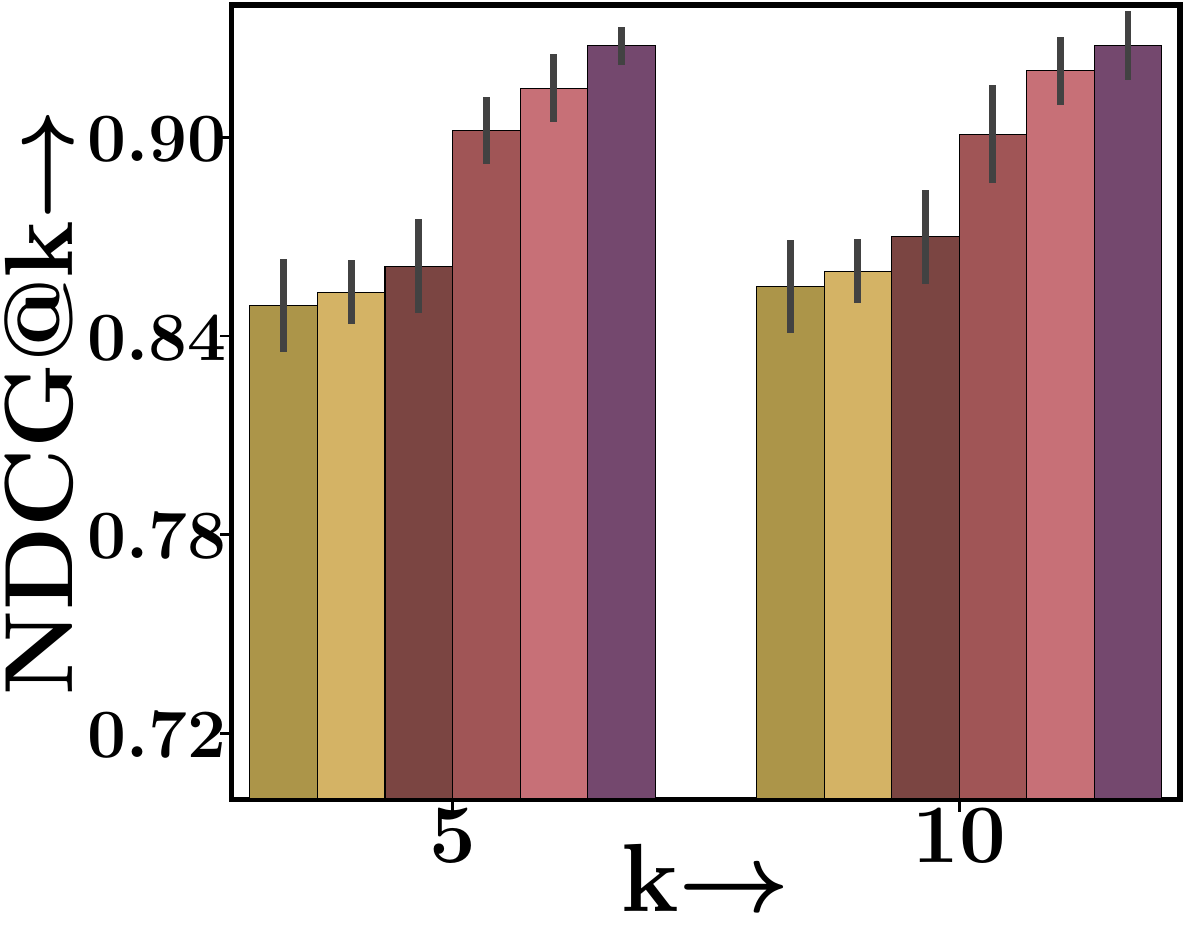}}
\caption{\tdk}
\end{subfigure}
{\includegraphics[width=0.6\linewidth]{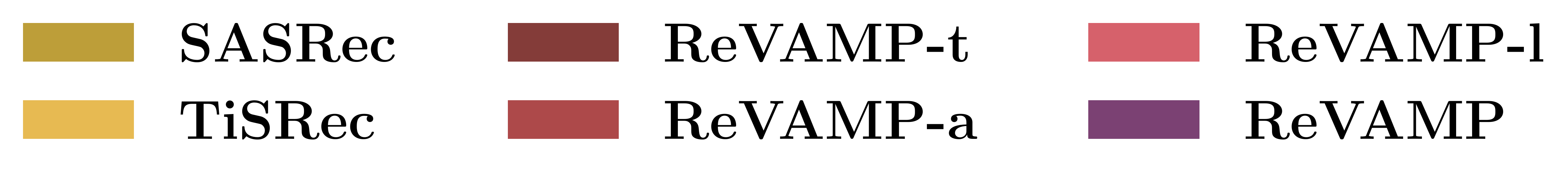}}
\vspace{-2mm}
\caption{Ablation study with different \textit{relative} positional encodings used in \revamp and their comparison with SASRec~\cite{sasrec} and TiSRec~\cite{tisasrec}.}
\vspace{-2mm}
\label{revfig:ablation}
\end{figure}

Figure \ref{revfig:ablation} summarizes our results where we observe that including relative positional encodings of any form, whether app-based or location-based, leads to better prediction performances. Interestingly, the contribution of location-based relative positional embeddings is more significant than the app-based and could be attributed to \textit{larger} variations in location-category than the app-category across an event sequence. For example, the difference between location categories of a university region and an office space will capture larger dynamics than the differences in smartphone app usage across these two regions. However, jointly learning all positional encoding leads to the best performance over both datasets. The improvements of \revamp-t over TiSRec~\cite{tisasrec} could be due to the inclusion of \textit{absolute} event encodings (both app and location).

\begin{figure}[t]
\centering
\begin{subfigure}{0.24\columnwidth}
  \centering
{\includegraphics[height=3cm]{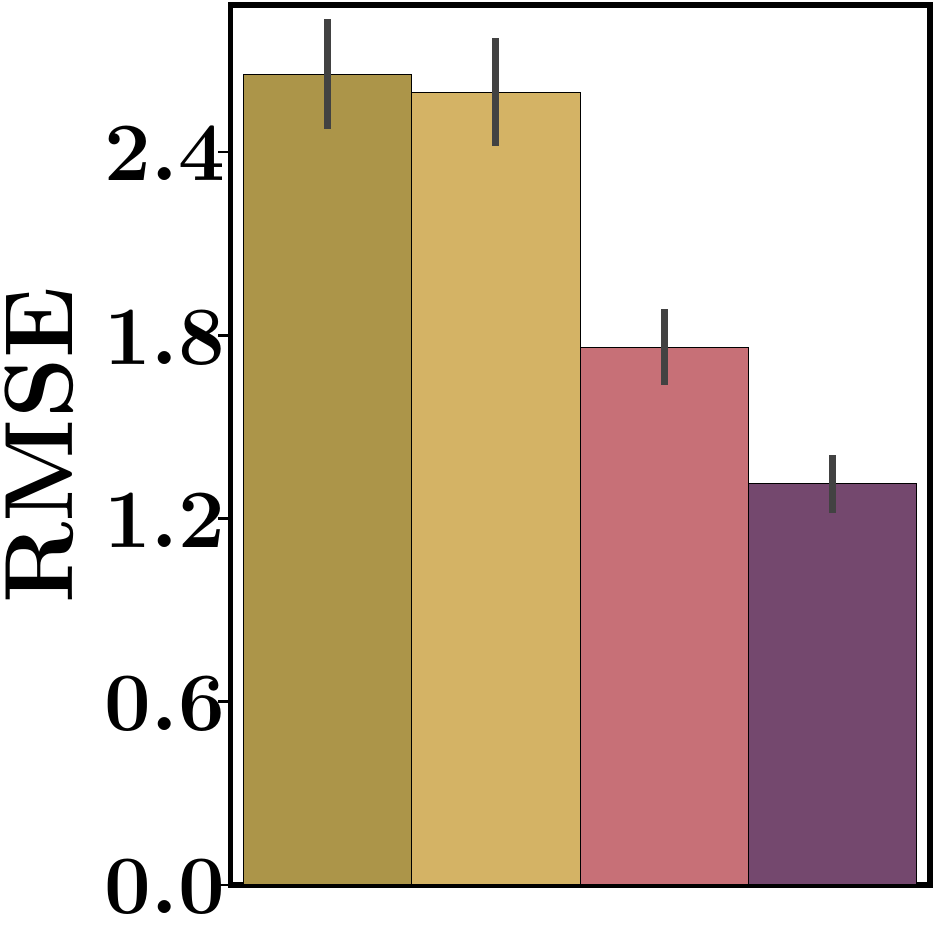}}
\caption{ST (App)}
\end{subfigure}
\hfill
\begin{subfigure}{0.24\columnwidth}
  \centering
{\includegraphics[height=3cm]{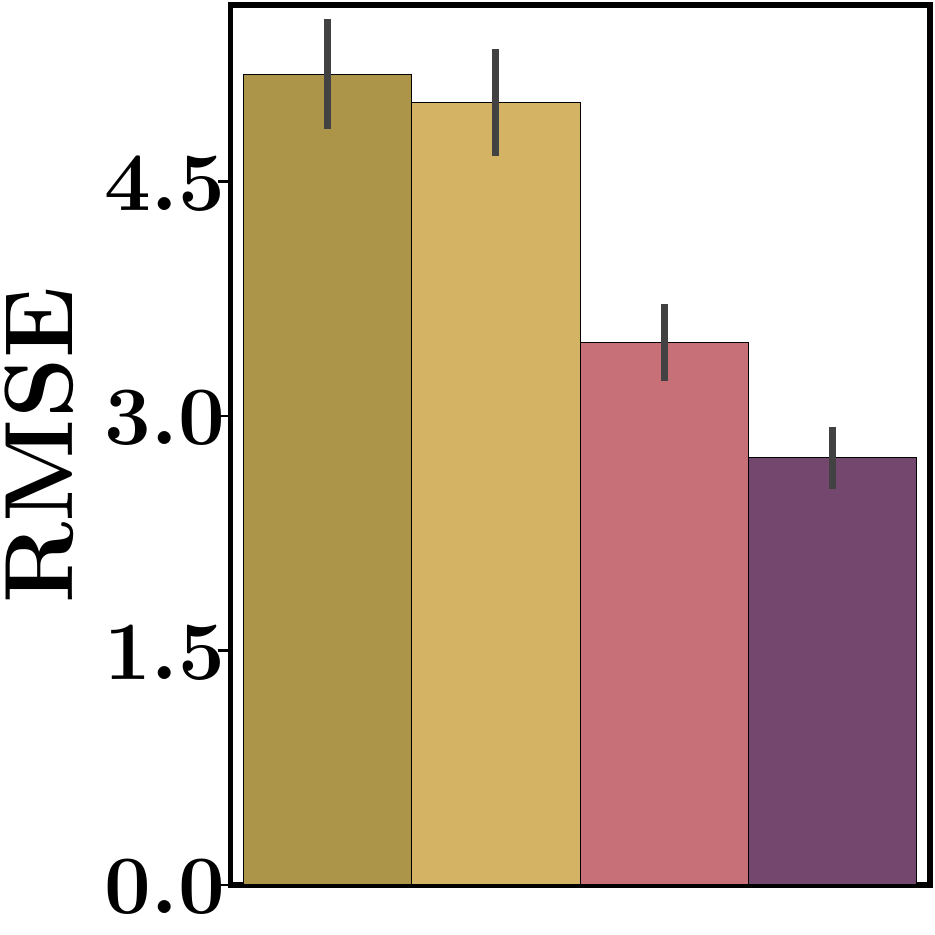}}
\caption{TD (App)}
\end{subfigure}
\hfill
\begin{subfigure}{0.24\columnwidth}
  \centering
{\includegraphics[height=3cm]{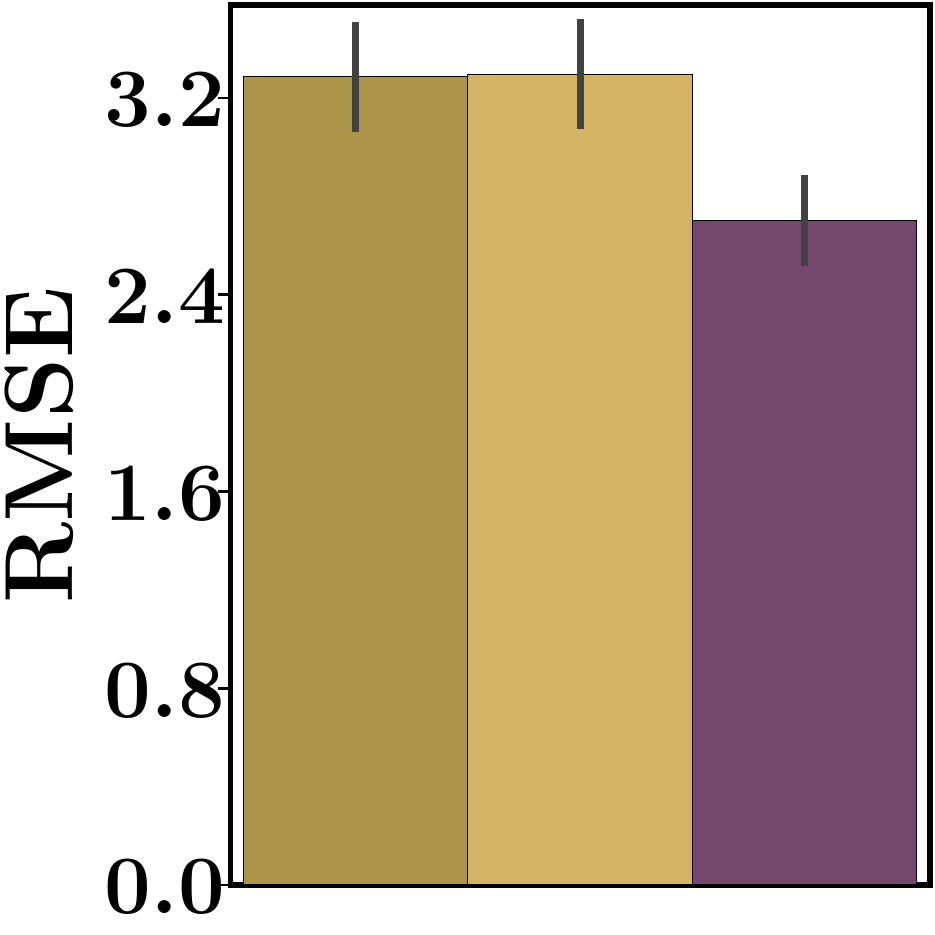}}
\caption{ST (POI)}
\end{subfigure}
\hfill
\begin{subfigure}{0.24\columnwidth}
  \centering
{\includegraphics[height=3cm]{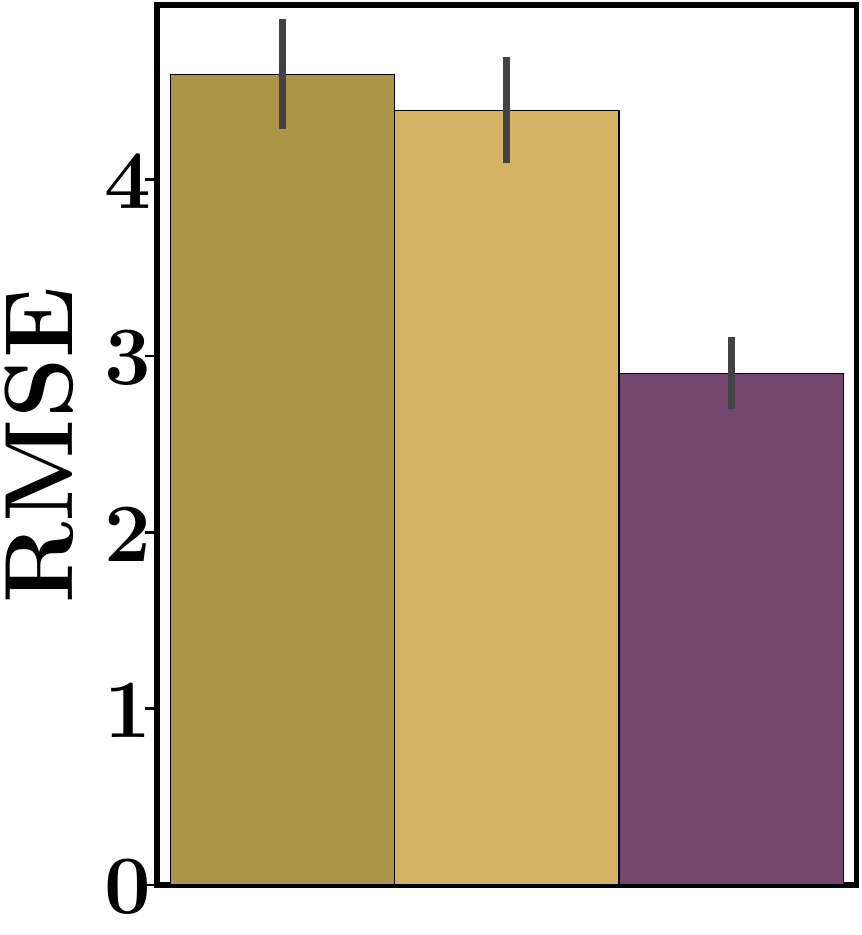}}
\caption{TD (POI)}
\end{subfigure}
{\includegraphics[width=0.8\linewidth]{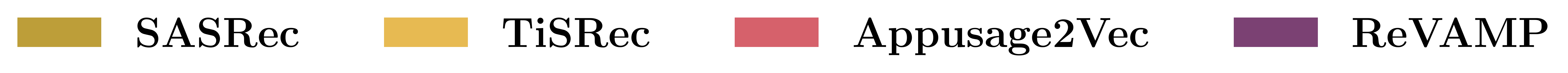}}
\vspace{-4mm}
\caption{Root-mean squared distance between the user-preference vector estimated by the model and the mean of the app- and location-category vectors of events in the test set. It shows that \revamp is the best performer for all the datasets.}
\vspace{-2mm}
\label{revfig:category}
\end{figure}

\subsection{App and Location Prediction Category} \label{revrq2}
Our goal via \revamp is to understand the smartphone activity of a user and correlate it with her mobile trajectories. Therefore, we perform an additional experiment to evaluate how effectively is \revamp able to predict the app- and the location category for the next user \cin. We also introduce an additional state-of-the-art smartphone-activity modeling baseline, Appusage2Vec~\cite{appusage2vec} which considers the category of the app and the time spent on the app by the user to learn an app-preference embedding for a user. We also compare with the state-of-the-art transformer-based models -- SASRec~\cite{sasrec} and TiSRec~\cite{tisasrec}. For an even comparison, we rank the models using the root-mean-squared (RMS) distance between the final user preference embedding obtained after learning on $N$ consecutive events of a user and the \textit{mean} of location and category embeddings of the $N+1$ event in the sequence. Accordingly, we also modify the architectures of SASRec and TiSRec to predict user affinity across the location and app category affinities. From the results in Figure~\ref{revfig:category}, we make the following observations: \begin{inparaenum}[(i)] \item  \revamp easily outperforms all other baselines for both apps and location category prediction. This illustrates the better user-preference modeling power of \revamp over other approaches, \item For app-category prediction, Appusage2Vec also outperforms both SASRec and TiSRec; however, \revamp easily outperforms Appusage2Vec across both the datasets.\end{inparaenum}

\subsection{Scalability of \revamp}
To address RQ3, \ie, determine the scalability of \revamp with different positional encodings -- absolute and relative, we present the epoch-wise time taken for training \revamp in Table \ref{revtab:scale}. Note that these running times exclude the time for pre-processing, where we calculate the inter-event app and location category-based differences. We note that the runtime of \revamp is linear with the number of users, and secondly, even for a large-scale dataset, like \tdk, we can optimize all parameters in \revamp well within 170 minutes. These run times are well within the range for designing recommender systems.
\begin{table}[t]
\centering
\caption{Run-time Statistics of training \revamp in minutes on a 32GB Tesla V100 GPU with 256 batch-size.}
\small
\begin{tabular}{lrrrrrr}
\toprule
\textbf{Epochs} & \textbf{20} & \textbf{60} & \textbf{100} & \textbf{160} & \textbf{200} & \textbf{300}\\
\midrule
\tel & 2.43 & 6.31 & 10.48 & 16.72 & 21.47 & 32.01\\
\tdk & 11.39 & 33.73 & 56.12 & 89.81 & 112.35 & 168.49\\
\bottomrule
\end{tabular}
\label{revtab:scale}
\end{table}

\begin{figure}[t]
\centering
\begin{subfigure}{0.45\columnwidth}
  \centering
{\includegraphics[height=4cm]{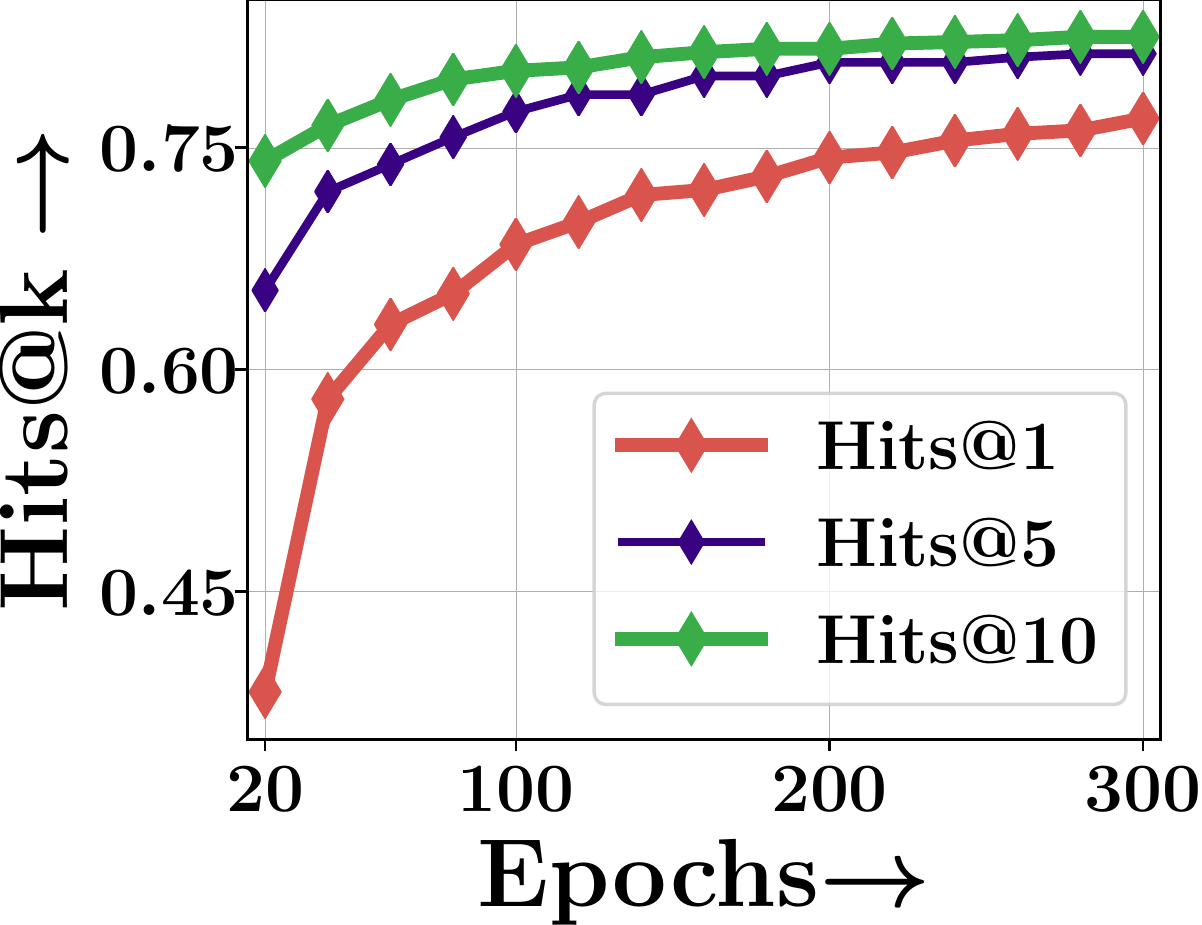}}
\caption{\tel}
\end{subfigure}
\hfill
\begin{subfigure}{0.45\columnwidth}
  \centering
{\includegraphics[height=4cm]{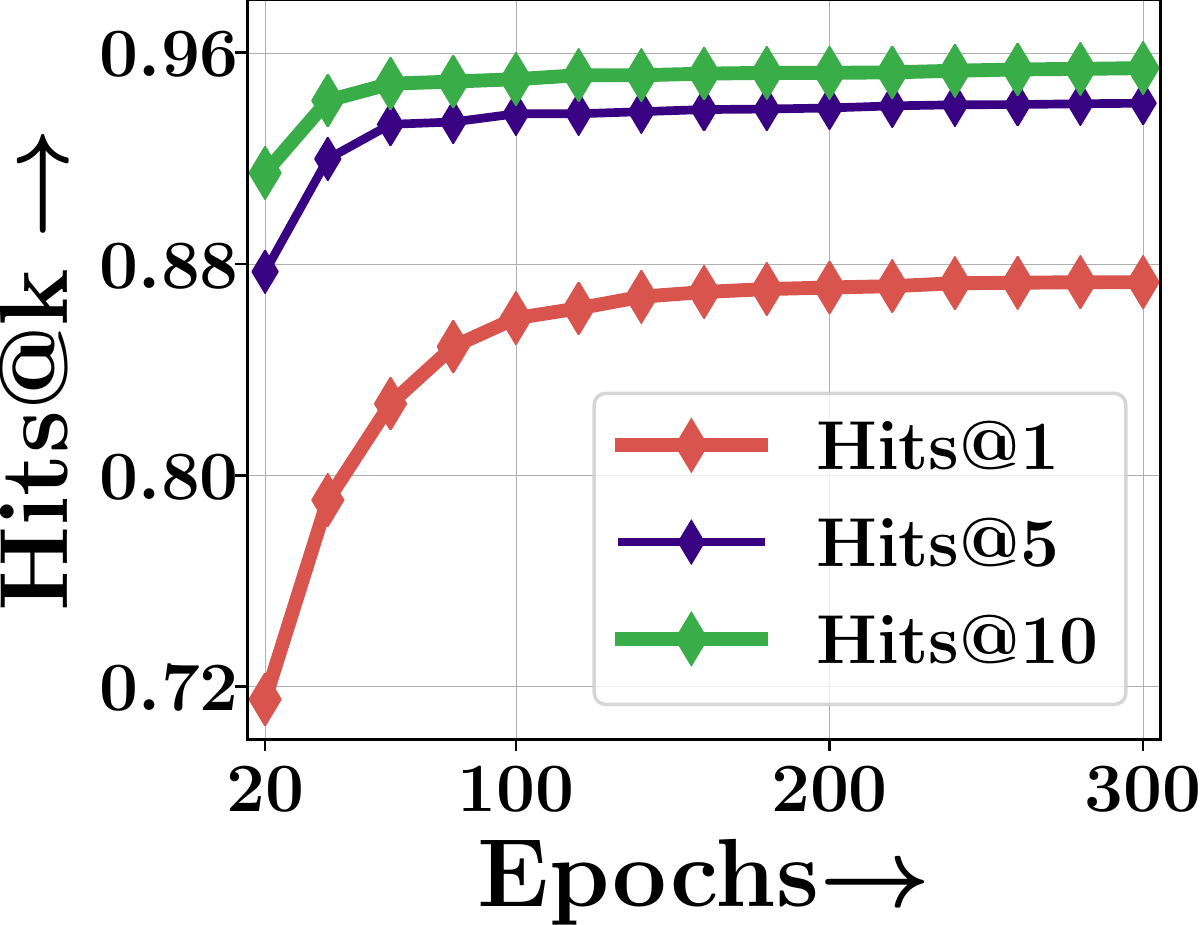}}
\caption{\tdk}
\end{subfigure}
\vspace{-2mm}
\caption{Epoch-wise recommendation performance of \revamp for both datasets in terms of Hits@1, 5, 10.}
\vspace{-2mm}
\label{revfig:runtime}
\end{figure}

\xhdr{Convergence of \revamp Training}
As we propose the first-ever application of the self-attention model for smartphones and human mobility, we also perform a convergence analysis during training \revamp. To emphasize the stability of \revamp training procedure, we plot the epoch-wise best prediction performance of \revamp across both datasets in Fig \ref{revfig:runtime}. From the results, we note that despite the multi-variate nature of the data and the disparate positional encodings, \revamp converges only in a few training iterations. It is also important to note that the \revamp significantly outperforms other RNN-based baselines even with limited training of 40 iterations.

\section{Conclusion}\label{revsec:conc}
In this chapter, we highlighted the drawbacks of modern POI recommender systems that ignore the smartphone usage characteristics of users. We also proposed a novel sequential POI recommendation model, called \revamp, that incorporates the smartphone usage details of a user while simultaneously maintaining user privacy. Inspired by the success of relative positional encodings and self-attention models, \revamp uses relative as well as absolute positional encodings determined by the inter-\cin variances in the smartphone app category, POI category, and time over the \cins in the sequence. Our experiments over two diverse datasets from China show that \revamp significantly outperforms other state-of-the-art baselines for POI recommendation. Moreover, we also show the contribution of each component in the \revamp architecture and analyze the learning stability of the model. 


%% file: chapters/008_neuroseqret.tex
\newcommand{\nsrf}{\textsc{SelfAttn-}\nsr}
\newcommand{\nsrs}{\textsc{CrossAttn-}\nsr}

\newcommand{\Tcal}{\mathcal{T}}
\newcommand{\Hcal}{\mathcal{H}}
\newcommand{\Mcal}{\mathcal{M}}
\newcommand{\Ucal}{\mathcal{U}}
\newcommand{\Dcal}{\mathcal{D}}
\newcommand{\Bcal}{\mathcal{B}}
\newcommand{\Qr}{\mathcal{Q}}
\newcommand{\Cr}{\mathcal{C}}
\newcommand{\II}{\mathbb{I}}

\newcommand{\sgn}{\mathop{\mathrm{sign}}}
\newcommand{\cp}{\backslash}

\newcommand{\hash}{\bm{\zeta}}
\newcommand{\zzc}{\bm{v}^c}
\newcommand{\ehdr}[1]{\vspace{1mm}\noindent\emph{---#1}}
\newcommand{\intensity}{\rho}
\newcommand{\reg}{\lambda}
\newcommand{\rel}{{+}}
\newcommand{\nrel}{{-}}
\newcommand{\unw}{U}
\newcommand{\pend}{{}}
\newcommand{\kernel}{\kappa}
\newcommand{\para}[1]{\vspace{1mm} \noindent{{#1}}}
\newcommand{\viz}{\emph{viz.}}
\newcommand{\tpprank}{\texttt{Rank}}

\newcommand{\bracex}[1]{\left(#1\right)}
\newcommand{\newq}{{q{'}}}
\newcommand{\hq}{\zb^{(q,q)}}
\newcommand{\hc}{\zb^{(c,c)}}
\newcommand{\ha}{{\zb^{(a)}}}

\newcommand{\hnewq}{{\zb^{(\newq,\newq)}}}
\newcommand{\znc}{{\bm{\zeta}^{(c)}}}
\newcommand{\zc}{{\bm{\zeta}^{(c)}}}
\newcommand{\app}[1]{Appendix~\ref{app:#1}}
\newcommand{\znq}{\bm{\zeta}^{(\newq)}}
\newcommand{\zq}{\bm{\zeta}^{(q)}}

\newcommand{\eat}[1]{}

\newcommand{\yb}{\bm{y}}
\newcommand{\pb}{\bm{p}}
\newcommand{\kb}{\bm{k}}
\newcommand{\ub}{\bm{u}}
\newcommand{\Ib}{\bm{I}}
\renewcommand{\indicator}[1]{{\llbracket #1 \rrbracket }}

\newcommand{\querytt}{\texttt{query}}
\newcommand{\Querytt}{\texttt{Query}}

\newcommand{\keytt}{\texttt{key}}
\newcommand{\Keytt}{\texttt{Key}}

\newcommand{\valuett}{\texttt{value}}
\newcommand{\Valuett}{\texttt{Value}}
\newcommand{\bhbc}{\overline{\bm{h}} ^{(c,q)}}
\newcommand{\bhbq}{\overline{\bm{h}} ^{(q,q)}}
\newcommand{\bhbcc}{\overline{\bm{h}} ^{(c,c)}}
\newcommand{\bhb}{\overline{\bm{h}} ^{(\bullet,q)}}


\section{Introduction}
Recent developments in predictive modeling using marked temporal point processes (MTPP) have enabled an accurate characterization of several real-world applications involving continuous-time event sequences (CTESs). However, the retrieval problem of such sequences remains largely unaddressed in literature. In this chapter, we address the problem of retrieving temporal event sequences from a large corpus using neural MTPP models. This is a first-of-its-kind application of MTPP as the earlier developments in MTPP models focused on improving the predictive analytics in several real-world applications--- from information diffusion in social networks to healthcare--- by characterizing them with continuous-time event sequences (CTESs)~\cite{Valera2014,rizoiu_hip,wang2017human,daley2007introduction,gupta2021learning, initiator,du2015dirichlet,srijan,de2016learning,rmtpp,farajtabar2017fake,jing2017neural}. Therefore, in this context, given a query sequence, retrieval of \emph{relevant} CTESs from a corpus of sequences is a challenging problem having a wide variety of search-based applications. For example, in audio or music retrieval, one may like to search sequences having different audio or music signatures; the retrieval of ECG sequences relevant to one pathological query ECG sequence can help in the early detection of cardiac disease; in social networks, retrieval of trajectories of information diffusion, relevant to a given trajectory can assist in viral marketing, fake news detection, \etc\ Despite having a rich literature on searching similar time-series~\cite{blondel2021differentiable, gogolou2020data, alaee2020matrix, timegan, cai2019dtwnet, shen2018accelerating, cuturi2017soft, paparrizos2015k}, the problem of designing retrieval models specifically for CTES has largely been unaddressed in the past. Moreover, as shown in our experiments, the existing search methods for time sequences are largely ineffective for a CTES retrieval task since the underlying characterization of the sequences varies across these two domains. 
 
\subsection{Our Contribution}
In this chapter, we first introduce \nsr, a family of supervised retrieval models for continuous-time event sequences, and then develop a trainable locality-sensitive hashing (LSH) based method for efficient retrieval over very large datasets~\cite{neuroseqret}. Specifically, our contributions are as follows:

\xhdr{Query unwarping} The notion of relevance between two sequences varies across applications. A relevant sequence pair can share very different individual attributes, which can mislead the retrieval model if the sequences are compared as-it-is. In other words, an observed sequence may be a warped transformation of a hidden sequence~\cite{ido,gervini2004self}. To tackle this problem, \nsr\ first applies a trainable unwarping function on the query sequence before the computation of a relevance score. Such an unwarping function is a monotone transformation, which ensures that the chronological order of events across the observed and the unwarped sequences remains the same~\cite{ido}.

\xhdr{Neural relevance scoring model}
In principle, the relevance score between two sequences depends on their latent similarity. We measure such similarity by comparing the generative distribution between the query-corpus sequence pairs. In detail, we feed the unwarped query sequence and the corpus sequence into a neural MTPP-based relevance scoring model,  which computes the relevance score using a Fisher kernel~\cite{fisher} between the corpus and the unwarped query sequences. Such a kernel offers two key benefits over other distribution similarity measures, \eg, KL divergence or Wasserstein distance: (i) it computes a natural similarity score between query-corpus sequence pairs in terms of the underlying generative distributions; and, (ii) it computes a dot product between the gradients of log-likelihoods of the sequence pairs, which makes it compatible with locality-sensitive hashing for certain design choices and facilitates efficient retrieval. In this context, we provide two MTPP models, leading to two variants of \nsr, which allows a nice tradeoff between accuracy and efficiency.

\noindent{\underline{\nsrf}}: {Here, we use transformer Hawkes process~\cite{thp} which computes the likelihood of corpus sequences independently of the query sequence.} Such a design admits precomputable corpus likelihoods, which in turn allows for prior indexing of the corpus sequences before observing the unseen queries. This setup enables us to apply LSH for efficient retrieval.

\noindent{\underline{\nsrs}}: Here, we propose a novel cross attention-based neural MTPP model to compute the sequence likelihoods. Such a cross-attention mechanism renders the likelihood of corpus sequence dependent on the query sequence, making it a more powerful retrieval model. While \nsrs is not directly compatible with such a hashing-based retrieval, it can be employed in a telescopic manner--- where a smaller set of relevant candidates are first retrieved using LSH applied on top of \nsrf, and then reranked using \nsrs. Therefore, from the design perspective, these two models provide a tradeoff between accuracy and efficiency.

\noindent Having computed the relevance scores, we learn the unwarping function and the MTPP model by minimizing a pairwise ranking loss based on the ground truth relevance labels.

\xhdr{Scalable retrieval}
Next, we use the predictions made by \nsrf\ to develop a novel hashing method that enables efficient sequence retrieval. More specifically, we propose an optimization framework that compresses the learned sequence embeddings into binary hash vectors, while simultaneously limiting the loss due to compression. Then, we use locality-sensitive hashing~\cite{GionisIM1999hash} to bucketize the sequences into hash buckets so that sequences with similar hash representations share the same bucket. Finally, given a query sequence, we consider computing relevance scores only with the sequences within its bucket. Such a hashing mechanism combined with high-quality sequence embeddings achieves fast sequence retrieval with no significant loss in performance. Finally, our experiments with real-world datasets from different domains show that both variants of \nsr\ outperform several baselines including the methods for continuous-time series retrieval. Moreover, we observe that our hashing method applied on \nsrf\ can make a tradeoff between the retrieval accuracy and efficiency more effectively than baselines based on random hyperplanes as well as exhaustive enumeration.

\section{Preliminaries}
\subsection{Notations and MTPP}
An event $e$ in an MTPP is realized using a tuple $(t,x)$, where $t\in\mathbb{R}_+$ and $x\in \mathcal{C}$ are the arrival time and the mark of the event $e$. Then, we use $\Hcal(t)$ to denote a continuous time event sequence (CTES) where each event has arrived until and excluding time $t$, \ie, $\Hcal(t):=\set{e_i=(t_i,x_i)\given t_{i-1}<t_{i}<t}$. Moreover we use $\Tcal(t)$ and $\Mcal(t)$ to denote the sequence of arrival times $\set{t_i\given e_i\in\Hcal(t)}$  and the marks $\set{x_i\given e_i\in\Hcal(t)}$.  Finally, we denote the counting process $N(t)$ as counts of the number of events that happened until and excluding time $t$, encapsulating the generative mechanism of the arrival times.

\xhdr{Generative model for CTES}  
The underlying MTPP model consists of two components -- (i) the dynamics of the arrival times and (ii) the dynamics of the distribution of marks. Most existing works~\cite{rmtpp, sahp,mei_icml,nhp,shelton,thp} model the first component using an intensity function which explicitly models the likelihood of an event in the infinitesimal time window $[t,t+dt)$, \ie,  $\lambda^{\pend}(t )= \text{Pr} (dN(t)=1|\Hcal(t))$. In contrast, we use an intensity-free approach following the proposal by~\citet{intfree}, where we explicitly model the distribution of the arrival time $t$ of the next event $e$. Specifically, we denote the density $\intensity$ of the arrival time and  the distribution $m^{\pend}$ of the mark  of the next event as follows:
\begin{align}\label{eq:qm}
 \intensity(t) dt & = \text{Pr} (e \text{ in } [t,t+dt) \given \Hcal(t)), \\
 m^{\pend}(x)&=\text{Pr} (x\given \Hcal(t))
\end{align}
As discussed by~\citet{intfree}, such an intensity-free MTPP model enjoys several benefits over its intensity-based counterparts in terms of facilitating efficient training, scalable prediction, computation of expected arrival times, \etc\ Given a sequence of observed events $\Hcal(T)$ collected during the time interval $(0,T]$, the likelihood function is given by:
\begin{equation}
 p (\Hcal(T)) =\textstyle  \prod_{e_i=(t_i,x_i)\in\Hcal(T)}  \intensity(t_i) \times   m^{\pend}(x_i)
\end{equation}

\subsection{Problem setup}
Next, we set up our problem of retrieving a ranked list of sequences from a corpus of continuous-time event sequences (CTESs)
which are relevant to a given query CTES. 

\xhdr{Query and corpus sequences, relevance labels} We operate on a large corpus of sequences $\set{\Hcal_c(T_c)\given c\in \Cr}$, where $\Hcal_c(T_c)=\{(t^{(c)} _i, x^{(c)} _i)  \given t^{(c)} _i < T_c \}$. We are given a set of query sequences $\set{\Hcal_q(T_q) \given q\in\Qr}$ with $\Hcal_q(T_q)=\{(t^{(q)} _i, x^{(q)} _i) \given t^{(q)} _i < T_q\}$, as well as a query-specific relevance label for the set of corpus sequences. That is, for a given query sequence $\Hcal_q$, we have: $y(\Hcal_q,\Hcal_c)=+1$  if $\Hcal_c$ is marked as relevant to $\Hcal_q$ and $y(\Hcal_q,\Hcal_c)=-1$ otherwise. 

We define $\Cr_{q\rel} =\{c\in\Cr \given y(\Hcal_q,\Hcal_c)=+1  \}, \text{and, }
\Cr_{q\nrel} =\{c\in\Cr \given y(\Hcal_q,\Hcal_c)=-1  \}$, with $\Cr = \Cr_{q\rel}  \cup \Cr_{q\nrel} $. Finally, we denote $T=\max\{T_q,T_c\given q\in\Qr,c\in\Cr\}$ as the maximum time of the data collection.

\xhdr{Our Goal} We aim to design an efficient CTES retrieval system, which would return a list of sequences from a known corpus of sequences, relevant to a given query sequence $\Hcal_q$. Therefore, we can view a sequence retrieval task as an instance of the ranking problem. Similar to other information retrieval algorithms, a CTES retrieval algorithm first computes the estimated relevance $s(\Hcal_q,\Hcal_c)$ of the corpus sequence $\Hcal_c$ for a given query sequence $\Hcal_q$ and then provides a ranking of $\Cr$ in the decreasing order of their scores.

\section{\nsr Model}
In this section, we describe \nsr family of MTPP-based models that we propose for the retrieval of continuous-time event sequences (CTES). We begin with an outline of its two key components.

\subsection{Components of \nsr} \label{subsec:components}
\nsr\  models the relevance scoring function between query and corpus
sequence pairs. However, the relevance of a corpus sequence to the query is latent and varies widely across applications. To accurately characterize this relevance measure, \nsr\ works in two steps. First, it unwarps the query sequences to make them compatible for comparison with the corpus sequences. Then, it computes the pairwise relevance score between the query and corpus sequences using neural MTPP models.

\xhdr{Unwarping query sequences} Direct comparison between a query and a corpus sequence can provide misleading outcomes, since
they also contain their own individual idiosyncratic factors in addition to sharing some common attributes. In fact, a corpus sequence can be highly relevant to the query, despite greatly varying in timescale, initial time, etc. In other words, it may have been generated by applying a warping transformation on a latent sequence~\cite{ido,gervini2004self}. Thus, a direct comparison between a relevant sequence pair may give a poor relevance score. To address this challenge, we first apply a trainable unwarping function~\cite{ido} $U(\cdot)$ on the arrival times of a query sequence, which enhances its compatibility for comparing it with the corpus sequences\footnote{\scriptsize We only apply the unwarping function on the times. Since marks belong to a fixed discrete set, we believe marks are directly comparable.}. More specifically, we define $\unw (\Hcal_q):=\{(\unw(t^{(q)} _i), x^{(q)} _i) \}$. In general, $\unw$  satisfies two properties~\cite{ido,gervini2004self}\eat{ which are}: \emph{unbiasedness}, \ie, having a small value of $\left\|\unw(t )]-t\right\|$ and \emph{monotonicity}, \ie, $ {d\, \unw(t  ) }/{dt} \ge 0$. These properties ensure that the chronological order of the events across both the warped observed sequence and the unwarped sequence remains the same.

Such a sequence transformation learns to capture the similarity between two sequences, even if it is not apparent due to different individual factors, as we shall later in our experiments (Figure~\ref{fig:UU}). 

\xhdr{Computation of relevance scores}
Given a query sequence $\Hcal_q$ and a corpus sequence $\Hcal_c$, we compute the relevance score $s(\Hcal_q,\Hcal_c)$ 
using two similarity scores, \eg,  (i) a \emph{model independent} sequence similarity score and (ii) a \emph{model based} sequence similarity score.

\ehdr{Model independent similarity score:} 
Computation of model-independent similarity score between two sequences is widely studied in literature~\cite{xiao2017wasserstein,mueen2016extracting,su2020survey}. They are computed using different distance measures between two sequences, \eg, DTW, Wasserstein distance, \etc, and therefore, can be immediately derived from data without using the underlying MTPP model. In this work, we compute the model-independent similarity score, $\textsc{Sim}_{\unw}(\Hcal_q,\Hcal_c)$, between $\Hcal_q$ and $\Hcal_c$ as follows:
\begin{align}
\textsc{Sim}_{\unw}(\Hcal_q,\Hcal_c)& = -\Delta_{t}(\unw(\Hcal_q),\Hcal_c)-\Delta_{x}(\Hcal_q,\Hcal_c)
\end{align}
where, $\Delta_{t}$ and $\Delta_{x}$ are defined as:
\begin{equation}
\Delta_{t}(\unw(\Hcal_q),\Hcal_c) = \sum_{i=0}^ {H_{\min}} \left| \unw(t^{(q)} _i)- t^{(c)} _i\right| + \hspace{-2mm}\sum_{\substack{t_i \in \Hcal_c\cup\Hcal_q\\ i > |H_{\min}| }} (T -t  _i),
\end{equation}
\begin{equation}
\Delta_{x}(\Hcal_q,\Hcal_c) = \sum_{i=0}^{H_{\min}} \II[x^{(q)} _i \neq x^{(c)}_i] + \big||\Hcal_c|-|\Hcal_q|\big|.
\end{equation}
Here, $H_{\min} = \min\{|\Hcal_q|,|\Hcal_c|\}$, $T=\max\{T_q,T_c\}$ where the events of $\Hcal_q$ and $\Hcal_c$ are gathered until time $T_q$ and $T_c$ respectively; $\Delta_{t}(\unw(\Hcal_q),\Hcal_c)$ is the Wasserstein distance between the unwarped arrival time sequence $\unw(\Hcal_q)$ and the corpus sequence~\cite{xiao2017wasserstein} and, $\Delta_{x}(\Hcal_q,\Hcal_c)$ measures the matching error for the marks, wherein the last term penalizes the marks of last $|\Hcal_c|-|\Hcal_q|$ events of $|\Hcal_c|$.

\ehdr{Model-based similarity score using Fisher kernel:}
We hypothesize that the relevance score $s(\Hcal_q,\Hcal_c)$ also depends on a latent similarity that may not be immediately evident from the observed query and corpus sequences even after unwarping. Such similarity can be measured by comparing the generative distributions of the query-corpus sequence pairs. To this end, we first develop an MTPP-based generative model $p_{\theta}(\Hcal)$ parameterized by $\theta$ and then compute a similarity score using the Fisher similarity kernel between the unwarped query and corpus sequence pairs $(\unw(\Hcal_q),\Hcal_c)$~\cite{fisher}. Specifically, we compute the relevance score between the unwarped query sequence $\unw(\Hcal_q)$ and the corpus sequence $\Hcal_c$ as follows:
\begin{align}
\kernel_{p_{\theta}}(\Hcal_q,\Hcal_c) = \vb_{p_{\theta}}(\unw(\Hcal_q)) ^\top \vb_{p_{\theta}}(\Hcal_c),   \label{eq:fisher}
\end{align}
where $\theta$ is the set of trainable parameters; $\vb_{p_{\theta}} (\cdot)$ is given by
\begin{align}
 \vb_{p}(\Hcal) =  \Ib^{-1/2} _{\theta} \nabla_{\theta} \log p_{\theta}(\Hcal)/||\Ib^{-1/2} _{\theta} \nabla_{\theta} \log p_{\theta}(\Hcal)||_2,
\end{align}
$\Ib_{\theta}$ is the Fisher information matrix~\cite{fisher}, \ie,
$\Ib_{\theta} = \mathbb{E}_{\Hcal\sim p_{\theta}(\bullet)}\left[\nabla_{\theta} \log p_{\theta}(\Hcal)\nabla_{\theta} \log p_{\theta}(\Hcal) ^\top\right]$. We would like to highlight that $ \kernel_{p_{\theta}}(\Hcal_q,\Hcal_c)$ in Eq.~\eqref{eq:fisher} is a normalized version of Fisher kernel since $||\vb_{p_{\theta}}(\cdot)||=1$. Thus, $\kernel_{p_{\theta}}(\Hcal_q,\Hcal_c)$ measures the cosine similarity between $ \vb_{p_{\theta}}(\unw(\Hcal_q))$ and $\vb_{p_{\theta}}(\Hcal_c) $. 

Note that, KL divergence or Wasserstein distance could also serve our purpose of computing the latent similarity between the generative distributions. However, we choose the Fisher similarity kernel because of two reasons: (i) it is known to be a natural similarity measure that allows us to use the underlying generative model in a discriminative learning task~\cite{fisher,sewell2011fisher}; and, (ii) unlike KL divergence or other distribution (dis)similarities, it computes the cosine similarity between $ \vb_{p_{\theta}}(\unw(\Hcal_q))$ and $\vb_{p_{\theta}}(\Hcal_c) $, which makes it compatible with locality-sensitive hashing~\cite{charikar2002similarity}. 

\ehdr{Net relevance score:}
Finally, we compute the relevance score as:
\begin{align}\label{eq:relevance-score-function}
s_{p,\unw}\left(\Hcal_q,\Hcal_c\right) & =   \kernel_{p}\left(\Hcal_q,\Hcal_c\right) 
  +\gamma \textsc{Sim}_U\left(\Hcal_q,\Hcal_{c}\right)
\end{align}
where $\gamma$ is a hyperparameter. 
 
\subsection{Neural Parameterization of \nsr}
Here, we first present the neural architecture of the unwarping function and then describe the MTPP models used to compute the model-based similarity score in Eq.~\eqref{eq:fisher}. As we describe later, we use two MTPP models with different levels of modeling sophistication, \viz, \nsrf\ and \nsrs. In \nsrf, the likelihood of a corpus sequence is computed independently of the query sequence using self attention-based MTPP model, \eg, Transformer Hawkes Process~\cite{thp}. As a result, we can employ a locality-sensitive hashing-based efficient retrieval-based \nsrf. In \nsrs, on the other hand, we propose a more expressive and novel cross-attention MTPP model, where the likelihood of a corpus sequence is dependent on the query sequence. Thus, our models can effectively tradeoff between accuracy and efficiency.

\xhdr{Neural architecture of $U(\cdot)$}  As discussed in Section~\ref{subsec:components}, the unwarping function $U(\cdot)$ should be unbiased, \ie, a small value of $\left\|\unw(t)-t\right\|$ and monotonic, \ie, $d \unw(t) /dt > 0$. To this end, we model $U(\cdot)\approx U_{\phi}(\cdot)$ using a nonlinear monotone function which is computed using an unconstrained monotone neural network (UMNN)~\cite{umnn},\ie,
\begin{equation}
 U_{\phi}(t) = \int_0 ^t u_{\phi}(\tau) d\tau + \eta,\label{eq:umnn}
\end{equation}
where $\phi$ is the parameter of the underlying neural network $u_{\phi}(\cdot)$, $\eta\in \mathcal{N}(0,\sigma)$ and $u_{\phi}:\mathbb{R}\to\mathbb{R}_+$ is a non-negative non-linear function. Since the underlying monotonicity can be achieved only by enforcing the non-negativity of the integrand $u_{\phi}$, UMNN admits an unconstrained, highly expressive parameterization of monotonic functions. Therefore, any complex unwarping function $U_{\phi}(\cdot)$ can be captured using Eq.~\eqref{eq:umnn} by integrating a suitable neural model augmented with ReLU$(\cdot)$ in the final layer. In other words, if $u_{\phi}$ is a universal approximator for a positive function, then $U_{\phi}$  can capture any differentiable unwarping function. We impose an additional regularizer $\frac{1}{\sigma^2}\int_0 ^T \left\|u_{\phi}(t) -1\right\|^2 dt $ on our training loss  which ensures that $\left\|\unw(t)-t\right\|$ remains small.

\xhdr{Neural architecture of MTPP model $p_{\theta}(\cdot)$}
We provide two variants of $p_{\theta}(\cdot)$, which leads to two retrieval models, \viz, \nsrf\ and \nsrs. These two models offer a nice tradeoff between accuracy and efficiency.

\para{\nsrf}: Here, we use the Transformer Hawkes process~\cite{thp}, which applies a self-attention mechanism to model the underlying generative process. In this model, the gradient of corpus sequences $\vb_{\theta}(\Hcal_c) =\nabla_{\theta} \log p_{\theta}(\Hcal_c)$ are computed independently of the query sequence $\Hcal_q$. Once we train the retrieval model,  $\vb_{\theta}(\Hcal_c)$ can be precomputed and bucketized before observing the test query. Such a model, together with the Fisher kernel-based cosine similarity scoring model, allows us to apply locality-sensitive hashing for efficient retrieval. 
 
\para{\nsrs}: The above self-attention-based mechanism models a query agnostic likelihood of the corpus sequences. Next, we introduce a \emph{cross attention} based MTPP model which explicitly takes into account the underlying query sequence while modeling the likelihood of the corpus sequence. Specifically, we measure the latent \eat{cross-attention} {relevance score} between $\Hcal_q$ and $\Hcal_c$ \eat{using the likelihood of a corpus sequence} via a query-induced MTPP model built using the cross-attention between the generative process of both the sequences. 

Given a query sequence $\Hcal_q$ and the first $r$ events of  the corpus sequence $\Hcal_c$, we parameterize the generative model for $(r+1)$-th event, \ie, $p (e^{(c)}_{r+1} \given  \Hcal({t_r}))$ as $ p_{\theta  \pend}(\cdot)$, where $p_{\theta  \pend} (e^{(c)} _{r+1}) = \intensity _{\theta  \pend} (t^{(c)} _{r+1})\, m^{\pend}_{\theta  \pend} (x^{(c)} _{r+1})$, where $\intensity$ and $m^{\pend}$ are the density and distribution functions for the arrival time and the mark respectively, as described in Eq.~\eqref{eq:qm}. 

\ehdr{Input Layer:} For each event $e_i ^{(q)}$ in the query sequence $\Hcal_q$ and each event $e_j ^{(c)}$ in the first $r$ events in the corpus sequence $\Hcal_c$,  the input layer computes the initial embeddings $\yb^{(q)} _i$ and $\yb^{(c)} _j$ as follows:
\begin{equation}
\yb^{(q)}_i = \wb_{y, x} x^{(q)}_i + \wb_{y, t}\unw(t^{(q)}_i) + \wb_{y, \Delta t} \left(\unw(t^{(q)}_i) - \unw(t^{(q)}_{i-1})\right) + \bb_{y}, \forall i\in[|\Hcal_q|],
\end{equation}
\begin{equation}
\yb^{(c)}_j = \wb_{y, x} x^{(c)}_j + \wb_{y, t}t^{(c)}_j + \wb_{y, \Delta t} \left(t^{(c)}_j - t^{(c)}_{j-1}\right)   + \bb_{y}, \forall j\in[|\Hcal_c(t_{r})|-1]
\end{equation}
where $\wb_{\bullet,\bullet}$ and $\bb_{y}$ are trainable parameters.

\ehdr{Attention layer:}
The second layer models the interaction between all the query events and \emph{the past corpus events}, \ie, $\Hcal_q$ and $\Hcal_c(t_{r})$ using an attention mechanism. Specifically, following the existing attention models~\cite{transformer,sasrec,tisasrec} it first adds a trainable position embedding $\pb$ with $\yb$--- the output from the previous layer. As compared to fixed positional encodings~\cite{transformer}, these learnable encodings demonstrate better performances~\cite{sasrec, tisasrec}. More specifically, we have the updates: $\yb^{(q)}_i \leftarrow \yb^{(q)}_i + \pb_i$ and $ \yb^{(c)}_j \leftarrow \yb^{(c)}_j + \pb_j$. Where, $\pb_\bullet\in \mathbb{R}^D$. Next, we apply  two linear transformations on the vectors  $[\yb^{(q)}_i]_{i\in [|\Hcal_q|]}$ and one linear transformation on  $[\yb^{(c)}_j]_{j\in[r]}$, \ie, $ \sb_j = \Wb^S  \yb^{(c)}_j,   \kb_i = \Wb^K  \yb^{(q)}_i,   \vb_i = \Wb^V  \yb^{(q)}_i$. The state-of-the-art works on attention models~\cite{transformer,thp,sasrec,tisasrec} often refer $\sb_\bullet$, $\kb_\bullet$ and $\vb_\bullet$ as \querytt,\footnote{\scriptsize The term  \querytt\ in this attention model is different from ``query'' sequence}, \keytt\ and \valuett\ vectors respectively. Similarly, we call the trainable weights $\Wb^S, \Wb^K$ and $\Wb^V$ as the \Querytt, \Keytt\ and \Valuett\ matrices, respectively. Finally, we use the standard attention recipe~\cite{transformer} to compute the final embedding vector $\hb_j ^{(c,q)}$ for the event $e^{(c)} _j$, induced by query $\Hcal_q$. Such a recipe adds the values weighted by the outputs of a softmax function induced by the \querytt\ and \keytt, \ie,
\begin{align}
\hb^{(c,q)} _j = 
 \sum_{i\in [|\Hcal_q|]} \frac{\exp\left( \sb_j ^\top \kb_i /\sqrt{D} \right)}{\sum_{i'\in [|\Hcal_q|]}\exp\left( \sb_j ^\top \kb_{i'} /\sqrt{D} \right)} \vb_i, \label{eq:attn}
\end{align}

\ehdr{Output layer:}
Given the vectors $\hb^{(c,q)} _j$ provided by the attention mechanism~\eqref{eq:attn}, we first apply a feed-forward neural network on them to compute  $\bhbc_r$ as follows:
\begin{equation*}
\bhbc _{r} = \sum_{j=1}^r \left[\wb_{\overline{h}}\odot\text{ReLU}(\hb^{(c,q)} _j \odot \wb_{h, f} + \bb_{h, o})  + \bb_{\overline{h}}\right],
\end{equation*}
where $\wb_{\bullet,\bullet}$, $\wb_\bullet$ and $\bb_{v}$. Finally, we use these vectors to compute the probability density of the arrival time of the next event $e^{(c)} _{r+1}$, \ie,  $\intensity_{\theta  \pend}(t_{r+1})$ and the mark distribution $m^{\pend} _{\theta  \pend}(x_{r+1})$. In particular, we realize $\intensity_{\theta  \pend}(t_{r+1})$ using a lognormal distribution of inter-arrival times, \ie,
\begin{align}
 \hspace{-2mm}t^{(c)}_{r+1}- t^{(c)}_r \sim \textsc{LogNormal}\left(\mu_e\left(\bhbc _{r} \right),\sigma^2 _e\left(\bhbc _{r} \right) \right)\nn,
\end{align}
where, $\left[\mu_e\left(\bhbc _{r} \right),\sigma_e\left(\bhbc _{r} \right)\right] = \Wb_{t,q}  \bhbc _{r} \hspace{-2mm}+\bb_{t,q}$. Similarly, we model the mark distribution as,
\begin{align}
\hspace{-2mm} m^{\pend} _{\theta  \pend}(x_{r+1}) =  \dfrac{\exp\left(\wb_{x,m}^\top \bhbc +b_{x,m} \right)}{\sum_{x'\in \mathcal{C}}\exp\left(\wb_{x',m}^\top \bhbc +b_{x',m}\right)}, \label{eq:mark-mtpp-x}
\end{align}
where $\Wb_{\bullet,\bullet}$ are the trainable parameters. Therefore the set of trainable parameters for the underlying MTPP models is $\theta=\set{\Wb^\bullet, \Wb_{\bullet,\bullet}, \wb_\bullet,\wb_{\bullet,\bullet}, \bb_{\bullet},\bb_{\bullet,\bullet}}$.

\xhdr{Tradeoff between accuracy and efficiency} Note that\eat{, unlike \nsrf, } the likelihoods of corpus sequences in \nsrs\ depend on the query sequence, making it more expressive than \nsrf. However, these query-dependent corpus likelihoods cannot be precomputed before observing the query sequences and thus cannot be directly used for hashing. However, it can be deployed on top of \nsrf, where a smaller set of relevant candidates is first selected using LSH applied on \nsrf\ and then reranked using this \nsrf. Thus, our proposal offers a nice tradeoff between accuracy and efficiency.

\subsection{Parameter estimation}
Given the query sequences $\set{\Hcal_q}$, the corpus sequences $\set{\Hcal_c}$ along with their relevance labels $\set{y(\Hcal_q,\Hcal_c)}$, we seek to find $\theta$ and $\phi$ which ensure that:
\begin{align}
 s_{p_{\theta},\unw_{\phi}}(\Hcal_q,\Hcal_{c_+}) \gg s_{p_{\theta},\unw_{\phi}}(\Hcal_q, \Hcal_{c_-})\, \forall\, c_{\pm} \in \Cr_{q \pm}.
\end{align}
To this aim, we minimize the following pairwise ranking loss~\cite{Joachims2002ranksvm} to estimate the parameters $\theta,\phi$:
\begin{align}
\underset{\theta,\phi} {\text{min}}\sum_{q\in\Qr}\sum_{\substack{c_+\in \Cr_{q\rel},\\ c-\in \Cr_{q\nrel} }}&    \big[s_{p_{\theta},\unw_{\phi}}(\Hcal_q,\Hcal_{c_-})- \nn  s_{p_{\theta},\unw_{\phi}}(\Hcal_q,\Hcal_{c_+}) + \delta \big]_+, 
\end{align}
where, $\delta$ is a tunable margin.

\section{Scalable Retrieval with Hashing}
Once we learn the model parameters $\theta$ and $\phi$, we can rank the set of corpus sequences $\Hcal_c$ in the decreasing order of $s_{p_{\theta},\unw_{\phi}}(\Hcal_{\newq},\Hcal_c)$ for a new query $\Hcal_{\newq}$ and return top$-K$ sequences. Such an approach requires $|\Ccal|$ comparisons per each test query, which can take a huge amount of time for many real-life applications where $|\Cr|$ is high. However, for most practical query sequences, the number of relevant sequences is a very small fraction of the entire corpus of sequences. Therefore, the number of comparisons between query-corpus sequence pairs can be reduced without significantly impacting the retrieval quality by selecting a small number of candidate corpus sequences that are more likely to be relevant to a query sequence. We first briefly describe the traditional random hyperplane-based hashing method and highlight its limitations before presenting our retrieval approach that uses a specific trainable hash code with desirable properties. 

\subsection{Random hyperplane based hashing method} \label{sec:rh}
Since the relevance between the query and corpus sequence pairs $(\Hcal_q,\Hcal_c)$ is measured using the cosine similarity between the gradient vectors, \ie, $\kernel_{p_{\theta}}(\Hcal_q,\Hcal_c)$, one can use random hyperplane based locality sensitive hashing method for hashing the underlying gradient vectors $\vb_{p_{\theta}} (\Hcal_c)$~\cite{charikar2002similarity}.
Towards this goal, after training \nsrf\, we generate $R$ unit random vectors $\ub_r\in \mathbb{R}^D$ from i.i.d. Normal distributions and then compute a binary hash code, $\hash^c = [\sgn (\ub_1 ^\top \vb_{p_{\theta}}(\Hcal_c)),\dots, \sgn (\ub_R ^\top \vb_{p_{\theta}}(\Hcal_c))] $ for each $c\in \Ccal$. This leads to $2^R$ possible hash buckets $\set{\Bcal}$, where
each corpus sequence is assigned to one hash bucket using the algorithm proposed by~\citet{GionisIM1999hash}.

When we encounter an unseen test query $\Hcal_q$,  we compute the corresponding hash code $\hash^q$, assign it to a bucket $\Bcal$ and finally return \emph{only those sequences} $\Hcal_c$ which were assigned to this bucket $\Bcal$. Thus, for each query, the number of comparisons is reduced from $|\Ccal|$ to $|\Bcal|$, \ie, the number of corpus sequences in the bucket $\Bcal$.
Thus, if the corpus sequences are assigned uniformly across the different buckets, then the expected number of comparisons becomes $|\Ccal|/2^R$, which provides a significant improvement for  $R>2$. 

\xhdr{Limitations of Binary Hash Codes} In practice, binary hash codes are not trained from data, and consequently, they are not optimized to be uniformly distributed across different hash buckets. Consequently, the assignment of corpus sequences across different buckets may be quite skewed, leading to inefficient sequence retrieval.

\subsection{Trainable Hashing for Retrieval}
Responding to the limitations of the random hyperplane-based hashing method, we propose to learn the hash codes from data so that they can be optimized for performance.

\xhdr{Computation of a trainable hash code}
We first apply a trainable nonlinear transformation $\Lambda_{\psi}$ with parameter $\psi$ on the gradients $\zzc = \vb_{p_{\theta}}(\Hcal_c)$ and then learn the binary hash vectors $\hash^c = \sgn\bracex{\Lambda_{\psi}\bracex{\zzc}}$  by solving the following optimization, where we use $\tanh\left(\Lambda_{\psi}(\cdot)\right)$ as a smooth approximation of $\sgn\left(\Lambda_{\psi}(\cdot)\right)$.
\begin{align}\label{eq:hash}
 & \min_{\psi} \frac{\eta_1}{|\Cr|}\sum_{c\in\Cr}   \left|\bm{1}^\top \tanh\bracex{\Lambda_{\psi}\bracex{\zzc}}\right| + \frac{\eta_2}{|\Cr|}\sum_{c\in\Cr}   \left\| \left|\tanh\bracex{\Lambda_{\psi}\bracex{\zzc}}\right|-\bm{1}\right\|_1  \\  
 &\qquad + \frac{2\eta_3}{{D \choose 2}}  \cdot \bigg| \sum_{\substack{c\in \Ccal \\ i\neq j\in [D]}}  \tanh\bracex{\Lambda_{\psi}\bracex{\zzc}[i]} \cdot  \tanh\bracex{\Lambda_{\psi} \bracex{\zzc}[j]}\bigg| \nn
\end{align}

\begin{algorithm}[t]
\small
        \caption{Efficient retrieval with hashing}
        \begin{algorithmic}[1]
          \REQUIRE  Trained corpus embeddings
          $\set{\vb^c = \vb_{p_{\theta}} (\Hcal_c)} $ using \nsrf; \\
          new query sequences $\set{\Hcal_{\newq}}$, $K$:
          \# of corpus sequences to return; trained models for \nsrf\ and \nsrs.
          \STATE \textbf{Output:} $\set{L_\newq}$: top-$K$ relevant sequences from $\set{\Hcal_c}$.
          \STATE $\psi\leftarrow\textsc{TrainHashNet}\left(\Lambda_{\psi},[\vb^c]_{c\in\Cr} \right)$
          \STATE \textsc{InitHashBuckets}$(\cdot )$ 
          \FOR{$c\in \Cr$} 
          \STATE $\hash^c\leftarrow \textsc{ComputeHashCode}\left(\zzc;\Lambda_{\psi}\right)$
          \STATE $\Bcal\leftarrow\textsc{AssignBucket}(\hash^c)$
          \ENDFOR
          \FOR{each new query $\Hcal_{\newq}$} 
          \STATE $\vb^{q'}\leftarrow  \nsrf(\Hcal_{\newq})$ 
          \STATE $\hash^{q'}\leftarrow \textsc{ComputeHashCode} (\vb^{q'};  {\Lambda_{\psi}} )$ 
          \STATE  $\Bcal\leftarrow\textsc{AssignBucket}(\hash^{\newq})$ 
            \FOR{$c\in \Bcal$} 
          \STATE $\vb^{\newq} _{\text{cross}}, \vb^{c} _{\text{cross}} \leftarrow \nsrs(\Hcal_{\newq},\Hcal_c)$
          \STATE $s_{p_{\theta},U_{\phi}} (\Hcal_{q'},\Hcal_c) \leftarrow \textsc{Score}(\vb^{\newq} _{\text{cross}}, \vb^{c} _{\text{cross}}, \Hcal_q,\Hcal_c)$
          \ENDFOR
          \STATE $L_{\newq}\leftarrow \textsc{Rank}(\set{s_{p_{\theta},U_{\phi}} (\Hcal_{q'},\Hcal_c)},K)$
        \ENDFOR
        \STATE \textbf{Return}  $\set{L_{\newq}}$
        \end{algorithmic}      \label{alg:key}
\end{algorithm}

Here, $\sum_{i=1}^3 \eta_i=1$. Moreover, different terms in Eq.~\eqref{eq:hash} allow the hash codes, $\hash^c$, to have a set of four desired properties: (i) the first term ensures that the numbers of $+1$ and $-1$ are evenly distributed in the hash vectors $\hash^c=  \tanh\bracex{\Lambda_{\psi}\bracex{\zzc}}$; (ii) the second term encourages the entries of $\hash^c$ to become as close to $\pm 1$ as possible so that $\tanh(\cdot)$ gives an accurate approximation of $\sgn(\cdot)$ ; (iii) the third term ensures that the entries of the hash codes, $\hash^c$, contain independent information and therefore they have no redundant entries.  Trainable hashing has been used in other domains of information retrieval, including graph hashing~\cite{qin2020ghashing,roy2020adversarial}, document retrieval~\cite{zhang2010self, salakhutdinov2009semantic}. However, to the best of our knowledge, such an approach has never been proposed for continuous-time sequence retrieval.

\xhdr{Outline of our retrieval method}
We summarize our retrieval procedure in Algorithm~\ref{alg:key}. We are given gradient vectors $\vb^c = \vb_{p_{\theta}}(\Hcal_c)$ obtained by training \nsrf.  Next, we train an additional neural network $\Lambda_{\psi}$ parameterized by $\psi$ (\textsc{TrainHashNet}$()$, line 2), which is used to learn a binary hash vector $\hash^c$ for each sequence $\Hcal_c$. Then these hash codes are used to arrange corpus sequences in different hash buckets (for-loop in lines 4--7) using the algorithm proposed by~\citet{GionisIM1999hash}, so that two sequences $\Hcal_c, \Hcal_{c'}$ lying in the same hash bucket have a very high value of cosine similarity $\cos(\vb^c,\vb^{c'})$. Finally, once a new query $\Hcal_\newq$ comes, we first compute  $\vb^{\newq}$ using the trained \nsrf\ model and then compute the binary hash codes $\hash^{\newq}$ using the trained hash network $\Lambda_\psi$ (lines 9--10). Next, we assign an appropriate bucket $\Bcal$ to it (line 11) and finally compare it with only the corpus sequences in the same bucket, \ie, $\Hcal_c\in\Bcal$ (lines 12--15) using our model. 

\xhdr{Bucket Assignment}
As suggested in~\cite{GionisIM1999hash}, we design multiple hash tables and assign a bucket to the hashcode of a sequence using only a set of bits selected randomly. More specifically, let the number of hash tables be $M$. Given a query sequence, we calculate its hashcode using the procedure described in Algorithm~\ref{alg:key}, $\hash^{\newq} = \sgn\left(\Lambda_{\psi}(\vb^{\newq})\right)$. The hash code is a $R$ dimension vector with $\hash^{\newq} \in \{-1, 1\}^{R}$ and from this vector, we consider $L$ bits at random positions to determine the bucket to be assigned to the sequence. Here, $\hash^{\newq} $ represents the numbers between $\{0, 2^{L}-1\}$, \ie,  one of the $2^{L}$ different buckets in a hash table. Correspondingly, we assign $\hash^{\newq}$ into a bucket. However, such a procedure is dependent on the specific set of bits --that were selected randomly-- used for deciding the bucket ID. Therefore, we use $M$ hash-tables and repeat the procedure of sampling $L$ bits and bucket assignment for each table. We follow a similar bucket assignment procedure for corpus sequences. As described in Algorithm~\ref{alg:key}, for an incoming query sequence in the test set, we use the above bucket assignment procedure and compute the relevance score for only the corpus sequences within the same buckets. For all our experiments, we set $H$ the same as the hidden dimension $D$, $M = 10$, and $L=12$.

Recall that we must use \nsrf\ to compute gradient vectors for subsequent hash code generation (in lines 9--10). However, at the last stage for final score computation and ranking,  we can use any variant of \nsr (in line 13), preferably \nsrs, since the corpus sequences have already been indexed by our LSH method.

\section{Experiments}
In this section, we provide a comprehensive evaluation of \nsr and our hashing method.
\newcommand{\seq}{\texttt{seq}}
\begin{table}[t]
\caption{Statistics of the search corpus for all datasets. ${|\Cr_{q\rel}|}/{|\Cr|}$ denotes the ratio of positive corpus sequences to the total sequences sampled for training. The ratio is kept the same for all queries.}
\small
    \centering
    \begin{tabular}{l|ccccc}
    \toprule
    \textbf{Dataset} & \textbf{Audio} & \textbf{Celebrity} & \textbf{Electricity} & \textbf{Health} & \textbf{Sports}\\ \hline
    $  {|\Cr_{q\rel}|}/{|\Cr|} $ & 0.25 & 0.23 & 0.20 & 0.28 & 0.30\\
    Total Events & 1M & 50M & 60M & 60M & 430k \\
    \# Marks & 5 & 16 & 5 & 5 & 21 \\
    \bottomrule
    \end{tabular}
    \label{tab:dset_details}
    \vspace{-4mm}
\end{table}

\subsection{Experimental Setup} \label{sec:exp-setup}
\xhdr{Datasets} We use five real-world datasets containing event sequences from various domains: (i) Audio~\cite{coucke2018snips}, (ii) Sports~\cite{multithumos}, (iii) Celebrity~\cite{nagrani17}, (iv) Electricity~\cite{refit}, and (v) Health~\cite{ecg}. The statistics of all datasets are given in Table~\ref{tab:dset_details}. Across all datasets, $|\Hcal_q| = 5K$ and $|\Hcal_c| = 200K$. We partition the set of queries into 50\% training, 10\% validation, and the rest as test sets. During training, we negatively sample 100 corpus sequences for each query.

\begin{compactitem}
\item \textbf{Audio:} The dataset contains audio files for spoken commands to a smart-light system and the demographics(age, nationality) of the speaker.  Here, a query corpus sequence pair is relevant if they are from an audio file with a common speaker.

\item \textbf{Sports:} The dataset contains actions (\eg, run, pass, shoot) taken while playing different sports. We consider the time of action and action class as time and mark of sequence, respectively. Here, a query corpus sequence pair is relevant if they are from a common sport.

\item \textbf{Celebrity:} In this dataset, we consider the series of frames extracted from youtube videos of multiple celebrities as event sequences where event-time denotes the video-time, and the mark is decided upon the coordinates of the frame where the celebrity is located. Here, a query corpus sequence pair is relevant if they are from a video file having a common celebrity.

\item \textbf{Electricity:}  This dataset contains the power-consumption records of different devices across smart homes in the UK. We consider the records for each device as a sequence with an event mark as the \textit{normalized} change in the power consumed by the device and the time of recording as event time. Here, a query corpus sequence pair is relevant if they are from a  similar appliance.

\item \textbf{Health:} The dataset contains ECG records for patients suffering from heart-related problems. Since the length of the ECG record for a single patient can be up to 10 million, we generate smaller individual sequences of length 10,000 and consider each such sequence as an independent sequence. The marks and times of events in a sequence are determined using a similar procedure as in Electricity. Here, a query corpus sequence pair is relevant if they are from a common patient.
\end{compactitem}

For Health, Celebrity, and Electricity datasets, we lack the true ground-truth labeling of relevance between sequences. Therefore, we adopt a heuristic in which, given a dataset $\Dcal$, from each sequence $\seq_q\in\Dcal$ with $q\in[|\Dcal|]$, we first sample a set of sub-sequences $\mathcal{U}_q=\set{\Hcal\subset \texttt{seq}_q}$ with $|\Ucal_q|\sim \text{Unif}\,[200,300]$. For each such collection $\Ucal_q$, we draw exactly one query $\Hcal_q$ uniformly at random from $\Ucal_q$, \ie, $\Hcal_q\sim\Ucal_q$. Then, we define $\Cr=\cup_{q\in[|\Dcal|]}\Ucal_q\cp \Hcal_q$, $\Cr_{q\rel}=\Ucal_q\cp \Hcal_q$ and $\Cr_{q\nrel}=\cup_{c\neq q}\big(\Ucal_c\cp \Hcal_c\big)$.

\xhdr{System Configuration}
All our models were implemented using Pytorch v1.6.0~\footnote{\scriptsize https://pytorch.org/}. We conducted all our experiments on a server running Ubuntu 16.04, CPU: Intel(R) Xeon(R) Gold 6248 2.50GHz, RAM: 377GB, and GPU: NVIDIA Tesla V100. 

\xhdr{Hyperparameters setup}
We set the hyper-parameters values of \nsr\ as follows: \begin{inparaenum}[(i)] \item contribution of model-independent similarity score in Eq.~\eqref{eq:relevance-score-function}, $\gamma = 0.1$; \item margin parameters for parameter estimation, $\delta \in \{0.1, 0.5, 1\}$ and weight for constraint violations, $\lambda \in \{0.1, 0.5, 1\}$; \item weight parameters for hashing objective~\eqref{eq:hash} $\eta_1, \eta_2, \eta_3 \in \{0.1, 0.2, 0.25\}$ and correspondingly $\eta_4 \in \{0.25, 0.4, 0.7\}$. 
\end{inparaenum}

\begin{table}[tbh]
\caption{Hyper-parameter values used for different datasets. The values are determined by fine-tuning the performance on the validation set.}
\small
\centering
\resizebox{\textwidth}{!}{
    \begin{tabular}{lccccc}
    \toprule
    \multirow{2}{*}{\textbf{Parameters}} & \multicolumn{5}{c}{\textbf{Datasets}} \\
    \cmidrule(lr){2-6}
    & \textbf{Audio} & \textbf{Celebrity} & \textbf{Electricity} & \textbf{Health} & \textbf{Sports}\\ \midrule
    $\gamma$ & 0.1 & 0.1 & 0.5 & 0.1 & 0.1 \\
    $\delta$ & 0.5 & 0.5 & 0.1 & 0.1 & 0.5 \\
    $\{\eta_1, \eta_2, \eta_3\}$ & \{0.4, 0.3, 0.3\} & \{0.4, 0.3, 0.3\} & \{0.4, 0.3, 0.3\} & \{0.5, 0.25, 0.25\} & \{0.5, 0.25, 0.25\} \\
    Batch-size $\mathcal{B}$ & 32 & 32 & 32 & 16 & 16 \\
    $D$ & 64 & 64 & 48 & 32 & 32 \\
    \bottomrule
    \end{tabular}
    }
    \vspace{-4mm}
\label{tab:hyperparameters} 
\end{table}

Moreover, the values of training specific parameter values are: \begin{inparaenum}[(i)] \item batch-size, $\mathcal{B}$ is selected from $\{16, 32\}$, \ie, for each batch we select $\mathcal{B}$ query sequences and all corresponding corpus sequences; \item hidden-layer dimension for cross-attention model, $D \in \{32, 48, 64\}$; \item number of attention blocks $N_b = 2$; \item number of attention heads $N_h = 1$ and \item UMNN network as a two-layer feed-forward network with dimension $\{128, 128\}$. \end{inparaenum} We also add a dropout after each attention layer with probability $p = 0.2$ and an $l_2$ regularizer over the trainable parameters with the coefficient set to $0.001$. All our parameters are learned using the Adam optimizer. We summarize the details of hyperparameters across different datasets in Table~\ref{tab:hyperparameters}.

\xhdr{Evaluation metrics}
We evaluate \nsr and the baselines using mean average precision (MAP), NDCG@k, and mean reciprocal rank (MRR). We calculate these metrics as follows:
\begin{equation}
\text{MAP} = \frac{1}{|\Hcal_{q'}|}\sum_{q' \in \Hcal_{q'}} \text{AP}_{q'}, \quad \text{NDCG@k} = \frac{\text{DCG}_k}{\text{IDCG}_k}, \quad \text{MRR} = \frac{1}{|\Hcal_{q'}|} \sum_{q' \in \Hcal_{q'}}\frac{1}{r_{q'}},
\end{equation}
where $\text{AP}_{q'}, \text{DCG}_k, \text{IDCG}_k$, and $r_{q'}$ denote the average precision, discounted cumulative gain at top-$k$ position, ideal discounted cumulative gain (at top-$k$), and the topmost rank of a related corpus sequence respectively. For all our evaluations, we follow a standard evaluation protocol~\cite{sasrec, tisasrec} for our model and all baselines wherein for each query sequence in the test set, we rank all relevant corpus sequences and 1000 randomly sampled non-relevant sequences. All confidence intervals and standard deviations are calculated after five independent runs. For all metrics -- MAP, NDCG, and MRR, we report results in terms of percentages with respect to maximum possible value \ie, 1.

\xhdr{Baseline Implementations}
For all the baselines, we use the official python implementations released by the authors of MASS, UDTW, and Sharp. The other implementations are same as in Chapter~\ref{chapter:imtpp}. For MASS and UDTW, we report the results using the default parameter values. For Sharp, we tune the hyper-parameter `\textit{gamma}' (for more details see~\cite{blondel2021differentiable}) based on the validation set. In RMTPP, we set the BPTT length to 50, the RNN hidden layer size to 64, and the event embedding size 16. These are the parameter values recommended by the authors. For SAHP and THP, we set the dimension to 128 and the number of heads to 2. The values for all other transformer parameters are similar to the ones we used for the attention part in \nsr.

\newcommand{\secbest}{\underline}
\newcommand{\thbest}{\fbox}
\begin{table}[t!]
\centering
\caption{Retrieval quality in terms of Mean Average Precision (MAP in \%) of all the methods across five datasets on the test set. Numbers with bold font (underlined) indicate the best (second best) performer. Boxed numbers indicate the best-performing state-of-the-art baseline. Results marked \textsuperscript{$\dagger$} are statistically significant (two-sided Fisher's test with $p \le 0.1$) over the best performing state-of-the-art baseline (\tpprank-THP or Sharp). The standard deviation for MASS and UDTW are zero since they are deterministic retrieval algorithms.}
\footnotesize
\begin{tabular}{l|ccccc}
\toprule
\textbf{Dataset} & \multicolumn{5}{c}{\textbf{Mean Average Precision (MAP) in \%}} \\ \hline 
 & Audio & Celebrity & Electricity & Health & Sports\\ \hline \hline
MASS~\cite{mass} & 51.1$\pm$0.0 & 58.2$\pm$0.0 & 19.3$\pm$0.0 & 26.4$\pm$0.0 & 54.7$\pm$0.0 \\
UDTW~\cite{ucrdtw} & 50.7$\pm$0.0 & 58.7$\pm$0.0 & 20.3$\pm$0.0 & 28.1$\pm$0.0 & 54.5$\pm$0.0 \\
Sharp~\cite{blondel2021differentiable} & 52.4$\pm$0.2 & 59.8$\pm$0.5 & 22.8$\pm$0.2 & 28.6$\pm$0.2 & \thbest{56.8$\pm$0.3} \\
RMTPP~\cite{rmtpp} & 48.9$\pm$2.3 & 57.6$\pm$1.8 & 18.7$\pm$0.8 & 24.8$\pm$1.2 & 50.3$\pm$2.5 \\
\texttt{Rank}-RMTPP & 52.6$\pm$2.0 & 60.3$\pm$1.7 & 23.4$\pm$0.7 & 29.3$\pm$0.6 & 55.8$\pm$2.1 \\
SAHP~\cite{sahp} & 49.4$\pm$3.2& 57.2$\pm$2.9 & 19.0$\pm$1.8 & 26.0$\pm$2.1 & 53.9$\pm$3.6 \\
\texttt{Rank}-SAHP & 52.9$\pm$1.8& 61.8$\pm$2.3 & 26.5$\pm$1.2 & 31.6$\pm$1.1 & 55.1$\pm$2.3 \\
THP~\cite{thp} & 51.8$\pm$2.3& 60.3$\pm$1.9 & 21.3$\pm$0.9 & 27.9$\pm$0.9 & 54.2$\pm$2.1 \\
\texttt{Rank}-THP & \thbest{54.3$\pm$1.7}& \thbest{63.1$\pm$2.1} & \thbest{29.4$\pm$0.9} & \thbest{33.6$\pm$1.3} &  56.3$\pm$1.9 \\
\hline
\nsrf & \secbest{55.8$\pm$1.8} & \secbest{64.4$\pm$1.9} & \secbest{30.7$\pm$0.7} & \secbest{35.9$\pm$0.9} & \secbest{57.6$\pm$1.9} \\
\nsrs & \textbf{56.2$\pm$2.1}\textsuperscript{$\dagger$} & \textbf{65.1$\pm$1.9}\textsuperscript{$\dagger$} & \textbf{32.4$\pm$0.8}\textsuperscript{$\dagger$} & \textbf{37.4$\pm$0.9}\textsuperscript{$\dagger$} & \textbf{58.7$\pm$2.1}\\
\bottomrule
\end{tabular}
\label{tab:exp_map}
\vspace{3mm}
\caption{Retrieval quality in terms of NDCG@10 (in \%) of all the methods.}
\begin{tabular}{l|ccccc}
\toprule
\textbf{Dataset} & \multicolumn{5}{c}{\textbf{NDCG@10}} \\ \hline 
 & Audio & Celebrity & Electricity & Health & Sports\\ \hline \hline
MASS~\cite{mass} & 20.7$\pm$0.0 & 38.7$\pm$0.0 & 9.1$\pm$0.0 & 13.6$\pm$0.0 & 22.3$\pm$0.0 \\
UDTW~\cite{ucrdtw} & 21.3$\pm$0.0 & 39.6$\pm$0.0 & 9.7$\pm$0.0 & 14.7$\pm$0.0 & 22.9$\pm$0.0 \\
Sharp~\cite{blondel2021differentiable} & 21.9$\pm$0.2 & 40.6$\pm$0.5 & 11.7$\pm$0.1 & 16.8$\pm$0.1 & 23.7$\pm$0.2 \\
RMTPP~\cite{rmtpp} & 20.1$\pm$1.9 & 39.4$\pm$2.1 & 8.3$\pm$0.8 & 12.3$\pm$0.5 & 19.1$\pm$1.8 \\
\tpprank-RMTPP & 22.4$\pm$1.3 & 41.2$\pm$1.3 & 11.4$\pm$0.4 & 15.5$\pm$0.5 & 23.9$\pm$1.4 \\
SAHP~\cite{sahp} & 20.4$\pm$2.3 & 39.0$\pm$3.1 & 8.7$\pm$1.2 & 13.2$\pm$1.4 & 22.6$\pm$2.5 \\
\tpprank-SAHP & 23.3$\pm$1.4 & 42.1$\pm$1.7 & 13.3$\pm$0.7 & 17.5$\pm$0.9 & 25.4$\pm$1.8 \\
THP~\cite{thp} & 22.1$\pm$1.1 & 40.3$\pm$1.2 & 10.4$\pm$0.6 & 14.4$\pm$0.3 & 22.9$\pm$1.1 \\
\tpprank-THP & \thbest{25.4$\pm$0.9} & \thbest{44.2$\pm$1.0} & \thbest{15.3$\pm$0.4} & \thbest{19.7$\pm$0.4} & \thbest{26.5$\pm$0.9} \\ \hline
\nsrf & \secbest{25.9$\pm$1.1} & \secbest{45.8$\pm$1.0} & \secbest{16.5$\pm$0.5} & \secbest{20.4$\pm$0.4} & \textbf{27.8$\pm$1.1} \\
\nsrs & \textbf{28.3$\pm$1.1}\textsuperscript{$\dagger$} & \textbf{46.9$\pm$1.2}\textsuperscript{$\dagger$} & \textbf{18.1$\pm$0.7}\textsuperscript{$\dagger$} & \textbf{22.0$\pm$0.4}\textsuperscript{$\dagger$} & \secbest{27.9$\pm$1.2} \\
\bottomrule
\end{tabular}
\vspace{-4mm}
\label{tab:exp_n10}
\end{table}

\begin{table}[t!]
\footnotesize
\centering
\caption{Retrieval quality in terms of mean reciprocal rank (MRR in \%).}
\begin{tabular}{l|ccccc}
\toprule
\textbf{Dataset} & \multicolumn{5}{c}{\textbf{Mean Reciprocal Rank (MRR)}} \\ \hline 
 & Audio & Celebrity & Electricity & Health & Sports\\ \hline \hline
MASS~\cite{mass} & 57.3$\pm$0.0 & 63.7$\pm$0.0 & 17.6$\pm$0.0 & 27.2$\pm$0.0 & 61.6$\pm$0.0\\
UDTW~\cite{ucrdtw} & 58.5$\pm$0.0 & 64.8$\pm$0.0 & 18.7$\pm$0.0 & 29.2$\pm$0.0 & 61.2$\pm$0.0\\
Sharp~\cite{blondel2021differentiable} & 58.7$\pm$1.7 & 65.4$\pm$2.6 & 19.8$\pm$0.6 & 30.4$\pm$0.7 & 61.1$\pm$2.3\\
RMTPP~\cite{rmtpp} & 54.2$\pm$3.9 & 64.5$\pm$4.6 & 15.8$\pm$1.2 & 25.2$\pm$1.8 & 56.2$\pm$4.8\\
\texttt{Rank}-RMTPP & 60.8$\pm$3.7 & 65.6$\pm$4.0 & 23.3$\pm$1.5 & 30.7$\pm$1.7 & 62.7$\pm$3.9\\
SAHP~\cite{sahp} & 56.9$\pm$4.3 & 64.2$\pm$4.8 & 17.2$\pm$1.3 & 26.7$\pm$2.1 & 59.4$\pm$4.9\\
\texttt{Rank}-SAHP & 60.3$\pm$2.5 & 66.1$\pm$2.9 & 25.9$\pm$1.2 & 32.6$\pm$1.4 & 62.1$\pm$2.9\\
THP~\cite{thp} & 58.6$\pm$2.6 & 65.0$\pm$2.9 & 20.1$\pm$1.1 & 28.2$\pm$1.3 & 60.9$\pm$3.3\\
\texttt{Rank}-THP & 62.2$\pm$2.8 & 68.9$\pm$3.0 & 31.4$\pm$1.4 & 36.2$\pm$1.7 & 63.4$\pm$2.9\\
\hline
\nsrf & 63.6$\pm$2.7 & 69.3$\pm$3.1 & 33.4$\pm$1.6 & 37.9$\pm$1.7 & 64.3$\pm$3.1\\
\nsrs & \textbf{64.5$\pm$2.9} & \textbf{70.1$\pm$3.3} & \textbf{35.2$\pm$1.7} & \textbf{40.3$\pm$1.9} & \textbf{66.7$\pm$3.1}\\
\bottomrule
\end{tabular}
\label{tab:exp_mrr}
\vspace{3mm}
\centering
\caption{Retrieval quality in terms of NDCG@20 (in \%) of all the methods.}
\begin{tabular}{l|ccccc}
\toprule
\textbf{Dataset} & \multicolumn{5}{c}{\textbf{NDCG@20}} \\ \hline 
 & Audio & Celebrity & Electricity & Health & Sports\\ \hline \hline
MASS~\cite{mass} & 17.5$\pm$0.0 & 31.4$\pm$0.0 & 8.1$\pm$0.0 & 13.5$\pm$0.0 & 16.3$\pm$0.0\\
UDTW~\cite{ucrdtw} & 17.9$\pm$0.0 & 32.5$\pm$0.0 & 8.8$\pm$0.0 & 14.4$\pm$0.0 & 16.0$\pm$0.0\\
Sharp~\cite{blondel2021differentiable} & 18.2$\pm$0.5 & 33.6$\pm$0.7 & 11.9$\pm$0.3 & 15.9$\pm$0.4 & 17.2$\pm$0.7\\
RMTPP~\cite{rmtpp} & 16.0$\pm$1.1 & 32.2$\pm$1.6 & 7.1$\pm$0.5 & 12.1$\pm$0.7 & 17.1$\pm$1.1\\
\texttt{Rank}-RMTPP & 20.2$\pm$1.0 & 33.4$\pm$1.4 & 10.5$\pm$0.5 & 15.4$\pm$0.6 & 21.8$\pm$1.2\\
SAHP~\cite{sahp} & 19.8$\pm$1.4 & 31.7$\pm$2.1 & 7.8$\pm$0.5 & 13.1$\pm$0.9 & 19.2$\pm$1.5\\
\texttt{Rank}-SAHP & 21.5$\pm$0.9 & 34.1$\pm$1.0 & 12.3$\pm$0.4 & 17.4$\pm$0.6 & 22.9$\pm$1.0\\
THP~\cite{thp} & 19.7$\pm$0.8 & 32.9$\pm$1.0 & 9.5$\pm$0.4 & 14.4$\pm$0.5 & 20.8$\pm$0.9\\
\texttt{Rank}-THP & 21.8$\pm$0.9 & 37.7$\pm$1.2 & 14.4$\pm$0.6 & 19.3$\pm$0.6 & 23.3$\pm$1.1\\
\hline
\nsrf & 22.9$\pm$1.3 & 40.3$\pm$1.5 & 16.3$\pm$0.7 & 20.9$\pm$1.0 & 23.8$\pm$1.6\\
\nsrs & \textbf{24.2$\pm$1.5} & \textbf{42.0$\pm$1.8} & \textbf{17.6$\pm$0.8} & \textbf{22.3$\pm$1.0} & \textbf{25.7$\pm$1.7}\\
\bottomrule
\end{tabular}
\vspace{-4mm}
\label{tab:exp_n20}
\end{table}

\xhdr{Baselines} We consider three continuous time-series retrieval models: (i) MASS~\cite{mass}, (ii) UDTW~\cite{ucrdtw}, and (iii) Sharp~\cite{blondel2021differentiable}; and, three MTPP models (iv) RMTPP~\cite{rmtpp}, (v) SAHP~\cite{sahp}, and (vi) THP~\cite{thp}. For sequence retrieval with MTPP models, we first train them across all the sequences using maximum likelihood estimation. Then, given a test query $\Hcal_{\newq}$, this MTPP method ranks the corpus sequences $\set{\Hcal_c}$ in decreasing order of their cosine similarity  $\text{CosSim}(\texttt{emb}^{(\newq)},\texttt{emb}^{(c)})$, where $\texttt{emb}^{(\bullet)}$ is the corresponding sequence embedding provided by the underlying MTPP model. In addition, we build supervised ranking models over these approaches, \emph{viz.,} \tpprank-RMTPP, \tpprank-SAHP and \tpprank-THP corresponding to RMTPP, SAHP, and THP. Specifically, \tpprank-MTPP formulates a ranking loss on the query-corpus pairs based on the cosine similarity scores along with the likelihood function to get the final training objective. Therefore, the vanilla MTPP models are used as unsupervised models and the corresponding \tpprank-MTPP models work as supervised models.  

\xhdr{Evaluation protocol} We partition the set of queries $\Qr$ into 50\% training, 10\% validation, and the rest as test sets. First, we train a retrieval model using the set of training queries. Then, for each test query $\newq$, we use the trained model to obtain a top-$K$ ranked list from the corpus sequences. Next,  we compute the average precision (AP) and discounted cumulative gain (DCG) of each top-$K$ list, based on the ground truth. Finally, we compute the mean average precision (MAP) and NDCG@$K$ by averaging AP and DCG values across all test queries. We set $K \in \{10, 20\}$. 

\subsection{Results on retrieval accuracy}
\xhdr{Comparison with baselines}
First, we compare the retrieval performance of our model against the baselines. Tables~\ref{tab:exp_map}, \ref{tab:exp_n10}, \ref{tab:exp_mrr}, and \ref{tab:exp_n20} summarizes the results in terms of MAP, NDCG@10, MRR, and NDCG@20 respectively. From the results, we make the following observations:
\begin{compactitem}
\item Both \nsrs\ and \nsrf\ outperform all the baselines by a substantial margin.
\item \nsrs outperforms \nsrf, since the former has a higher expressive power.
\item The variants of baseline MTPP models trained for sequence retrieval, \ie, \tpprank-RMTPP, \tpprank-SAHP, and \tpprank-THP outperform the vanilla MTPP models.
\item The performances of vanilla MTPPs and the time series retrieval models (MASS, UDTW, and Sharp) are comparable.
\end{compactitem}

\begin{table}[t]
    \caption{Ablation study of \nsrs and its variants in terms of Mean Average Precision (MAP) in \%.}
    \centering
    \resizebox{\textwidth}{!}{
    \begin{tabular}{l|ccccc}
    \toprule
    \textbf{Variation of} $s_{p_{\theta},U_{\phi}} (\Hcal_q,\Hcal_c)$ & \textbf{Audio} & \textbf{Celebrity} & \textbf{Health} & \textbf{Electricity} & \textbf{Sports} \\ \hline
    $-\Delta_x (\Hcal_q,\Hcal_c) - \Delta_t (U_{\phi}(\Hcal_q),\Hcal_c)$ & 36.1$\pm$0.0 & 43.7$\pm$0.0 & 18.9$\pm$0.0 & 18.9$\pm$0.0 & 41.3$\pm$0.0\\
    $ \kernel_{p_{\theta}}(\Hcal_q,\Hcal_c)$ & 53.9$\pm$1.9 & 62.5$\pm$1.3 & 33.6$\pm$0.7 & 30.6$\pm$0.9 & 56.3$\pm$2.1\\
    $\kernel_{p_{\theta}}(\Hcal_q,\Hcal_c) -\gamma \Delta_x (\Hcal_q,\Hcal_c)$ & 54.6$\pm$1.9 & 63.1$\pm$1.4 & 33.7$\pm$0.7 & 30.8$\pm$0.8 & 55.6$\pm$2.0\\
    $\kernel_{p_{\theta}}(\Hcal_q,\Hcal_c) -\gamma \Delta_t (U_{\phi}(\Hcal_q),\Hcal_c)$ & 55.7$\pm$2.0 & 63.7$\pm$1.8 & 35.9$\pm$0.8 & 31.3$\pm$0.8 & 58.1$\pm$2.0\\
    \nsrs Without $U_{\phi}(\cdot)$ & 55.2$\pm$2.2 & 62.9$\pm$2.0 & 34.3$\pm$0.9 & 29.7$\pm$1.3 & 56.2$\pm$2.3\\
    \nsrs & 56.2$\pm$2.1 & 65.1$\pm$1.9 & 37.4$\pm$0.9 & 32.4$\pm$0.8 & 58.7$\pm$2.1\\
    \bottomrule
    \end{tabular}
    }
    \vspace{-4mm}
    \label{tab:model_ablation}
\end{table}

\xhdr{Ablation Study}
Next, we compare the retrieval performance across four model variants: 
\begin{inparaenum}[(i)] 
\item our model with only model-independent score, \ie, $s_{p_{\theta},U_{\phi}} (\Hcal_q,\Hcal_c) = -\Delta_x (\Hcal_q,\Hcal_c) - \Delta_t (U_{\phi}(\Hcal_q),\Hcal_c)$;
 \label{var:withoutmodel}
\item our model with only model-dependent score, \ie, $s_{p_{\theta},U_{\phi}}(\Hcal_q,\Hcal_c) = \kernel_{p_{\theta}}(\Hcal_q,\Hcal_c)$;
 \label{var:withmodel}
\item our model without any model-independent time similarity, \ie, $s_{p_{\theta},U_{\phi}}(\Hcal_q,\Hcal_c) = \kernel_{p_{\theta}}(\Hcal_q,\Hcal_c) -\gamma \Delta_x (\Hcal_q,\Hcal_c)$;
 \label{var:withouttime}
\item our model without any model-independent mark similarity, \ie, $s_{p_{\theta},U_{\phi}}(\Hcal_q,\Hcal_c) = \kernel_{p_{\theta}}(\Hcal_q,\Hcal_c) -\gamma \Delta_t (U_{\phi}(\Hcal_q),\Hcal_c)$;
 \label{var:withoutmark}
\item  our model without unwarping function $U_{\phi}(\cdot)$;
 \label{var:withoutU}
 and
\item the complete design of our model.
\label{var:our}
\end{inparaenum}

\begin{figure}[t]
 \centering
\begin{subfigure}{0.45\columnwidth}
  \centering
  \includegraphics[height=4cm]{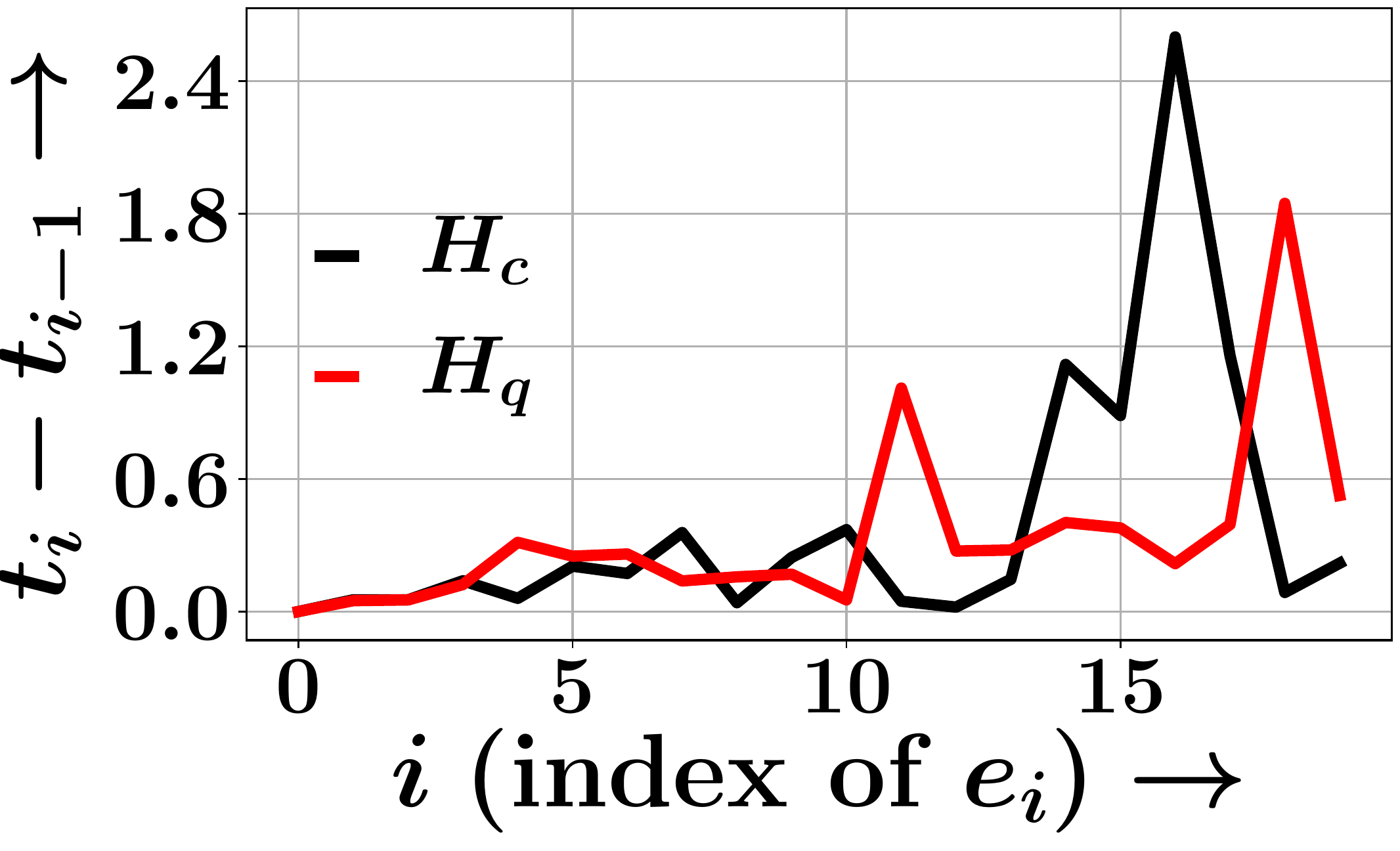}
  \caption{$\Hcal_q, \Hcal_c$}
\end{subfigure}
\hspace{2mm}
\begin{subfigure}{0.45\columnwidth}
  \centering
  \includegraphics[height=4cm]{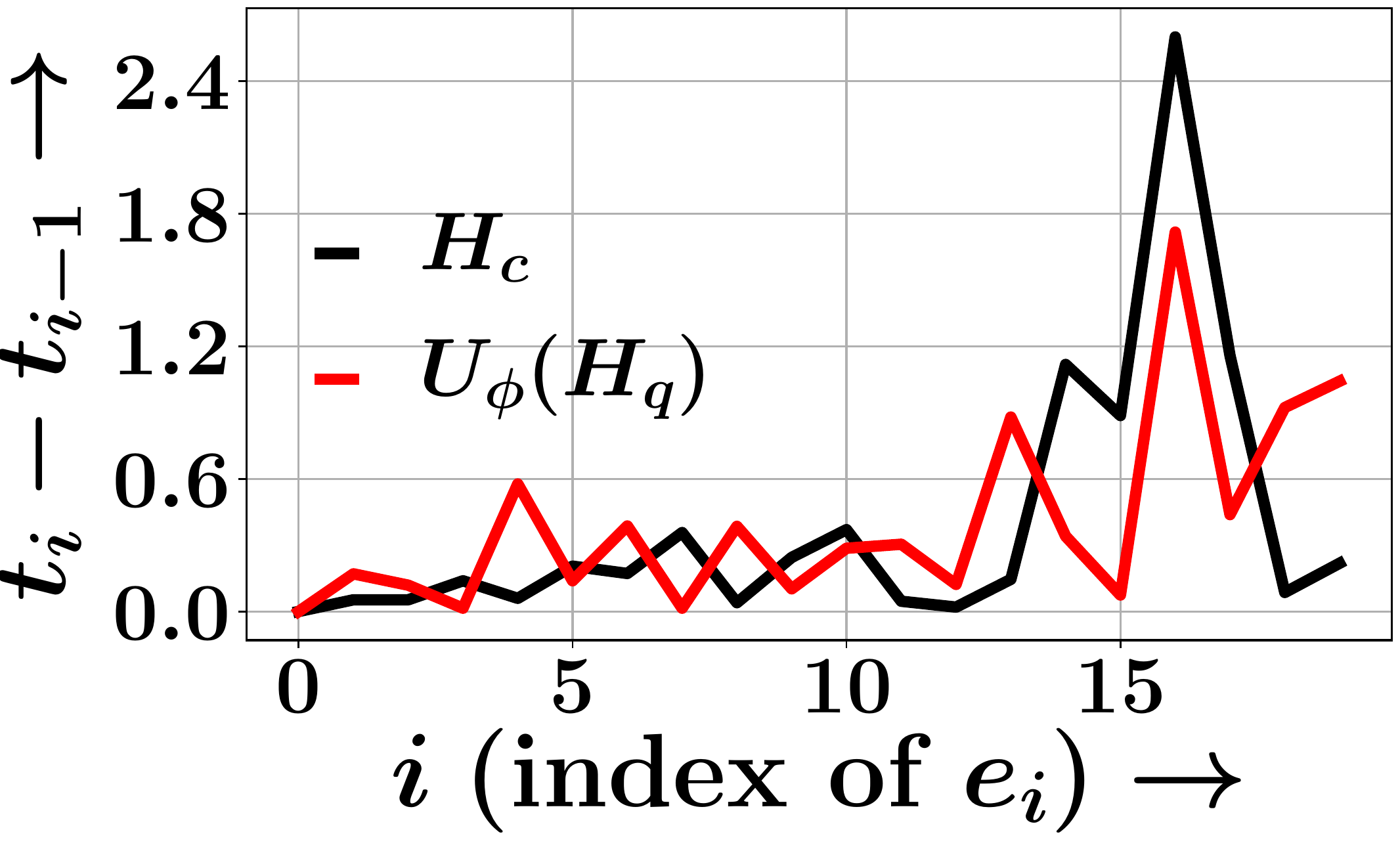}
  \caption{$U_{\phi}(\Hcal_q), \Hcal_c$}
\end{subfigure}
\caption{Effect of unwarping on a \emph{relevant} query-corpus pair in Audio. $U_{\phi}(\cdot)$ learns to transform $\Hcal_q$ in order to capture a high value of its latent similarity with $\Hcal_c$.}
\vspace{-4mm}
\label{fig:UU}
\end{figure}

In all cases, we used \nsrs. Table~\ref{tab:model_ablation} shows that the complete design of our model~(variant \eqref{var:our})  achieves the best performance. We further note that removing $\kernel_{p_\theta}$ from the score (variant \eqref{var:withoutmodel}) leads to significantly poor performance. Interestingly, our model without any mark-based similarity~(variant \eqref{var:withoutmark}) leads to better performance than the model without time similarity  (variant~\eqref{var:withouttime})--- this could be attributed to the larger variance in query-corpus time distribution than the distribution of marks. 

\xhdr{Effect of $U_{\phi}(\cdot)$}
Finally, we observe that the performance deteriorates if we do not use an unwarping function $U_{\phi}(\cdot)$ (variant~\eqref{var:withoutU}). Figure~\ref{fig:UU} illustrates the effect of $U_{\phi}(\cdot)$. It shows that $U_{\phi}(\cdot)$ is able to learn a suitable transformation of the query sequence, which encapsulates the high value of latent similarity with the corpus sequence.

\begin{figure}[t]
\centering
\begin{subfigure}{0.3\columnwidth}
  \centering
  \includegraphics[height=3.5cm]{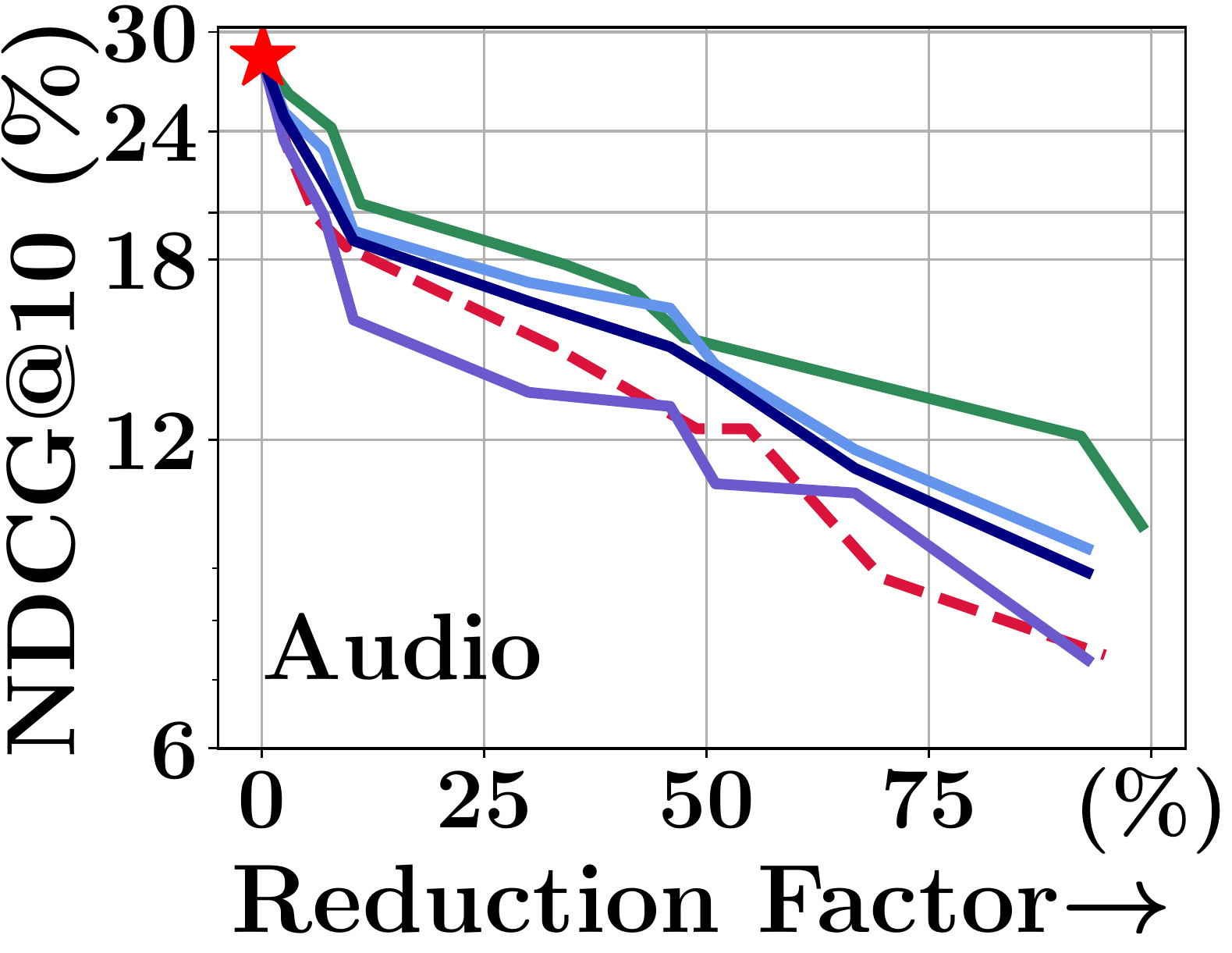}
  \caption{$\Hcal_q, \Hcal_c$}
\end{subfigure}
\hfill
\begin{subfigure}{0.3\columnwidth}
  \centering
  \includegraphics[height=3.5cm]{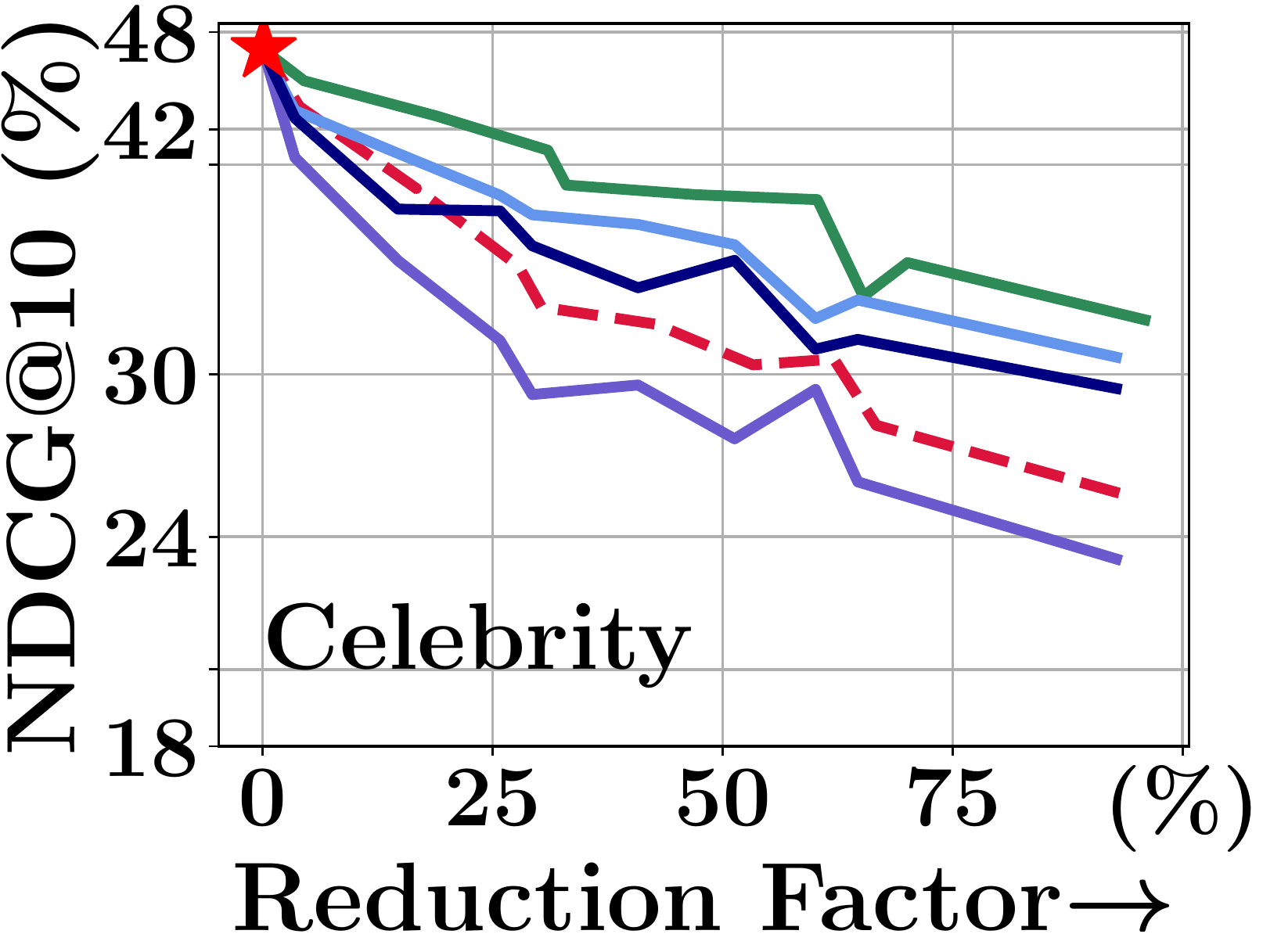}
  \caption{$\Hcal_q, \Hcal_c$}
\end{subfigure}
\hfill
\begin{subfigure}{0.3\columnwidth}
  \centering
  \includegraphics[height=3.5cm]{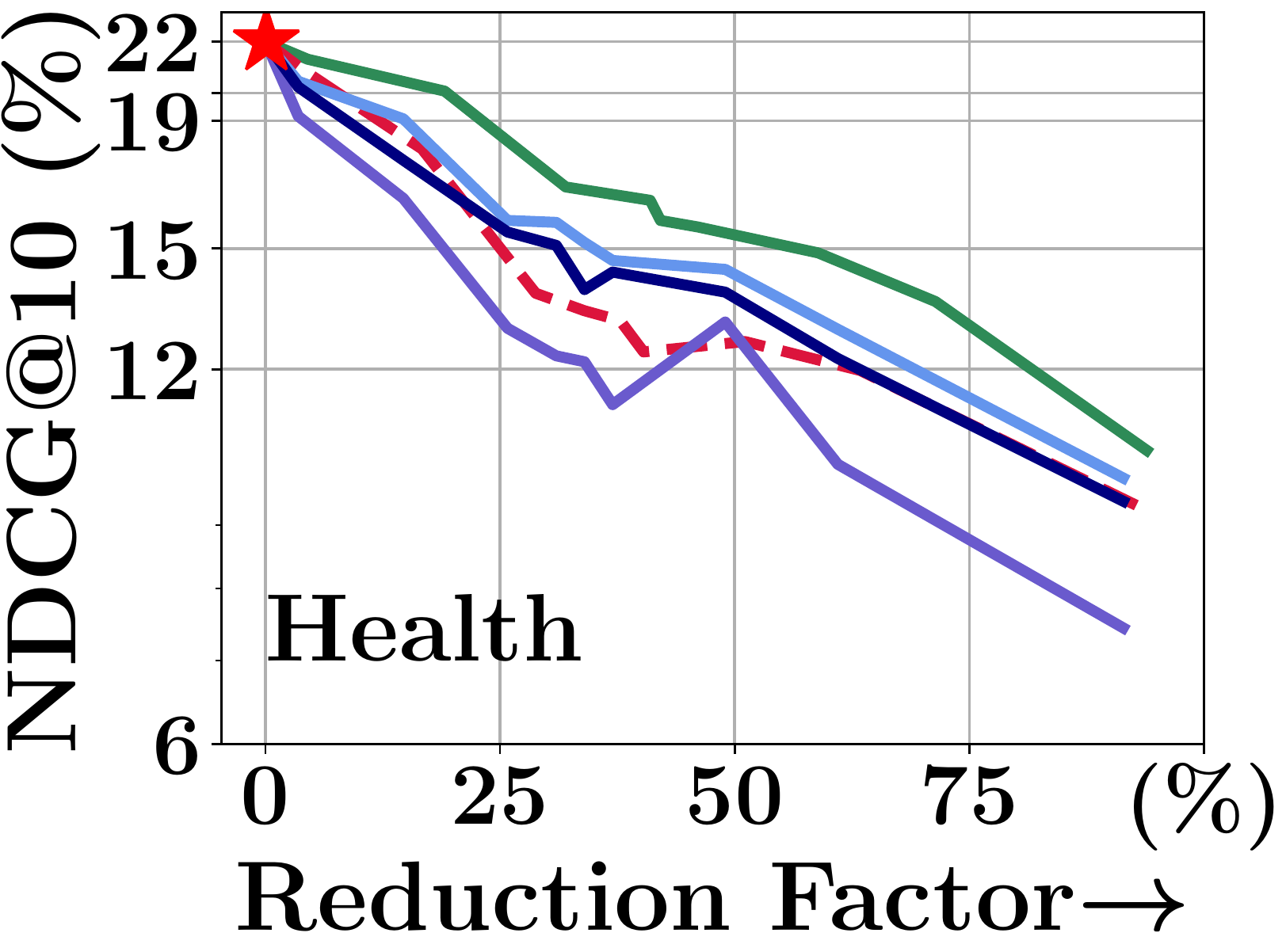}
  \caption{$\Hcal_q, \Hcal_c$}
\end{subfigure}
\begin{subfigure}{0.45\columnwidth}
  \centering
  \includegraphics[height=3.5cm]{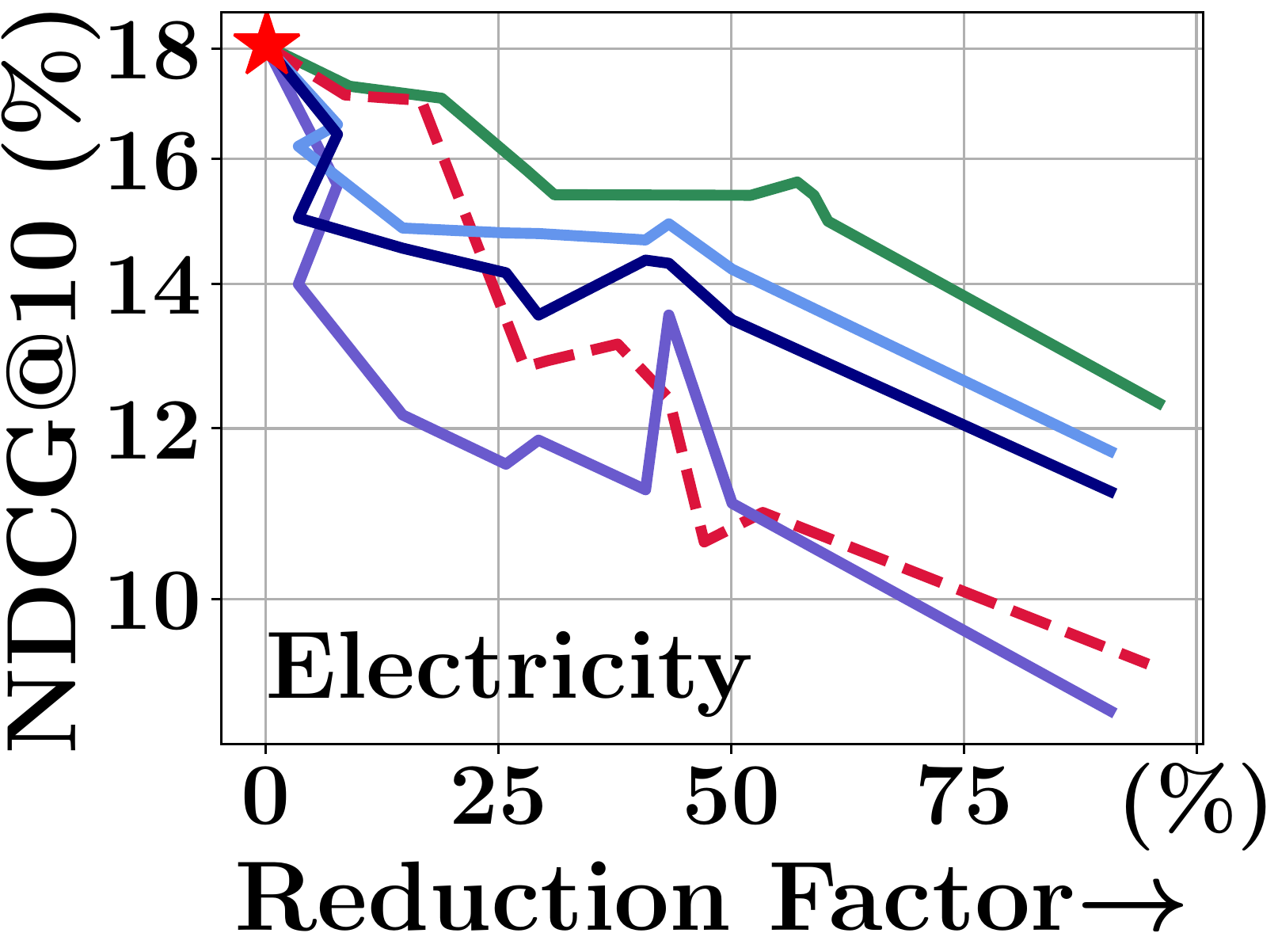}
  \caption{Health}
\end{subfigure}
\hspace{-1cm}
\begin{subfigure}{0.45\columnwidth}
  \centering
  \includegraphics[height=3.5cm]{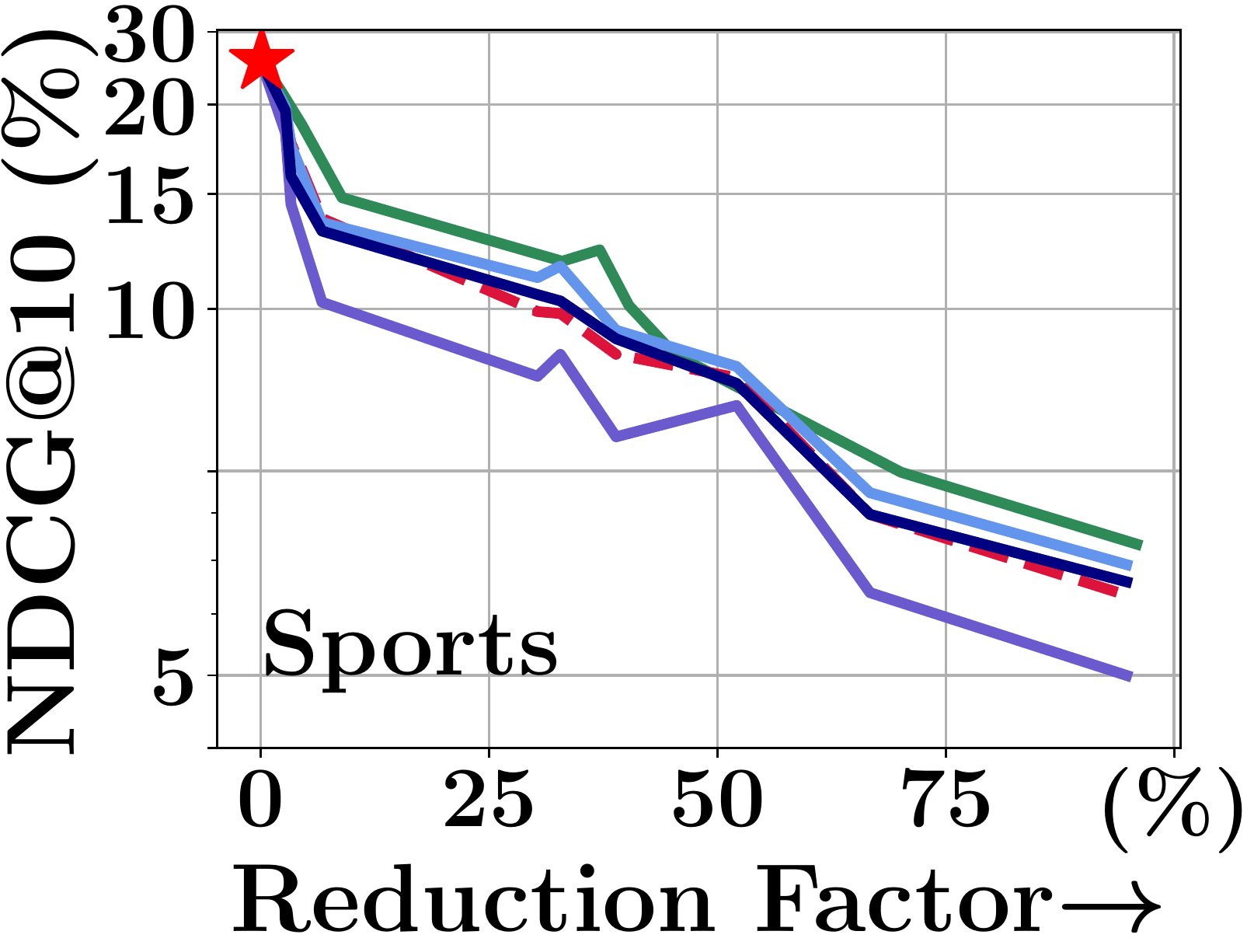}
  \caption{Health}
\end{subfigure}
{\includegraphics[height=1.2cm]{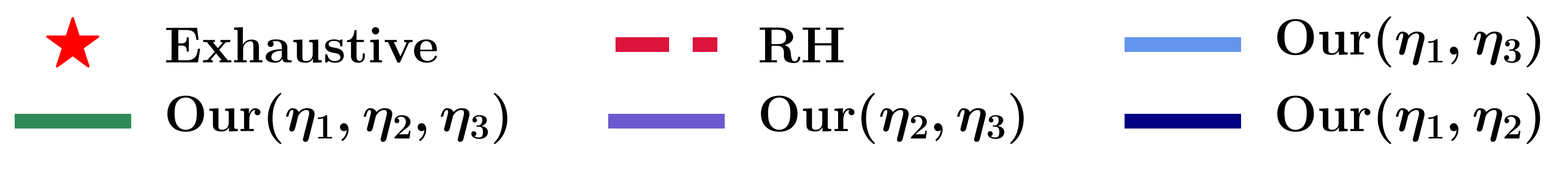}}
\caption{Tradeoff between NDCG@10 vs. Reduction factor, \ie, \% reduction in number of comparisons between query-corpus pairs w.r.t. the exhaustive comparisons for different hashing methods. The point marked as $ \color{red} {\star}$ indicates the case with exhaustive comparisons on the set of corpus sequences.}
\vspace{-4mm}
\label{fig:hash}
\end{figure}

\subsection{Results on Retrieval Efficiency}
We compare our efficient sequence retrieval method given in Algorithm~\ref{alg:key} against the random hyperplane (RH) method and three variants of our proposed training problem in Eq.~\eqref{eq:hash}. 
\begin{inparaenum}[(i)] 
\item Our $(\eta_2, \eta_3)$ which sets $\eta_1=0$ and thus does not enforce even distribution of $\pm 1 $ in $\hash^c$;
\item Our $(\eta_1, \eta_3)$ which sets $\eta_2=0$ and thus $\tanh$ does not accurately approximate $\sgn$;
\item Our $(\eta_1, \eta_2)$ which sets $\eta_3=0$ and thus does not enforce $\hash^c$ to be compact and free of redundancy. 
\end{inparaenum}
Our $(\eta_1,\eta_2,\eta_3)$ is the complete design that includes all trainable components. Figure~\ref{fig:hash} summarizes the results.

\xhdr{Comparison with random hyperplane}   Figure~\ref{fig:hash}  shows that our method (Our$(\eta_1,\eta_2,\eta_3)$) demonstrates better Pareto efficiency than RH. This is because RH generates hash code in a data-oblivious manner, whereas our method learns the hash code on top of the trained embeddings.

\xhdr{Ablation study on different components of Eq.~\eqref{eq:hash}}
Figure~\ref{fig:hash} summarizes the results, which shows that (i) the first three variants are outperformed by Our$(\eta_1,\eta_2,\eta_3)$; (ii) the first term having $\eta_1 \neq 0$, which enforces an even distribution of $\pm 1$, is the most crucial component for the loss function--- as the removal of this term causes significant deterioration of the performance.

\subsection{Analysis at a Query Level} 
\begin{figure}[t]
\centering
\begin{subfigure}{0.3\columnwidth}
  \centering
  \includegraphics[height=3.5cm]{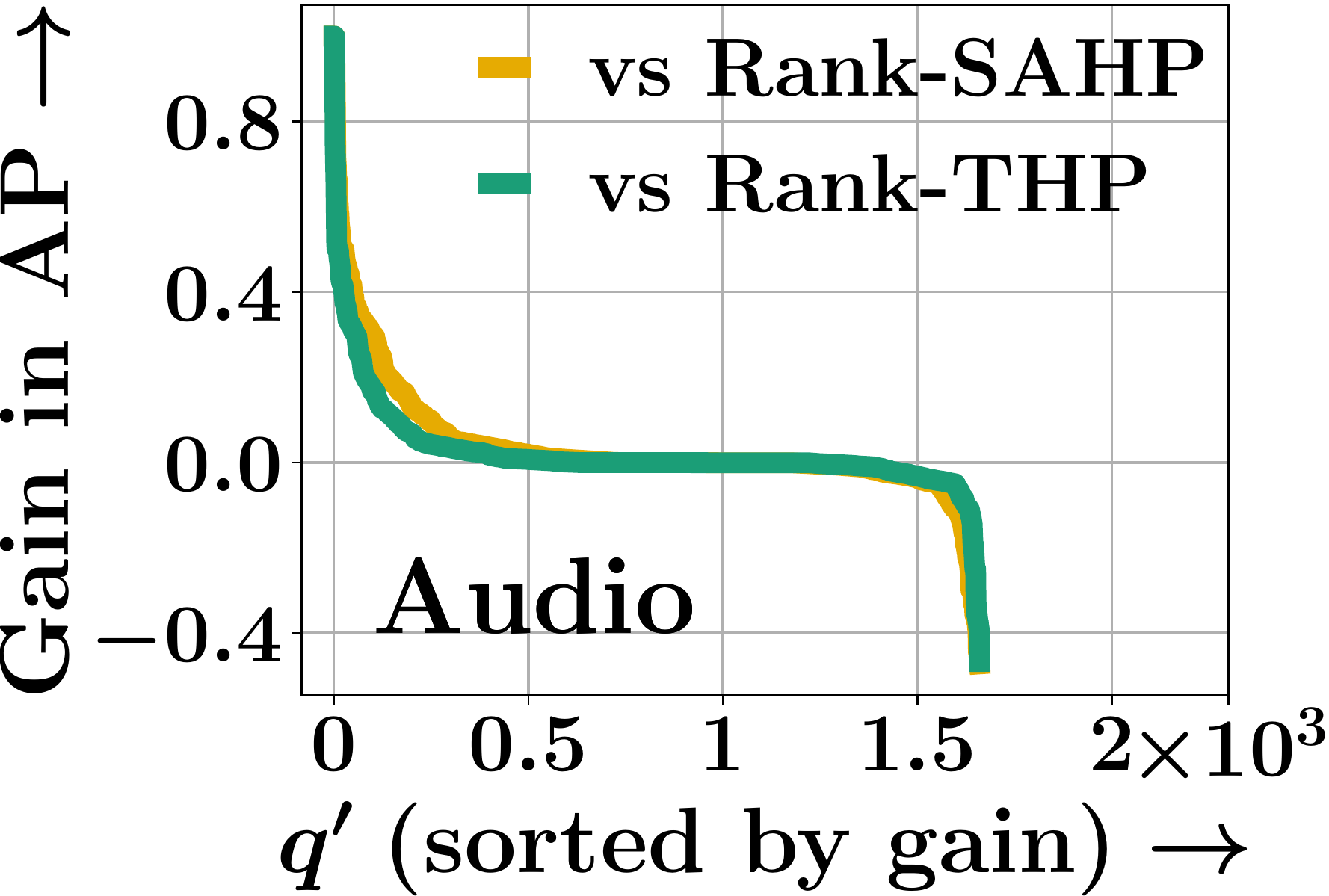}
  \caption{Audio}
\end{subfigure}
\hfill
\begin{subfigure}{0.3\columnwidth}
  \centering
  \includegraphics[height=3.5cm]{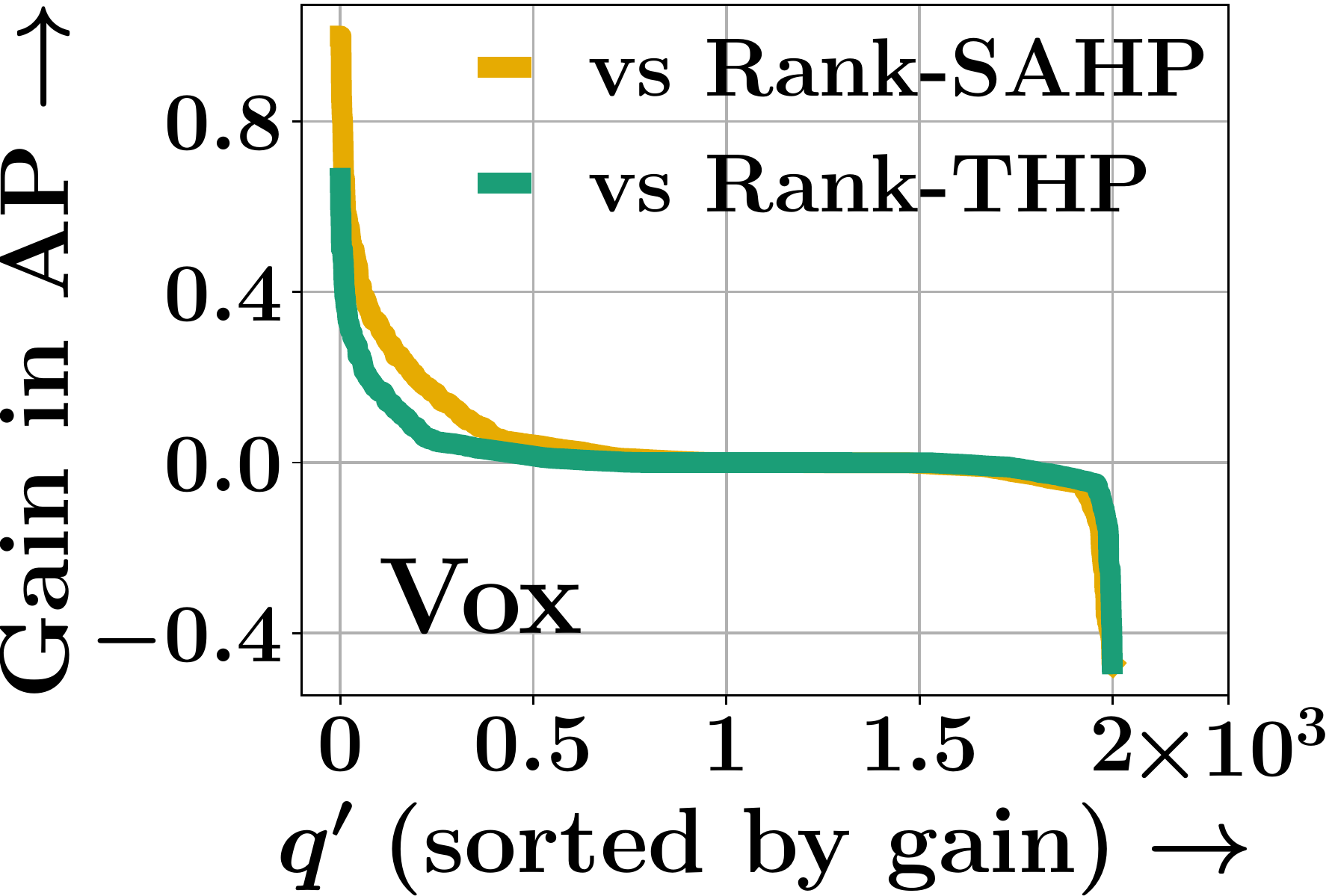}
  \caption{Celebrity}
\end{subfigure}
\hfill
\begin{subfigure}{0.3\columnwidth}
  \centering
  \includegraphics[height=3.5cm]{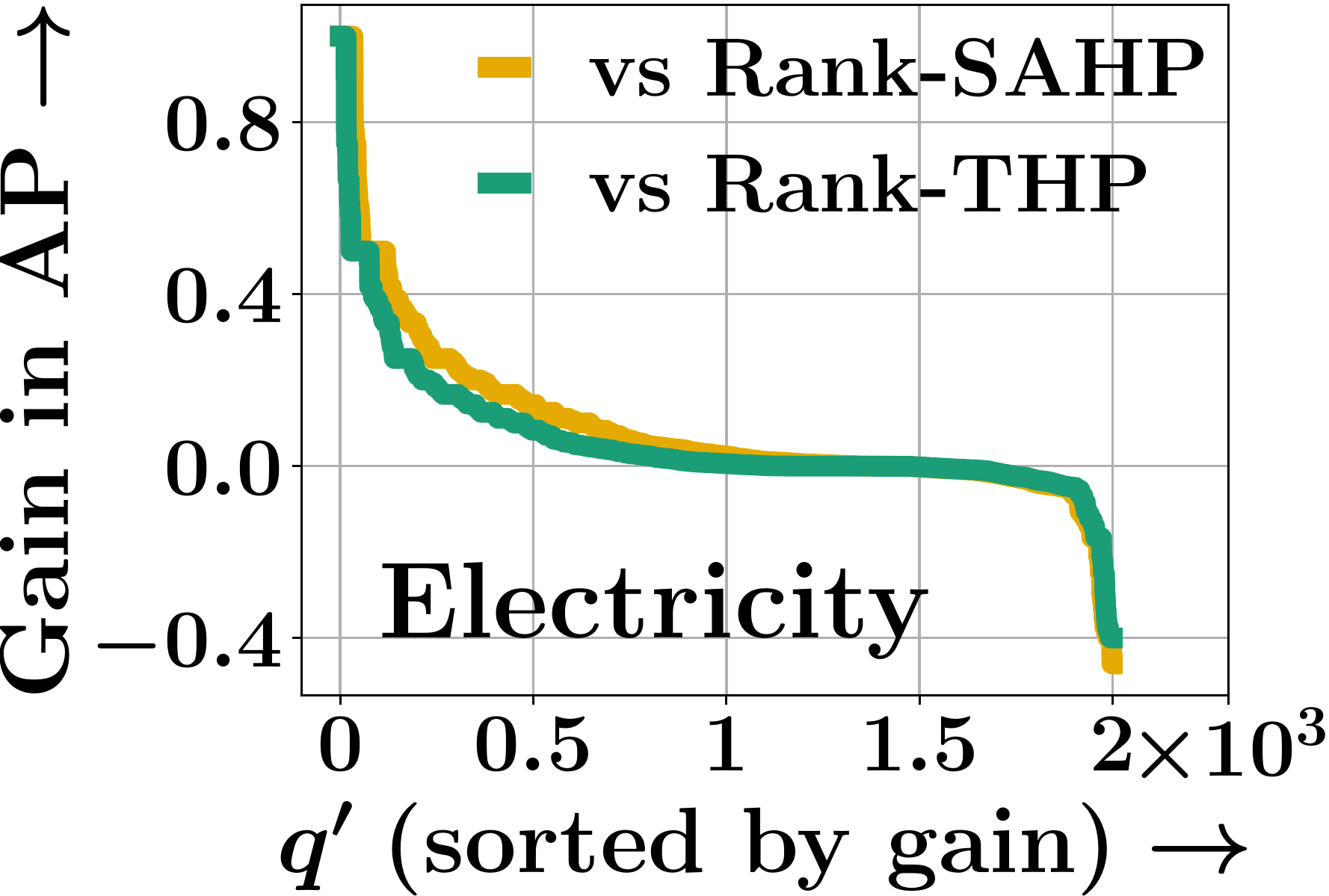}
  \caption{Electricity}
\end{subfigure}

\begin{subfigure}{0.3\columnwidth}
  \centering
  \includegraphics[height=3.5cm]{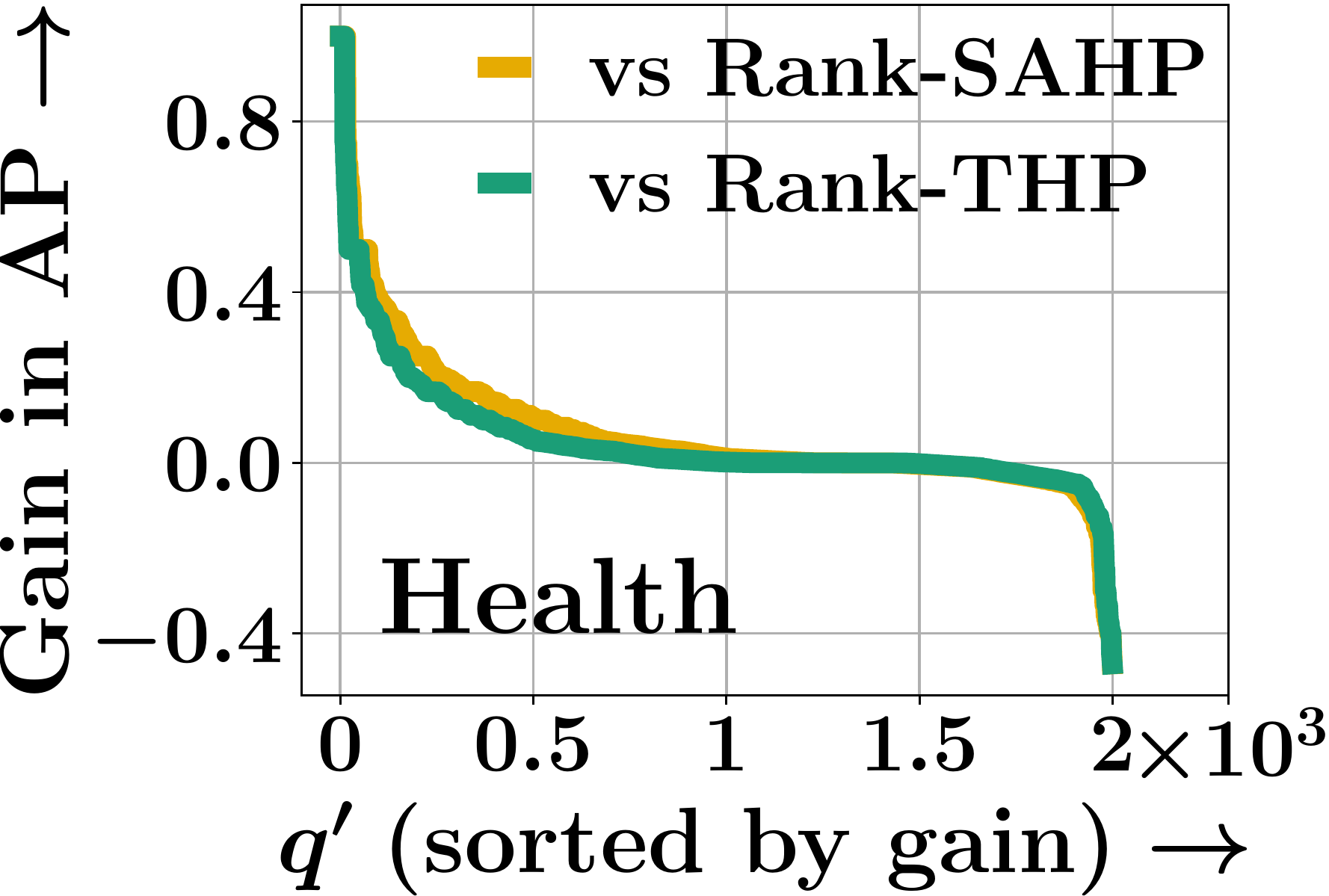}
  \caption{Health}
\end{subfigure}
\hspace{1cm}
\begin{subfigure}{0.3\columnwidth}
  \centering
  \includegraphics[height=3.5cm]{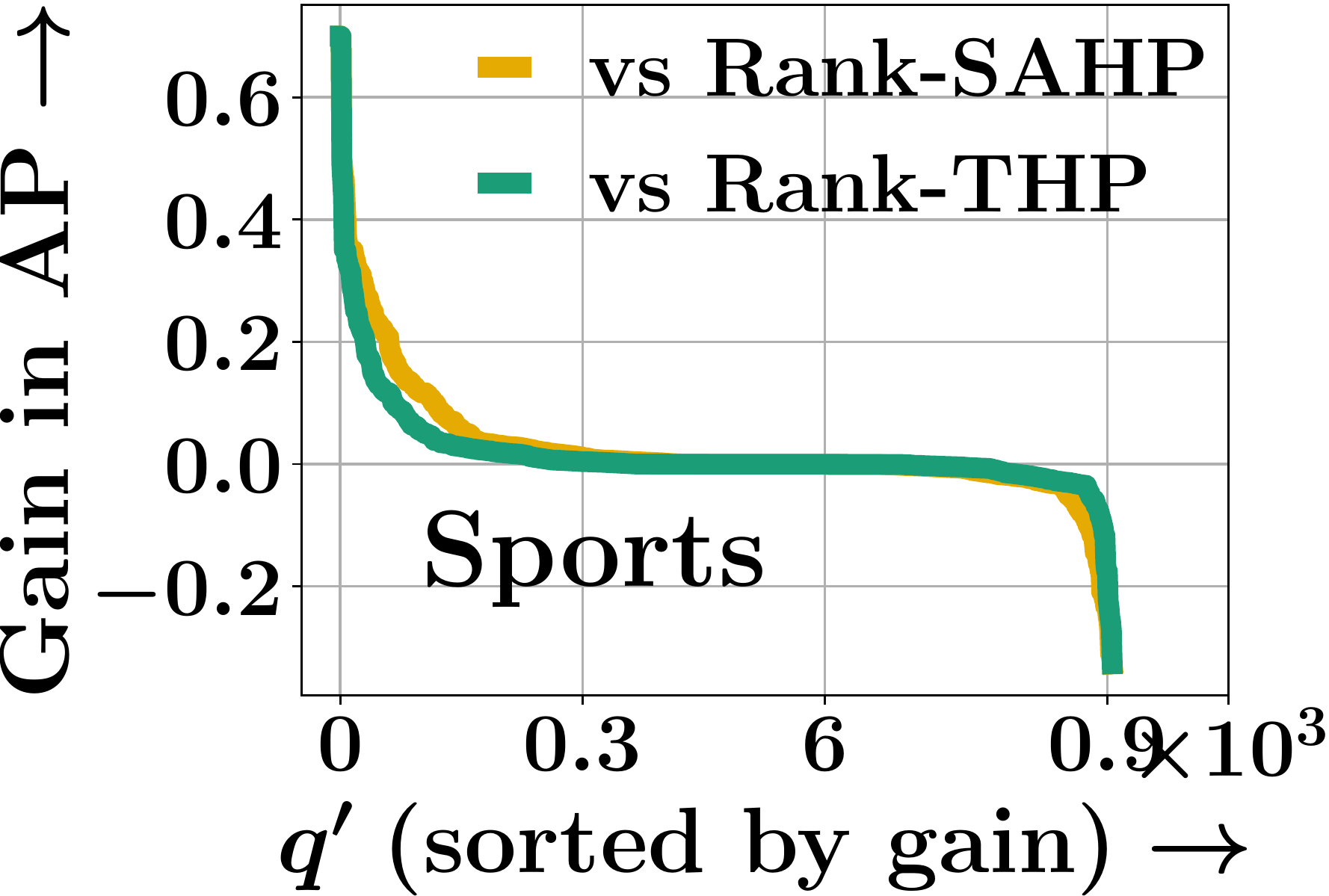}
  \caption{Sports}
\end{subfigure}
\caption{Query-wise performance comparison between \nsr and best baseline methods -- \tpprank-THP, \tpprank-SAHP. Queries are sorted by the decreasing gain in AP.}
\label{fig:app_drill}
\vspace{-4mm}
\end{figure}

Next, we compare the performance between \nsr and other state-of-the-art methods at a query level. Specifically, for each query $\Hcal_q$ we compute the advantage of using \nsr in terms of gain in average precision, \ie, AP(\nsr) $-$ AP(baseline) for two most competitive baselines -- \tpprank-SAHP and \tpprank-THP. We summarize the results in Figure~\ref{fig:app_drill}, which show that for at least 70\% of the queries, \nsr outperforms or fares competitively with these baselines across all datasets.

\begin{table}[t!]
    \centering
    \caption{Training-times of \nsr for all datasets.}
    \small
    \begin{tabular}{c|ccccc}
    \toprule
    \textbf{Run Time} & \textbf{Audio} & \textbf{Celebrity} & \textbf{Electricity} & \textbf{Health} & \textbf{Sports}\\ \midrule
    \nsr & $\le$ 3hr & $\le$ 5hr & $\le$ 6hr & $\le$ 6hr & $\le$ 2hr\\ 
    \midrule
    \end{tabular}
    \vspace{-4mm}
    \label{tab:runtime}
\end{table}

\subsection{Runtime Analysis}
Next, we calculate the run-time performance of \nsr. With this experiment, our goal is to determine if the training times of \nsr are suitable for designing solutions for real-world problems. From the results in Table~\ref{tab:runtime}, we note that even for datasets with up to 60 million events, the training times are well within the feasible range for practical deployment.

\subsection{Query Length}
We perform an additional experiment of sequence retrieval with varying query lengths. Specifically, we sample queries of lengths $|\mathcal{H}_q| \sim Unif(10, 50)$ and $|\mathcal{H}_q| \sim Unif(50, 100)$ and report the sequence retrieval results in Table~\ref{tab:length1} and Table~\ref{tab:length2} respectively. The results show that the performance of all models deteriorates significantly as we reduce the length of query sequences. They also show that even with smaller query lengths, \nsrs significantly outperforms the other state-of-the-art baseline \texttt{Rank}-THP.

\begin{table}[t!]
    \caption{Retrieval quality in terms of mean average precision (MAP) for query sequence lengths sampled between 10 and 50.}
    \small
    \centering
    \begin{tabular}{l|ccccc}
    \hline
    \textbf{$|\mathcal{H}_q| = (10,20)$} & \textbf{Audio} & \textbf{Celebrity} & \textbf{Health} & \textbf{Electricity} & \textbf{Sports}\\ \hline
    \texttt{Rank}-THP & 18.97 & 22.06 & 9.48 & 12.27 & 26.58\\
    \nsrs & 21.30 & 25.77 & 15.83 & 10.79 & 27.63\\
    \hline
    \end{tabular}
    \label{tab:length1}
    \centering
    \caption{Retrieval quality in terms of mean average precision (MAP) for query sequence lengths sampled between 50 and 100.}
    \begin{tabular}{l|ccccc}
    \hline
    \textbf{$|\mathcal{H}_q| = (50,100)$} & \textbf{Audio} & \textbf{Celebrity} & \textbf{Health} & \textbf{Electricity} & \textbf{Sports}\\ \hline
    \texttt{Rank}-THP & 27.94 & 37.85 & 17.58 & 21.61 & 31.26\\
    \nsrs & 28.58 & 41.92 & 19.67 & 23.54 & 36.92\\
    \hline
    \end{tabular}
    \vspace{-4mm}
    \label{tab:length2}
\end{table}

\subsection{Qualitative Analysis}
\begin{figure}
\centering
\begin{subfigure}{\columnwidth}
  \centering
 {\includegraphics[height=3cm]{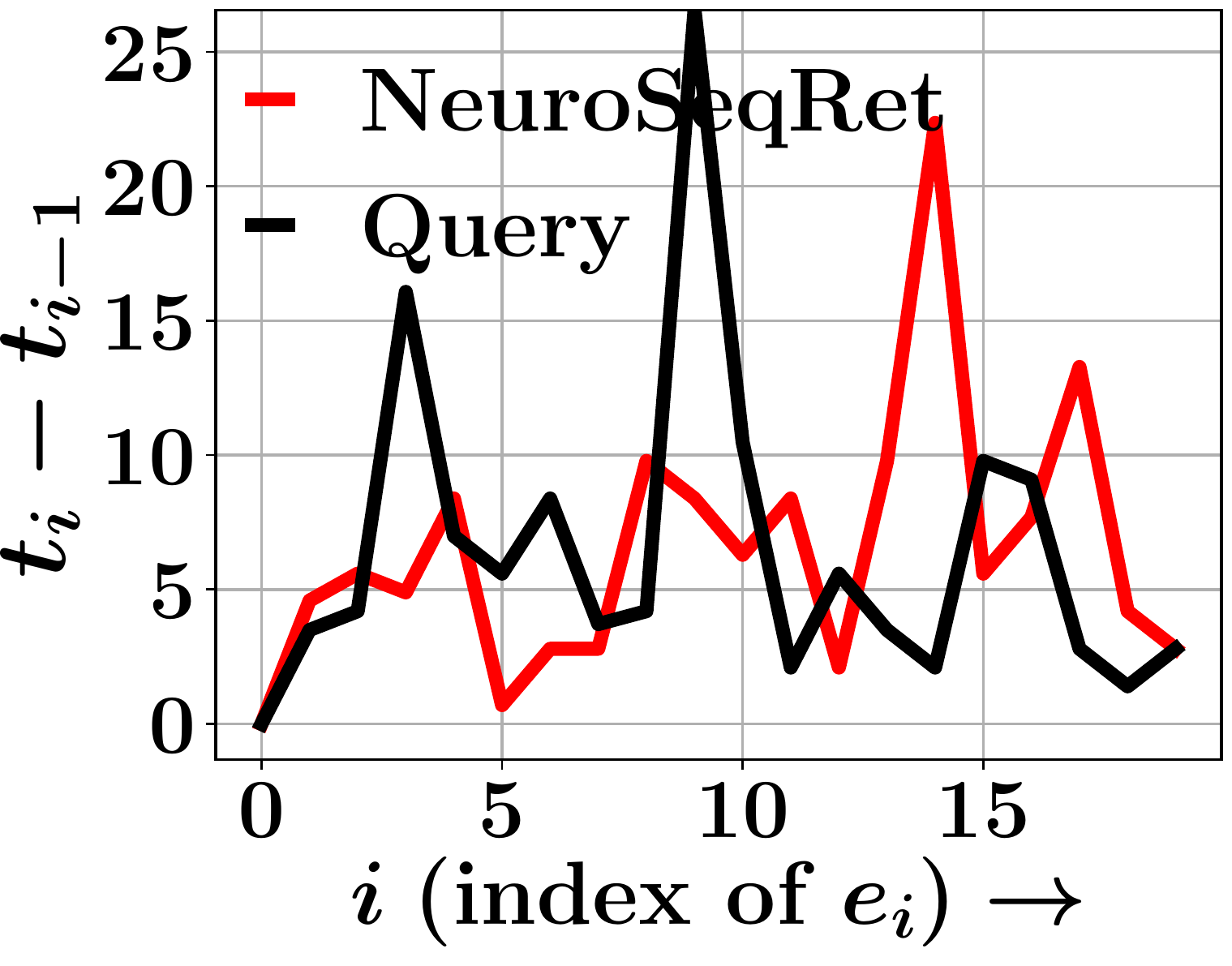}}
 {\includegraphics[height=3cm]{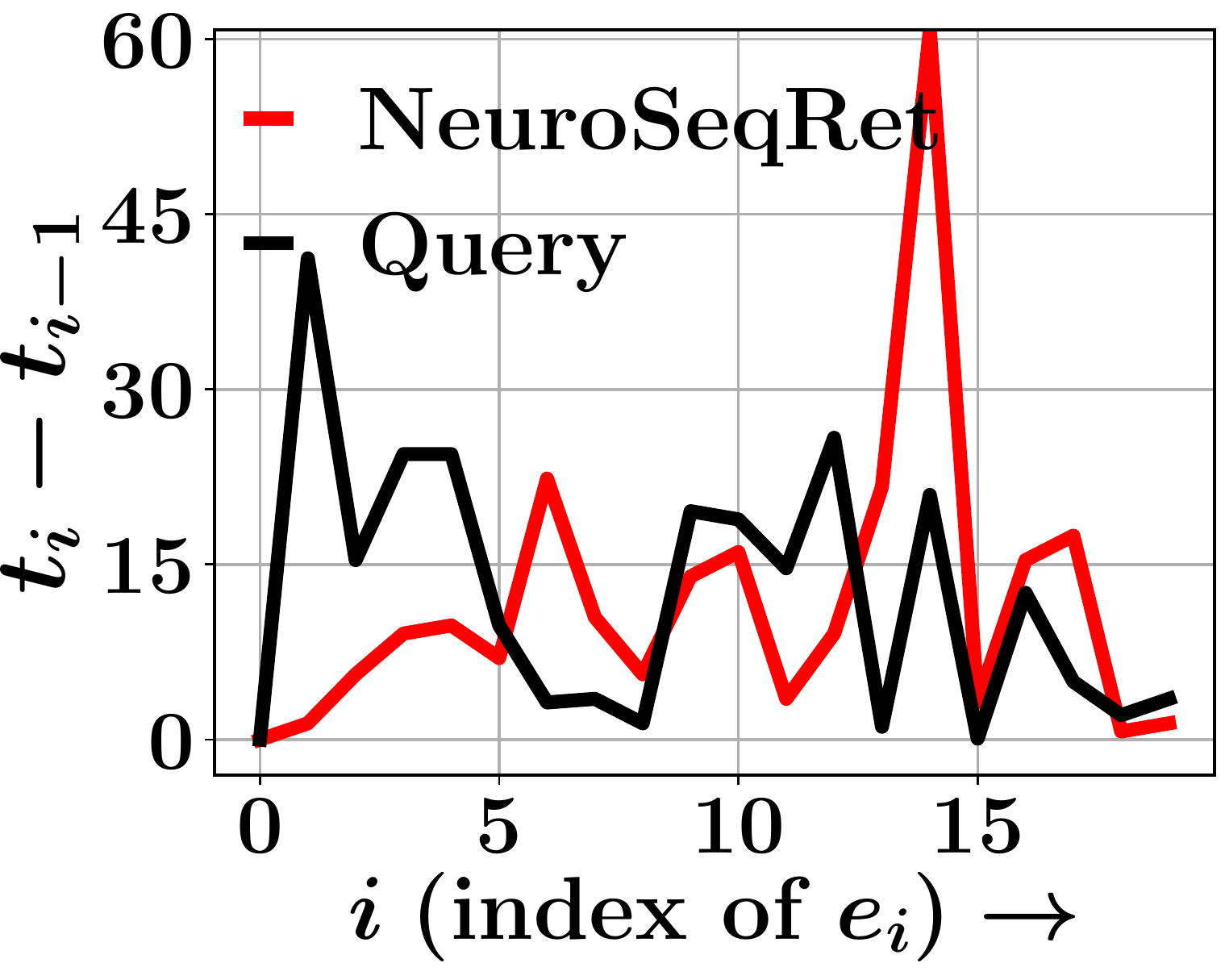}}
 {\includegraphics[height=3cm]{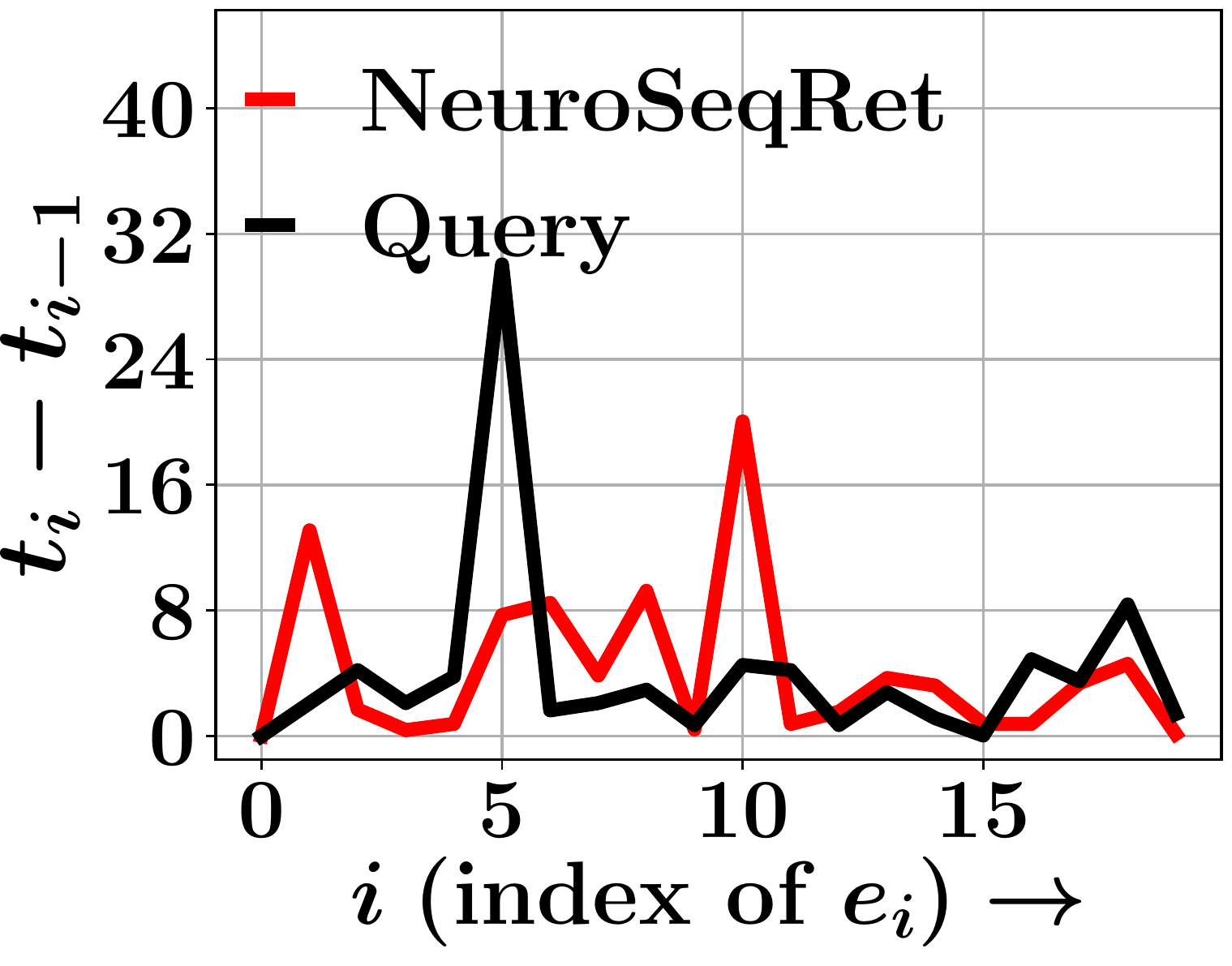}}
 {\includegraphics[height=3cm]{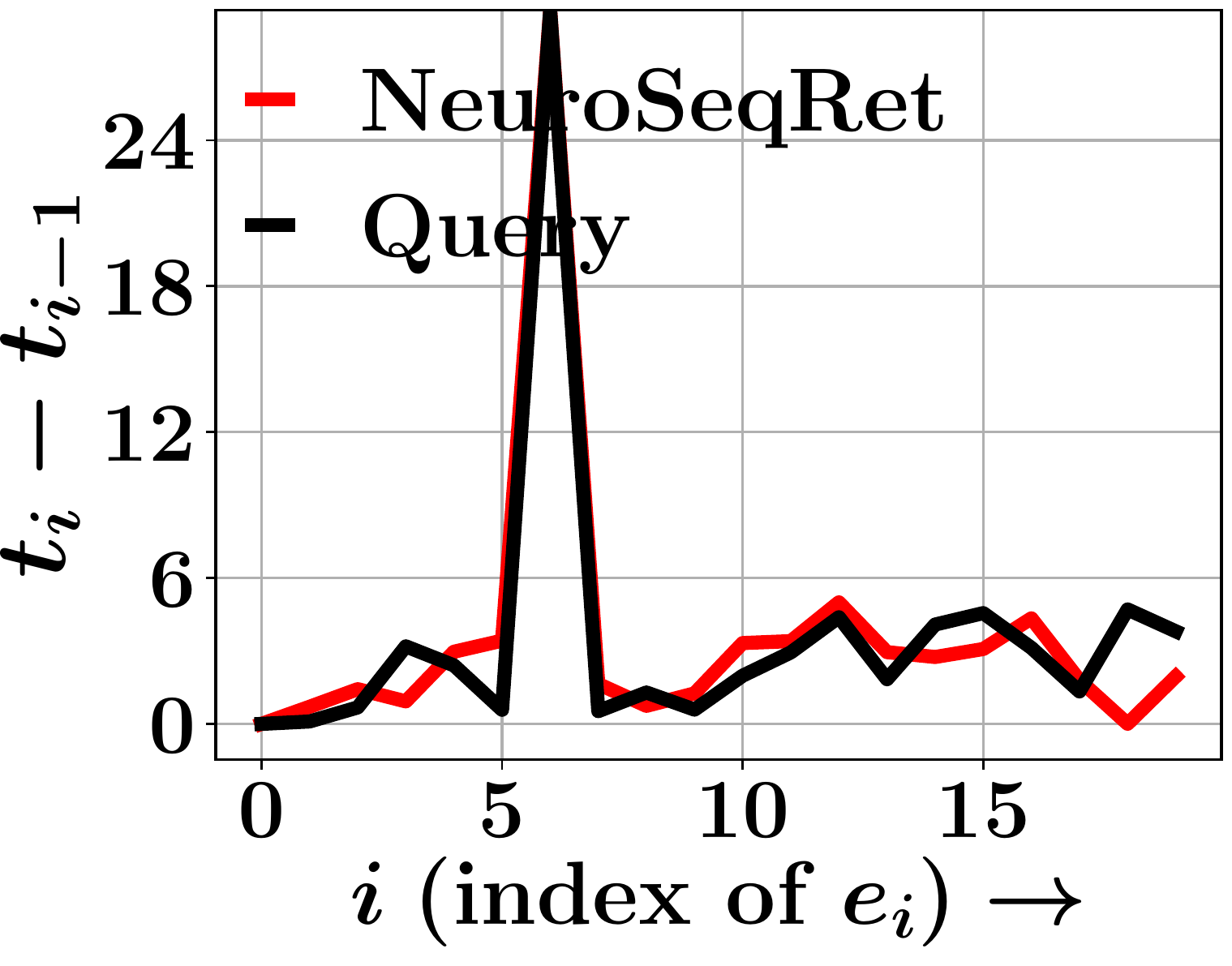}}
  \caption{Audio}
\end{subfigure}

\begin{subfigure}{\columnwidth}
  \centering
{\includegraphics[height=3cm]{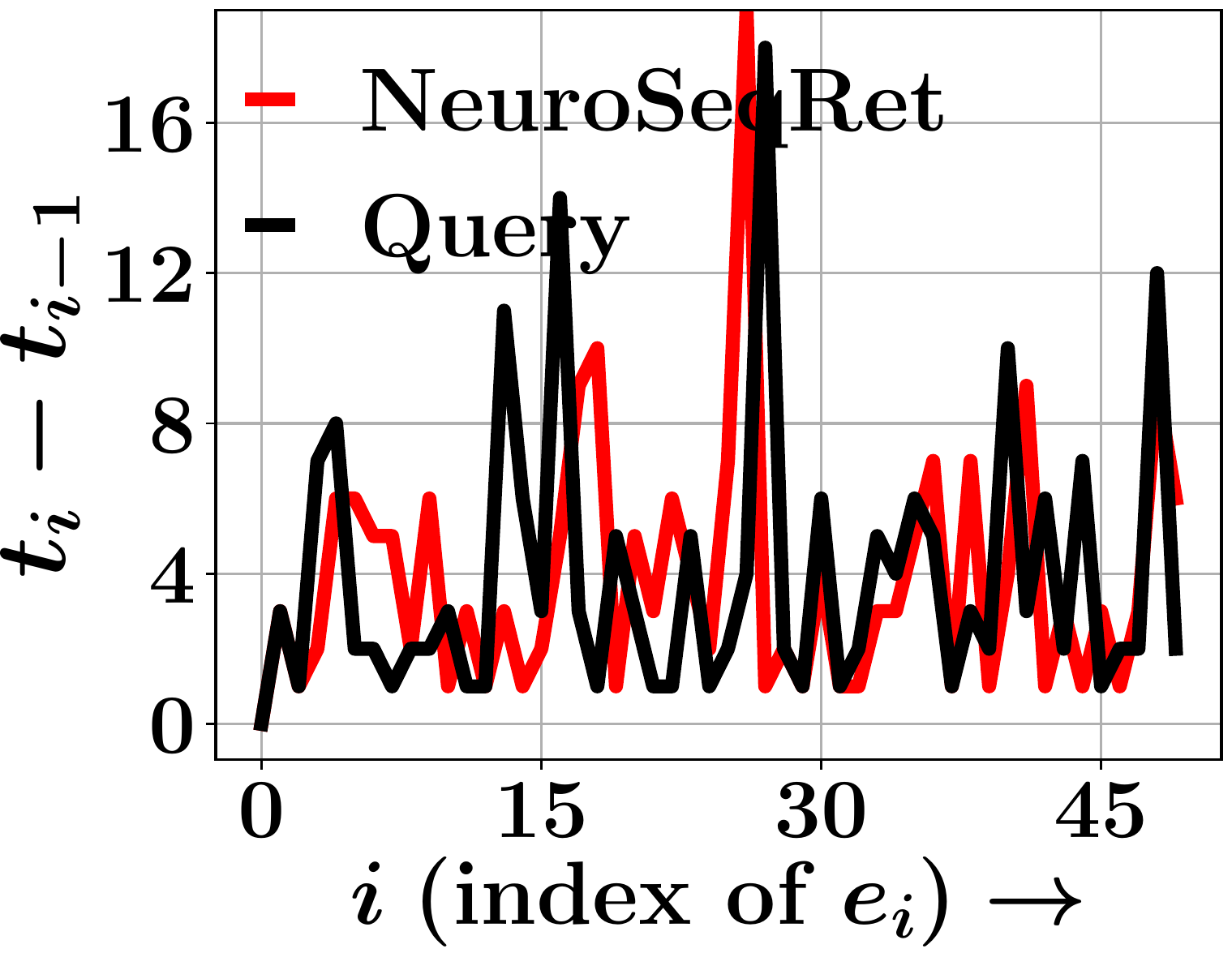}}
{\includegraphics[height=3cm]{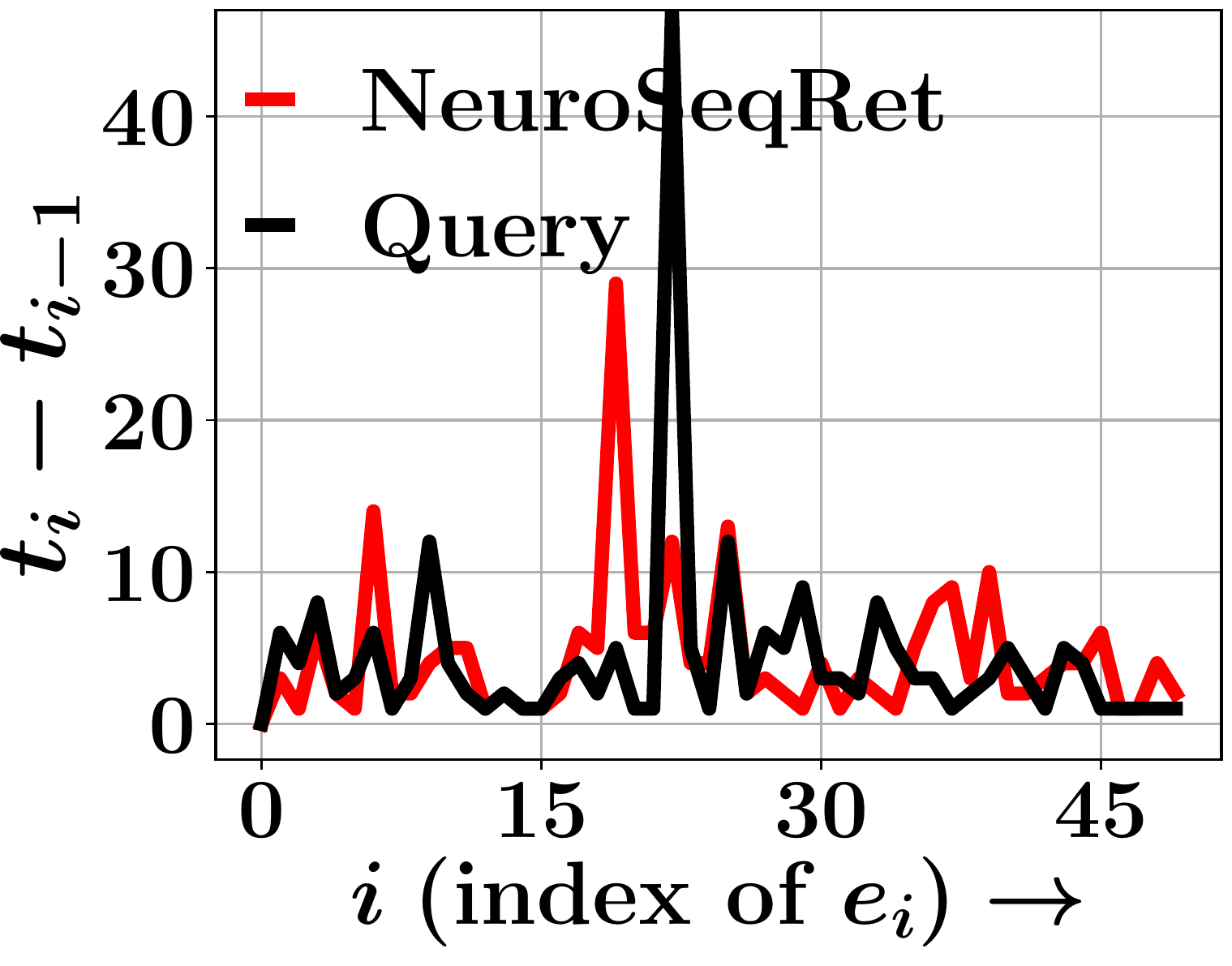}}
{\includegraphics[height=3cm]{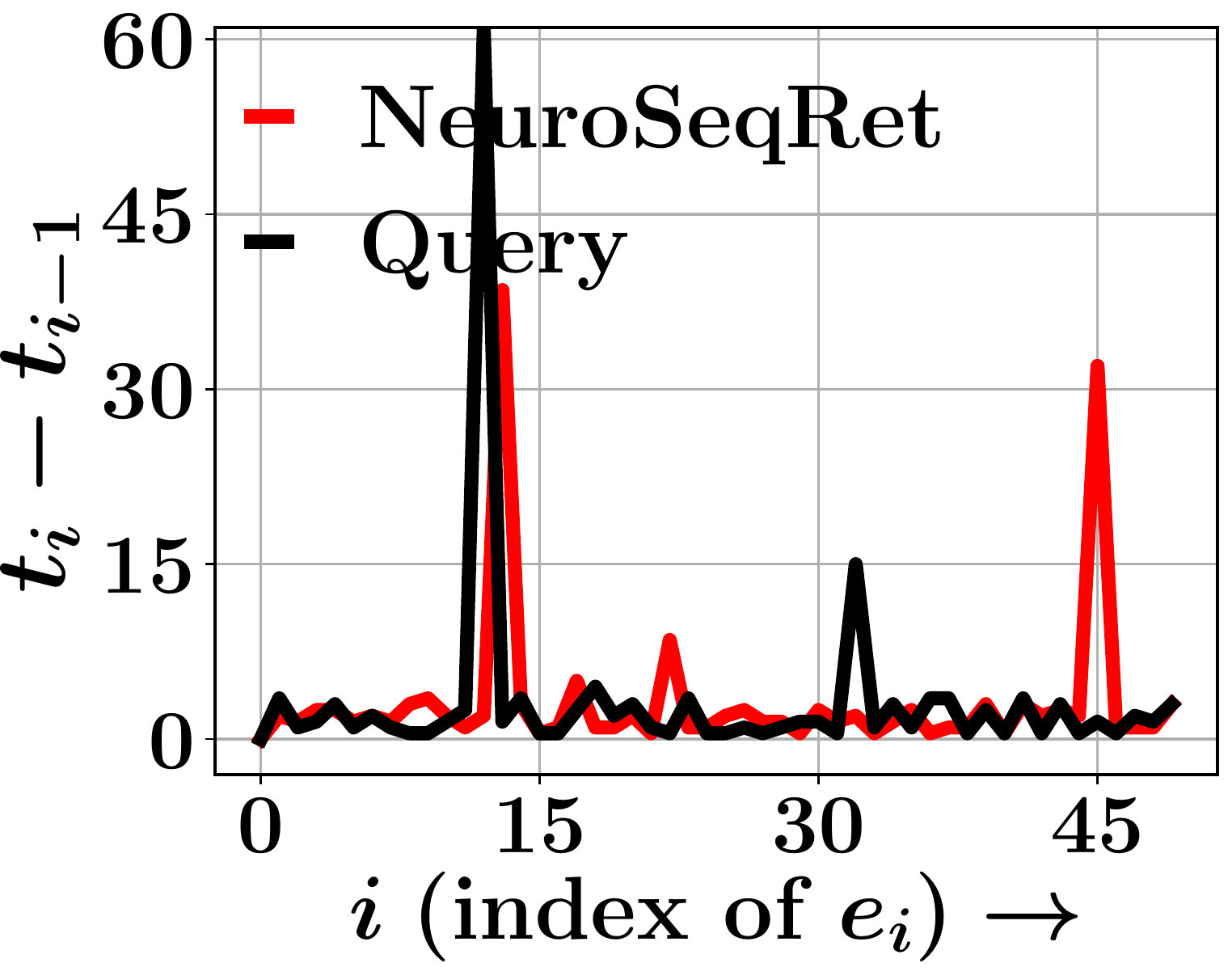}}
{\includegraphics[height=3cm]{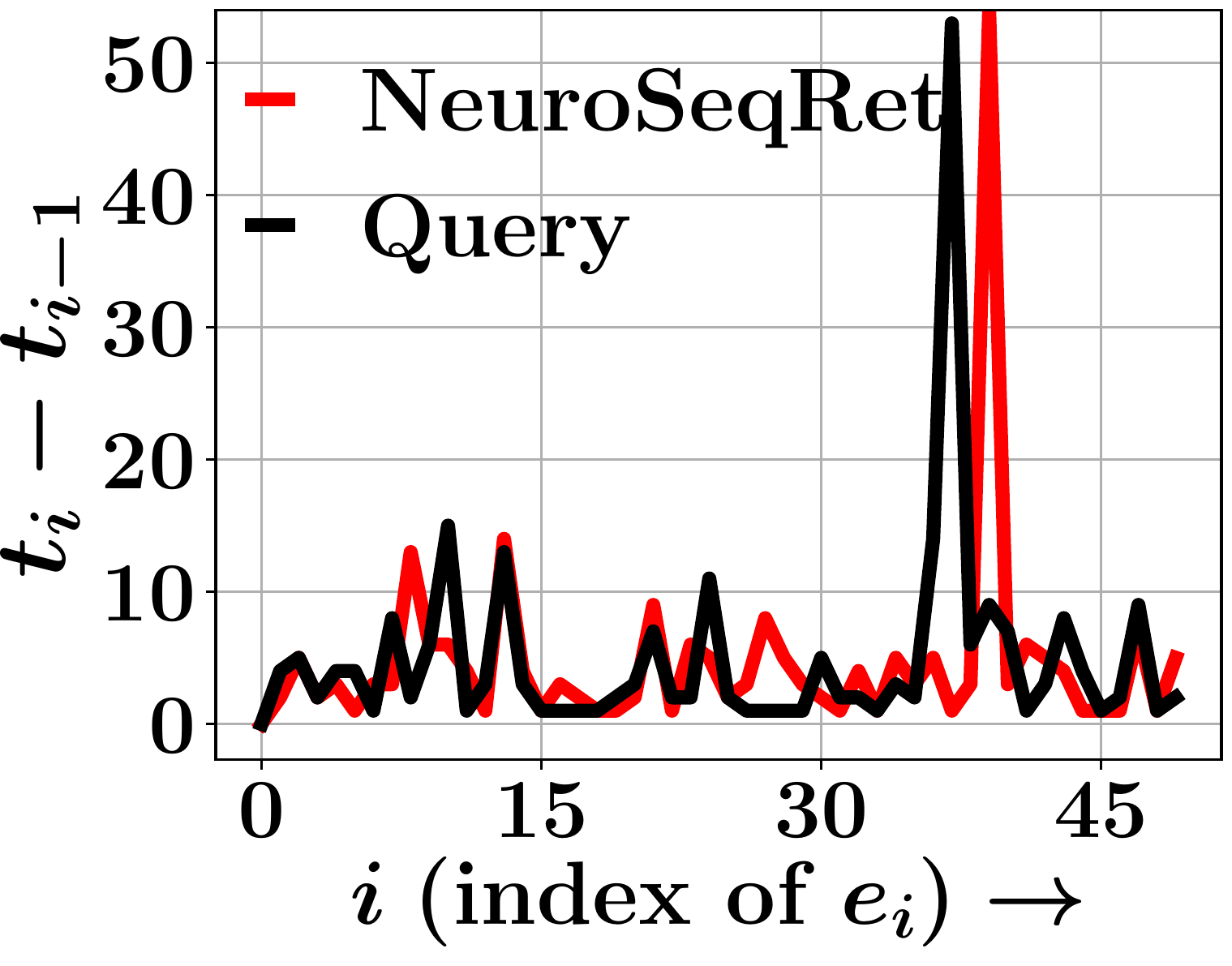}}
  \caption{Celebrity}
\end{subfigure}

\begin{subfigure}{\columnwidth}
  \centering
{\includegraphics[height=3cm]{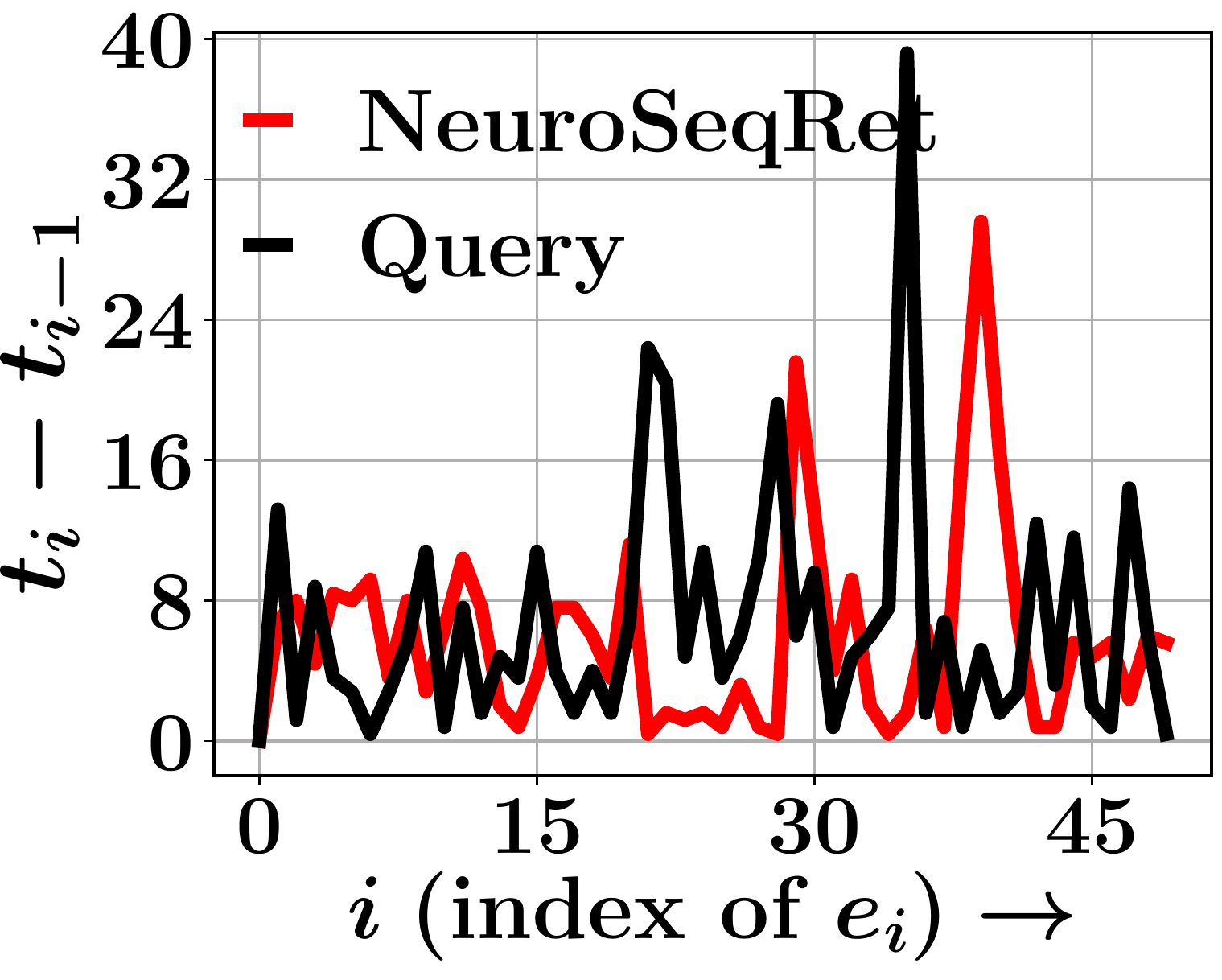}}
{\includegraphics[height=3cm]{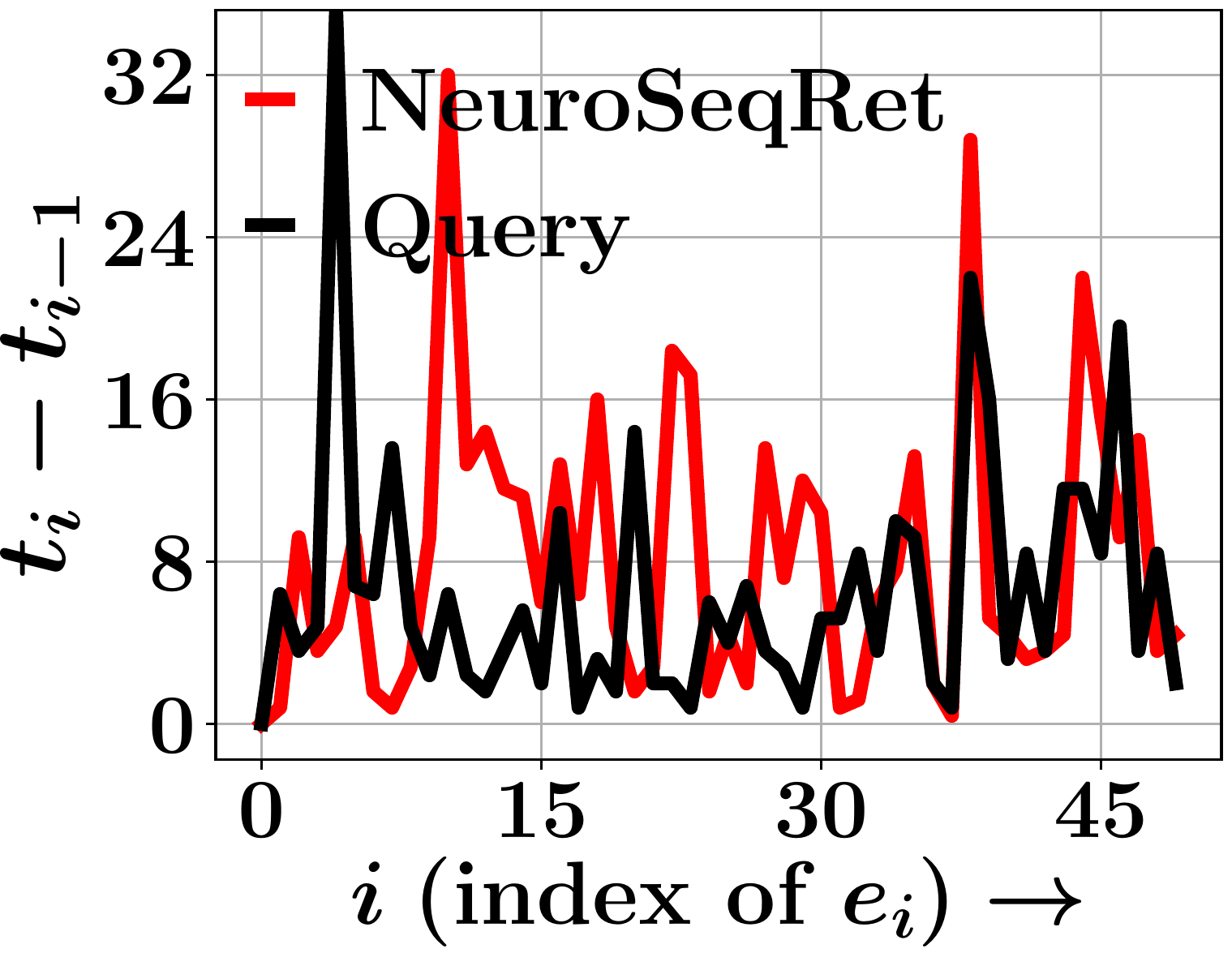}}
{\includegraphics[height=3cm]{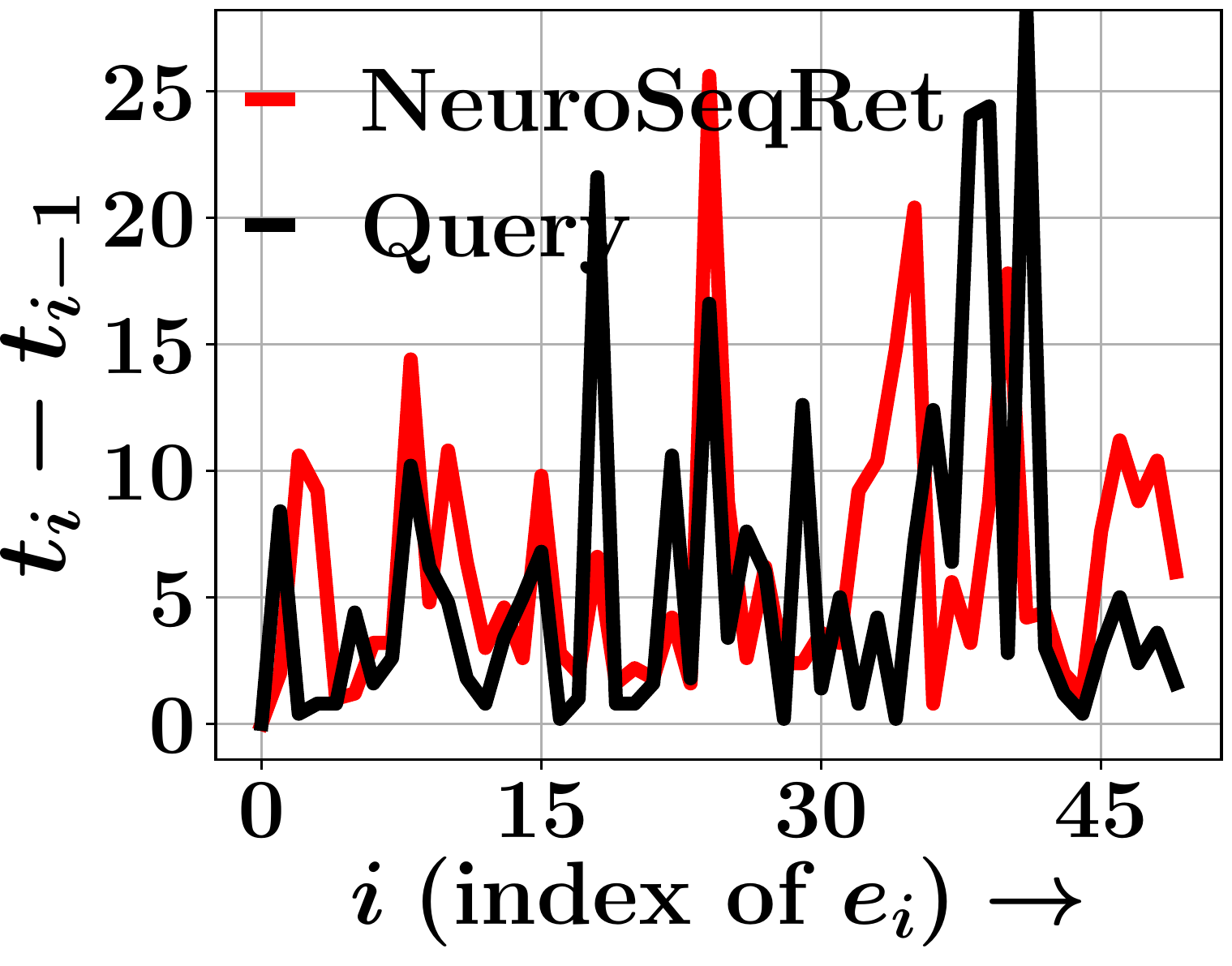}}
{\includegraphics[height=3cm]{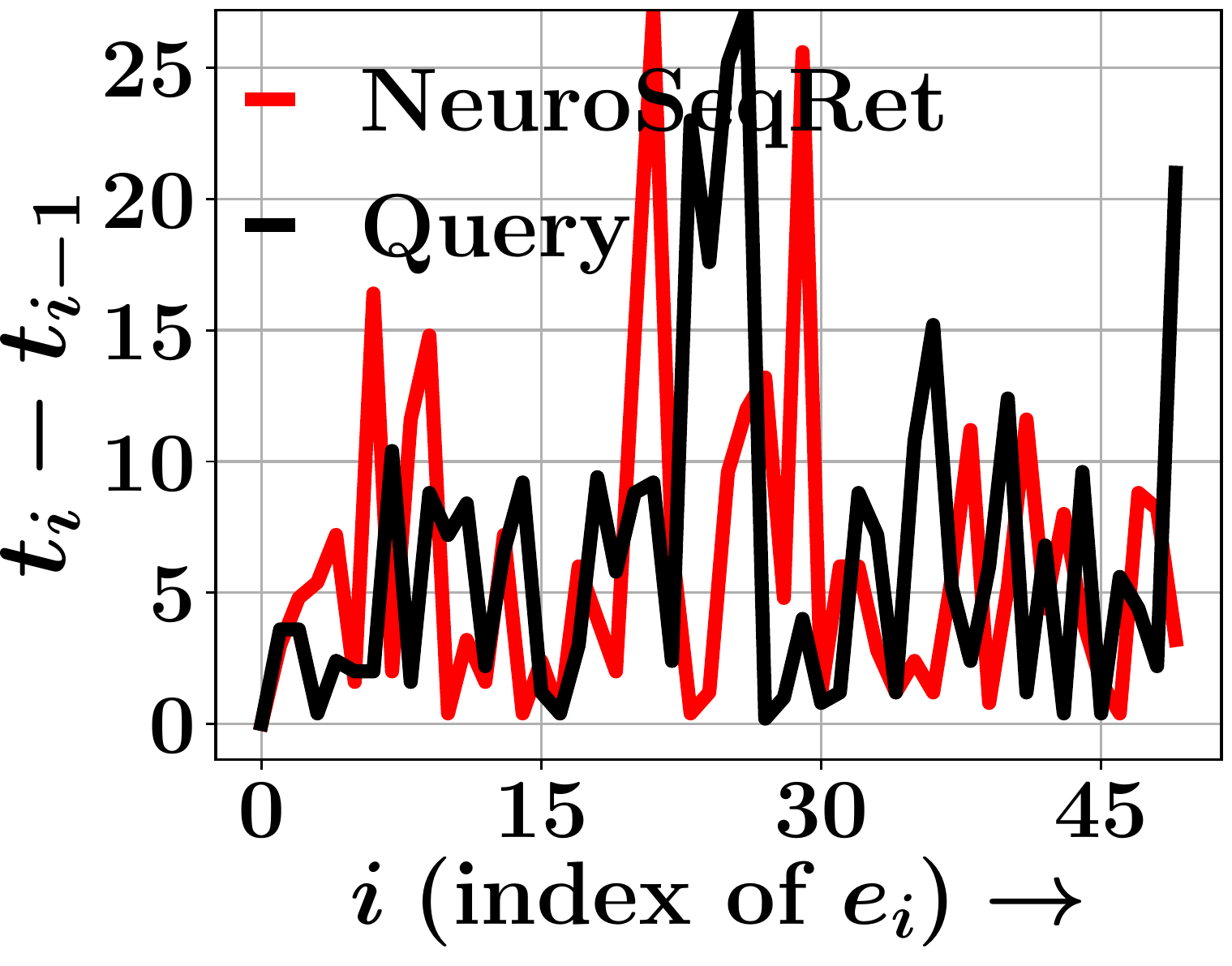}}
  \caption{Electricity}
\end{subfigure}

\begin{subfigure}{\columnwidth}
  \centering
{\includegraphics[height=3cm]{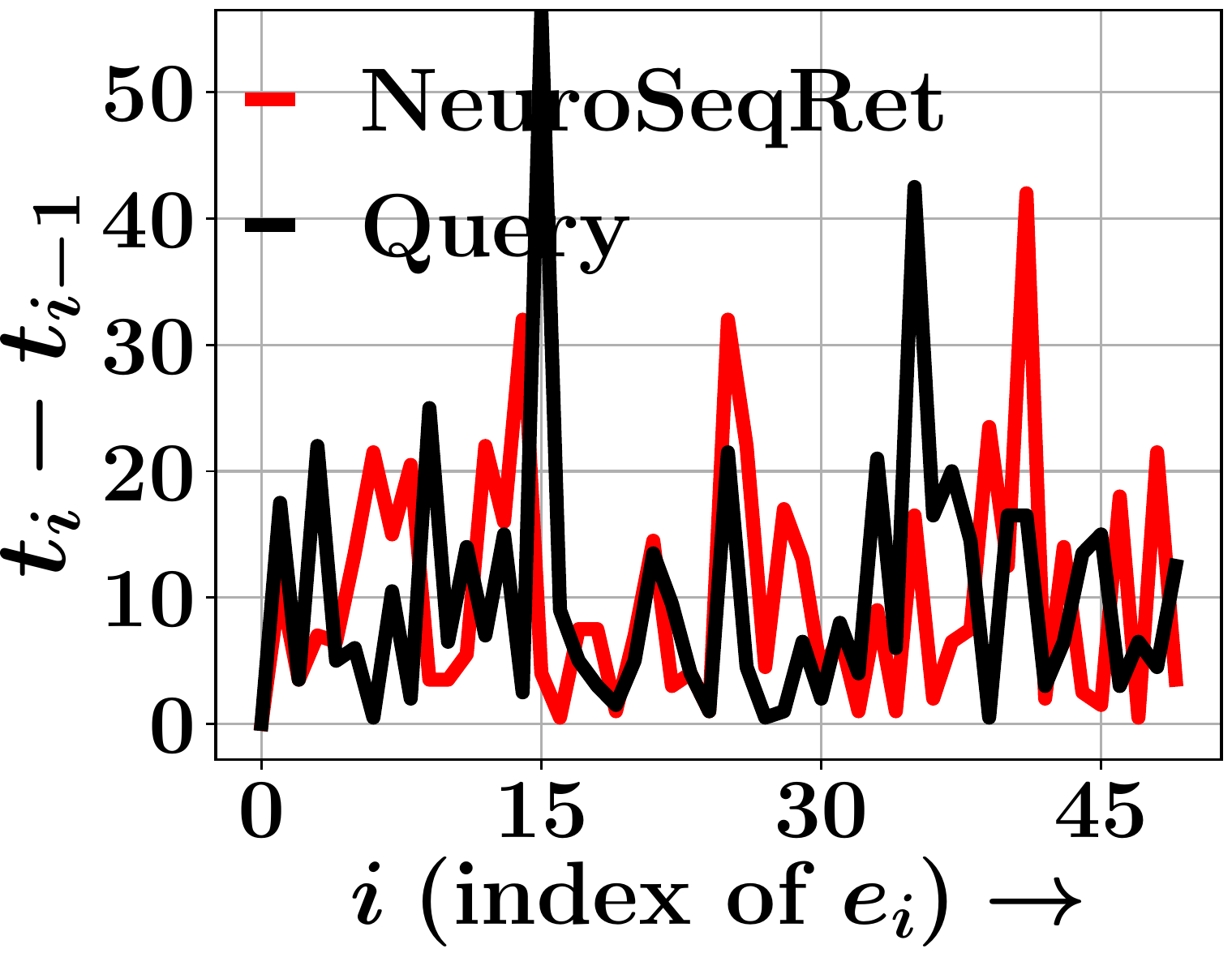}}
{\includegraphics[height=3cm]{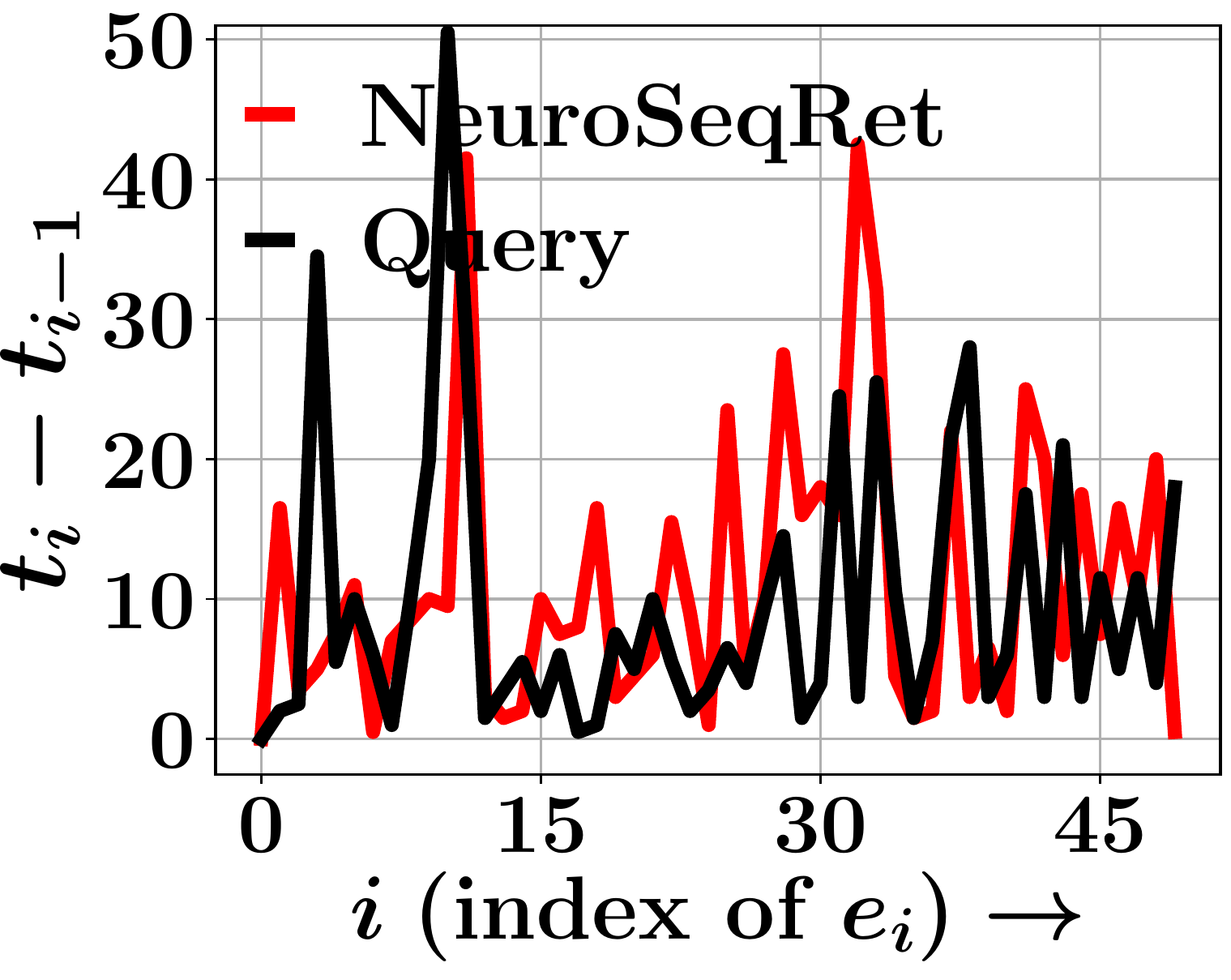}}
{\includegraphics[height=3cm]{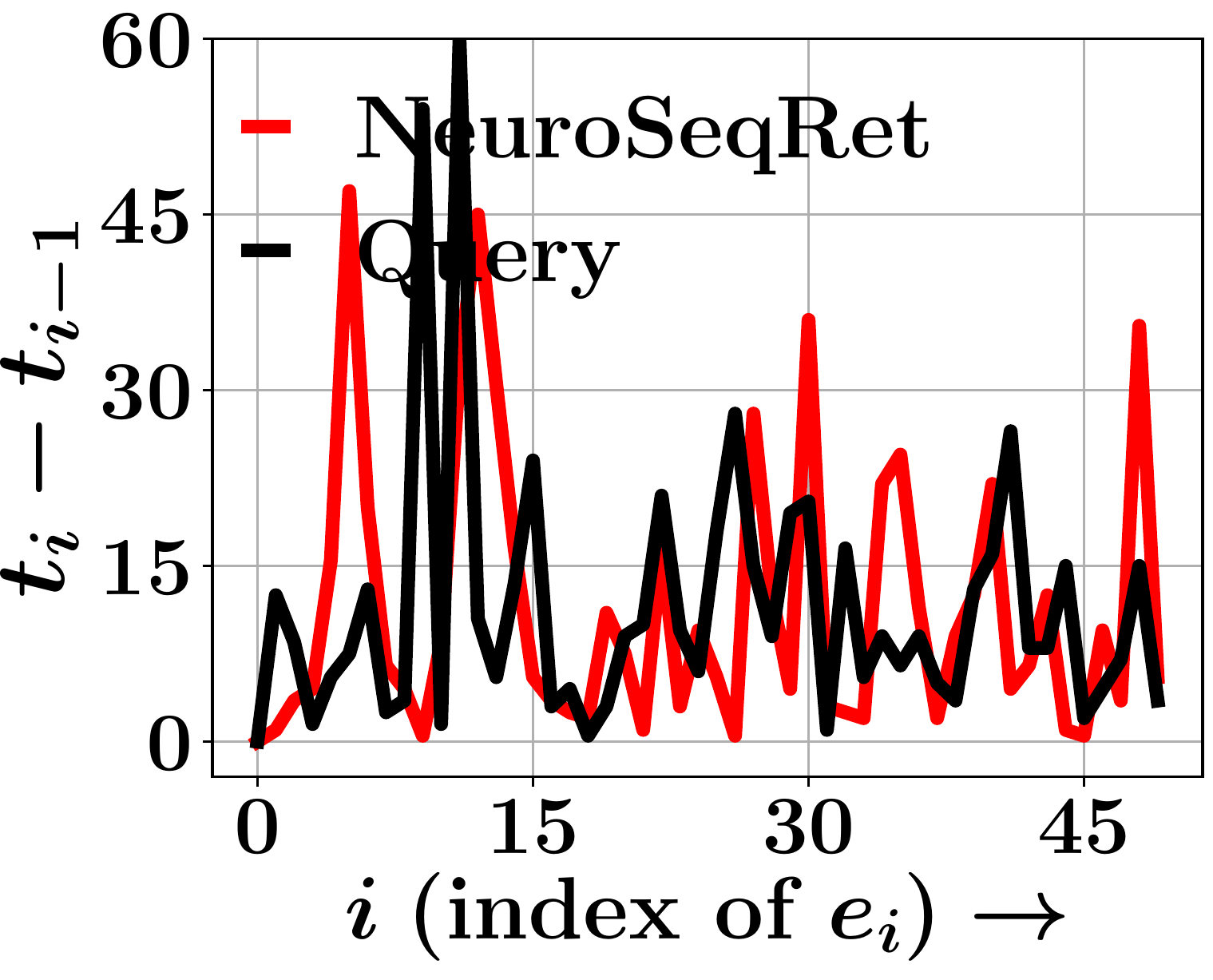}}
{\includegraphics[height=3cm]{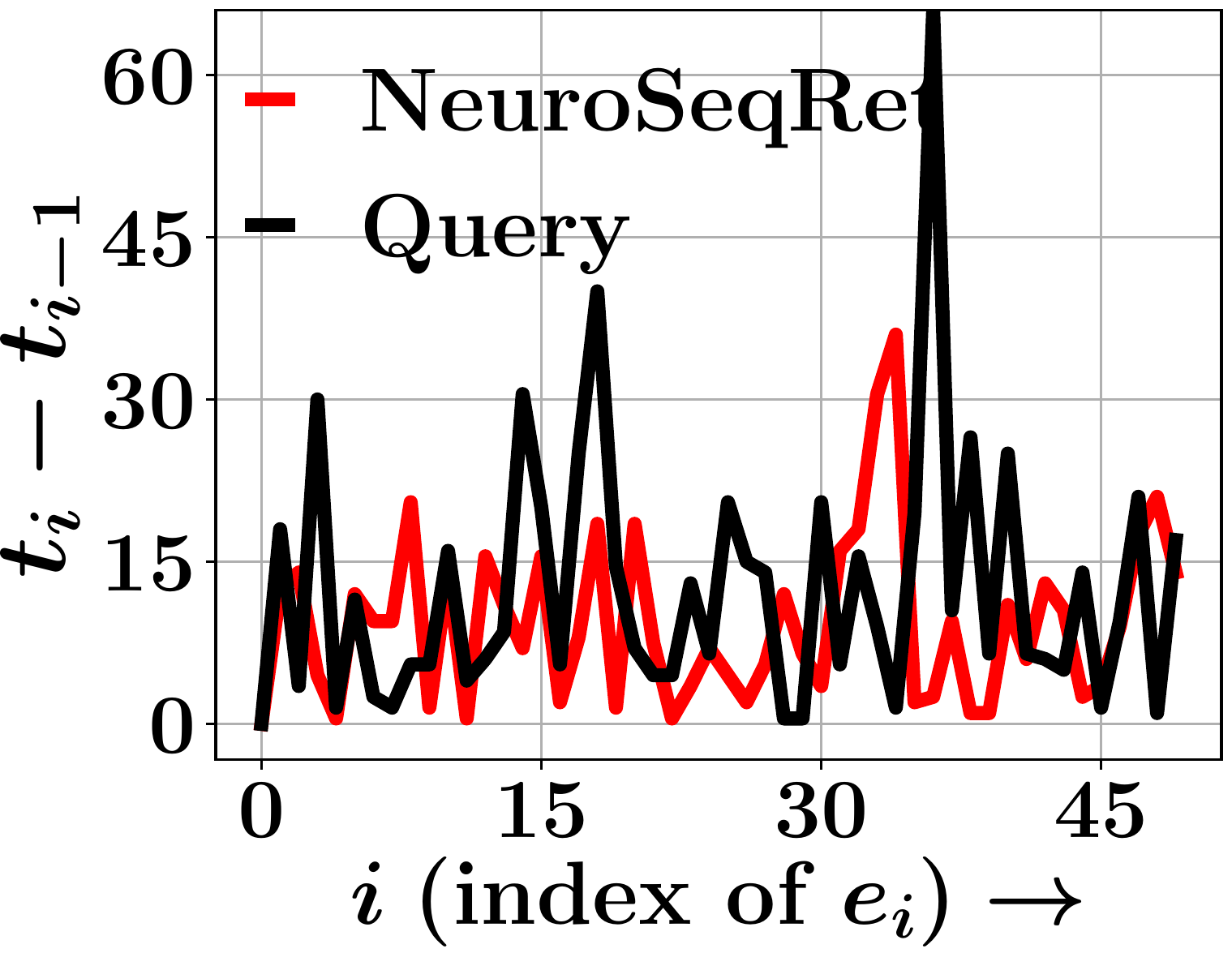}}
  \caption{Health}
\end{subfigure}

\begin{subfigure}{\columnwidth}
  \centering
 {\includegraphics[height=3cm]{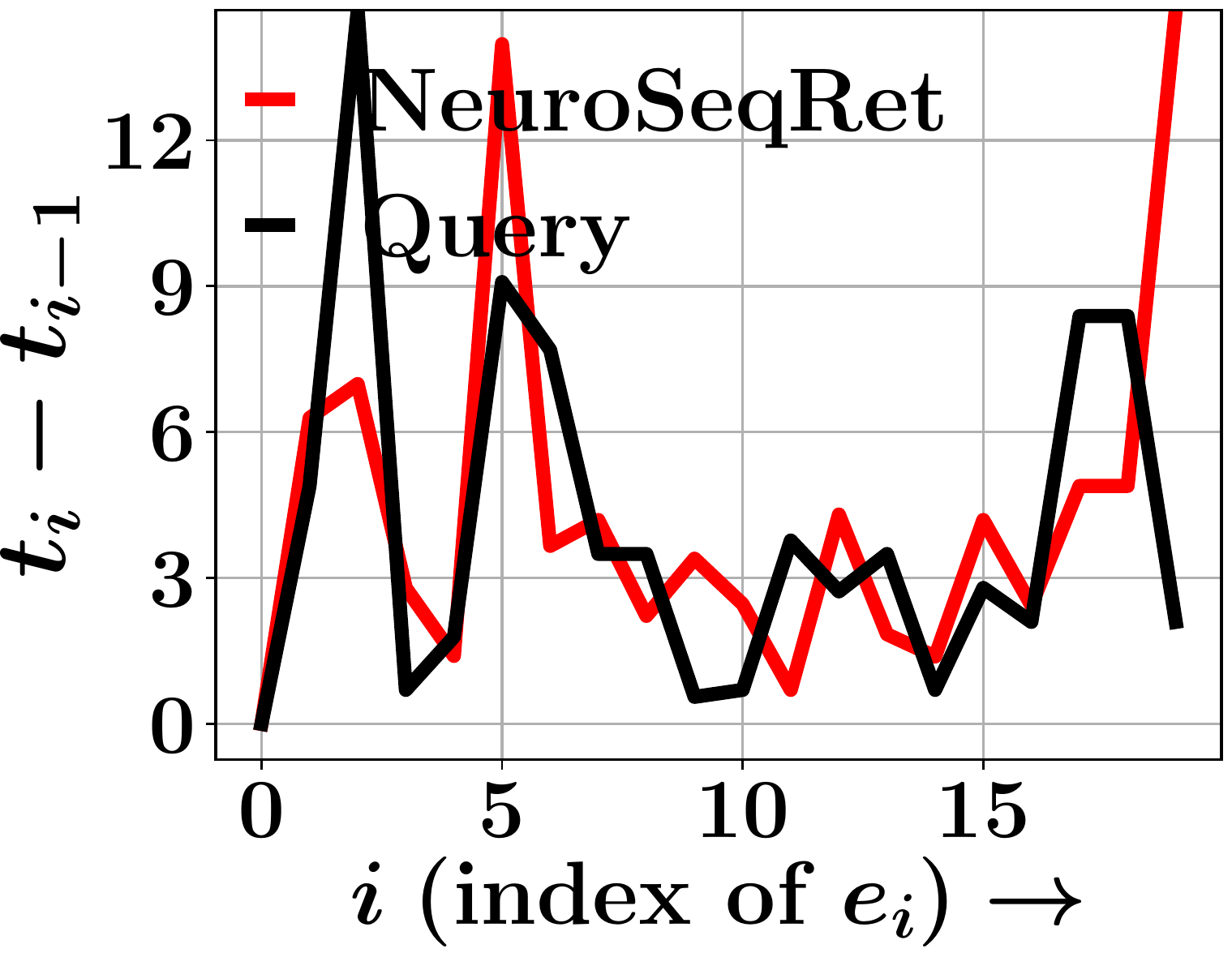}}
 {\includegraphics[height=3cm]{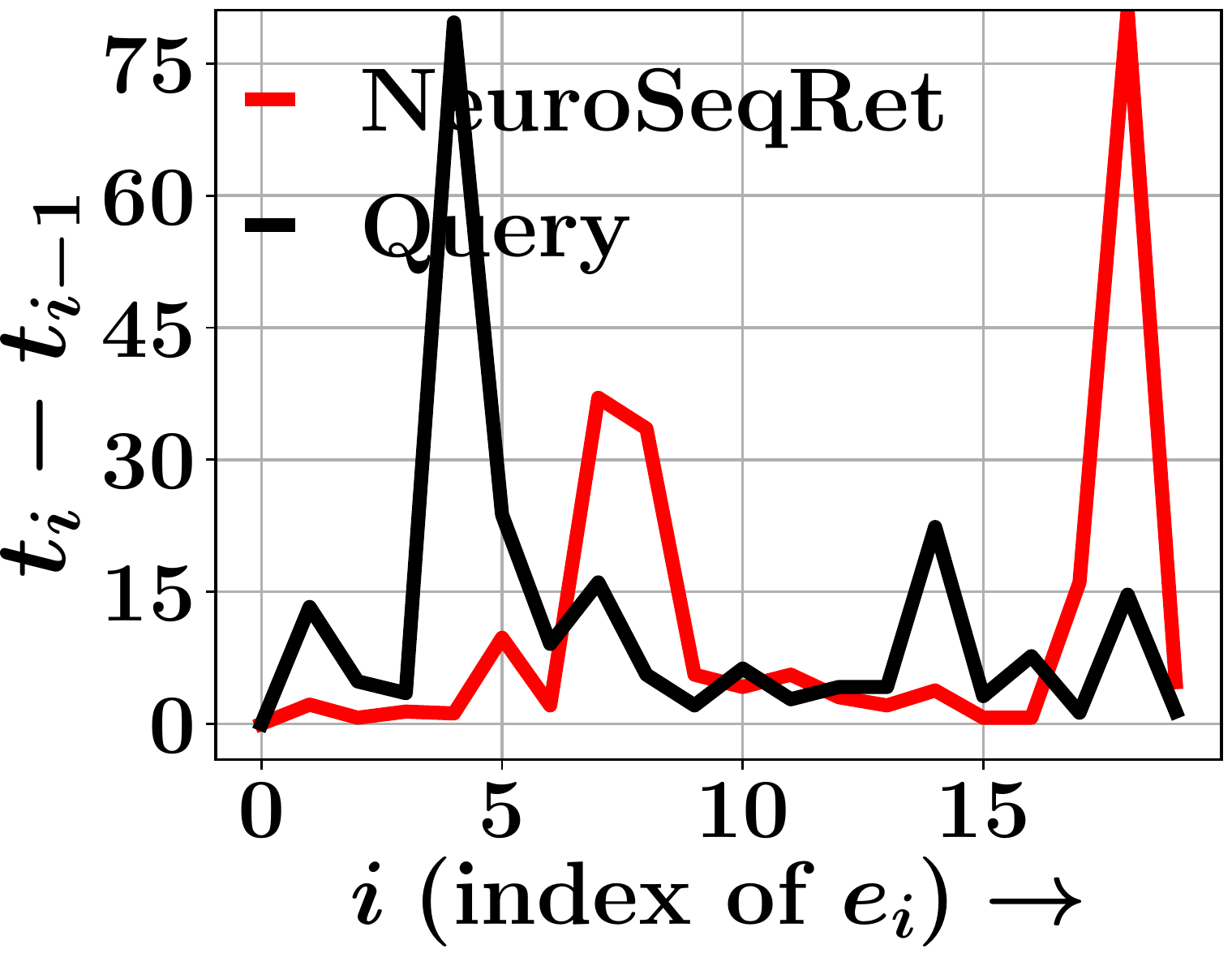}}
 {\includegraphics[height=3cm]{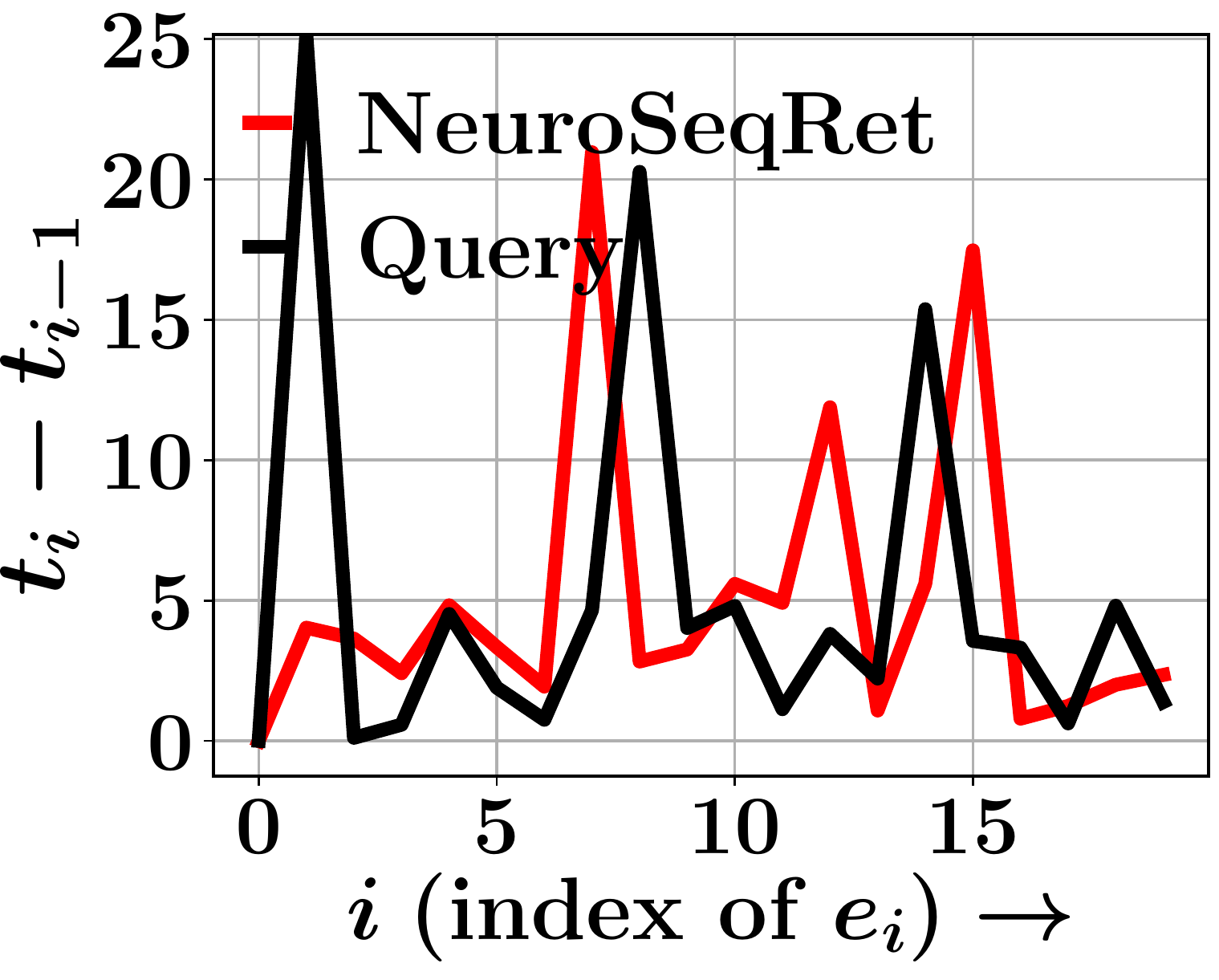}}
 {\includegraphics[height=3cm]{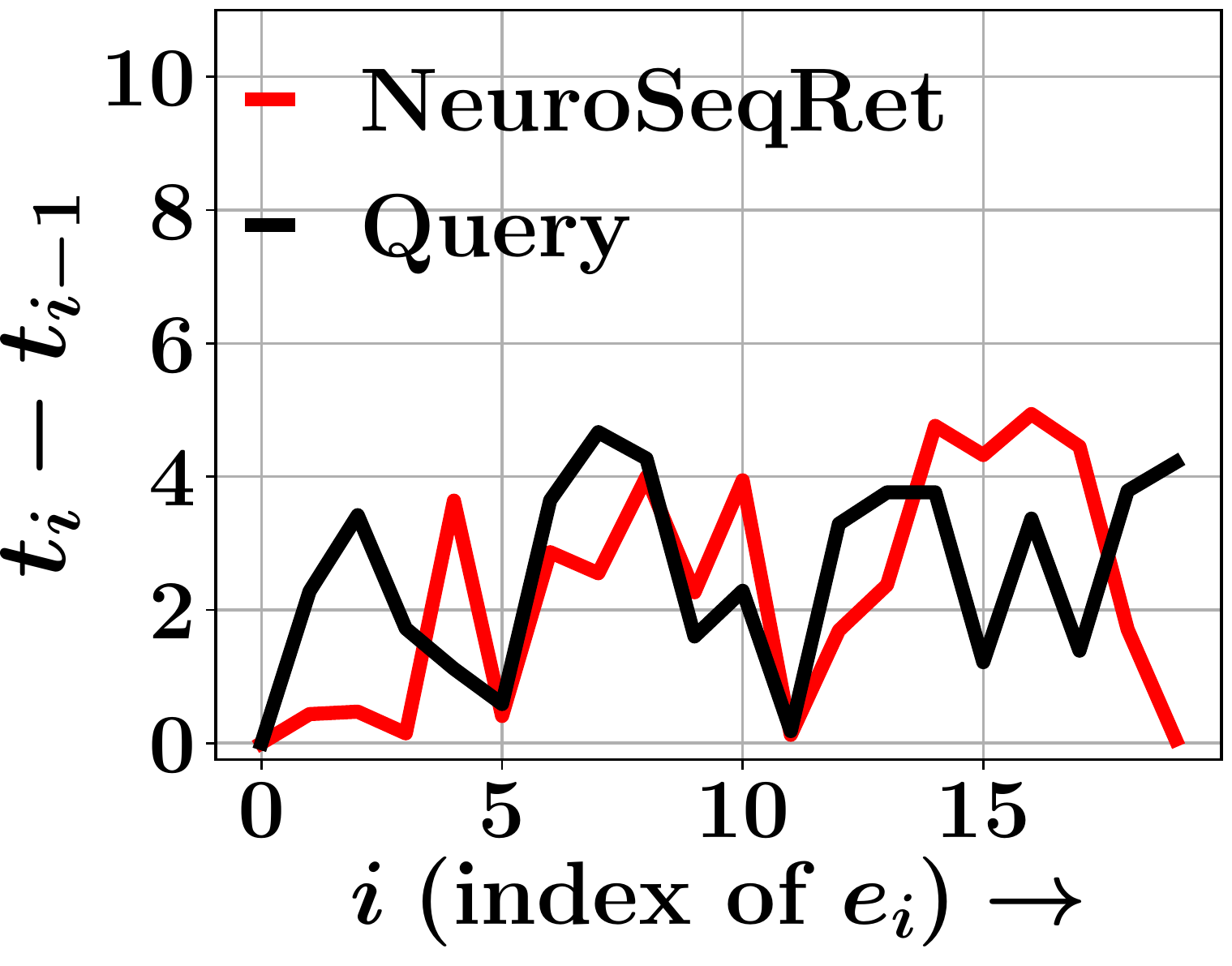}}
  \caption{Sports}
\end{subfigure}
\caption{Qualitative examples of inter-event times of events in a query sequence and the top-search results by \nsr for all datasets.}
\label{fig:qualitative}
\end{figure}

To get deeper insights into the working of our model, we perform a qualitative analysis between a query sequence from the dataset and the sequence retrieved by \nsr. More specifically, we aim to understand the similar patterns between query and corpus sequences that \nsr searches for in the corpus and plot the query sequence and the corresponding top-ranked relevant corpus sequence retrieved by \nsr. The results across all datasets in Figure~\ref{fig:qualitative} show that the inter-arrival times of the CTES retrieved by \nsr closely matches the query inter-arrival times.

\section{Conclusion}
In this chapter, we proposed a novel supervised continuous-time event sequence retrieval system called \nsr\ using neural MTPP models. To achieve efficient retrieval over a very large corpus of sequences, we also propose a trainable hash-coding of corpus sequences which can be used to narrow down the number of sequences to be considered for similarity score computation. Our experiments with real-world datasets from a diverse range of domains show that our retrieval model is more effective than several baselines. Our work opens several avenues for future work, including the design of generative models for relevance sequences and counterfactual explanations for relevance label predictions. 


%% file: chapters/009_proactive.tex
\newcommand{\bfast}{Breakfast}
\newcommand{\mult}{Multi-THUMOS}
\newcommand{\act}{Activity-Net}
\renewcommand{\ours}{\textsc{ProAct}\xspace}

\section{Introduction}
A majority of the data generated via human activities such as running, playing basketball, cooking, \etc, can be represented as a sequence of actions over a continuous time. These actions denote a step taken by a user towards achieving a certain goal and vary in their start and completion times, depending on the user and the surrounding environment~\cite{mehrasa2017learning,breakfast,avae}. Therefore, unlike synthetic time series, these continuous-time action sequences (CTAS) can vary significantly even if they consist of the same set of actions. For \eg, one person making omelets may take a longer time to cook eggs while another may prefer to cook for a short time\footnote{\scriptsize https://bit.ly/3F5aEwX (Accessed October 2022)}; or in a football game, Xavi may make a faster pass than Pirlo, even though the goals and the sequence of actions are the same. In addition, modeling the dynamics of CTAS becomes increasingly challenging due to the limited ability of the current neural frameworks, recurrent or self-attention-based, in capturing the continuous nature of action times~\cite{transformer,sasrec}. This situation is further exacerbated due to the large variance in action-\textit{times} and \textit{types}. Therefore, the problem of modeling a CTAS has been overlooked by the past literature.

In recent years, neural marked temporal point processes (MTPP) have shown significant promise in modeling a variety of continuous-time sequences in healthcare~\cite{rizoiu_sir,rizoiu_hip}, finance~\cite{sahp,bacry}, education~\cite{sahebi}, and social networks~\cite{retweet_data,nhp,thp}. However, standard MTPP have a limited modeling ability for CTAS as: (i) they assume a homogeneity among sequences, \ie, they cannot distinguish between two sequences of similar actions but with different time duration; (ii) in a CTAS, an action may finish before the start of the next action and thus, to model this empty time interval an MTPP must introduce a new action type, \eg, \textit{NULL} or \textit{end-action}, which may lead to an unwarranted increase in the types of actions to be modeled; and (iii) they cannot encapsulate the additional features associated with an action, for \eg, the minimum time for completion, necessary previous actions, or can be extended to sequence generation.

\subsection{Our Contribution}
In this chapter, we present \proactive (\textbf{P}oint P\textbf{ro}cess flows for \textbf{Activ}ity S\textbf{e}quences), a normalizing flow-based neural MTPP framework, designed specifically to model the dynamics of a CTAS~\cite{proactive}. Specifically, \proactive addresses three challenging problems -- (i) action prediction, (ii) goal detection, and (iii) the first-of-its-kind task of \textit{end-to-end} sequence generation. We learn the distribution of actions in a CTAS using temporal normalizing flows (NF)~\cite{shakir,ppflows} conditioned on the dynamics of the sequence as well as the action features (\eg, minimum completion time, \etc). Such a flow-based formulation provides \proactive flexibility over other similar frameworks~\cite{karishma,intfree} to better model the inter-action dynamics within and across the sequences. Moreover, our model is designed for \textit{early} detection of the sequence goal, \ie, identifying the result of a CTAS without traversing the complete sequence. We achieve this by using a time-bounded optimization procedure, \ie, by incrementally increasing the probability of identifying the true goal via a \textit{margin}-based and a weighted factor-based learning~\cite{margin,earlyijcv}. Such an optimization procedure allows \proactive to model the goal-action hierarchy within a sequence, \ie, the \textit{necessary} set of actions in a CTAS towards achieving a particular goal, and simultaneously, the order of actions in CTAS with similar goals.

To the best of our knowledge, in this paper, we present the first-ever application of MTPP via \textit{end-to-end} action sequence generation. Specifically, given the resultant goal, \proactive can generate a CTAS with the necessary set of actions and their occurrence times. Such a novel ability for MTPP models can reinforce their usage in applications related to bio-signals~\cite{haradal}, sensor-data~\cite{sensegen}, \etc, and overcome the modeling challenge due to scarcity of activity data~\cite{donahue2020user,luo2020database,axolotl}. Buoyed by the success of attention models in sequential applications~\cite{transformer}, we use a self-attention architecture in \proactive to model the inter-action influences in a CTAS. In summary, the key contributions we make in this paper via \proactive are:
\begin{compactitem}
\item We propose \proactive, a novel temporal flow-based MTPP framework designed specifically for modeling human activities with a time-bounded optimization framework for early detection of CTAS goal. 
\item Our normalizing flow-based modeling framework incorporates the sequence and individual action dynamics along with the action-goal hierarchy. Thus, \proactive introduces the first-of-its-kind MTPP application of end-to-end action CTAS generation with just the sequence-goal as input.
\item Finally, we empirically show that \proactive outperforms the state-of-the-art models for all three tasks --  action prediction, goal detection, and sequence generation.
\end{compactitem}

\section{Preliminaries}\label{proact_sec:basics}
In this section, we present a background on activity prediction and the problems addressed in this chapter.

\subsection{Activity Prediction}
Activity modeling in videos is a widely used application with recent approaches focusing on frame-based prediction.~\citet{conc1} predicts the future actions via hierarchical representations of short clips,~\citet{conc2} jointly predicts future activity ad the starting time by capturing different sequence features, and a similar procedure is adopted by~\cite{conc3} that predicts the action categories of a sequence of future activities as well as their starting and ending time.~\citet{margin} propose a method for early classification of a sequence of frames extracted from a video by maximizing the margin-based loss between the correct and the incorrect categories, however, it is limited to visual data and cannot incorporate the action times. This limits its ability for use in CTAS, and sequence generation. A recent approach~\cite{avae} proposed to model the dynamics of action sequences using a variational auto-encoder built on top of a temporal point process. We consider their work as most relevant to \proactive as it also addressed the problem of CTAS modeling. However, as shown in our experiments \proactive was able to easily outperform it across all metrics. This could be attributed to the limited modeling capacity of VAE over normalizing flows. Moreover, their sampling procedure could not be extended to sequence generation. Therefore, in contrast to the past literature, \proactive is the first application of MTPP models for CTAS modeling and end-to-end sequence generation.

\subsection{Problem Formulation}
As mentioned earlier, we represent an activity via a continuous-time action sequence, \ie, a series of actions undertaken by users and their corresponding time of occurrences. We derive each CTAS from annotated frames of videos consisting of individuals performing certain activities. Specifically, for every video, we have a sequence of activity labels being performed in the video along with timestamps for each activity. Therefore, each CTAS used in our experiments is derived from these sequences of a video. Formally, we provide a detailed description of a CTAS in Definition~\ref{proact_def: ctas}.
\begin{definition}[Continuous Time Action Sequence]
\label{proact_def: ctas}
\textit{We define a continuous-time action sequence (CTAS) as a series of action events taken by a user to achieve a particular goal. Specifically, we represent a CTAS as an MTPP $\cm{S}_k=\{e_i=(c_i, t_i) | i \in[k] , t_i<t_{i+1}\}$, where $t_i \in \mathbb{R}^+$ is the start time of the action, $c_i\in \cm{C}$ is the discrete category or mark of the $i$-th action, and $\cm{C}$ is the set of all categories. Each CTAS has an associated result, $g \in \cm{G}$, that signifies the goal of the CTAS. Here, $\cm{G}$ denotes the set of all possible sequence goals.}
\end{definition}

To highlight the relationship between sequence goal and actions, consider the example of a CTAS with the goal of \cat{making-coffee}, would comprise of actions -- \cat{take-a-cup}, \cat{pour-milk}, \cat{add-coffee-powder}, \cat{add-sugar}, and \cat{stir} -- at different time intervals. Given the aforementioned definitions, we formulate the tasks of next action, sequence goal prediction, and sequence generation as:

\xhdr{Input}
A CTAS of all actions, $\cm{S}_k$, consisting of categories and times of different actions that lead to a goal $g$.

\xhdr{Output}
A probabilistic prediction model with three distinct tasks -- (i) to estimate the likelihood of the next action $e_{k+1}$ along with the action category and occurrence time; (ii) to predict the goal of the CTAS being modeled, \ie, $\widehat{g}$; and (iii) a generative model to sample a sequence of actions, $\widehat{\cm{S}}$ given the true sequence goal, $g$.

\section{\proactive Model}\label{proact_sec:model}
In this section, we first present a high-level overview of the \proactive model and then describe the neural parameterization of each component in detail. Lastly, we provide a detailed description of its optimization and sequence generation procedure.

\subsection{High Level Overview}\label{proact_sec:overview}
We use an MTPP denoted by $p_{\theta}(\cdot)$, to learn the generative mechanism of a continuous-time action sequence. Moreover, we design the sequence modeling framework of $p_{\theta}(\cdot)$ using a self-attention-based encoder-decoder model~\cite{transformer}. Specifically, we embed the actions in a CTAS, \ie, $\cm{S}_k$, to a vector embedding, denoted by $\bs{s}_k$, using a weighted aggregation of all past actions. Therefore, $\bs{s}_k$ signifies a compact neural representation of the sequence history, \ie, all actions till the $k$-th index and their marks and occurrence times. Recent research~\cite{thp,sahp} has shown that an attention-based modeling choice can better capture the long-term dependencies as compared to RNN-based MTPP models~\cite{rmtpp,nhp,intfree,fullyneural}. A detailed description of the embedding procedure is given in Section~\ref{proact_sec:detail}.

We use our MTPP $p_{\theta}(\cdot)$ to estimate the generative model for the $(k+1)$-th action conditioned on the past, \ie, $p(e_{k+1})$ as:
\begin{equation}
p_{\theta}(e_{k+1} | \bs{s}_k) = \mathbb{P}_{\theta}(c_{k+1}|\bs{s}_k) \cdot \rho_{\theta}(\Delta_{t, k+1}|\bs{s}_k),
\end{equation}
where $\mathbb{P}_{\theta}(\cdot)$ and $\rho_{\theta}(\cdot)$ denote the probability distribution of marks and the density function for inter-action arrival times, respectively. Note that both the functions are conditioned on $\bs{s}_k$ and thus \proactive requires a joint optimizing procedure for both -- action time and mark prediction. Next, we describe the mechanism used in \proactive to predict the next action and goal detection in a CTAS.

\xhdr{Next Action Prediction}
We determine the most probable mark and time of the next action, using $p_{\theta} (\cdot)$ via standard sampling techniques over $\mathbb{P}_{\theta}(\cdot)$ and $\rho_{\theta}(\cdot)$ respectively~\cite{rmtpp,intfree}.
\begin{equation}
\widehat{e_{k+1}} \sim p_{\theta}(e_{k+1} | \bs{s}_k),
\end{equation}
In addition, to keep the history embedding up-to-date with all past actions, we iteratively update $\bs{s}_k$ to $\bs{s}_{k+1}$ by incorporating the details of action $e_{k+1}$.

\xhdr{Goal Detection}
Since the history embedding, $\bs{s}_k$, represents an aggregation of all past actions in a sequence, it can also be used to capture the influences between actions and thus, can be extended to detect the goal of the CTAS. Specifically, to detect the CTAS goal, we use a non-linear transformation over $\bs{s}_k$ as:
\begin{equation}
\widehat{g} \sim \mathbb{P}_{g' \in \cm{G}}(\Phi(s_k)),
\label{proact_eq:goal}
\end{equation}
where $\mathbb{P}_{\bullet}$ denotes the distribution over all sequence goals and $\Phi(\cdot)$ denotes the transformation via a fully-connected MLP layer.

\subsection{Neural Parameterization}\label{proact_sec:detail}
Here, we present a detailed description of the architecture of our MTPP, $p_{\theta}(\cdot)$, and the optimization procedure in \proactive. Specifically, we realize $p_{\theta}(\cdot)$ via a three-layer architecture:

\begin{compactitem}
\item[(1)] \textbf{Input Layer}
As mentioned in Section~\ref{proact_sec:basics}, each action $e_i \in \cm{S}_k$ is represented by a mark $c_i$ and time $t_i$. Therefore, we embed each action as a combination of all these features as:
\begin{equation}
\bs{y}_i = \bs{w}_{y, c} c_i + \bs{w}_{y, t} t_{i} + \bs{w}_{y, \Delta} \Delta_{t,i} + \bs{b}_y,
\end{equation}
where $\bs{w}_{\bullet, \bullet}, \bs{b}_{\bullet}$ are trainable parameters and $\bs{y}_i \in \mathbb{R}^D$ denotes the vector embedding for the action $e_i$ respectively. In other sections as well, we denote weight and bias as $\bs{w}_{\bullet, \bullet}$ and $\bs{b}_{\bullet, \bullet}$ respectively. 

\item[(2)] \textbf{Self-Attention Layer}
We use a \textit{masked} self-attention layer to embed the past actions to $\bs{s}_k$ and to interpret the influence between the past and the future actions. In detail, we follow the standard attention procedure~\cite{transformer} and first add a trainable positional encoding, $\bs{p}_i$, to the action embedding, \ie, $\bs{y}_i \leftarrow \bs{y}_i + \bs{p}_i$. Such trainable encodings are shown to be more scalable and robust for long sequence lengths as compared to those based on a fixed function~\cite{sasrec,tisasrec}. Later, to calculate an attentive aggregation of all actions in the past, we perform three independent linear transformations on the action representation to get the \textit{query}, \textit{key}, and \textit{value} embeddings, \ie, 
\begin{equation}
\bs{q}_i = \bs{W}^Q  \bs{y}_i, \quad \bs{k}_i = \bs{W}^K  \bs{y}_i, \quad \bs{v}_i = \bs{W}^V  \bs{y}_i,
\end{equation}
where $\bs{q}_{\bullet}, \bs{k}_{\bullet}, \bs{v}_{\bullet}$ denote the query, key, and value vectors, respectively. Following standard self-attention model, we represent $\bs{W}^{Q}$, $\bs{W}^{K}$ and $\bs{W}^{V}$ as trainable \textit{Query}, \textit{Key}, and \textit{Value} matrices respectively. Finally, we compute $\bs{s}_k$ conditioned on the history as:
\begin{equation}
\bs{s}_k =  \sum_{i=1}^{k} \frac{\exp\left(\bs{q}_k^{\top} \bs{k}_i /\sqrt{D} \right)}{\sum_{i'=1}^{k}\exp\left( \bs{q}_k^{\top} \bs{k}_{i'} /\sqrt{D} \right)} \bs{v}_i, \label{proact_eq:attn}
\end{equation}
where $D$ denotes the number of hidden dimensions. Here, we compute the attention weights via a \textit{softmax} over the interactions between the query and key embeddings of each action in the sequence and perform a weighted sum of the value embeddings.

Now, given the representation $\bs{s}_k$, we use the attention mechanism in Eqn.~\eqref{proact_eq:attn} and apply a feed-forward neural network to incorporate the necessary non-linearity to the model as:
\begin{equation*}
\bs{s}_k \leftarrow \sum_{i=1}^k \big[ \bs{w}_{s, m}\odot\textsc{ReLU}(\bs{s}_i \odot \bs{w}_{s, n} + \bs{b}_{s, n})  + \bs{b}_{s, m} \big],
\end{equation*}
where $\bs{w}_{s, m}, \bs{b}_{s, m}$ and $\bs{w}_{s, n}, \bs{b}_{s, n}$ are trainable parameters of the outer and inner layer of the point-wise feed-forward layer.  

To support faster convergence and training stability, we employ (i) layer normalization; (ii) stacking multiple self-attention blocks; and (iii) multi-head attention. Since these are standard techniques~\cite{ba2016layer,transformer}, we omit their mathematical descriptions.

\item[(3)] \textbf{Output Layer}
At every index $k$, \proactive outputs the next action and the most probable goal of the CTAS. We present the prediction procedure for each of them as follows:

\noindent \textit{\underline{Action Prediction:}}
We use the output of the self-attention layer, $\bs{s}_k$ to estimate the mark distribution and time density of the next event, \ie, $\mathbb{P}_{\theta}(e_{k+1})$ and $\rho_{\theta}(e_{k+1})$ respectively. Specifically, we model the $\mathbb{P}_{\theta}(\cdot)$ as a softmax over all other marks as:
\begin{equation}
\mathbb{P}_{\theta}(c_{k+1}) = \frac{\exp\left(\bs{w}_{c, s}^{\top} \bs{s}_i + \bs{b}_{c, s} \right)}{\sum_{c'=1}^{|\cm{C}|}\exp\left( \bs{w}_{c', s}^{\top} \bs{s}_i + \bs{b}_{c', s} \right)},
\label{proact_eqn: samplemark}
\end{equation}
where, $\bs{w}_{\bullet, \bullet}$ and $\bs{b}_{\bullet, \bullet}$ are trainable parameters.
\end{compactitem}

In contrast to standard MTPP approaches that rely on an intensity-based model~\cite{rmtpp,nhp,thp,sahp}, we capture the inter-action arrival times via a \textit{temporal} normalizing flow (NF). In detail, we use a \textit{lognormal} flow to model the temporal density $\rho_{\theta}(\Delta_{t, k+1})$. Moreover, standard flow-based approaches~\cite{intfree,ppflows} utilize a common NF for all events in a sequence, \ie, the arrival times of each event are determined from a single or mixture of flows trained on all sequences. We highlight that such an assumption restricts the ability to model the dynamics of a CTAS, as unlike standard events, an action has three distinguishable characteristics -- (i) every action requires a minimum time for completion; (ii) the time taken by a user to complete an action would be similar to the times of another user; and (iii) similar actions require similar times to complete. For example, the time taken to complete the action \cat{add-coffee} would require a certain minimum time of completion, and these times would be similar for all users. Intuitively, the time for completing the action \cat{add-coffee} would be similar to those for the action \cat{add-sugar}. 

To incorporate these features in \proactive, we identify actions with similar completion times and model them via independent temporal flows. Specifically, we cluster all actions $c_i \in \cm{C}$ into $\cm{M}$ non-overlapping clusters based on the \textit{mean} of their times of completion, and for each cluster, we define a trainable embedding $\bs{z}_r \in \mathbb{R}^{D} \, \forall r \in \{1, \cdots, \cm{M}\}$. Later, we sample the start-time of the future action by conditioning our temporal flows on the cluster of the current action as:
\begin{equation}
\widehat{\Delta_{t, k+1}} \sim \textsc{LogNormal}\left(\bs{\mu}_k , \bs{\sigma}^2_k \right),
\label{proact_eqn: sampletime}
\end{equation}
where, $[\bs{\mu}_k ,\bs{\sigma}^2_k]$, denote the mean and variance of the lognormal temporal flow and are calculated via the sequence embedding and the cluster embedding as:
\begin{equation}
\bs{\mu}_k = \sum_{r=1}^{\cm{M}} \cm{R}(e_k, r) \big(\bs{w}_{\mu} \left(\bs{s}_{k}\odot\bs{z}_{c, i} \right) + \bs{b}_{\mu}\big),
\end{equation}
\begin{equation}
\bs{\sigma}^2_k  = \sum_{r=1}^{\cm{M}} \cm{R}(e_k, r)  \big( \bs{w}_{\sigma} \left(\bs{s}_{k}\odot\bs{z}_{c, i} \right) + \bs{b}_{\sigma}\big),
\end{equation}
where $\bs{w}_{\bullet}, \bs{b}_{\bullet}$ are trainable parameters, $\cm{R}(e_k, r)$ is an indicator function that determines if event $e_k$ belongs to the cluster $r$ and $\bs{z}_r$ denotes the corresponding cluster embedding. Such a cluster-based formulation facilitates the ability of the model to assign similar completion times for events in the same cluster. To calculate the time of the next action, we add the sampled time difference to the time of the previous action $e_k$, \ie,
\begin{equation}
\widehat{t_{k+1}} = t_k + \widehat{\Delta_{t, k+1}}
\end{equation}
where, $\widehat{t_{k+1}}$ denotes the predicted time for the action $e_{k+1}$. 

\noindent \textit{\underline{Goal Detection:}}
In contrast to other MTPP approaches~\cite{rmtpp,nhp,thp,sahp,intfree}, an important feature of \proactive is identifying the goal of a sequence, \ie, a hierarchy on top of the actions in a sequence, based on the past sequence dynamics. To determine the goal of a CTAS, we utilize the history embedding $\bs{s}_k$ as it encodes the inter-action relationships of all actions in the past. Specifically, we use a non-linear transformation via a feed-forward network, denoted as $\Phi(\cdot)$ over $\bs{s}_k$ and apply a softmax over all possible goals.
\begin{equation}
\Phi(\bs{s}_k) = \textsc{ReLU} (\bs{w}_{\Phi, s} \bs{s}_k + \bs{b}_{\Phi, s}),
\end{equation}
where, $\bs{w}_{\bullet, \bullet}, \bs{b}_{\bullet, \bullet}$ are trainable parameters. We sample the most probable goal as in Eqn.~\eqref{proact_eq:goal}. 

We highlight that we predict the CTAS goal at each interval, though a CTAS has only one goal. This is to facilitate \textit{early} goal detection in comparison to detecting the goal after traversing the entire CTAS. More details are given in Section~\ref{proact_sec:early} and Section~\ref{proact_sec:optimization}.

\subsection{Early Goal Detection and Action Hierarchy}\label{proact_sec:early}
Here, we highlight the two salient features of \proactive -- early goal detection and modeling the goal-action hierarchy.

\xhdr{Early Goal Detection}
Early detection of sequence goals has many applications ranging from robotics to vision~\cite{earlyiccv,earlyijcv}. To facilitate early detection of the goal of a CTAS in \proactive, we devise a ranking loss that forces the model to predict a \textit{non-decreasing} detection score for the correct goal category. Specifically, the detection score of the correct goal at the $k$-th index of the sequence, denoted by $p_k(g| \bs{s}_k, \Phi)$, must be more than the scores assigned the correct goal in the past. Formally, we define the ranking loss as:
\begin{equation}
\cm{L}_{k, g} = \max \big(0, p^*_k(g) - p_k(g| \bs{s}_k, \Phi)\big),
\label{proact_eqn: margin}
\end{equation}
where $p^*_k(g)$ denotes the maximum probability score given to the correct goal in all past predictions.
\begin{equation}
p^*_k(g) = \max_{j \in \{1, k-1\}} p_j(g| \bs{s}_j, \Phi),
\label{proact_eqn: pastmax}
\end{equation}
where $p_k(g)$ denotes the probability score for the correct goal at index $k$. Intuitively, the ranking loss $\cm{L}_{k,g}$ would penalize the model for predicting a smaller detection score for the correct CTAS goal than any previous detection score for the same goal.

\xhdr{Action Hierarchy}
Standard MTPP approaches assume the category of marks as independent discrete variables, \ie, the probability of an upcoming mark is calculated independently~\cite{rmtpp,nhp,sahp,thp}. Such an assumption restricts the predictive ability while modeling CTAS, as in the latter case, there exists a hierarchy between goals and actions that lead to the specific goal. Specifically, actions that lead to a common goal may have similar dynamics and it is also essential to model the relationships between the actions of different CTAS with a common goal. We incorporate this hierarchy in \proactive along with our next action prediction via an action-based ranking loss. In detail, we devise a loss function similar to Eqn.~\ref{proact_eqn: margin} where we restrict the model to assign non-decreasing probabilities to all actions leading to the goal of CTAS under scrutiny. 
\begin{equation}
\cm{L}_{k, c} = \sum_{c' \in \cm{C}^*_g} \max \big(0, p^*_k(c') - p_k(c'| \bs{s}_k)\big),
\label{proact_eqn: cat_margin}
\end{equation}
where $\cm{C}^*_g, p_k(c'| \bs{s}_k)$ denote a set of all actions in CTAS with the goal $g$ and the probability score for the action $c' \in \cm{C}^*_g$ at index $k$ respectively. Here, $p^*_k(c')$ denotes the maximum probability score given to action $c'$ in all past predictions and is calculated similar to Eqn.~\ref{proact_eqn: pastmax}. We regard  $\cm{L}_{k, g}$ and $\cm{L}_{k, c}$ as \textit{margin} losses, as they aim to increase the difference between two prediction probabilities.

\subsection{Optimization}\label{proact_sec:optimization}
We optimize the trainable parameters in \proactive, \ie, the weight and bias tensors ($\bs{w}_{\bullet, \bullet}$ and $\bs{b}_{\bullet, \bullet}$) for our MTPP $p_{\theta} (\cdot)$, using a two channels of training consisting of action and goal prediction. Specifically, to optimize the ability of \proactive for predicting the next action, we maximize the joint likelihood for the next action and the lognormal density distribution of the temporal flows.
\begin{equation}
\mathscr{L} = \sum_{k = 1}^{|\cm{S}|} \log \big( \mathbb{P}_{\theta}(c_{k+1}|\bs{s}_k) \cdot \rho_{\theta} (\Delta_{t, k+1}| \bs{s}_k) \big ),
\label{proact_eqn:likelihood}
\end{equation}
where $\mathscr{L}$ denotes the joint likelihood, which we represent as the sum of the likelihoods for all CTAS. In addition to action prediction, we optimize the \proactive parameters for \textit{early} goal detection via a temporally weighted cross entropy (CE) loss over all sequence goals. Specifically, we follow a popular reinforcement recipe of using a time-varying \textit{discount} factor over the prediction loss as:
\begin{equation}
\cm{L}_g = \sum_{k = 1}^{|\cm{S}|} \gamma^k \cdot \cm{L}_{\textsc{CE}} \big(p_k(g| \bs{s}_k)\big),
\label{proact_eqn: discount}
\end{equation}
where $\gamma \in [0,1], \cm{L}_{\textsc{CE}} \big(p_k(g| \bs{s}_k)\big)$ denote the decaying factor and a standard softmax-cross-entropy loss respectively. Such a recipe is used exhaustively for faster convergence of reinforcement learning models~\cite{rl_book}. Here, the discount factor penalizes the model for taking longer times to detect the CTAS goal by decreasing the gradient updates to the loss.

\xhdr{Margin Loss}
In addition, we minimize the margin losses given in Section~\ref{proact_sec:early} with the current optimization procedure. Specifically, we minimize the following loss:
\begin{equation}
\cm{L}_m = \sum_{k=1}^{|\cm{S}|} \cm{L}_{k, g} + \cm{L}_{k, c},
\end{equation}
where $\cm{L}_{k, g}$ and $\cm{L}_{k, c}$ are margin losses defined in Eqn.~\ref{proact_eqn: margin} and Eqn.~\ref{proact_eqn: cat_margin} respectively. We learn the parameters of \proactive using an Adam~\cite{adam} optimizer for both likelihood and prediction losses.

\subsection{Sequence Generation}\label{proact_sec:generation}
An important ability of \proactive is the end-to-end generation of action sequences. Specifically, given the CTAS goal as input, we can generate a most probable sequence of actions that may lead to that specific goal. Such a feature has a range of applications from sports analytics~\cite{mehrasa2017learning}, forecasting~\cite{prathamesh}, identifying the duration of an activity~\cite{avae}, \etc

A standard approach for training a sequence generator is to sample future actions in a sequence and then compare them with the true actions~\cite{timegan}. However, such a procedure has multiple drawbacks as it is susceptible to noises during training and deteriorates the scalability of the model. Moreover, we highlight that such sampling-based training cannot be applied to a self-attention-based model as it requires a fixed-sized sequence as input~\cite{transformer}. Therefore, we resort to a two-step generation procedure that is defined below:
\begin{compactitem}
\item[1] \textbf{Pre-Training:} The first step requires training all \proactive parameters for action prediction and goal detection. This step is necessary to model the relationships between actions and goals and we represent the set of optimized parameters as $\theta^*$ and the corresponding MTPP as $p_{\theta^*}(\cdot)$ respectively.
\item[2] \textbf{Iterative Sampling:} We iteratively sample events and update parameters via our trained MTPP till the model predicts the correct goal for CTAS or we encounter an \texttt{<EOS>} action. Specifically, using $p_{\theta^*}(\cdot)$ and the first \textit{real} action ($e_1$) as input, we calculate the detection score for the correct goal, \ie, $p_1(g| \bs{s}_k)$ and while its value is highest among all probable goals, we sample the mark and time of next action using Eqn.~\ref{proact_eqn: samplemark} and Eqn.~\ref{proact_eqn: sampletime} respectively.
\end{compactitem}
Such a generation procedure harnesses the fast sampling of temporal normalizing flows and simultaneously is conditioned on the action and goal relationships. A detailed pseudo-code of the sequence generation procedure used in \proactive is given in Algorithm~\ref{proact_axoalgo}.

\begin{algorithm}[t!]
\textbf{Input:} $g$: Goal of CTAS \\ $e_1$: First Action \\ $p_{\theta^*}(\cdot)$: Trained MTPP\\
\textbf{Output:} $\widehat{S}$: Generated CTAS
$\cm{S}_1 \leftarrow e_1$ \\
$k = 1$\\ 
  \While {$k < \mathtt{max\_len}$}
  {
    Sample the mark of next action: $\widehat{c_{k+1}} \sim \mathbb{P}_{\theta^*}(\bs{s}_k)$\\
    Sample the time of next action: $\widehat{t_{k+1}} \sim \rho_{\theta^*}(\bs{s}_k)$\\
    Add to CTAS: $\cm{S}_{k+1} \leftarrow \cm{S}_{k} + e_{k+1}$\\
    Update the MTPP parameters $\bs{s}_{k+1} \leftarrow p(\bs{s}_{k}, e_{k+1})$\\
    Calculate most probable goal: $\widehat{g}_k = \max_{\forall g'} \big(p_k(g'| \bs{s}_k)\big)$\\
    \uIf{$\widehat{g_i} != g \, \mathrm{or} \, \widehat{c_{k+1}} == \mathtt{<EOS>}$}
  {
  Add EOS mark: $\widehat{\cm{S}} \leftarrow \cm{S}_{k+1} + \mathtt{<EOS>}$\\
  Exit the sampling procedure: $\textsc{Break}$\\
  }
  Increment iteration: $k \leftarrow k + 1$\\
  
  }\label{proact_endfor}
  Return generated CTAS: return $\widehat{\cm{S}}$\\
\caption{Sequence Generation with \proactive}\label{proact_axoalgo}
\end{algorithm}

\section{Experiments}\label{proact_sec:exp}
In this section, we present the experimental setup and the empirical results to validate the efficacy of \proactive. Through our experiments, we aim to answer the following research questions: 
\begin{compactitem}
\item[\textbf{RQ1}] What is the action-mark and time prediction performance of \proactive in comparison to the state-of-the-art baselines?
\item[\textbf{RQ2}] How accurately and quickly can \proactive identify the goal of an activity sequence?
\item[\textbf{RQ3}] How effectively can \proactive generate an action sequence?
\end{compactitem}

\subsection{Datasets}
To evaluate \proactive, we need timestamped action sequences and their goals. Therefore, we derive CTAS from three activity modeling datasets sourced from different real-world applications -- cooking, sports, and collective activity. The datasets vary significantly in terms of origin, sparsity, and sequence lengths. We highlight the details of each of these datasets below: 
\begin{compactitem}
\item \textbf{\bfast~\cite{breakfast}.} This dataset contains CTAS derived from 1712 videos of different people preparing breakfast. The actions in a CTAS and sequence goals can be classified into 48 and 9 classes, respectively. These actions are performed by 52 different individuals in 18 different kitchens.

\item \textbf{\mult~\cite{multithumos}.} A sports activity dataset that is designed for action recognition in videos. We derive the CTAS using 400 videos of individuals involved in different sports such as discus throw, baseball, \etc\ The actions and goals can be classified into 65 and 9 classes, respectively and on average, there are 10.5 action class labels per video.

\item \textbf{\act~\cite{activitynet}.} This dataset comprises activity categories collected from 591 YouTube videos with a total of 49 action labels and 14 goals.
\end{compactitem}
We highlight that in \act, many of the videos are shot by amateurs in many uncontrolled environments, the variances within the CTAS of the same goal are often large, and the lengths of CTAS vary and are often long and complex.

\subsection{Baselines}
We compare the action prediction performance of \proactive with the following state-of-the-art methods: 
\begin{compactitem}
\item \textbf{NHP~\cite{nhp}}: Models an MTPP using continuous-time LSTMs for capturing the temporal evolution of sequences.
\item \textbf{RMTPP~\cite{rmtpp}}: A recurrent neural network that models time differences to learn a representation of past events.
\item \textbf{AVAE~\cite{avae}}: A variational autoencoder-based MTPP framework designed specifically for activities in a sequence.
\item \textbf{SAHP~\cite{sahp}}: A self-attention model to learn the temporal dynamics using an aggregation of historical events. 
\item \textbf{THP~\cite{thp}}: Extends the transformer model~\cite{transformer} to include the \textit{conditional} intensity of event arrival and the inter-mark influences.
\end{compactitem}
We omit comparison with other continuous-time models~\cite{fullyneural,intfree,xiao2017wasserstein,xiaointaaai,hawkes} as they have already been outperformed by these approaches. 

\subsection{Evaluation Criteria}
Given the dataset $\cm{D}$ of $N$ action sequences, we split them into training and test sets based on the goal of the sequence. Specifically, for each goal $g \in \cm{G}$, we consider 80\%  of the sequences as the training set and the other last 20\% as the test set. We evaluate \proactive and all baselines on the test set in terms of (i) mean absolute error (MAE) of predicted times of action, and (ii) action prediction accuracy (APA) described as:
\begin{equation}
\mathrm{MAE} = \frac{1}{|\cm{S}|}\sum_{e_i\in \cm{S}}[|t_i-\widehat{t}_i|], \quad \mathrm{APA} = \frac{1}{|\cm{S}|}\sum_{e_i\in \cm{S}} \#(c_i=\widehat{c}_i),
\end{equation}
where, $\widehat{t_i}$ and $\widehat{c_i}$ are the predicted time and type the $i$-th action in test set. Moreover, we follow a similar protocol to evaluate the sequence generation ability of \proactive and other models. For goal prediction, we report the results in terms of accuracy (ratio) calculated across all sequences. We calculate confidence intervals across five independent runs.

\subsection{Experimental Setup}
\xhdr{System Configuration}
All our experiments were done on a server running Ubuntu 16.04. CPU: Intel(R) Xeon(R) Gold 5118 CPU @ 2.30GHz , RAM: 125GB and GPU: NVIDIA Tesla T4 16GB DDR6. 

\xhdr{Parameter Settings}
For our experiments, we set $D=16$, $\cm{M}=8$, $\gamma=0.9$ and weigh the margin loss $\cm{L}_m$ by 0.1. In addition, we set a $l_2$ regularizer over the parameters with coefficient value $0.001$.

\begin{table}[t!]
\small
\caption{Performance of all the methods in terms of action prediction accuracy (APA). Bold (underline) fonts indicate the best performer (baseline). Results marked \textsuperscript{$\dagger$} are statistically significant (i.e. two-sided Fisher's test with $p \le 0.1$) over the best baseline.}
\centering
\begin{tabular}{l|ccc}
\toprule
\textbf{Dataset} & \multicolumn{3}{c}{\textbf{Action Prediction Accuracy (APA)}} \\ \hline 
 & \bfast & \mult & \act \\ \hline \hline
NHP~\cite{nhp} & 0.528$\pm$0.024 & 0.272$\pm$0.019 & 0.684$\pm$0.034 \\
AVAE~\cite{avae} & 0.533$\pm$0.028 & 0.279$\pm$0.022 & 0.678$\pm$0.036 \\
RMTPP~\cite{rmtpp} & 0.542$\pm$0.022 & 0.274$\pm$0.017 & 0.683$\pm$0.034 \\
SAHP~\cite{sahp} & 0.547$\pm$0.031 & 0.287$\pm$0.023 & 0.688$\pm$0.042 \\
THP~\cite{thp} & \underline{0.559$\pm$0.028} & \underline{0.305$\pm$0.018} & \underline{0.693$\pm$0.038} \\
\ours-c & 0.561$\pm$0.027 & 0.297$\pm$0.020 & 0.698$\pm$0.038 \\
\ours-t & 0.579$\pm$0.025 & 0.306$\pm$0.018 & 0.722$\pm$0.035 \\
\proactive & \textbf{0.583$\pm$0.027}\textsuperscript{$\dagger$} & \textbf{0.316$\pm$0.019} & \textbf{0.728$\pm$0.037}\textsuperscript{$\dagger$} \\
\bottomrule
\end{tabular}
\vspace{3mm}
\label{pro_tab:apa}
\caption{Performance of all the methods in terms of mean absolute error (MAE).}
\begin{tabular}{l|ccc}
\toprule
\textbf{Dataset} & \multicolumn{3}{c}{\textbf{Mean Absolute Error (MAE)}} \\ \hline 
 & \bfast & \mult & \act \\ \hline \hline
NHP~\cite{nhp} & 0.411$\pm$0.019 & \underline{0.017$\pm$0.002} & 0.796$\pm$0.045 \\
AVAE~\cite{avae} & 0.417$\pm$0.021 & 0.018$\pm$0.002 & 0.803$\pm$0.049 \\
RMTPP~\cite{rmtpp} & \underline{0.403$\pm$0.018} & \underline{0.017$\pm$0.002} & \underline{0.791$\pm$0.046} \\
SAHP~\cite{sahp} & 0.425$\pm$0.031 & 0.019$\pm$0.003 & 0.820$\pm$0.072 \\
THP~\cite{thp} & 0.413$\pm$0.023 & 0.019$\pm$0.002 & 0.806$\pm$0.061 \\ \hline
\ours-c & 0.415$\pm$0.027 & 0.015$\pm$0.002 & 0.774$\pm$0.054 \\
\ours-t & 0.407$\pm$0.025 & 0.015$\pm$0.002 & 0.783$\pm$0.058 \\
\proactive & \textbf{0.364$\pm$0.028}\textsuperscript{$\dagger$} & \textbf{0.013$\pm$0.002}\textsuperscript{$\dagger$} & \textbf{0.742$\pm$0.059}\textsuperscript{$\dagger$} \\
\bottomrule
\end{tabular}
\label{pro_tab:mae}
\end{table}

\begin{figure}[b]
\centering
\hfill
\begin{subfigure}[b]{0.45\columnwidth}
\includegraphics[height=4cm]{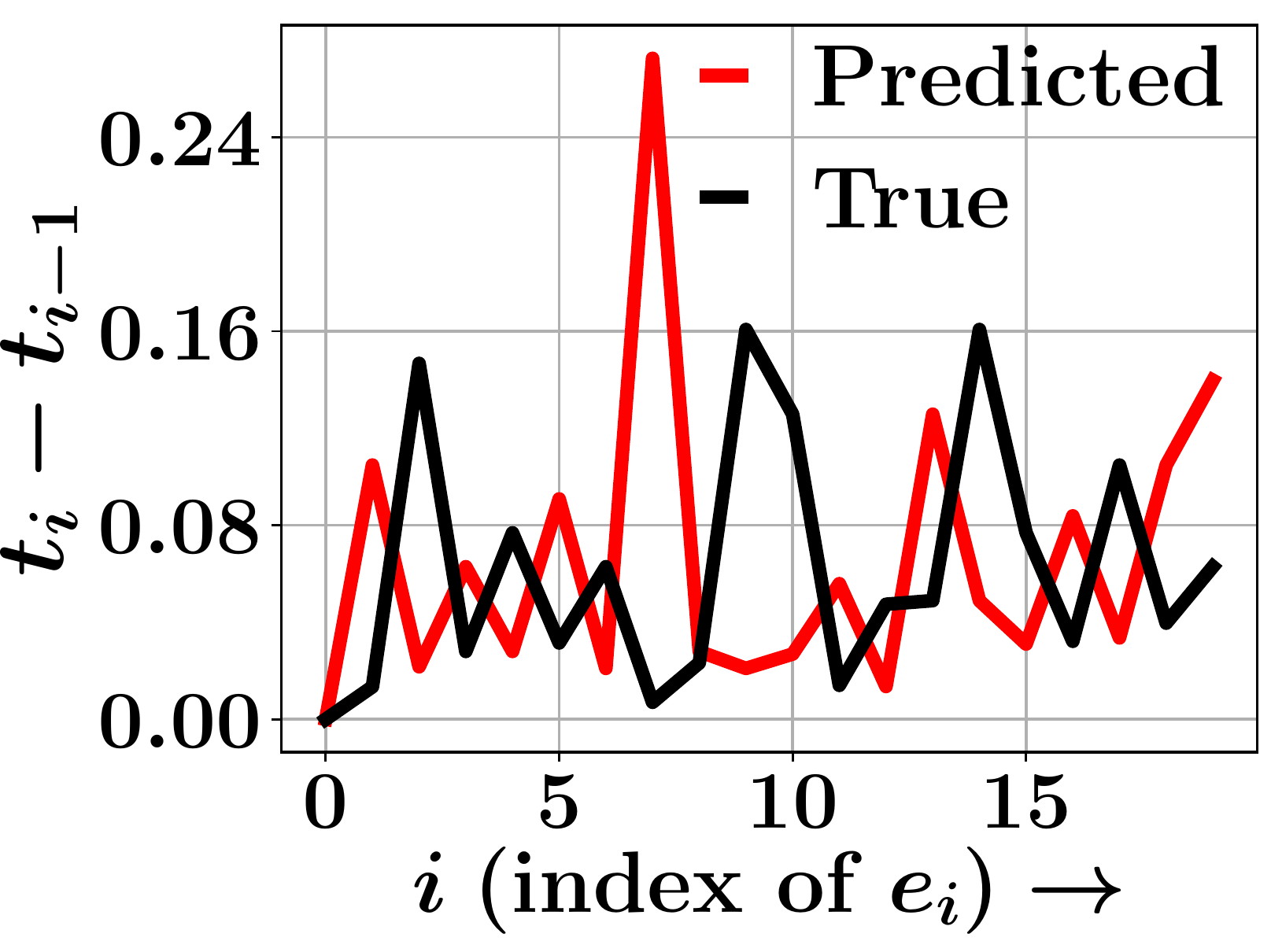}
\caption{\mult}
\end{subfigure}
\hfill
\begin{subfigure}[b]{0.45\columnwidth}
\includegraphics[height=4cm]{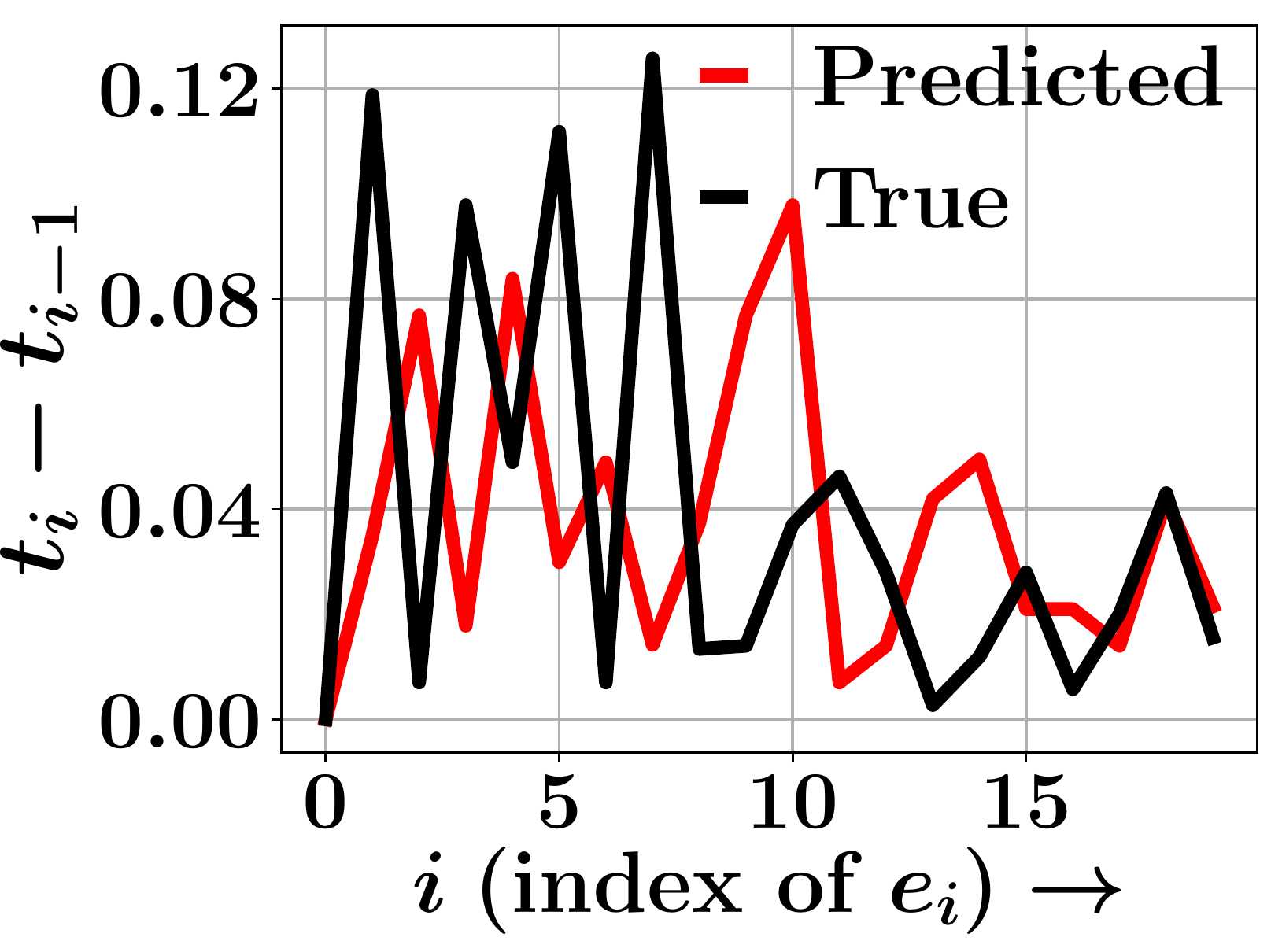}
\caption{\act}
\end{subfigure}
\hfill
\vspace{-0.3cm}
\caption{\label{proact_fig:qualitative} Real life \textit{true} and \textit{predicted} inter-arrival times $\Delta_{t,k}$ of different events $e_k$ for (a) \mult\ and (b) \act\ datasets. The results show that the true arrival times match with the times predicted by \proactive.}
\end{figure}

\begin{figure*}[t]
\centering
\begin{subfigure}[b]{0.6\linewidth}
{\includegraphics[height=4cm]{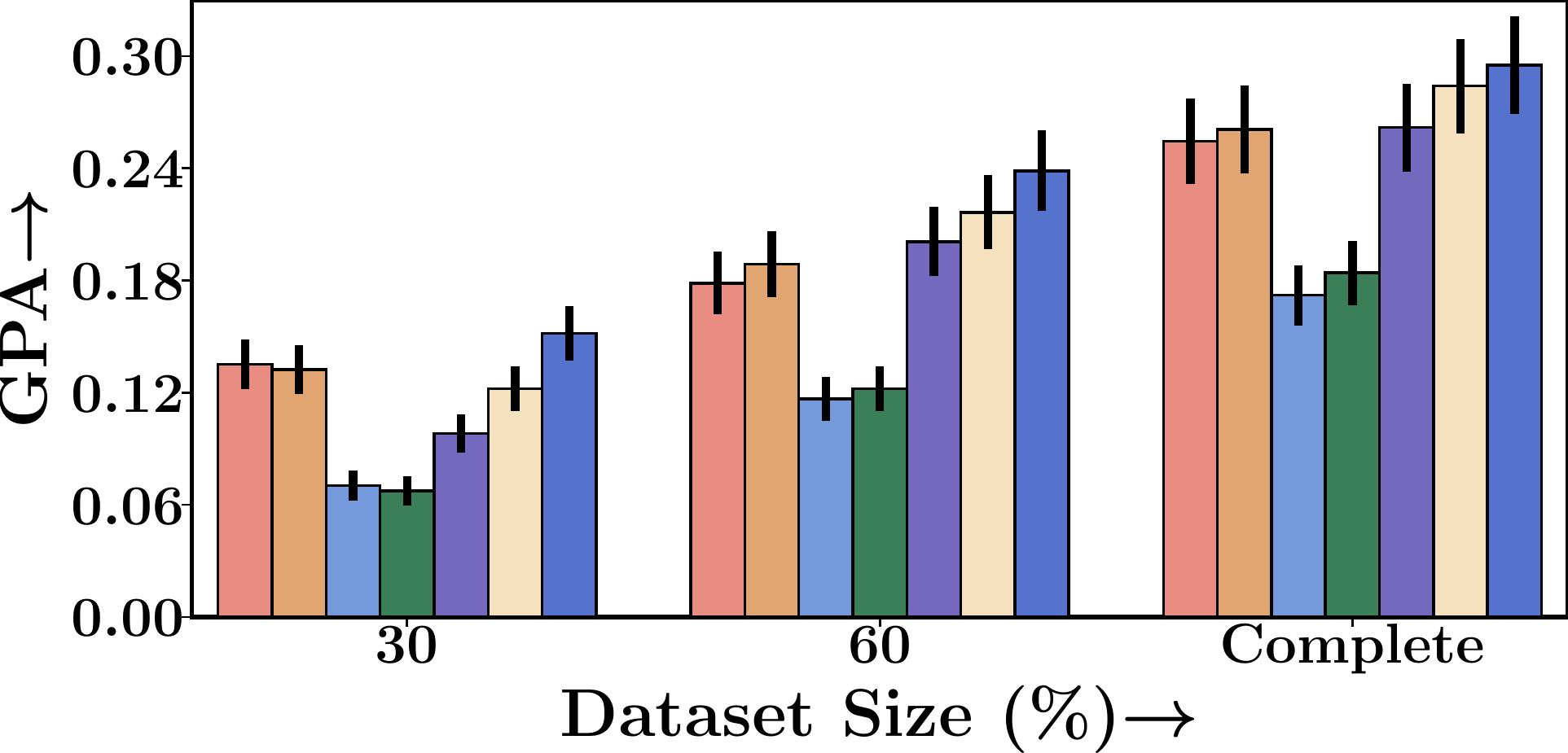}}
\caption{\bfast}
\end{subfigure}
\begin{subfigure}[b]{0.6\linewidth}
{\includegraphics[height=4cm]{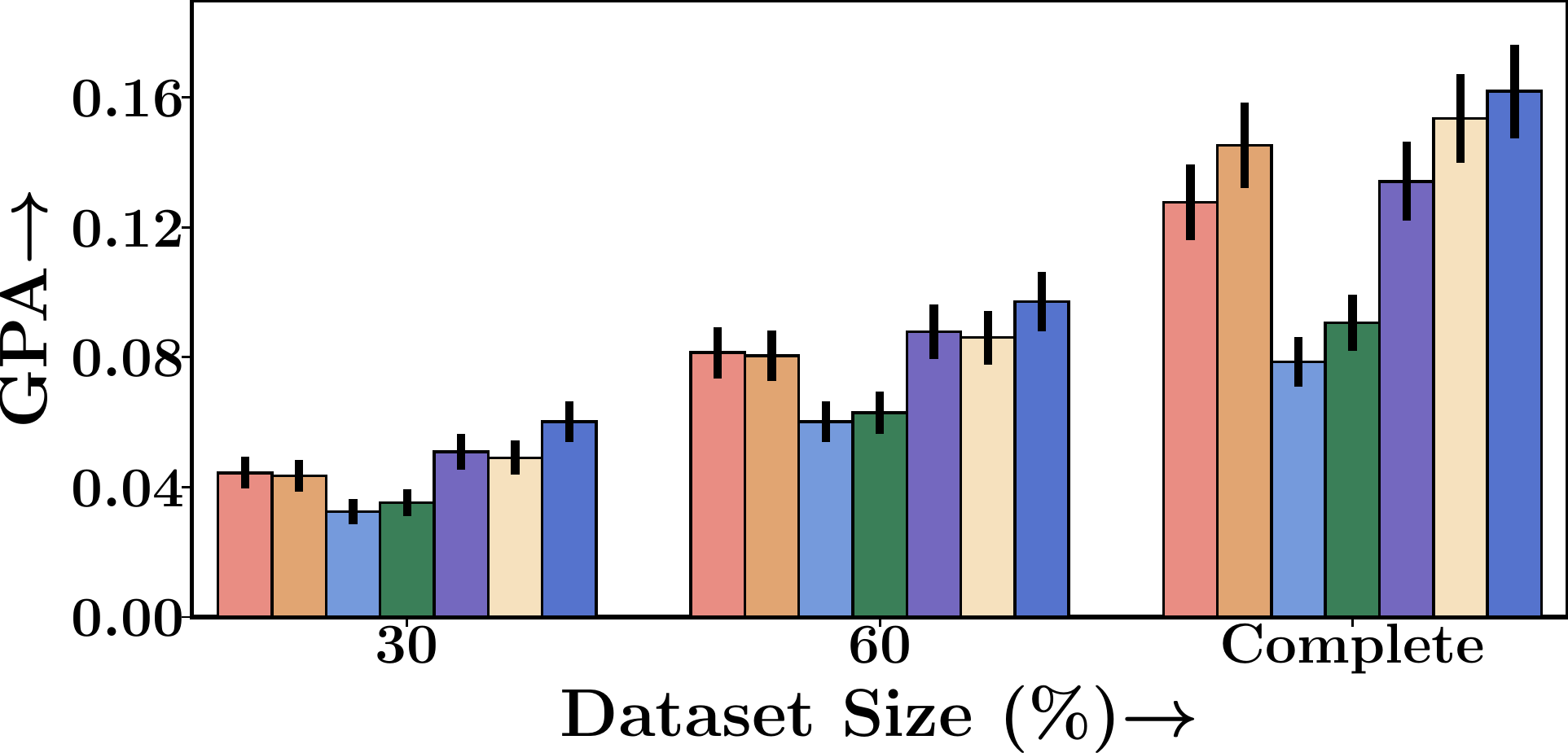}}
\caption{\mult}
\end{subfigure}
\begin{subfigure}[b]{0.6\linewidth}
{\includegraphics[height=4cm]{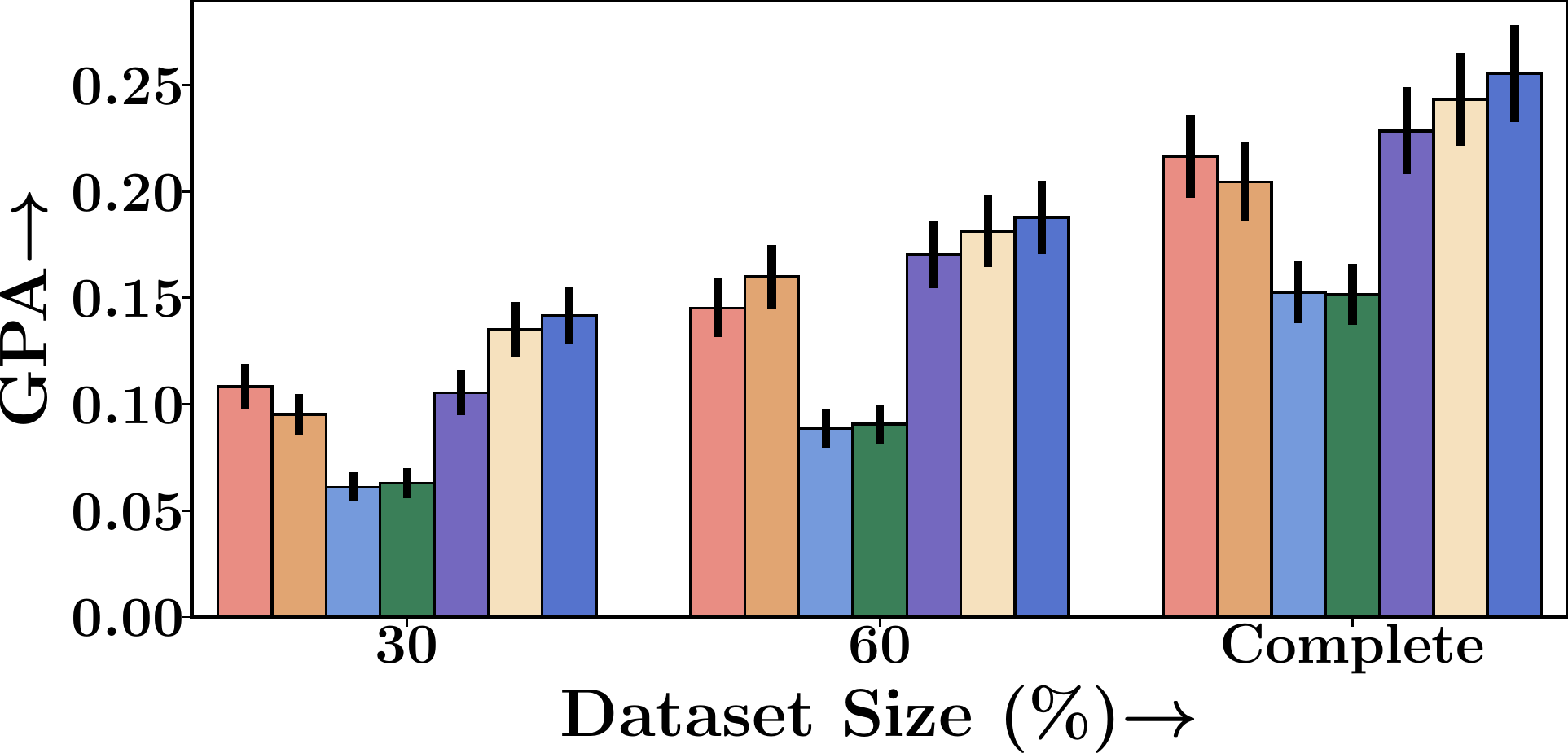}}
\caption{\act}
\end{subfigure}
{\includegraphics[height=0.45cm]{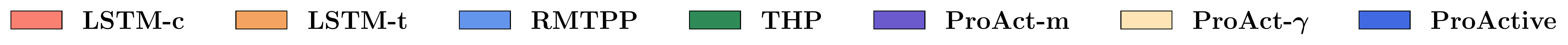}}
\vspace{-0.3cm}
\caption{Sequence goal prediction performance of \proactive, its variants -- \ours-m and \ours-$\gamma$, and other baseline models. The results show that \proactive can effectively detect the CTAS goal even with smaller test sequences as input.}
\label{proact_fig:UU}
\end{figure*}

\subsection{Action Prediction Performance}
To address RQ1, we report on the performance of action prediction of different methods in terms of APA and MAE in Table~\ref{pro_tab:apa} and Table~\ref{pro_tab:mae}, respectively. In addition, we include two variants of \proactive -- (i) \ours-c, which represents our model without the goal-action hierarchy loss and cluster-based flows, and (ii) \ours-t, which represents our model without cluster-based flows. From the results, we note the following:
\begin{compactitem}
\item \proactive consistently yields the best prediction performance on all the datasets. In particular, it improves over the strongest baselines by 8-27\% for time prediction and by 2-7\% for action prediction. These results signify the drawbacks of using standard sequence approaches for modeling a temporal action sequence. 

\item RMTPP~\cite{rmtpp} is the second-best performer in terms of MAE of time prediction in almost all the datasets. We also note that for \act dataset, THP~\cite{thp} outperforms RMTPP for action category prediction. However, \proactive still significantly outperforms these models across all metrics.

\item Neural MTPP methods that deploy a self-attention for modeling the distribution of action -- namely THP, SAHP, and \proactive, achieve better performance in terms of category prediction.

\item Despite AVAE~\cite{avae} being a sequence model designed specifically for activity sequences, other neural methods that incorporate complex structures using self-attention or normalizing flows easily outperform it.
\end{compactitem}
\noindent To sum up, our empirical analysis suggests that \proactive can better model the underlying dynamics of a CTAS as compared to all other baseline models. Moreover, the performance gain over \ours-c and \ours-t highlights the need for modeling action hierarchy and cluster-based flows. 

\xhdr{Qualitative Assessment}
We also perform a qualitative analysis to highlight the ability of \proactive to model the inter-arrival times for action prediction. Specifically, we plot the actual inter-action time differences and the time-difference predicted by \proactive in Figure~\ref{proact_fig:qualitative} for \mult\ and \act\ datasets. From the results, we note that the predicted inter-arrival times closely match with the true inter-arrival times, and \proactive is even able to capture large time differences (peaks).

\subsection{Goal Prediction Performance}
To address RQ2, we evaluate the goal detection performance of \proactive along with other baselines. To highlight the \textit{early} goal detection ability of our model,  we report the results across different variants of the test set, \ie, with the initial 30\% and 60\% of the actions in the CTAS in terms of goal prediction accuracy (GPA). In addition, we introduce two novel baselines, LSTM-c, and LSTM-t, that detect the CTAS goal using just the types and the times of actions, respectively. We also compare with the two best-performing MTPP baselines -- RMTPP and THP which we extend for the task of goal detection by a $k$-means clustering algorithm. In detail, we obtain the sequence embedding, say $\bs{s}_k$ using the MTPP models and then cluster them into $|\cm{G}|$ clusters based on their cosine similarities and perform a maximum \textit{polling} across each cluster, \ie, predict the most common goal for each cluster as the goal for all CTAS in the same cluster. In addition, we introduce two new variants of our model to analyze the benefits of early goal detection procedures in \proactive -- (i) \proactive-m, which represents our model without the goal-based margin loss given in Eqn.~\ref{proact_eqn: margin} and (ii) \proactive-$\gamma$, is our model without the discount-factor weight in Eqn.~\ref{proact_eqn: discount}. We also report the results for the complete model \proactive.

The results for goal detection in Figure \ref{proact_fig:UU}, show that the complete design of \proactive achieves the best performance among all other models. We also note that the performance of MTPP-based models deteriorates significantly for this new task which shows the unilateral nature of the prediction prowess of MTPP models, unlike \proactive. Interestingly, the variant of \proactive without the margin loss \proactive-m performs poorly as compared to the one without the discount factor, \proactive-$\gamma$. This could be attributed to better convergence guarantees with a margin-based loss over the latter. Finally, we observe that standard LSTM models are easily outperformed by our model, thus reinforcing the need for joint training of types and action times.

\begin{figure}[t]
\centering
\hfill
\begin{subfigure}[b]{0.30\columnwidth}
\includegraphics[height=3cm]{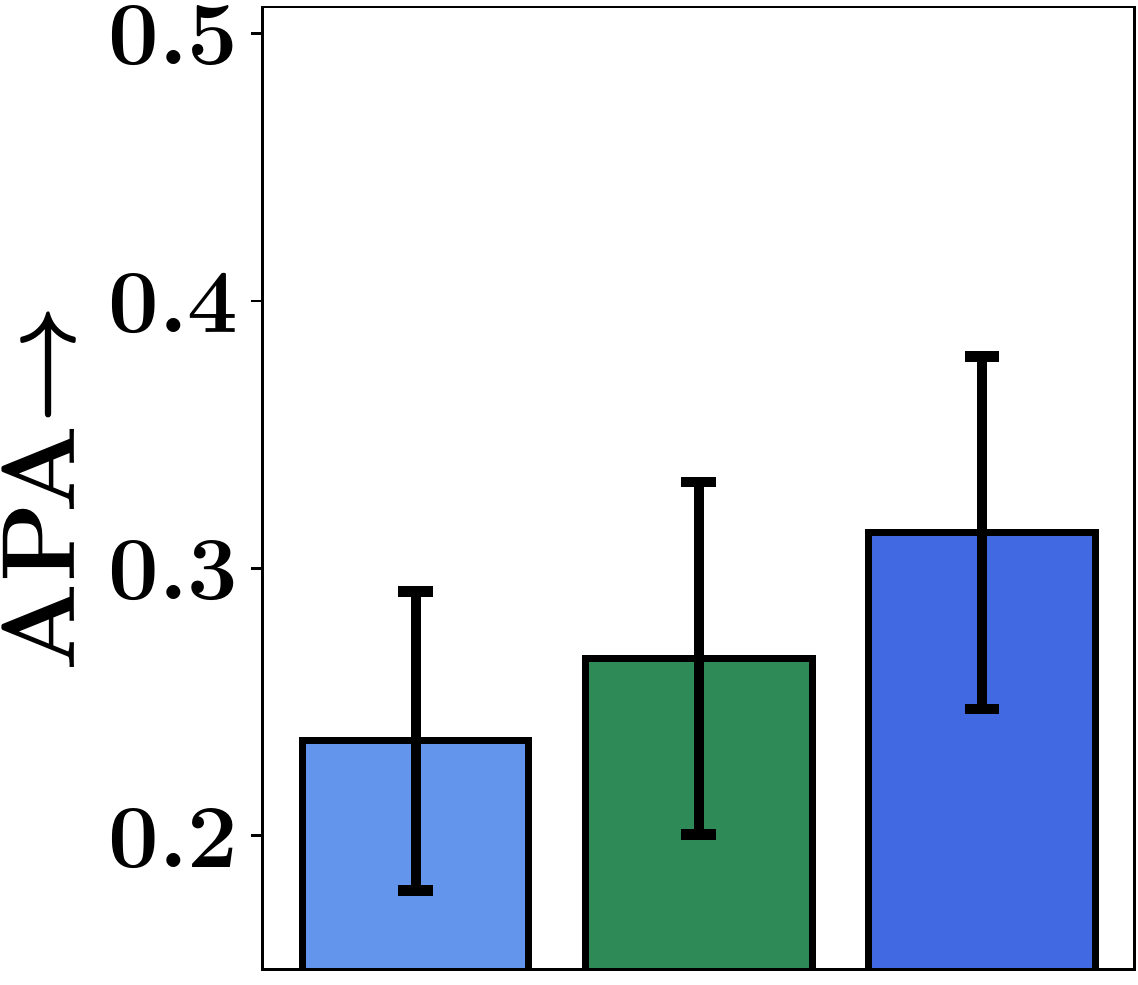}
\caption{\bfast}
\end{subfigure}
\hfill
\begin{subfigure}[b]{0.30\columnwidth}
\includegraphics[height=3cm]{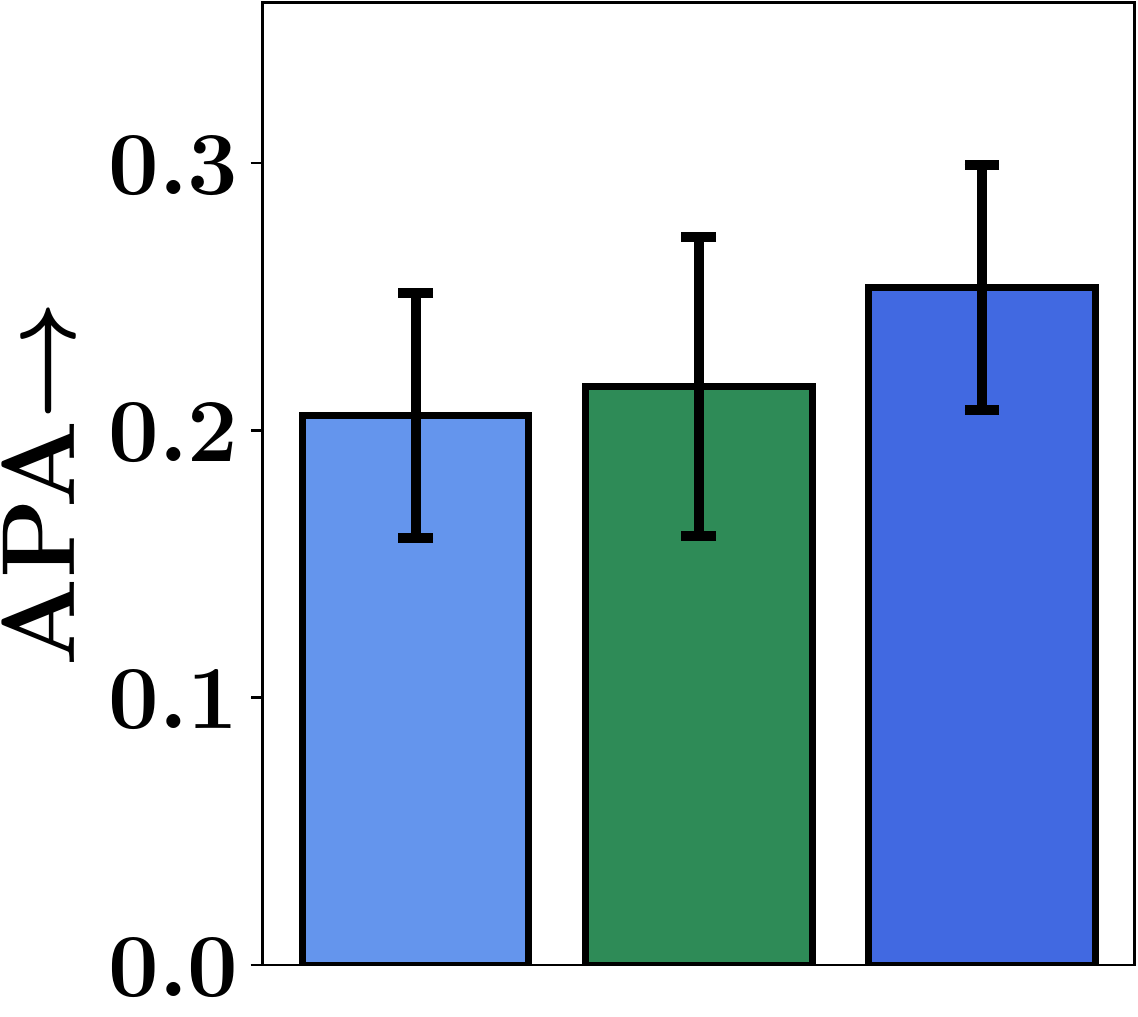}
\caption{\mult}
\end{subfigure}
\hfill
\begin{subfigure}[b]{0.30\columnwidth}
\includegraphics[height=3cm]{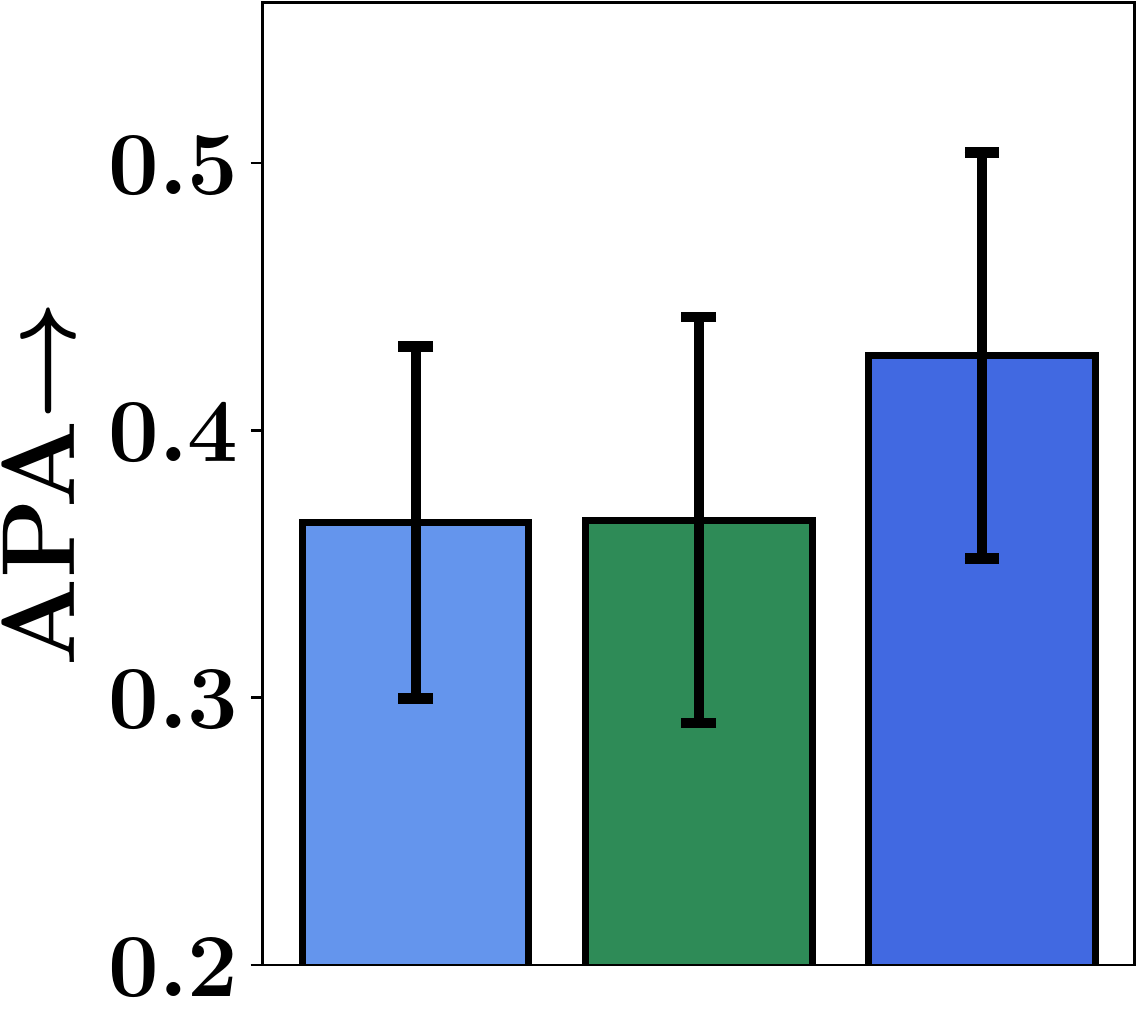}
\caption{\act}
\end{subfigure}
\hfill
{\includegraphics[height=0.5cm]{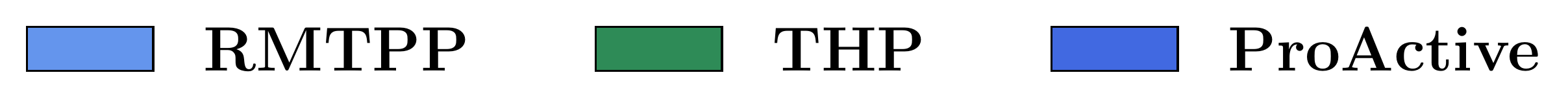}}
\vspace{-0.3cm}
\caption{\label{proact_fig:mpa} Sequence Generation results for \proactive and other baselines in terms of APA for action prediction. \vspace{0.3cm}}
\end{figure}

\begin{figure}[t]
\centering
\hfill
\begin{subfigure}[b]{0.30\columnwidth}
\includegraphics[height=3cm]{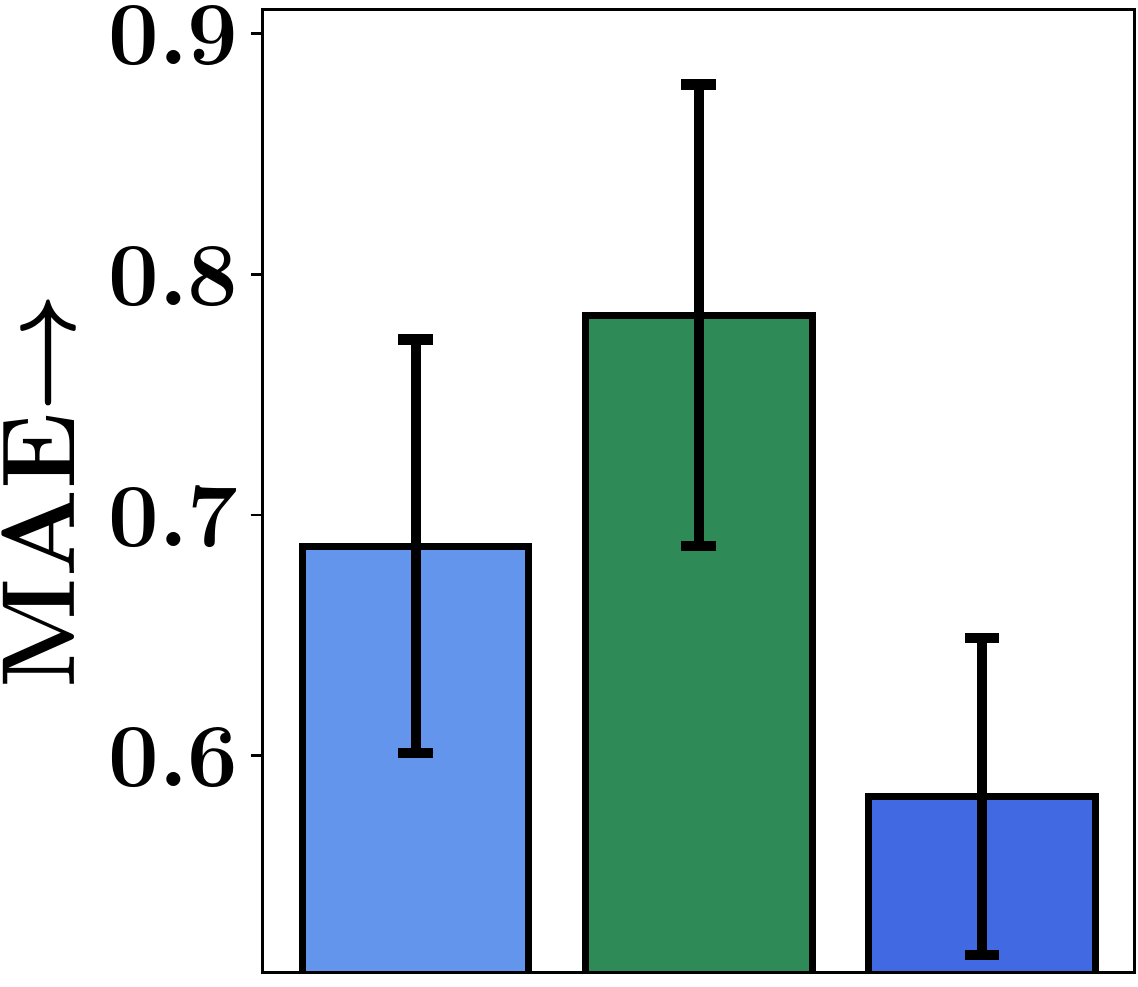}
\caption{\bfast}
\end{subfigure}
\hfill
\begin{subfigure}[b]{0.30\columnwidth}
\includegraphics[height=3cm]{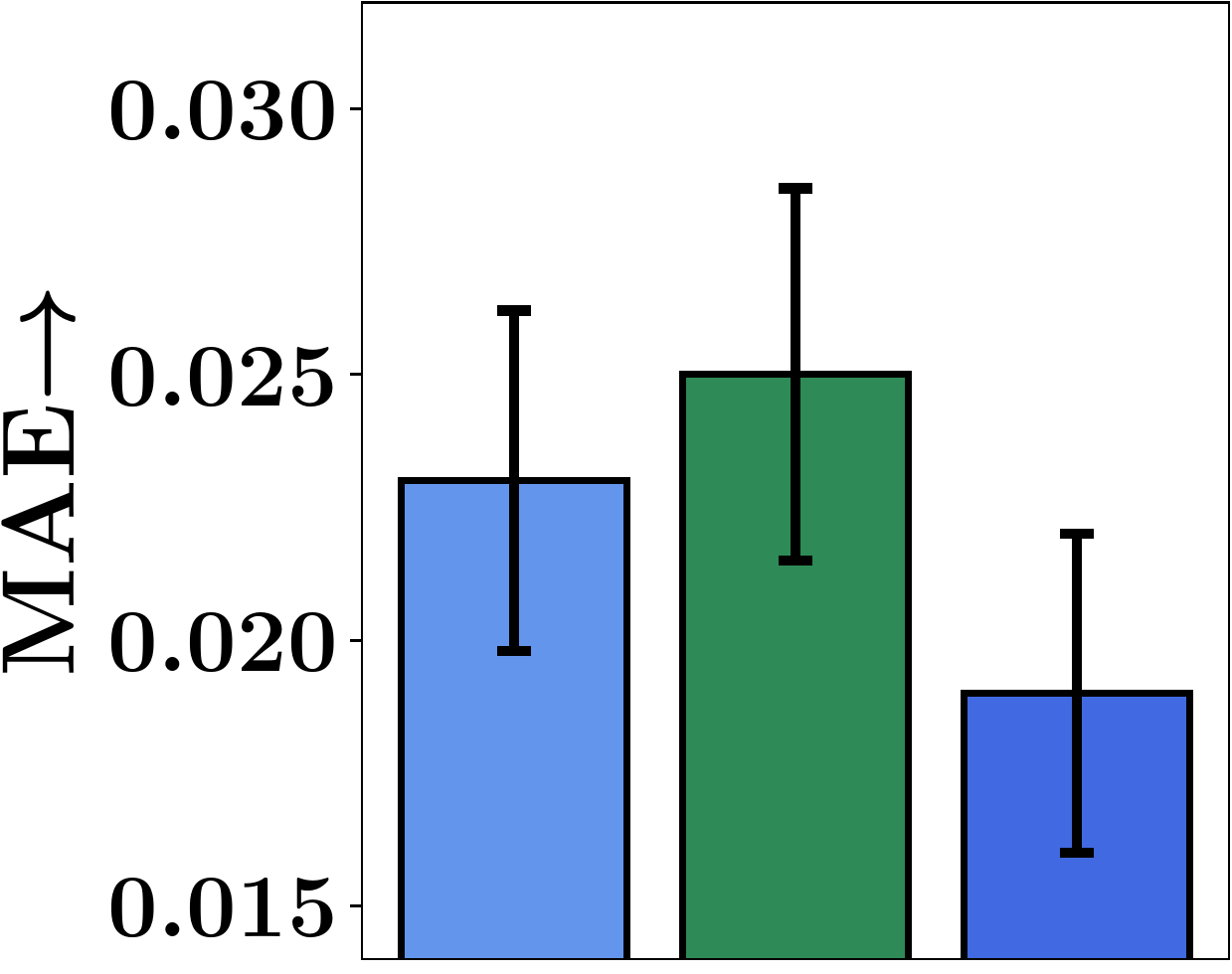}
\caption{\mult}
\end{subfigure}
\hfill
\begin{subfigure}[b]{0.30\columnwidth}
\includegraphics[height=3cm]{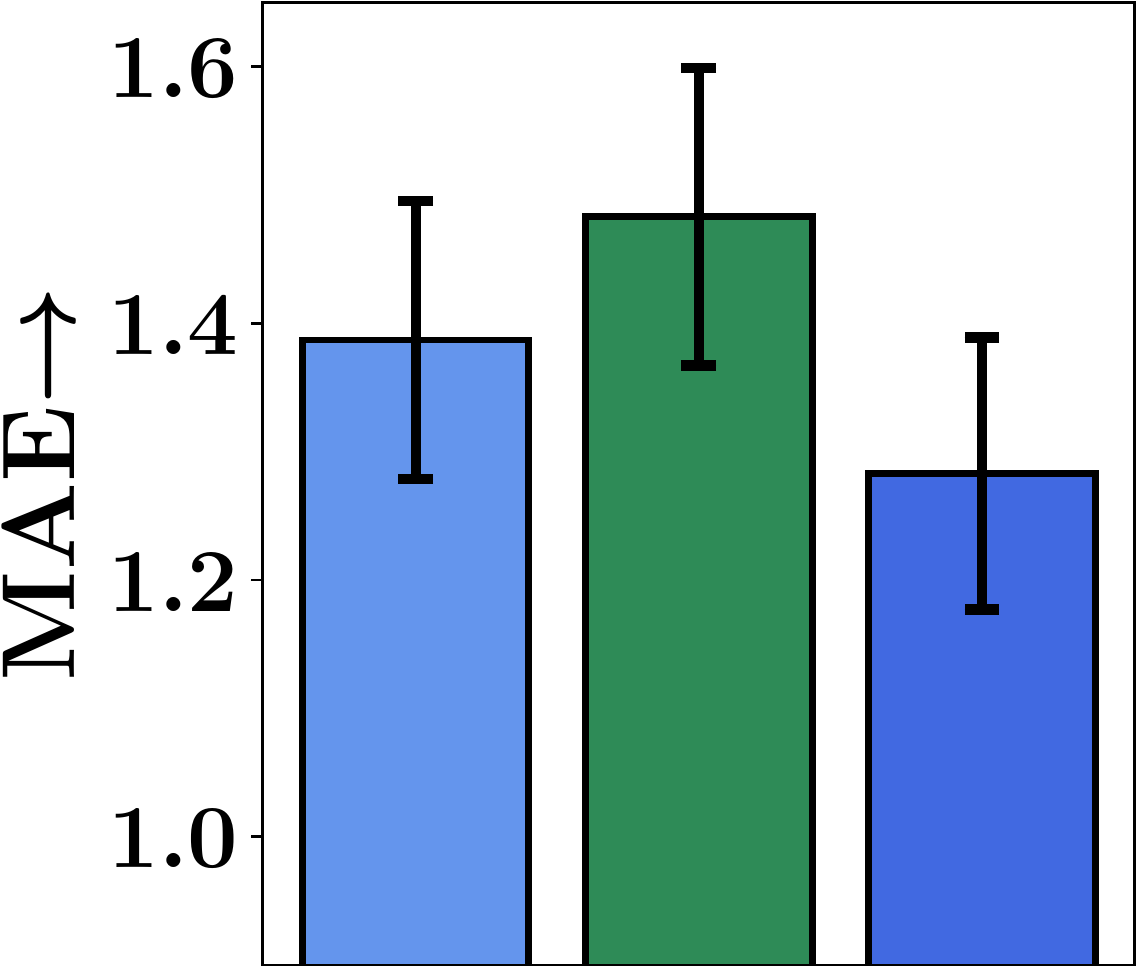}
\caption{\act}
\end{subfigure}
\hfill
{\includegraphics[height=0.5cm]{proactive/legend_cat.pdf}}
\vspace{-0.3cm}
\caption{\label{proact_fig:mae} Sequence Generation results for \proactive and other baselines in terms of MAE for time prediction.}
\end{figure}

\subsection{Sequence Generation Performance}
To address RQ3, we evaluate the sequence generation ability of \proactive. Specifically, we generate all the sequences in the test set by giving the \textit{true} goal of the CTAS and the first action as input to the procedure described in Section~\ref{proact_sec:generation}. However, there may be differences in the lengths of the generated and true sequences, \ie, the length of generated sequences is usually greater than the true CTAS. Therefore, we compare the actions in the true sequence with the initial $|\cm{S}|$ generated actions. Such an evaluation procedure provides us the flexibility of comparing with other MTPP models such as RMTPP~\cite{rmtpp} and THP~\cite{thp}. As these models cannot be used for end-to-end sequence generation, we alter their underlying model for \textit{forecasting} future actions given the first action and then iteratively update and sample from the MTPP parameters. We report the results in terms of APA and MAE for action and time prediction in Figure~\ref{proact_fig:mpa} and Figure~\ref{proact_fig:mae}, respectively. The results show that \proactive can better capture the generative dynamics of a CTAS in comparison to other MTPP models. We also note that the prediction performance deteriorates significantly in comparison to the results given in Tables~\ref{pro_tab:apa} and ~\ref{pro_tab:mae}. This could be attributed to the error that gets compounded in further predictions made by the model. Interestingly, the performance advantage that \proactive has over THP and RMTPP is further widened during sequence generation. 

\xhdr{Length Comparison}
Here, we report the results for the length comparison of the generated sequence and the true sequence. Specifically, we identify the count of instances where \proactive was able to effectively capture the generative mechanism of a sequence as:
\begin{equation}
\mathrm{CL} = \frac{1}{N}\sum_{\forall \cm{S}} \#(|\cm{S}|=|\widehat{\cm{S}}|),
\end{equation}
where CL denotes the \textit{Correct-Length} ratio with values $0.21$, $0.11$, and $0.16$ for datasets \bfast, \mult, and \act\ respectively. On a coarse level, these results might seem substandard; however, given the difficulty associated with the problem of sequence generation using just the CTAS goal, we believe these values are satisfactory. Moreover, we believe that the sequence generation procedure of \proactive opens up new frontiers for generating action sequences.

\xhdr{Scalability} For all datasets, the run-times for training \proactive are within 1 hour and thus are within the practical range for deployment in real-world scenarios. These running times further reinforce our design choice of using a neural MTPP due to their faster learning and closed-form sampling~\cite{intfree,ppflows}.

\section{Conclusion} \label{proact_sec:conc}
Standard deep-learning models are not designed for modeling sequences of actions localized in continuous time. However, neural MTPP models overcome this drawback but have limited ability to model the events performed by a human. Therefore, we developed a novel point process flow-based architecture called \proactive for modeling the dynamics of a CTAS. \proactive solves the problems associated with action prediction and goal prediction, and for the first time, we extend MTPP for end-to-end CTAS generation. Our experiments on three large-scale, diverse datasets reveal that \proactive can significantly improve over the state-of-the-art baselines across all metrics. Moreover, the results also reinforce the novel ability of \proactive to generate a CTAS. We hope that such an application will open many horizons for using MTPP in a wide range of tasks. 

%% file: chapters/010_conclusion.tex
To conclude, traditional recommender systems have limited ability to model user-item interactions with changing dynamics. In this thesis, we present the research directions, possible data-related problems, and solutions for learning recommender systems that utilize user-item interactions in the form of continuous-time event sequences (CTES). This thesis has been divided into three parts based on the issues addressed and the type of solutions proposed. We also identify the sequence modeling prowess of our proposed models and design solutions for applications beyond recommender systems. Here, we summarize the contribution of each part of this thesis:

In Chapter~\ref{chapter:imtpp}, we address the problem of modeling missing events in temporal sequences. The presence of missing events can significantly deteriorate the quality of the CTES data and, consecutively, the performance of recommender systems trained on these datasets. The traditional sequence models assume that the underlying sequence is complete -- an ideal setting. Thus, we proposed \imtpp, a novel MTPP model that learns the generative processes of observed events and missing events with missing events as latent random variables. Then, we devise an unsupervised training method to jointly learn the parameters of the observed and missing MTPP models using variational inference. \imtpp can effectively impute the missing data among the observed events, which in turn enhances its predictive prowess. Later, its enhanced version, \imtppp, can impute and identify the optimal position of missing events in a sequence. However, one unaddressed aspect of the missing data problem is partially missing events \ie, events with either the time \textit{or} mark missing. This is a challenging problem, as standard MTPP models are not designed to handle such sequences or functions in the absence of event times. In addition, the constrained optimization procedure in \imtppp can be improved by using Lagrangian multipliers. This would prevent the two-step procedure mentioned in \imtppp while simultaneously facilitating faster convergence.

In the second part of the thesis, we address the problems associated with designing better sequence modeling frameworks for the application of spatial recommendations. Specifically, given the continuous-time mobility records of a user, we use external data or features to overcome the data scarcity in the region and better recommend better POI candidates to a user in the spatial network. This part is divided into three chapters based on the three different real-world applications: (i) top-$k$ POI recommendation; (ii) sequential POI recommendation and \cin time prediction; and (iii) using smartphone usage to predict the mobility of a user.

In Chapter~\ref{chapter:axolotl}, we design solutions for the top-$k$ POI recommendation and propose \axolotl (\textbf{A}utomated \textit{cross} \textbf{Lo}cation-network \textbf{T}ransfer \textbf{L}earning), a graph neural network-based framework that transfers mobility knowledge across regions via -- (i) a novel meta-learning procedure derived from spatial as well as a social network, and (ii) a \textit{lightweight} unsupervised cluster-based transfer technique. We also address the problems associated with an extremely data-scarce region and devise a suitable cluster-based alignment loss that enforces similar embeddings for communities of users and locations with similar dynamics. In future work, we plan to modify the transfer procedure of \axolotl to incorporate novel meta-learning approaches, namely ProtoMAML~\cite{protomaml} and Reptile~\cite{reptile}. Additionally, we plan to experiment with approaches~\cite{sagpool, diffpool, spat} that automatically identify the number of clusters within a region while simultaneously maintaining the scalability of our model.

In Chapter~\ref{chapter:reformd}, we design solutions to overcome the problems due to data scarcity for developing sequential POI recommendation frameworks. Thus, we propose \reformd (\textbf{Re}usable \textbf{F}lows f\textbf{or} \textbf{M}obility \textbf{D}ata), a point process framework that learns the dynamics of user-specific check-in sequences in a region using normalizing flows to learn the inter-check-in time and geo-distributions. \reformd achieved better performance in data-scarce regions without any prerequisite of overlapping users. In future work, we plan to explore the relationship between MTPP-based transfer for product recommendation and incorporate complex transfer techniques such as meta-learning~\cite{maml}.

In Chapter~\ref{chapter:revamp}, we propose \revamp (\textbf{Re}lative position \textbf{V}ector for \textbf{A}pp-based \textbf{M}obility \textbf{P}rediction. This sequential POI recommendation approach uses smartphone app-usage logs to identify the mobility preferences of a user while simultaneously maintaining the user's privacy. Inspired by the success of relative positional encodings and self-attention models, \revamp uses relative and absolute positional encodings determined by the inter-\cin variances in the smartphone app category, POI category, and time over the \cins in the sequence. Through \revamp, we aim to understand the relationship between the smartphone usage of users and their mobility preferences. A drawback of the current formulation is that it requires the entire dataset to train its parameters. However, modern privacy-conscious techniques use a federated learning approach to train the model parameters with decentralized data. In future work, we plan to expand \revamp using such an efficient architecture.

In the third part, we highlight that the sequence modeling ability of the proposed models can have applications beyond recommender systems. Thus, we extend these models for two novel applications -- (i) efficient retrieval of event sequences; and (ii) human activity prediction. In Chapter~\ref{chapter:nsr}, we propose \nsr, a novel MTPP model to retrieve and rank a relevant set of continuous-time event sequences for a given query sequence from a large corpus of sequences. \nsr has many advantages over the previous models, such as a trainable unwarping function that makes it easier to compare query-corpus sequence pairs, especially when they have individually different attributes. Moreover, it has a learnable hashing for efficient retrieval of sequences with millions of events. Currently, in high-impact applications like healthcare, our method requires additional care since an incorrect prediction made by our model would have an adverse impact. In the end, one may consider a human-in-loop retrieval system that will mitigate such risk via human intervention. Moreover, event sequences from some domains, \eg, mobility records, can contain user-specific data and their use may lead to privacy violations. To that aim, it would be interesting to consider designing a privacy-preserving retrieval system for continuous-time event sequences. Empirically we show that our hash-code-based retrieval is Pareto-efficient, offering a sweet spot in trading off the computational cost and the retrieval performance.

In Chapter~\ref{chapter:proact}, we address the problem of modeling action sequences and first highlight the dissimilarities between standard time series and human-performed action sequences. To efficiently model these action sequences, we propose \proactive (\textbf{P}oint P\textbf{ro}cess flows for \textbf{Activ}ity S\textbf{e}quences), a neural MTPP framework for modeling the continuous-time distribution of actions in activity sequences while simultaneously addressing three real-world applications -- next action prediction, sequence-goal prediction, and \textit{end-to-end} sequence generation. \proactive brings forward the first-of-its-kind application of MTPP models via sequence generation, \ie, given the goal of an action sequence, it can generate a sequence with actions towards achieving that particular goal. In future work, we plan to incorporate a generative adversarial network~\cite{gan,xiao2017wasserstein} with action sampling and train it simultaneously with the MTPP.

Therefore, this thesis addresses some critical limitations of the current approaches for modeling continuous-time event sequences. To achieve this, we proposed robust and scalable neural network-based models that show their prowess in their respective applications. In addition, we design MTPP frameworks for two novel and high-impact applications. However, the research directions addressed in this have led to several open questions and future works. Here, we list out a few of them that were unaddressed in this thesis and, to the best of our knowledge, have been overlooked by the past literature as well.
\begin{compactitem}
    \item \textbf{Evaluating Time Prediction Performance.} A significant advantage of the neural MTPP over traditional sequence modeling frameworks is their ability to effectively understand the time of a future event. To evaluate the time-prediction performance of an MTPP model, the current procedures utilize one of the two popular metrics -- mean absolute error (MAE) or root mean squared error (RMSE)~\cite{rmtpp, sahp, thp}. However, both of these metrics evaluate the time-prediction ability in an \textit{event}-wise setting, \ie, perform a one-vs-one matching between the predicted and the actual data and then take mean overall events. This metric provides limited insights into the generative ability of MTPP models and the ability to results for the entire sequence. An interesting future direction for this thesis will be to utilize sequence-to-sequence comparison metrics such as dynamic time warping (DTW) to evaluate the prediction performance, or if possible, use a trainable DTW~\cite{cuturi2017soft} as a loss in the optimization procedure. Such an optimization procedure can give deeper insights into the generative and long-tail forecasting ability of MTPP models, which can be further used to design better MTPP frameworks. 

    \item \textbf{Repetitiveness in Recommendations.} Recent research has shown that the purchase records of users on an e-commerce platform may have repetitive patterns, \ie, there is a temporal dependency between the consecutive purchase of similar items~\cite{buyitagain}. Incorporating this dependency into a continuous-time model has been studied with traditional MTPP models. However, there has been limited research in exploring periodicity in a neural MTPP model. This is a challenging task as the neural MTPP models do not have an \textit{explicit} influence function, and these influences are approximated by a neural layer. Thus, an interesting future work could be to incorporate the repetitive purchasing nature of a user into the mathematical framework of a neural MTPP. A possible solution could be to optimize the ability of the model to predict similar purchases in the next time window, \ie, the likelihood that a user will re-buy the same product in the near future.

    \item \textbf{Learning Neural MTPP at Scale.} A drawback of using MTPP for modeling the dynamics of a CTES is their larger inference and sampling times for long sequences. This problem arises due to the inherently sequential nature of MTPP, which restricts their scalability even in the presence of highly-parallel modern hardware. However, recent research has shown that the conditional intensity function of an MTPP can be replaced by a normalizing flow based on triangular maps that allows parallel sampling and likelihood computation~\cite{shchur2020fast}. In a similar context, researchers have also \textit{decoupled} the neural MTPP to gain better scalability, \ie, rather than a joint neural MTPP, they use deep recurrent neural networks to capture complex temporal dependency patterns, and the self-excitation dynamics are modeled with a traditional Hawkes processes~\cite{turkmen2020fastpoint}. These techniques can be used to design scalable MTPP models, however, a detailed analysis of how these additions result in the overall improvement in scalability in the context of recommender systems has not been performed in the past. Thus, another interesting future work for this thesis will be a detailed study that compares the training and inference times of MTPP models while plugging-in several heuristics that will be a breakthrough in scaling these techniques and their applicability in large-scale recommender systems. 
    
    \item \textbf{Human in the Loop.} Lastly, we highlight the limited ability of sequence modeling frameworks to incorporate feedback given by users along with the training data. Incorporating additional feedback given by a user, such as a ranked list of desired items, is necessary for modern recommender systems as recent research has shown that the recommendations produced by a majority of ML models do not correlate with actual human preferences~\cite{hil_rec, ferrari2019we}. This is a major drawback of the current models and highlights their limited practicability even when they consist of sophisticated ML techniques. Thus, a crucial problem that needs to be addressed will be training neural models over CTES data with additional aids from other domains, such as crowd-sourcing, with the goal of incorporating real human feedback on the ranked recommendations.
\end{compactitem}

To summarize, in this thesis, we have designed solutions for a few real-world problems associated with recommender systems, and have discussed some open questions and directions for improvements.

%% file: rest/pub.tex
\chapter*{List of Publications}
\addcontentsline{toc}{chapter}{List of Publications}
Here we list out all \textit{first}-author articles accepted/published during the Ph.D tenure.\\

\xhdr{Publications based on this Thesis}
{
\begin{enumerate}
\item \textbf{Vinayak Gupta}, Abir De, Sourangshu Bhattacharya, and Srikanta Bedathur. \textquotedblleft Learning Temporal Point Processes with Intermittent Observations\textquotedblright, In \textit{Proc. of the 24th International Conference on Artificial Intelligence and Statistics (AISTATS)}, 2021.

\item \textbf{Vinayak Gupta} and Srikanta Bedathur. \textquotedblleft Region Invariant Normalizing Flows for Mobility Transfer\textquotedblright, In \textit{Proc. of the 30th ACM Intl. Conference on Information and Knowledge Management (CIKM)}, 2021.

\item \textbf{Vinayak Gupta}, Abir De, and Srikanta Bedathur. \textquotedblleft Learning Temporal Point Processes for Efficient Retrieval of Continuous Time Event Sequences\textquotedblright, In \textit{Proc. of the 36th AAAI Conference on Artificial Intelligence (AAAI)}, 2022.

\item \textbf{Vinayak Gupta} and Srikanta Bedathur. \textquotedblleft ProActive: Self-Attentive Temporal Point Process Flows for Activity Sequences\textquotedblright, In \textit{Proc. of the 28th ACM SIGKDD Conference on Knowledge Discovery and Data Mining (KDD)}, 2022.

\item \textbf{Vinayak Gupta} and Srikanta Bedathur. \textquotedblleft Doing More with Less: Overcoming Data Scarcity for POI Recommendation via Cross-Region Transfer\textquotedblright, \textit{ACM Transactions on Intelligent Systems and Technology (ACM TIST)}, 2022.

\item \textbf{Vinayak Gupta}, Abir De, Sourangshu Bhattacharya, and Srikanta Bedathur. \textquotedblleft Modeling Continuous Time Sequences with Intermittent Observations using Marked Temporal Point Processes\textquotedblright, \textit{ACM Transactions on Intelligent Systems and Technology (ACM TIST)}, 2022.

\item \textbf{Vinayak Gupta} and Srikanta Bedathur. \textquotedblleft Modeling Spatial Trajectories using Coarse-Grained Smartphone Logs\textquotedblright, \textit{IEEE Transactions on Big Data (IEEE TBD)}, 2022.
\end{enumerate}
\par}

\xhdr{Other Publications}
{
\begin{enumerate}
\item Ankita Likhyani*, \textbf{Vinayak Gupta}*, P. K. Srijith, Deepak, and Srikanta Bedathur. \textquotedblleft Modeling Implicit Communities from Geo-tagged Event Traces using Spatio-Temporal Point Processes\textquotedblright, In \textit{Proc. of the 21st International Conference on Web Information Systems Engineering (WISE)}, 2020.
\end{enumerate}
\par}

\begin{center}
\begin{tabular}{|l|}
\hline
\textbf{This work received the \textit{Outstanding Doctoral Paper} award at}\\
\textbf{The First International Conference on AI-ML-Systems 2021.}\\
\hline
\end{tabular}
\end{center}

%% file: rest/bio.tex
\chapter*{Biography}
\addcontentsline{toc}{chapter}{Biography}

Vinayak Gupta is a Ph.D. student at the Department of Computer Science and Engineering, Indian Institute of Technology (IIT) Delhi. He obtained a Bachelors' degree in Computer Science and Engineering from the Indian Institute of Information Technology (IIIT) Jabalpur. His research interests are in data mining. Specifically, his research addresses the problems related to -- limited training data, asynchronous data collection in sequences, event-imputation, end-to-end sequence generation, activity recognition, and spatial recommendation for users -- in temporal sequences using point processes and graph neural networks.